\PassOptionsToPackage{unicode}{hyperref}
\PassOptionsToPackage{hyphens}{url}
\PassOptionsToPackage{dvipsnames,svgnames,x11names}{xcolor}
\documentclass[
  12pt]{article}

\usepackage{amsmath,amssymb}
\usepackage{iftex}
\ifPDFTeX
  \usepackage[T1]{fontenc}
  \usepackage[utf8]{inputenc}
  \usepackage{textcomp} %
\else %
  \usepackage{unicode-math}
  \defaultfontfeatures{Scale=MatchLowercase}
  \defaultfontfeatures[\rmfamily]{Ligatures=TeX,Scale=1}
\fi
\usepackage{lmodern}
\ifPDFTeX\else  
\fi
\IfFileExists{upquote.sty}{\usepackage{upquote}}{}
\IfFileExists{microtype.sty}{%
  \usepackage[]{microtype}
  \UseMicrotypeSet[protrusion]{basicmath} %
}{}
\makeatletter
\@ifundefined{KOMAClassName}{%
  \IfFileExists{parskip.sty}{%
    \usepackage{parskip}
  }{%
    \setlength{\parindent}{0pt}
    \setlength{\parskip}{6pt plus 2pt minus 1pt}}
}{%
  \KOMAoptions{parskip=half}}
\makeatother
\usepackage{xcolor}
\makeatletter
\ifx\paragraph\undefined\else
  \let\oldparagraph\paragraph
  \renewcommand{\paragraph}{
    \@ifstar
      \xxxParagraphStar
      \xxxParagraphNoStar
  }
  \newcommand{\xxxParagraphStar}[1]{\oldparagraph*{#1}\mbox{}}
  \newcommand{\xxxParagraphNoStar}[1]{\oldparagraph{#1}\mbox{}}
\fi
\ifx\subparagraph\undefined\else
  \let\oldsubparagraph\subparagraph
  \renewcommand{\subparagraph}{
    \@ifstar
      \xxxSubParagraphStar
      \xxxSubParagraphNoStar
  }
  \newcommand{\xxxSubParagraphStar}[1]{\oldsubparagraph*{#1}\mbox{}}
  \newcommand{\xxxSubParagraphNoStar}[1]{\oldsubparagraph{#1}\mbox{}}
\fi
\makeatother

\usepackage{longtable,booktabs,array}
\usepackage{calc} %
\usepackage{etoolbox}
\makeatletter
\patchcmd\longtable{\par}{\if@noskipsec\mbox{}\fi\par}{}{}
\makeatother
\IfFileExists{footnotehyper.sty}{\usepackage{footnotehyper}}{\usepackage{footnote}}
\makesavenoteenv{longtable}
\usepackage{graphicx}
\makeatletter
\def\maxwidth{\ifdim\Gin@nat@width>\linewidth\linewidth\else\Gin@nat@width\fi}
\def\maxheight{\ifdim\Gin@nat@height>\textheight\textheight\else\Gin@nat@height\fi}
\makeatother
\setkeys{Gin}{width=\maxwidth,height=\maxheight,keepaspectratio}
\makeatletter
\def\fps@figure{htbp}
\makeatother

\makeatletter
\@ifpackageloaded{caption}{}{\usepackage{caption}}
\AtBeginDocument{%
\ifdefined\contentsname
  \renewcommand*\contentsname{Table of contents}
\else
  \newcommand\contentsname{Table of contents}
\fi
\ifdefined\listfigurename
  \renewcommand*\listfigurename{List of Figures}
\else
  \newcommand\listfigurename{List of Figures}
\fi
\ifdefined\listtablename
  \renewcommand*\listtablename{List of Tables}
\else
  \newcommand\listtablename{List of Tables}
\fi
\ifdefined\figurename
  \renewcommand*\figurename{Figure}
\else
  \newcommand\figurename{Figure}
\fi
\ifdefined\tablename
  \renewcommand*\tablename{Table}
\else
  \newcommand\tablename{Table}
\fi
}
\@ifpackageloaded{float}{}{\usepackage{float}}
\floatstyle{ruled}
\@ifundefined{c@chapter}{\newfloat{codelisting}{h}{lop}}{\newfloat{codelisting}{h}{lop}[chapter]}
\floatname{codelisting}{Listing}

\makeatother
\makeatletter
\@ifpackageloaded{caption}{}{\usepackage{caption}}
\@ifpackageloaded{subcaption}{}{\usepackage{subcaption}}
\makeatother

\ifLuaTeX
  \usepackage{selnolig}  %
\fi
\usepackage[]{natbib}
\usepackage{bookmark}

\IfFileExists{xurl.sty}{\usepackage{xurl}}{} %
\hypersetup{
  pdftitle={Deformed Q-learning for High-Dimensional Dynamic Treatment Regimes},
  pdfauthor={Anonymous},
  pdfkeywords={Generalized linear models; Multi-stage inference; Penalized regression; Policy regret},
  colorlinks=true,
  linkcolor={blue},
  filecolor={Maroon},
  citecolor={Blue},
  urlcolor={Blue},
  pdfcreator={LaTeX}}

\usepackage{amsthm,bm,multirow,tabularx,enumitem,placeins}
\usepackage{authblk}

\newtheorem{thm}{Theorem}[section]
\newtheorem{ass}{Assumption}[section]
\newtheorem{pro}{Proposition}[section]
\newtheorem{cor}{Corollary}[section]
\newtheorem{example}{Example}[section]
\newtheorem{lemma}{Lemma}[section]
\newtheorem{remark}{Remark}[section]

\graphicspath{{figures/}}
\newcommand{\anon}{1}

\begin{document}

\def\spacingset#1{\renewcommand{\baselinestretch}%
{#1}\small\normalsize} \spacingset{1}

\if1\anon
{
  \title{\bf Deformed Q-learning for High-Dimensional Dynamic Treatment Regimes}
  \author[1]{Ming Cheng}
  \author[1]{Zhao Chen}
  \author[2]{Christina Dan Wang}
  \affil[1]{School of Data Science, Fudan University}
  \affil[2]{Business Division, New York University Shanghai}

  \date{}
  \hypersetup{pdfauthor={Ming Cheng; Zhao Chen; Christina Dan Wang}}
  \maketitle
} \fi

\if0\anon
{
  \bigskip
  \bigskip
  \bigskip
  \begin{center}
    {\LARGE\bf Deformed Q-learning for High-Dimensional Dynamic Treatment Regimes}
  \end{center}
  \medskip
} \fi

\bigskip
\begin{abstract}
	In high-dimensional dynamic treatment regimes, nonlinear terminal outcome models complicate upstream conditional-mean modeling and inference based on recursively estimated continuation values. We introduce a Deformed Q-learning procedure for general action spaces that accommodates a terminal-stage GLM, random designs, conditional heteroscedasticity, and fine-grained action discretization. The procedure changes the recursive criterion by propagating the terminal deformed optimal Q-function through sparse block-wise linear conditional-mean models. Under the stated models and regularity conditions, we establish consistent and asymptotically normal estimators, accounting for downstream estimation uncertainty in inference for optimal Q-values, deformed optimal Q-values, and their contrasts. We bound population regret relative to the original terminal mean-outcome objective without explicit stage-wise accumulation. Estimation and inference also extend to a growing number of stages under conditions uniform across stages. Simulation studies demonstrate accurate estimation and variable selection. In an empirical application to sepsis management using the MIMIC-III database, the proposed framework supports stage-specific value inference, while an auxiliary action-alignment analysis provides an exploratory signal that early-stage concordance with the learned recommendation may be associated with improved survival.
\end{abstract}

\noindent%
\vfill

\newpage
\spacingset{1.8}

\section{Introduction}\label{intro}
Personalized decision-making is an important topic in modern statistical decision theory. It focuses on selecting optimal treatments for individuals based on covariates and plays a crucial role in applications such as precision medicine. In single-stage settings, this problem is formalized through an individualized treatment rule (ITR), which maps covariates to a treatment that maximizes the expected outcome. In multi-stage settings, it extends to dynamic treatment regimes (DTRs), which assign treatments at each stage based on the whole history to optimize long-term outcomes.

For multi-stage decisions with binary action space, extensive methodologies have been developed based on recursive regression such as Q-learning \citep{Watkins1989,Watkins1992}, including its interactive extensions \citep{Laber2014IQ,Linn2017IQ}, and A-learning \citep{Murphy2003,Robins2004}. \citet{Chakraborty2010} develop a two-stage Q-learning framework with thresholding and bootstrap techniques to address nonregularity in inference. Building on this, \citet{Song2015} introduce penalized Q-learning (PQ-learning) using nonconvex penalties such as SCAD \citep{Fan2001} and MCP \citep{Zhang2010}, yielding a stage-wise recursive structure. High-dimensional extensions deliver consistent variable selection and valid inference, including \citet{Shi2018}, who develop high-dimensional A-learning via Dantzig-selector-type estimating equations \citep{Candes2007}, preserve the double robustness of A-learning, and allow robustness to nuisance-model misspecification, and \citet{Zhu2018}, who study high-dimensional Q-learning with folded-concave penalization and hard thresholding for valid value-function inference.

Beyond binary actions, a parallel line of research studies continuous-action decision-making. In biomedical applications, such rules are often described as individualized dosing rules (IDRs). Representative approaches include outcome-weighted learning \citep{Chen2016}, kernel-based extensions \citep{Kallus2018}, and semiparametric formulations \citep{Chernozhukov2019}. \citet{Cai2023} propose Jump Interval-Learning to obtain IDRs, offering interpretability and clinical flexibility. In the high-dimensional continuous-action setting, \citet{Ding2024} propose a single-stage statistical learning method based on action discretization and the DROVE penalty, which targets adjacent action-level discontinuity and enables optimal Q-value inference.

Beyond the above, statistical learning foundations for DTRs continue to broaden. \citet{Zhao2015} introduce BOWL and SOWL to optimize long-term outcomes via weighted classification. For infinite-horizon settings, \citet{Shi2022,Shi20241} study policy evaluation and advantage learning with asymptotic guarantees, with further contributions on unmeasured confounding \citep{Shi20243} and robust or heterogeneous treatment decision learning \citep{Shi2018maximin}. These directions broaden the statistical learning foundation for DTRs and provide complementary perspectives for the finite-horizon, model-based Q-learning problem studied here.

Taken together, these literatures point to a recursive inference problem in multi-stage Q-learning. Sequential sepsis dosing provides a concrete example: treatment decisions are repeated over time, each action can be represented as a clinically structured fluid-vasopressor block \citep{Komorowski2018,choi2024}, and the treatment trajectory is evaluated through a terminal binary reward such as survival status, or equivalently through a loss such as final mortality. Under this structure, the upstream pseudo-outcome depends on downstream high-dimensional estimation, block-wise action maximization, and a transformed continuation value induced by the nonlinear terminal mean model. Its asymptotic distribution therefore carries within-stage variation, downstream estimation effects, and cross-stage covariance terms through backward recursion. The methodological challenge is to build valid stage-specific value and value-difference inference for this generated pseudo-outcome, with explicit uncertainty propagation.

The main contributions of this paper are as follows. First, we formulate high-dimensional multi-stage Q-learning with block-structured actions and deformed recursive pseudo-outcomes as a cross-stage inference problem. The procedure propagates the terminal deformed optimal Q-function and imposes sparse block-wise linear models on its upstream conditional means. This formulation makes explicit how terminal GLM modeling, block-wise action maximization, and upstream regression jointly determine the target estimand and covariance structure.

Second, we establish a recursive oracle property and develop stage-wise inference accounting for terminal coefficient uncertainty in upstream inference. The resulting covariance separates stage-specific score variation from propagated estimation error, with a zero cross contribution in the asymptotic covariance. The theory retains the dependence among inherited and newly estimated coefficients and extends to a growing number of stages under uniform conditions.

We also characterize the population regret of the deformed-optimal regime relative to the regime maximizing the expected terminal outcome through a directional Wasserstein bound based on policy-induced distributions of the terminal deformed optimal Q-function, and identify structural conditions for zero regret.

Finally, we illustrate the proposed statistical approach through a MIMIC-III sepsis study with joint intravenous-fluid and vasopressor action blocks. The analysis reports stage-specific value inference as its primary inferential component. As an auxiliary assessment, a stage-wise action-alignment analysis adjusted for observed confounders yields a positive early-stage contrast, providing an exploratory signal that concordance with the learned recommendation during the early treatment window may be associated with improved survival.

The remainder of the paper is organized as follows. Section~\ref{method} presents the proposed methodology. Section~\ref{theory} develops the asymptotic theory. Sections~\ref{simu} and~\ref{real} report the simulation and empirical studies, respectively.

\section{Methodology}\label{method}

\subsection{Model Setup}

We consider an \(M\)-stage sequential decision problem. Let \(\mathbf{Z}\) denote a random trajectory from the population,
\[
\mathbf{Z} = (\mathbf{X}^{(1)},\mathbf{A}^{(1)},\ldots, \mathbf{X}^{(M)},\mathbf{A}^{(M)},Y^{(M)}).
\]
At stage \(m\), \(\mathbf{X}^{(m)}\in\mathbb{R}^{k_m}\) denotes the covariates and \(\mathbf{A}^{(m)}\in\mathcal{A}\subset\mathbb{R}^a\) the action. The action space \(\mathcal{A}\) is assumed common across stages, whereas the covariate dimension \(k_m\) may vary with \(m\). We focus on settings in which the treatment regime is evaluated solely by the terminal outcome \(Y^{(M)}\).
The history is defined recursively by \(\mathbf{H}^{(1)}=\mathbf{X}^{(1)}\), 
\[
\mathbf{H}^{(m)}
=
\left(
(\mathbf{H}^{(m-1)})^\top,\ 
(\mathbf{A}^{(m-1)})^\top,\ 
(\mathbf{X}^{(m)})^\top
\right)^\top,\qquad m\ge 2 .
\]
The observed sample consists of independent copies \(\{\mathbf{Z}_i\}_{i=1}^n\) of \(\mathbf Z\). Throughout, unsubscripted quantities such as \(\mathbf H^{(m)}\), \(\mathbf A^{(m)}\), \(Y^{(M)}\), and the design vectors below denote population-level random variables. A subscript \(i\) denotes the corresponding quantity for subject \(i\). 

Equivalently, this is a finite-horizon offline RL problem, with \(\mathbf H^{(m)}\), \(\mathbf A^{(m)}\), and the terminal \(Y^{(M)}\) serving as state, action, and reward, respectively.
The data consist of observed trajectories, as in the DTR literature. In the asymptotic analysis, the number of stages \(M\) and the model dimensions, including \(\sum_{m=1}^M k_m\), are allowed to grow with \(n\).

For clarity, we first present the construction for the two-stage case \((M=2)\). The stage-wise \textit{Q-functions} are recursively defined by
\begin{align}
	Q^{(2)}(\mathbf H^{(2)},\mathbf A^{(2)})
	&=
	\mathbb E\{Y^{(2)}\mid \mathbf H^{(2)},\mathbf A^{(2)}\},
	\notag\\
	Q^{(1)}(\mathbf H^{(1)},\mathbf A^{(1)})
	&=
	\mathbb E\{\mathcal C^{(2)}(\mathbf H^{(2)})
	\mid \mathbf H^{(1)},\mathbf A^{(1)}\}.
	\label{Q}
\end{align}
where \(\mathcal C^{(2)}\) denotes the continuation value from the terminal stage.

\subsection{Terminal-Stage Model}
In many sequential decision problems, the action is richer than a binary on--off choice. For example, in the sepsis application, the action at each stage consists of the joint doses of intravenous fluids and vasopressors and is therefore continuous and multidimensional. Since estimating a separate effect for every possible dose is infeasible, we instead divide the action space into a manageable number of categories, or \emph{blocks}, and allow the treatment effect to vary across blocks.

We represent the blocks by the indicator basis \(\{\varphi_k\}_{k=1}^{L_n}\) associated with a partition \(\mathcal A=\bigsqcup_{k=1}^{L_n}\mathcal A_{(k)}\), where \(\varphi_k(\mathbf a)=I_{\{\mathbf a\in\mathcal A_{(k)}\}}\). Thus exactly one block indicator is active for any action \(\mathbf a\). The number of blocks \(L_n\) may increase with \(n\), so the action space is resolved more finely as data accrue.

At the terminal stage, the reward is frequently non-Gaussian. In the sepsis application, for example, the final outcome is a binary survival indicator. To accommodate such outcomes while retaining a parsimonious linear predictor, we use a canonical generalized linear model. Let $
\bm{\beta}^{(2)}
=
\left(
c^{(2)},
(\bm{\eta}^{(2)}_0)^\top,
(\bm{\eta}^{(2)}_2)^\top,
\ldots,
(\bm{\eta}^{(2)}_{L_n})^\top
\right)^\top,$ 
where \(c^{(2)}\) is an intercept, \(\bm{\eta}^{(2)}_0\) represents the main effect part of the history, and \(\bm{\eta}^{(2)}_k\), \(k\geq2\), represents the history-dependent treatment effect associated with block \(k\). As usual, we take block 1 as the reference and set \(\bm{\eta}^{(2)}_1=\mathbf 0\).

Define $
\bm{\varphi}_{-1}(a)=(\varphi_2(a),\ldots,\varphi_{L_n}(a))^\top$ 
and let the terminal-stage design vector be
\begin{equation}
\mathbf S^{(2)}:=\mathbf S^{(2)}(\mathbf H^{(2)},\mathbf A^{(2)})=\left(1,(\mathbf H^{(2)})^\top,\left\{\bm{\varphi}_{-1}(\mathbf A^{(2)})\otimes \mathbf H^{(2)}\right\}^\top\right)^\top.
\label{eq:S2-pop}
\end{equation}
Here \(\otimes\) denotes the Kronecker product. The corresponding linear predictor is
\[
(\bm{\beta}^{(2)})^\top\mathbf S^{(2)}
=
c^{(2)}
+
(\bm{\eta}^{(2)}_0)^\top\mathbf H^{(2)}
+
\sum_{k=2}^{L_n}
\varphi_k(\mathbf A^{(2)})
(\bm{\eta}^{(2)}_k)^\top\mathbf H^{(2)}.
\]
The first two components specify the main effect between the history and the reward, while the remaining components allow this relationship to vary across action blocks.

Conditional on \(\mathbf S^{(2)}\), we assume that \(Y^{(2)}\) follows a canonical exponential-family model with density
\begin{equation}
\begin{aligned}
&f_{Y^{(2)}\mid\mathbf S^{(2)}}\!\left(Y^{(2)}\mid \mathbf S^{(2)};\bm{\beta}^{(2)},\phi^{(2)}\right)\\
&\quad =\exp\left\{
\frac{
Y^{(2)}(\bm{\beta}^{(2)})^\top\mathbf S^{(2)}
-b^{(2)}\left((\bm{\beta}^{(2)})^\top\mathbf S^{(2)}\right)
}{\phi^{(2)}}\right\}
h^{(2)}(Y^{(2)},\phi^{(2)}),
\end{aligned}
\label{glm}
\end{equation}
where \(b^{(2)}(\cdot)\) is the cumulant function and \(\phi^{(2)}\) is a dispersion parameter. Then the terminal stage Q-function defined in \eqref{Q} is
\begin{equation}
Q^{(2)}(\mathbf H^{(2)},\mathbf A^{(2)})=\mathbb E[Y^{(2)}\mid \mathbf H^{(2)},\mathbf A^{(2)}]=b^{(2)'}((\bm{\beta}^{(2)})^\top\mathbf S^{(2)}).
\label{Qglm}
\end{equation}
Here \(b^{(2)'}\), the first derivative of the cumulant function \(b^{(2)}\), is the GLM mean function. For a GLM, \(b^{(2)'}\) is monotone increasing on the natural-parameter space; hence its inverse exists on the relevant mean-parameter space and equals the canonical link.

\subsection{Recursive Deformation}\label{linkcriterion}

The nonlinear mean map in \eqref{Qglm} distinguishes the terminal-stage GLM from a linear Q-function model. Under the recursion in \eqref{Q}, the choice of \(\mathcal C^{(2)}\) determines the upstream target. Mean-scale Q-learning takes this continuation value to be
\(
\Gamma_0(\mathbf H^{(2)})
=
\max_{\mathbf a\in\mathcal A}
Q^{(2)}(\mathbf H^{(2)},\mathbf a).
\)
Under the terminal-stage GLM, however, \(\Gamma_0\) is obtained after applying the mean function \(b^{(2)'}\) to the terminal deformed optimal Q-function.

We map the terminal optimal Q-function back to the linear-predictor scale through the inverse mean function. Since \(b^{(2)'}\) is increasing, we define the deformed continuation value by
\begin{equation}
\begin{aligned}
\widetilde{\Gamma}(\mathbf H^{(2)})
&=
(b^{(2)\prime})^{-1}\{\Gamma_0(\mathbf H^{(2)})\}=
\max_{\mathbf a\in\mathcal A}
(\bm{\beta}^{(2)})^\top
\mathbf S^{(2)}(\mathbf H^{(2)},\mathbf a)
\\
&=
c^{(2)}
+
(\bm{\eta}^{(2)}_0)^\top\mathbf H^{(2)}
+
\max\Bigl\{
0,\,
\max_{2\le k\le L_n}
(\bm{\eta}^{(2)}_k)^\top\mathbf H^{(2)}
\Bigr\}.
\end{aligned}
\label{deform}
\end{equation}
We refer to this inverse-mean transformation as a \emph{deformation}. It preserves the terminal-stage maximizer while expressing the continuation value through the terminal deformed optimal Q-function. In the preceding stages, optimality is therefore defined with respect to the resulting deformed continuation criterion.

Substituting \(\widetilde{\Gamma}\) into the stage-1 recursion yields the upstream target
\[
Q_{\mathrm{dfm}}^{(1)}(\mathbf H^{(1)},\mathbf A^{(1)}) = \mathbb E\left[\widetilde{\Gamma}(\mathbf H^{(2)})\mid \mathbf H^{(1)},\mathbf A^{(1)}\right].
\]
The deformation removes the terminal GLM nonlinearity, so the continuation value enters the stage-1 target through blockwise linear predictors. We assume that the resulting conditional mean has the following sparse block-wise linear form:
\begin{equation}
Q_{\mathrm{dfm}}^{(1)}(\mathbf{H}^{(1)},\mathbf{A}^{(1)}) = c^{(1)} + (\bm{\eta}^{(1)}_0)^\top \mathbf{H}^{(1)} + \sum_{k=1}^{L_n} \varphi_k(\mathbf{A}^{(1)})(\bm{\eta}^{(1)}_k)^\top \mathbf{H}^{(1)} = (\bm{\beta}^{(1)})^\top\mathbf{S}^{(1)}.
\label{q1linear}
\end{equation}
where
\begin{align*}
\bm{\beta}^{(1)}
&=
\left(c^{(1)},(\bm{\eta}^{(1)}_0)^\top,
(\bm{\eta}^{(1)}_1)^\top,\ldots,
(\bm{\eta}^{(1)}_{L_n})^\top\right)^\top,\\
\mathbf{S}^{(1)}
&=
\left(1,(\mathbf{H}^{(1)})^\top,
\left\{(\varphi_k(\mathbf{A}^{(1)}))_{k=1}^{L_n}
\otimes\mathbf{H}^{(1)}\right\}^\top\right)^\top.
\end{align*}

For a logistic terminal model, \(Q_{\mathrm{dfm}}^{(1)}\) is the conditional mean of the optimized terminal log-odds, corresponding to the upstream regression on the log-odds scale discussed by \citet[Remark~2.4]{Linn2017IQ}. A conditional certainty-equivalent interpretation is given in Appendix~\ref{sec:ce-interpretation}.

\subsection{Deformed Q-learning}
We now introduce Deformed Q-learning, a backward regression procedure that uses the deformation to construct upstream deformed pseudo-outcomes, estimate stage-wise parameters and Q-functions, and induce treatment rules by maximizing the estimated Q-functions.

For each subject \(i\), let \(\mathbf S_i^{(2)}\) be the design vector in \eqref{eq:S2-pop}, and write \(\mathbf Y^{(2)}=(Y_1^{(2)},\ldots,Y_n^{(2)})^\top\), \(\bm\Psi^{(2)}=(\mathbf S_1^{(2)},\ldots,\mathbf S_n^{(2)})^\top\), and \(\bm\theta^{(2)}=\bm\Psi^{(2)}\bm\beta^{(2)}\). Up to constants and a dispersion factor, the log-likelihood under \eqref{glm} is
\[
\ell_n^{(2)}(\bm\beta^{(2)})
=\sum_{i=1}^n\left\{Y_i^{(2)}(\bm\beta^{(2)})^\top\mathbf S_i^{(2)}
-b^{(2)}\!\left((\bm\beta^{(2)})^\top\mathbf S_i^{(2)}\right)\right\}.
\]

We estimate \(\bm\beta^{(2)}\) by the penalized likelihood
\begin{equation}
	\widehat{\bm\beta}^{(2)}
	=
	\arg\max_{\bm\beta^{(2)}}
	\left[
	n^{-1}\ell_n^{(2)}(\bm\beta^{(2)})
	-
	p_{\lambda_n^{(2)}}(\bm\beta^{(2)})
	\right],
	\label{pGLM}
\end{equation}
where \(p_{\lambda_n^{(2)}}(\cdot)\) is a penalty with tuning parameter \(\lambda_n^{(2)}\). Denote the resulting estimates by
\(\widehat c^{(2)}\), \(\widehat{\bm\eta}^{(2)}_0\), and
\(\widehat{\bm\eta}^{(2)}_k\), with
\(\widehat{\bm\eta}^{(2)}_1=\mathbf 0\) for the reference block.

For the stage-1 deformed pseudo-outcome, write $
\mathbf H^{(2)}
=
\left(
(\mathbf H^{(1)})^\top,
(\mathbf A^{(1)})^\top,
(\mathbf X^{(2)})^\top
\right)^\top$ and $
\bm\eta^{(2)}_0
=
\left(
(\bm\eta^{(2)}_{0,1})^\top,
(\bm\eta^{(2)}_{0,2})^\top,
(\bm\eta^{(2)}_{0,3})^\top
\right)^\top. $
Then
\[
(\bm\eta^{(2)}_0)^\top\mathbf H^{(2)}
=
(\bm\eta^{(2)}_{0,1})^\top\mathbf H^{(1)}
+
(\bm\eta^{(2)}_{0,2})^\top\mathbf A^{(1)}
+
(\bm\eta^{(2)}_{0,3})^\top\mathbf X^{(2)}.
\]
The component \((\bm\eta^{(2)}_{0,1})^\top\mathbf H^{(1)}\) is already measurable at stage 1 and does not depend on the stage-1 action. We therefore use it as the inherited main effect and fix $
\bm\eta^{(1)}_0=\bm\eta^{(2)}_{0,1}.$
The remaining parameters are collected as $
\bm{\alpha}^{(2)}
=
\left(c^{(2)},(\bm{\eta}^{(2)}_{0,2})^\top,(\bm{\eta}^{(2)}_{0,3})^\top,
(\bm{\eta}^{(2)}_2)^\top,\ldots,
(\bm{\eta}^{(2)}_{L_n})^\top\right)^\top$ and 
$\bm{\alpha}^{(1)}
=
\left(c^{(1)},(\bm{\eta}^{(1)}_1)^\top,\ldots,
(\bm{\eta}^{(1)}_{L_n})^\top\right)^\top .$
This convention separates the inherited history main effect from the free stage-1 block coefficients and removes the usual non-uniqueness between treatment-independent history effects and block-specific effects. Consequently, the stage-1 design contains no separate main-effect term for \(\mathbf H^{(1)}\), so all \(L_n\) block-specific coefficients can be included without imposing a reference-block constraint.

Using \(\widehat{\bm{\beta}}^{(2)}\), define the stage-1 deformed pseudo-outcome by
\[
\widehat{Y}^{(1)}_{\mathrm{dfm}}
=
\widehat{c}^{(2)}
+(\widehat{\bm{\eta}}^{(2)}_{0,2})^\top \mathbf{A}^{(1)}
+(\widehat{\bm{\eta}}^{(2)}_{0,3})^\top \mathbf{X}^{(2)}
+ \max\Bigl\{
0,\,
\max_{2\le k\le L_n}
(\widehat{\bm{\eta}}^{(2)}_k)^\top \mathbf{H}^{(2)}
\Bigr\}.
\]
After subtracting the inherited main effect, we denote the regression target for \(\bm{\alpha}^{(1)}\) by \(q_{\mathrm{dfm}}^{(1)}\), reserving \(Q_{\mathrm{dfm}}^{(1)}\) for the full conditional mean:

\[
\begin{aligned}
q^{(1)}_{\mathrm{dfm}}(\mathbf{H}^{(1)}, \mathbf{A}^{(1)})
&:=Q^{(1)}_{\mathrm{dfm}}(\mathbf{H}^{(1)}, \mathbf{A}^{(1)})
-(\bm{\eta}^{(2)}_{0,1})^\top\mathbf H^{(1)}\\
&=c^{(1)}
+ \sum_{k=1}^{L_n}
\varphi_k(\mathbf{A}^{(1)})
(\bm{\eta}^{(1)}_k)^\top \mathbf{H}^{(1)}
=(\bm{\alpha}^{(1)})^\top \mathbf{S}^{(1)}_{\mathrm{dfm}}.
\end{aligned}
\]

Here \(\mathbf{S}^{(1)}_{\mathrm{dfm}}=(1,\allowbreak((\varphi_k(\mathbf{A}^{(1)}))_{k=1}^{L_n}\otimes\mathbf{H}^{(1)})^\top)^\top\). We estimate \(\bm{\alpha}^{(1)}\) by penalized LS
\begin{align}
	\widehat{\bm{\alpha}}^{(1)}
	=
	\arg \min_{\bm{\alpha}^{(1)}}
	{2}^{-1}\,
	\mathbb{P}_n
	\left(
	\widehat{Y}^{(1)}_{\mathrm{dfm}}
	- {q^{(1)}_{\mathrm{dfm}}(\mathbf{H}^{(1)}, \mathbf{A}^{(1)})}
	\right)^2
	+ p_{\lambda_n^{(1)}}(\bm{\alpha}^{(1)}),
\end{align}
where \(\mathbb{P}_n\) denotes the empirical measure and \(p_{\lambda_n^{(1)}}\) is the stage-1 penalty. Adding back the inherited main effect gives
\[
\widehat{Q}_{{\mathrm{dfm}}}^{(1)}(\mathbf{H}^{(1)}, \mathbf{A}^{(1)})
=
(\widehat{\bm{\eta}}^{(2)}_{0,1})^\top \mathbf{H}^{(1)}
+(\widehat{\bm{\alpha}}^{(1)})^\top \mathbf{S}^{(1)}_{\mathrm{dfm}}.
\]
The terminal-optimal and upstream deformed-optimal rules are estimated by
\[
\widehat{\pi}^{(2)}_*(\mathbf H^{(2)})
\in
\arg\max_{\mathbf a\in\mathcal A}
\widehat{Q}^{(2)}(\mathbf H^{(2)},\mathbf a),
\qquad
\widehat{\pi}^{(1)}_{{\mathrm{dfm}}}(\mathbf H^{(1)})
\in
\arg\max_{\mathbf a\in\mathcal A}
\widehat{Q}^{(1)}_{{\mathrm{dfm}}}(\mathbf H^{(1)},\mathbf a).
\]
The construction extends recursively to \(m=M,M-1,\ldots,1\): each upstream step forms a deformed pseudo-outcome, fits the corresponding Q-function, and maximizes it to obtain the deformed-optimal rule.

\section{Asymptotic Theory}\label{theory}

This section develops the asymptotic theory for the proposed Deformed Q-learning procedure. We present the main results in the two-stage setting, which already contains the key ingredients of the general recursion: terminal-stage GLM coefficient estimation, value inference after action maximization, and upstream regression based on generated deformed pseudo-outcomes. The diverging-stage analysis is handled through uniform versions of the assumptions. Most regularity conditions are collected in Appendix~A, while selected assumptions needed for the main statements are introduced below; proofs are provided in Appendix~C.

We impose the following standard conditions on the folded-concave penalty \citep{Fan2011}.
\begin{ass}\label{penal}
Let \(\rho(t;\lambda)=\lambda^{-1}p_\lambda(t)\). For each \(\lambda>0\), assume that:
\begin{enumerate}[label=\alph*),nosep,leftmargin=*]
\item \(\rho(t;\lambda)\) is nondecreasing and concave in \(t\in[0,\infty)\);
\item \(\rho(t;\lambda)\) is continuously differentiable in \(t>0\), with \(\rho'(0+;\lambda)>0\);
\item for fixed \(t>0\), \(\rho'(t;\lambda)\) is nondecreasing in \(\lambda\), and \(\rho'(0+;\lambda)\) is independent of \(\lambda\).
\end{enumerate}
\end{ass}
Prominent examples within this class include SCAD, proposed by \citet{Fan2001}, and MCP, introduced by \citet{Zhang2010}.

\subsection{Terminal Inference}\label{2-results}
Without loss of generality, assume the true terminal-stage coefficient vector is $\bm{\beta}^{(2)}_* = ((\bm{\beta}_{*,1}^{(2)})^\top, (\bm{\beta}_{*,2}^{(2)})^\top)^\top$ with $\bm{\beta}_{*,2}^{(2)} = \mathbf{0}$. Let $\bm{\Psi}^{(2)} = (\bm{\Psi}^{(2)}_1, \bm{\Psi}^{(2)}_2)$, where $\bm{\Psi}^{(2)}_1$ and $\bm{\Psi}^{(2)}_2$ denote the submatrices of $\bm{\Psi}^{(2)}$ corresponding to $\bm{\beta}_{*,1}^{(2)}$ and $\bm{\beta}_{*,2}^{(2)}$, respectively, with $\dim\{\bm{\beta}_*^{(2)}\}=p_n^{(2)},\dim\{\bm{\beta}_{*,1}^{(2)}\}=s_n^{(2)}$.  Define
$
g^{(2)}_n = {2}^{-1} \min \left\{ |\beta^{(2)}_{*,j}| : \beta^{(2)}_{*,j} \neq 0 \right\}
$ 
as half of the minimum signal.

Proposition~\ref{convergence2} establishes the oracle property of the terminal-stage penalized estimator \(\widehat{\bm\beta}^{(2)}\) under the stated random-design conditions, building on the folded-concave likelihood theory of \citet{Fan2011} and \citet{Ding2024}. The LLA conditions linking the computed solution to the oracle local solution are given in Remark~\ref{remark:lla-oracle}.

We consider inference for the terminal optimal Q-value. Let \( \{\mathcal{H}_i^{(2)}\}_{i=1}^N \) represent the testing samples with size $N$, independent of the training trajectory and drawn from the same distribution as \( \mathbf{H}^{(2)} \). We introduce the following assumption.
\begin{ass}\label{atensor}
	Assume
	\[
	\mathcal{A}= \mathcal{A}_1 \times \mathcal{A}_2 \times \cdots \times \mathcal{A}_a,
	\]
	where each component \(\mathcal{A}_l \subset \mathbb{R}\) is connected and attains its infimum.
	For each \(l=1,\dots,a\), let \(\mathcal A_l=\bigsqcup_{j=1}^{n_l}\mathcal A_{l,j}\) be a partition into nonempty blocks such that, for \(j=1,\dots,n_l-1\), \(x<y\) whenever \(x\in\mathcal A_{l,j}\) and \(y\in\mathcal A_{l,j+1}\).
	The \(L_n=\prod_{l=1}^a n_l\) basis functions are block indicators:
	\[
	\varphi_k(\mathbf a)=I_{\{\mathbf a\in\mathcal A_{(k)}\}},\qquad
	\mathcal{A}_{(k)} = \mathcal{A}_{1,j_1} \times \cdots \times \mathcal{A}_{a,j_a},
	\]
	where \(k\) indexes the tensor-block tuple \((j_1,\dots,j_a)\).
\end{ass}
Assumption~\ref{atensor} treats $\mathcal{A}$ as a Cartesian product of
one-dimensional spaces with inherent ordering. This construction parallels
the generation of Borel $\sigma$-algebras from semi-algebras of cylinder
sets. In practice, this requirement applies at the level of a
parameterization: more flexible action spaces can be represented through an appropriate block-structured coordinate system.
For example, a unit circle can be parameterized by an angle
$\theta \in [0,2\pi)$, and a unit disk by $(r,\theta)\in[0,1]\times[0,2\pi)$,
which satisfy Assumption~\ref{atensor} under such reparameterizations.

Fix an ordering of the action blocks and one representative \(\mathbf a_k\in\mathcal A_{(k)}\) for each block. Since the block-wise model is constant within each block, define the terminal rule by
\[
k_*(h)=\min\arg\max_{1\leq k\leq L_n}Q^{(2)}(h,\mathbf a_k),
\qquad \pi_*^{(2)}(h)=\mathbf a_{k_*(h)}.
\]
The estimated rule \(\widehat\pi_*^{(2)}\) uses the same convention with \(\widehat Q^{(2)}\). Write \(\mathcal S_{\pi^{(2)}}(h)=\mathbf S^{(2)}(h,\pi^{(2)}(h))\) for the design vector defined in Section~\ref{method}, and abbreviate \(b^{(2)}\) by \(b\). The deformed optimal Q-function and the optimal Q-function at the terminal stage, whose fitted versions are obtained by substituting \(\widehat{\bm\beta}^{(2)}\) and \(\widehat\pi_*^{(2)}\), are
\[
\widetilde Q_*^{(2)}(h)=(\bm\beta_*^{(2)})^\top\mathcal S_{\pi_*^{(2)}}(h),
\qquad Q_*^{(2)}(h)=b'\{\widetilde Q_*^{(2)}(h)\}.
\]
For the testing-history distribution, define
\[
\mu_*^{(2)}=\mathbb E\{Q_*^{(2)}(\mathcal H^{(2)})\},
\qquad
\widehat\mu_*^{(2)}=\frac1N\sum_{i=1}^N
b'\{(\widehat{\bm\beta}^{(2)})^\top\mathcal S_{\widehat\pi_*^{(2)}}(\mathcal H_i^{(2)})\}.
\]
Let \(\mathcal S_{\pi_*^{(2)},1}\) denote the design subvector corresponding to the nonzero coefficients \(\bm\beta_{*,1}^{(2)}\). The gradient of the terminal optimal Q-value with respect to \(\bm\beta_{*,1}^{(2)}\) is
\[
\mathbf g_*^{(2)}
=\mathbb E\!\left[b''\{\widetilde Q_*^{(2)}(\mathcal H^{(2)})\}
\mathcal S_{\pi_*^{(2)},1}(\mathcal H^{(2)})\right].
\]

\begin{thm}\label{2-opt}
Suppose all conditions in Proposition~\ref{convergence2} and Assumptions~A.7--A.9 hold. Let \(L_n=O(n^{\ell_1})\), \(s_n^{(2)}=O(n^{\ell_2^{(2)}})\), \(n^{\ell_3}=O(N)\), and \(N=O(n^{\ell_4})\), where \(\ell_2^{(2)}<1/3\), \(3\ell_2^{(2)}<\ell_3\), and \(2\ell_2^{(2)}+\ell_1<1/2\). As \(n,N\to\infty\), the terminal optimal Q-value estimator admits the expansion
\[
\begin{aligned}
\widehat\mu_*^{(2)}-\mu_*^{(2)}
&=(\mathbf g_*^{(2)})^\top
(\widehat{\bm\beta}^{(2)}_1-\bm\beta_{*,1}^{(2)})\\
&\quad+\frac1N\sum_{i=1}^N
\{Q_*^{(2)}(\mathcal H_i^{(2)})-\mu_*^{(2)}\}+o_p(n^{-1/2}).
\end{aligned}
\]
If \((\mathbf g_*^{(2)})^\top\mathbf V^{(2)}\mathbf g_*^{(2)}\to(\sigma_1^{(2)})^2\) and \((n/N)\operatorname{Var}\{Q_*^{(2)}(\mathcal H^{(2)})\}\to(\sigma_2^{(2)})^2\), then
\[
\sqrt n\,(\widehat\mu_*^{(2)}-\mu_*^{(2)})
\xrightarrow{d}\mathcal N\!\left(0,(\sigma_1^{(2)})^2+(\sigma_2^{(2)})^2\right).
\]
\end{thm}

The two leading terms correspond to terminal coefficient estimation and averaging over independent testing histories. The gradient \(\mathbf g_*^{(2)}\) converts coefficient uncertainty to the mean scale through \(b''\).

Let \(\widehat{\mathbf g}_*^{(2)}=\mathbb P_N[b''\{\widehat{\widetilde Q}_*^{(2)}(\mathcal H^{(2)})\}\mathcal S_{\widehat\pi_*^{(2)},1}(\mathcal H^{(2)})]\), using the design components corresponding to the nonzero coefficients in \(\widehat{\bm\beta}^{(2)}\). The asymptotic variance is estimated by
\[
(\widehat\sigma_*^{(2)})^2
=(\widehat{\mathbf g}_*^{(2)})^\top\widehat{\mathbf V}^{(2)}\widehat{\mathbf g}_*^{(2)}
+\frac nN\widehat{\operatorname{Var}}\{\widehat Q_*^{(2)}(\mathcal H^{(2)})\}.
\]

Appendix~B gives inference for the Q-value contrast
$\mathbb{E}[Q_*^{(2)}(\mathcal{H}^{(2)})] - \mathbb{E}[Q^{(2)}(\mathcal{H}^{(2)}, \pi^{(2)})]$
with a specified terminal policy $\pi^{(2)}$.

For a binary terminal outcome indicating survival, the inference is on the survival-probability scale, providing confidence intervals for differences in survival probability between terminal treatment rules.

\subsection{Recursive Inference}\label{1-results}
Assume that the stage-1 true coefficient takes the form \( \bm{\alpha}^{(1)}_* = ((\bm{\alpha}_{*,1}^{(1)})^\top, (\bm{\alpha}_{*,2}^{(1)})^\top)^\top \) with \( \bm{\alpha}_{*,2}^{(1)} = \mathbf{0} \). Let \( \bm{\Psi}^{(1)}_{\mathrm{dfm}} = \bigl((\mathbf S^{(1)}_{\mathrm{dfm},i})^\top\bigr)_{i=1}^n = (\bm{\Psi}^{(1)}_{\mathrm{dfm},1}, \bm{\Psi}^{(1)}_{\mathrm{dfm},2}) \) denote the design matrix, where \( \bm{\Psi}^{(1)}_{\mathrm{dfm},1} \) and \( \bm{\Psi}^{(1)}_{\mathrm{dfm},2} \) correspond to \( \bm{\alpha}_{*,1}^{(1)} \) and \( \bm{\alpha}_{*,2}^{(1)} \), respectively. Let $
g_n^{(1)} =2^{-1} \min \left\{ |\alpha_{*,j}^{(1)}| : \alpha_{*,j}^{(1)} \ne 0 \right\}$ 
denote half of the minimum signal,  \( p_n^{(1)} =\dim\{\bm{\alpha}_{*}^{(1)}\}\), and \( s_n^{(1)} =\dim\{\bm{\alpha}_{*,1}^{(1)}\}.\) 
We establish the recursive oracle property under the following upstream conditions.

In the two-stage terminal-outcome setting, \(Y^{(1)}=0\) and the stage-1 residual is \(\varepsilon^{(1)}=\widetilde\Gamma(\mathbf H^{(2)})-Q_{\mathrm{dfm}}^{(1)}(\mathbf H^{(1)},\mathbf A^{(1)})\). For the following assumptions, abbreviate \(\bm\Psi_{\mathrm{dfm}}^{(1)}\) by \(\bm\Psi=(\mathbf h_1,\ldots,\mathbf h_{p_n^{(1)}})=(\bm\Psi_1,\bm\Psi_2)\). For a terminal rule \(\pi\), let \(\mathbf W_\pi\) stack the terminal design vectors evaluated at \(\pi(\mathbf H_i^{(2)})\) and restricted to the components corresponding to the nonzero coefficients in \(\bm\alpha_*^{(2)}\).

\begin{ass}\label{upstream-design}
With probability \(1-o(1)\), for any terminal rule \(\pi\),
\[
\begin{alignedat}{2}
&\max_{1\leq j\leq p_n^{(1)}}\|\mathbf h_j\|_\infty
=O\left(\sqrt{n/\log p_n^{(1)}}\right),
&\qquad&\max_{1\leq j\leq p_n^{(1)}}\|\mathbf h_j\|_2=O(\sqrt n),\\
&\lambda_{\min}(\bm\Psi_1^\top\bm\Psi_1)\geq c n,
&&\|\bm\Psi_1^\top\bm\Psi_1\|_2=O(n),\\
&\|\bm\Psi_2^\top\bm\Psi_1\|_{2,\infty}=O(n),
&&\|\mathbf W_\pi^\top\mathbf W_\pi\|_2=O(n),\\
&\|\bm\Psi_2^\top\mathbf W_\pi\|_{2,\infty}=O(n),
&&\||\bm\Psi_2|^\top\mathbf W_\pi\|_{2,\infty}=O(n).
\end{alignedat}
\]
for a constant \(c>0\). Here, \(\|\cdot\|_{2,\infty}\) is the maximum Euclidean row norm, and \(|\cdot|\) denotes entrywise absolute values.
\end{ass}

\begin{ass}\label{upstream-signal}
The signal and penalty satisfy \(g_n^{(1)}\gg\lambda_n^{(1)}\gg\max\{\{(s_n^{(1)}+s_n^{(2)})/n\}^{1/2},\allowbreak(\log p_n^{(1)}/n)^{1/2}\}\), \(p'_{\lambda_n^{(1)}}(g_n^{(1)})=o\{(ns_n^{(1)})^{-1/2}\}\), and \(\lambda_n^{(1)}\kappa_0^{(1)}=o(1)\). Here, \(\kappa_0^{(1)}\) is the maximum local concavity of \(\rho(\cdot;\lambda_n^{(1)})\) over \(\|\bm\delta-\bm\alpha_{*,1}^{(1)}\|_\infty\leq g_n^{(1)}\), as defined in Appendix~A.
\end{ass}

\begin{ass}\label{upstream-moments}
The residual satisfies \(\max_{j\leq s_n^{(1)}}\mathbb E[\{\varepsilon^{(1)}(\mathbf S_{\mathrm{dfm},1}^{(1)})_j\}^2]<\infty\), and for constants \(b_\varepsilon,v_\varepsilon>0\),
\[
b_\varepsilon^2\,\mathbb E\!\left[
\exp\!\left\{\frac{|\varepsilon^{(1)}|}{b_\varepsilon}\right\}
-1-\frac{|\varepsilon^{(1)}|}{b_\varepsilon}
\,\middle|\,\mathbf H^{(1)},\mathbf A^{(1)}\right]
\leq\frac{v_\varepsilon}{2}.
\]
\end{ass}

Assumption~\ref{upstream-design} ensures identifiability of the nonzero coefficients and controls cross-stage design products. The bounds involving \(\mathbf W_\pi\) govern the transfer of terminal coefficient errors into the upstream regression. Assumption~\ref{upstream-signal} separates the minimum signal from the penalty level and the estimation error; the term \(\{(s_n^{(1)}+s_n^{(2)})/n\}^{1/2}\) accounts for uncertainty from both stages. Assumption~\ref{upstream-moments} imposes a conditional Bernstein bound on the residual induced by the state transition, allowing its variance to depend on history and action.

Appendix~A exploits the action-indicator and history-block structure to verify the design conditions for stage-wise coefficient estimation. Lemmas~\ref{lemmaA.1}--\ref{lemmaA.2} and~\ref{lemmaA.1-Mn-1}--\ref{lemmaA.2-Mn-1} give sufficient conditions under sub-Gaussian or bounded histories, together with restrictions on action-block probabilities, conditional second moments, cross-block dependence, and dimension growth.

\begin{thm}\label{convergence1}
In the two-stage terminal-outcome setting, assume all conditions in Proposition~\ref{convergence2}, Assumptions~A.7--A.9, and Assumptions~\ref{upstream-design}--\ref{upstream-moments}, with the history conditions applied to the training sample. Let \(L_n=O(n^{\ell_1})\) and \(s_n^{(2)}=O(n^{\ell_2^{(2)}})\), where \(2\ell_2^{(2)}+\ell_1<1/2\). Then there exists a strict local minimizer $
	\widehat{\bm{\alpha}}^{(1)} = \left((\widehat{\bm{\alpha}}^{(1)}_1)^\top, (\widehat{\bm{\alpha}}^{(1)}_2)^\top\right)^\top
$, such that as \(n\to\infty\),

	\(\mathbb P\{\widehat{\bm\alpha}^{(1)}_2=\mathbf0\}\to1\) and
	$
	\| \widehat{\bm{\alpha}}^{(1)} - \bm{\alpha}^{(1)}_* \|_2 = O_p(\sqrt{(s_n^{(2)}+s_n^{(1)})/n}).$
	If $3\ell^{(2)}_2+\ell_1<1/2$ and Assumption~\ref{normal1} further holds, then
	\[
	{\sqrt{n}} \,  \mathbf{C}^{(1)}(\mathbf{K}^{(1)})^{-1/2} \, \mathbf{D}^{(1)}(\widehat{\bm{\alpha}}^{(1)}_1 - \bm{\alpha}^{(1)}_{*,1}) \xrightarrow{d} \mathcal{N}(0, \mathbf{G}^{(1)}),
	\]
	where \( \mathbf{C}^{(1)} \in \mathbb{R}^{q \times s_n^{(1)}} \) for fixed \(q\), and \( \mathbf{C}^{(1)} (\mathbf{C}^{(1)})^\top \to \mathbf{G}^{(1)} \) is positive definite. Here
$$
\begin{alignedat}{2}
	&\mathbf{D}^{(1)} &=\; \mathbb{E}\left[\mathbf{S}^{(1)}_{\mathrm{dfm},1} (\mathbf{S}^{(1)}_{\mathrm{dfm},1})^\top\right], 
	&\quad \mathbf{F}^{(1)} = \mathbb{E}\left[(\varepsilon^{(1)})^2 \mathbf{S}^{(1)}_{\mathrm{dfm},1} (\mathbf{S}^{(1)}_{\mathrm{dfm},1})^\top\right],\\
	&\mathbf{J}^{(1)} &=\; \mathbb{E}\left[\mathbf{S}^{(1)}_{\mathrm{dfm},1} (\mathbf{S}_{\mathrm{dfm},\pi_*^{(2)},1})^\top\right],
	&\quad \mathbf{K}^{(1)} = \mathbf{F}^{(1)} + \mathbf{J}^{(1)} \mathbf{P}^{(2)} \mathbf{V}^{(2)} (\mathbf{P}^{(2)})^\top (\mathbf{J}^{(1)})^\top. 
\end{alignedat}
$$
	Here, \(\mathbf{S}^{(1)}_{\mathrm{dfm},1}\) is \(\mathbf{S}^{(1)}_{\mathrm{dfm}}\) restricted to \(\operatorname{supp}(\bm\alpha_*^{(1)})\), and
	\[
	\mathbf{S}_{\mathrm{dfm},\pi_*^{(2)},1}
	=\mathbf P^{(2)}\left\{\mathbf S^{(2)}\bigl(\mathbf H^{(2)},\pi_*^{(2)}(\mathbf H^{(2)})\bigr)\right\}_{\operatorname{supp}(\bm\beta_*^{(2)})},
	\]
	where \(\mathbf P^{(2)}\) extracts \(\bm\alpha_{*,1}^{(2)}\) from \(\bm\beta_{*,1}^{(2)}\). In \(\mathbf K^{(1)}\), \(\mathbf F^{(1)}\) is the covariance of the stage-1 regression score; the second term is the covariance contribution of terminal coefficient estimation error propagated through \(\mathbf J^{(1)}\mathbf P^{(2)}\). The score vectors from the two stages have zero cross-covariance, so there is no additional cross term.
\end{thm}

Theorem~\ref{convergence1} establishes upstream inference and an explicit covariance formula for estimated pseudo-outcomes under random designs and conditional heteroscedasticity, accounting for both terminal coefficient estimation and action maximization. Even when both stages use the same trajectories, the terminal score has conditional mean zero given the terminal history and action. The inherited and newly estimated coefficients share terminal estimation error, and their resulting dependence is retained in the full parameter covariance.

To estimate \(\mathbf V^{(1)}=(\mathbf D^{(1)})^{-1}\mathbf K^{(1)}(\mathbf D^{(1)})^{-1}\), the asymptotic covariance matrix of \(\sqrt n(\widehat{\bm\alpha}^{(1)}_1-\bm\alpha_{*,1}^{(1)})\), let \(\widehat{\bm\Psi}^{(1)}_{\mathrm{dfm},1}\) and \(\widehat{\bm\Psi}_{\mathrm{dfm},\widehat\pi_*^{(2)},1}\) stack the corresponding design vectors on the estimated supports, with the latter evaluated at the estimated terminal policy. Define
\[
\begin{aligned}
\widehat{\mathbf D}_n^{(1)}
&=(\widehat{\bm\Psi}^{(1)}_{\mathrm{dfm},1})^\top\widehat{\bm\Psi}^{(1)}_{\mathrm{dfm},1},\\
\widehat{\mathbf J}_n^{(1)}
&=(\widehat{\bm\Psi}^{(1)}_{\mathrm{dfm},1})^\top\widehat{\bm\Psi}_{\mathrm{dfm},\widehat\pi_*^{(2)},1},\\
\widehat{\mathbf F}_n^{(1)}
&=(\widehat{\bm\Psi}^{(1)}_{\mathrm{dfm},1})^\top\operatorname{diag}\{(\widehat\varepsilon_i^{(1)})^2\}_{i=1}^n\widehat{\bm\Psi}^{(1)}_{\mathrm{dfm},1},\\
\widehat{\mathbf K}_n^{(1)}
&=n^{-1}\widehat{\mathbf J}_n^{(1)}\widehat{\mathbf P}^{(2)}\widehat{\mathbf V}^{(2)}(\widehat{\mathbf P}^{(2)})^\top(\widehat{\mathbf J}_n^{(1)})^\top+\widehat{\mathbf F}_n^{(1)}.
\end{aligned}
\]
Here, \(\widehat\varepsilon_i^{(1)}=\widehat Y_{\mathrm{dfm},i}^{(1)}-(\widehat{\bm\alpha}^{(1)})^\top\mathbf S_{\mathrm{dfm},i}^{(1)}\), and \(\widehat{\mathbf P}^{(2)}\) is the corresponding selector on the estimated terminal support. The matrix \(\widehat{\mathbf V}^{(2)}\) estimates the asymptotic covariance of \(\sqrt n(\widehat{\bm\beta}^{(2)}_1-\bm\beta_{*,1}^{(2)})\), as specified in Appendix~B. The resulting covariance estimator is
\[
\widehat{\mathbf V}^{(1)}
=(\widehat{\mathbf D}_n^{(1)}/n)^{-1}
(\widehat{\mathbf K}_n^{(1)}/n)
(\widehat{\mathbf D}_n^{(1)}/n)^{-1}.
\]

We now turn to inference for the full parameter vector \( \bm{\beta}^{(1)} \). Define \( \bm{\gamma}^{(1)} = \bm{\eta}^{(2)}_{0,1} \) and \( \widehat{\bm{\gamma}}^{(1)} = \widehat{\bm{\eta}}^{(2)}_{0,1} \). Then \( \bm{\beta}^{(1)} \) is a permutation of the augmented vector \( \bm{\xi}^{(1)} = \left((\bm{\gamma}^{(1)})^\top, (\bm{\alpha}^{(1)})^\top\right)^\top \), and similarly, \( \bm{\beta}^{(2)} \) is a permutation of \( \bm{\xi}^{(2)} = \left((\bm{\gamma}^{(1)})^\top, (\bm{\alpha}^{(2)})^\top\right)^\top \).

\begin{cor}\label{convergence1full}
	Let \( \mathbf{P}^{(2)} \) and \( \mathbf P_{\gamma}^{(2)} \) be linear operators that extract 
	\( \bm{\alpha}^{(2)}_{*,1} \) and \( \bm{\gamma}^{(1)}_{*,1} \) from \( \bm{\beta}^{(2)}_{*,1} \), respectively. Under the normality conditions of Theorem~\ref{convergence1}, 
	$$ \sqrt{n} \widetilde{\mathbf{C}}^{(1)}(\mathbf V_{\xi}^{(1)})^{-1/2}(\widehat{\bm{\xi}}^{(1)}_1-\bm{\xi}^{(1)}_{*,1})\xrightarrow{d} \mathcal{N}(0,\widetilde{\mathbf{G}}^{(1)}),$$
	where $\dim\{\bm{\beta}^{(1)}_{*,1}\}=\dim\{\bm{\xi}^{(1)}_{*,1}\}=\tilde{s}_n^{(1)}$, $\widetilde{\mathbf{C}}^{(1)}\in \mathbb{R}^{q\times \tilde{s}^{(1)}_n}$ for a constant $q\leq \tilde{s}_n^{(1)}$, $\widetilde{\mathbf{C}}^{(1)}(\widetilde{\mathbf{C}}^{(1)})^\top\to \widetilde{\mathbf{G}}^{(1)}$ is positive definite, and
	\begin{align*}
		\mathbf V_{\xi}^{(1)}=	\left(\begin{matrix}
		\mathbf P_{\gamma}^{(2)} \,  {\mathbf{V}}^{(2)} (\mathbf P_{\gamma}^{(2)})^\top& \mathbf P_{\gamma}^{(2)} \,  {\mathbf{V}}^{(2)} ({\mathbf{P}^{(2)}})^\top {(\mathbf{J}^{(1)})^\top}(\mathbf{D}^{(1)})^{-1} \\ (\mathbf{D}^{(1)})^{-1}{\mathbf{J}^{(1)}}{\mathbf{P}^{(2)}} \,  {\mathbf{V}}^{(2)} (\mathbf P_{\gamma}^{(2)})^\top & (\mathbf{D}^{(1)})^{-1} \mathbf{K}^{(1)} (\mathbf{D}^{(1)})^{-1}
		\end{matrix}\right). 
	\end{align*}
\end{cor}
Similarly, the asymptotic covariance matrix of $\sqrt{n}(\widehat{\bm{\beta}}^{(1)}_{1}-\bm{\beta}^{(1)}_{*,1})$, denoted by $\widetilde{\mathbf{V}}^{(1)}$, is a permutation of $\mathbf V_{\xi}^{(1)}$, and can be estimated by $\widehat{\widetilde{\mathbf{V}}}^{(1)}$ using the sandwich formula.

Inference for stage-1 deformed optimal Q-values and value differences averages over the testing-history distribution; see Appendix~\ref{omitted}.

 More broadly, the two-stage framework naturally extends to multiple stages when \( M \) is finite.

We further allow the number of stages $M_n$ to grow with the sample size. Assumption~\ref{propagationm} bounds products of the coefficient-error propagation matrices uniformly over stages, controlling amplification along the backward recursion. Assumption~\ref{normalm}(b) requires accumulated remainders to vanish after the normalization used in the central limit theorem.

\begin{thm}\label{convergencem}
	Assume Assumptions~B.1--B.4 and B.7 hold for each stage.
	Then there exist strict local minimizers $
	\widehat{\bm{\alpha}}^{(m)} = \left((\widehat{\bm{\alpha}}^{(m)}_1)^\top, (\widehat{\bm{\alpha}}^{(m)}_2)^\top\right)^\top, 1\leq m \leq M_n,$
	such that \(\mathbb P\{\widehat{\bm\alpha}^{(m)}_2=\mathbf0,\ 1\leq m\leq M_n\}\to1\) and, uniformly over stages,

	$
	\| \widehat{\bm{\alpha}}^{(m)} - \bm{\alpha}^{(m)}_* \|_2 = O_p(\sqrt{\sum_{j\geq m}s_n^{(j)}\log n/n}),$
	where \( \widehat{\bm{\alpha}}^{(m)}_1 \) is the subvector of \( \widehat{\bm{\alpha}}^{(m)} \) formed by components in \( \operatorname{supp}(\bm{\alpha}^{(m)}_*) \).
	If Assumptions~B.5--B.6 further hold, then for $m<M_n$
	\[
	{\sqrt{n}} \,  \mathbf{C}^{(m)}(\mathbf{K}^{(m)})^{-1/2} \, \mathbf{D}^{(m)}(\widehat{\bm{\alpha}}^{(m)}_1 - \bm{\alpha}^{(m)}_{*,1}) \xrightarrow{d} \mathcal{N}(0, \mathbf{G}^{(m)}),
	\]
	where \( \mathbf{C}^{(m)} \in \mathbb{R}^{q \times s_n^{(m)}} \) for fixed \(q\), and \( \mathbf{C}^{(m)} (\mathbf{C}^{(m)})^\top \to \mathbf{G}^{(m)} \) is positive definite.
	
\end{thm}

The cumulative sparsity \(\sum_{j=m}^{M_n}s_n^{(j)}\) reflects estimation uncertainty from stage \(m\) onward; the \(\log n\) factor arises from simultaneous control across stages. The stage-specific covariance matrices follow the same backward recursion as in the two-stage setting; see Appendix~\ref{diverge}.

\subsection{Policy Regret}\label{sec:regret}

This section compares the deformed criterion with the expected terminal outcome criterion and quantifies the resulting policy regret.

For an upstream regime \(\pi\), let \(\mathbf H_{\pi}^{(M)}\) denote its induced terminal history and define
\[
\widetilde\Gamma(\mathbf H^{(M)})=\max_{\mathbf a\in\mathcal A}(\bm\beta_*^{(M)})^\top\mathbf S^{(M)}(\mathbf H^{(M)},\mathbf a).
\]
Since \(b^{(M)'}\) is increasing, the same terminal action optimizes both criteria. Define the expected terminal outcome \(V(\pi)\) and the deformed value \(V_{\mathrm{dfm}}(\pi)\) by
\[
V(\pi)
=
\mathbb E\!\left[
b^{(M)'}\!\left\{
\widetilde\Gamma(\mathbf H_{\pi}^{(M)})
\right\}
\right],
\qquad
V_{\mathrm{dfm}}(\pi)
=
\mathbb E\!\left\{
\widetilde\Gamma(\mathbf H_{\pi}^{(M)})
\right\}.
\]
Let \(\pi_*\) and \(\pi_{\mathrm{dfm}}\) maximize these two criteria over the same regime class, and write \(\mathcal R_{\mathrm{dfm}}=V(\pi_*)-V(\pi_{\mathrm{dfm}})\). Here, \(W_1\) denotes first-order Wasserstein distance.

\begin{pro}\label{prop:regret-multistage}
Suppose that the two policy-induced terminal deformed optimal Q-functions have finite first moments and that \(b^{(M)'}\) is increasing and \(L_b\)-Lipschitz on the convex hull of their combined supports. Then, for any fixed finite number of stages,
\begin{equation}
0\leq\mathcal R_{\mathrm{dfm}}\leq\frac{L_b}{2}\Big[W_1\!\left(\mathcal L\!\left\{\widetilde\Gamma(\mathbf H_{\pi_*}^{(M)})\right\},\mathcal L\!\left\{\widetilde\Gamma(\mathbf H_{\pi_{\mathrm{dfm}}}^{(M)})\right\}\right)-\left\{V_{\mathrm{dfm}}(\pi_{\mathrm{dfm}})-V_{\mathrm{dfm}}(\pi_*)\right\}\Big].
\label{eq:directional-regret-main}
\end{equation}
For the logistic mean function, \(L_b=1/4\).
\end{pro}

Let \(F_*\) and \(F_{\mathrm{dfm}}\) denote the distribution functions in Proposition~\ref{prop:regret-multistage}. The bracket in \eqref{eq:directional-regret-main} equals \(2\int_{\mathbb R}\{F_{\mathrm{dfm}}(t)-F_*(t)\}_+\,dt\). This quantity is bounded by the centered Wasserstein distance, so only adverse CDF crossings enter the bound. No explicit stage-count factor appears, although the induced distributions and \(W_1\) may depend on the horizon. See Appendix~C.

\begin{cor}\label{cor:regret-fosd}
Under the conditions of Proposition~\ref{prop:regret-multistage}, let \(U^{(m)}=Z_m-Q_{\mathrm{dfm}}^{(m)}(\mathbf H^{(m)},\mathbf A^{(m)})\), where \(Z_m\) is the terminal deformed optimal Q-function obtained by taking \(\mathbf A^{(m)}\) at stage \(m\) and following \(\pi_{\mathrm{dfm}}\) thereafter. Then
\[
\begin{aligned}
&U^{(m)}\perp\mathbf A^{(m)}\mid\mathbf H^{(m)}
\quad(m=1,\ldots,M-1)\\
&\qquad\Longrightarrow\quad F_{\mathrm{dfm}}(t)\leq F_*(t)
\quad(t\in\mathbb R)
\quad\Longrightarrow\quad \mathcal R_{\mathrm{dfm}}=0.
\end{aligned}
\]
\end{cor}

Conditional independence yields common location shifts and hence first-order stochastic dominance. Related error-independence assumptions appear in Q-learning \citep{Linn2017IQ,Zhu2018} and treatment-effect estimation \citep{Li2023RobustHTE}; here the condition is imposed on continuation residuals to ensure zero regret. The common two-point support example in Appendix~C gives this dominance without conditional independence.

\begin{pro}\label{prop:regret-second-order}
In the two-stage setting, suppose \(U^{(1)}=U_0^{(1)}+R^{(1)}\), where \(U_0^{(1)}\perp\mathbf A^{(1)}\mid\mathbf H^{(1)}\) and \(\mathbb E\{R^{(1)}\mid U_0^{(1)},\mathbf H^{(1)},\mathbf A^{(1)}\}=0\).
If the relevant second moments are finite and
\(\sup_t|b^{(2)'''}(t)|<\infty\), then
\[
\mathcal R_{\mathrm{dfm}}
\leq
\sup_t|b^{(2)'''}(t)|
\,
\mathbb E\!\left[
\sup_{\mathbf a\in\mathcal A}
\mathbb E\!\left\{
(R^{(1)})^2
\mid
\mathbf H^{(1)},
\mathbf A^{(1)}=\mathbf a
\right\}
\right].
\]
For the logistic mean function, \(\sup_t|b^{(2)'''}(t)|=1/(6\sqrt 3)\).
\end{pro}

When continuation functions are complex on the mean scale but deformed conditional means admit sparse models, Deformed Q-learning provides a tractable framework for multi-stage estimation and inference. Under conditional independence of the continuation residuals and actions given history, the deformed criterion preserves optimality for the expected terminal outcome. When conditional dependence is allowed, the regret bounds quantify the potential cost of changing the recursive criterion. The simulation study further illustrates a setting with conditional dependence and sparse deformed conditional means, in which the proposed method combines tractable recursive inference with a small loss in expected terminal outcome and achieves lower regret than the compared procedures.

\section{Simulation Study}\label{simu}

We consider a two-stage logistic model with random covariates, history-dependent treatment assignment, and conditionally heteroscedastic transitions. The first-stage reward is $Y^{(1)}\equiv0$, and the regime is evaluated by the binary terminal outcome $Y^{(2)}$. We use the indicator basis and discretize the action space $[0,1]$ into $L_n=11$ levels, $a_k=(k-1)/10$, $k=1,\ldots,11$.

The baseline history is $\mathbf H^{(1)}=(1,X,Z,V,\bm\xi_1^\top)^\top$, where $X$ takes values $-1$ and $1$ with equal probabilities, $Z$ is a standardized $\operatorname{Beta}(2,5)$ variable, and $V\sim N(0,1)$; these variables are mutually independent. The vectors $\bm\xi_1,\bm\xi_2\sim N_q(0,\Sigma_q)$, with $(\Sigma_q)_{ij}=0.5^{|i-j|}$, are mutually independent and independent of the remaining variables; they add irrelevant covariates at stages 1 and 2, respectively, with zero true coefficients. The key transition variables are $X_{21}\in\{-1,1\}$ and $W$, generated according to
\[
\begin{gathered}
\mathbb P(X_{21}=1\mid\mathbf H^{(1)},A^{(1)}=a_k)=p_{X,k},\\
W=g(A^{(1)})Z+\{1+0.8\tanh(V)\}\varepsilon,
\end{gathered}
\]
where $\varepsilon\sim N(0,1)$ is independent of the preceding variables and the binary transition innovation. The action-dependent loading function $g$ is specified in Appendix~\ref{app:sim-dgp}. The probabilities $p_{x,k}$ vary with the baseline state and first-stage action. The second-stage history augments $\mathbf H^{(1)}$ with $A^{(1)}$, $X_{21}$, $W$, and the additional noise vector $\bm\xi_2$. The terminal outcome follows
\[
\begin{aligned}
&Y^{(2)}\mid\mathbf H^{(2)},A^{(2)}=a_k\sim\operatorname{Bernoulli}\{\sigma(\eta_k)\},\\
&\eta_k=-\frac32-\frac32X+\frac32 A^{(1)}-\frac32X_{21}+W+d_k(X,X_{21}),
\end{aligned}
\]
where $\sigma(t)=(1+e^{-t})^{-1}$ and $d_k$ is an action-specific linear increment. For $h=(1,x,z,v,\bm\xi_1^\top)^\top$, the resulting deformed continuation is
\[
Q^{(1)}_{\mathrm{dfm}}(h,a_k)=\frac32+\frac32a_k+g(a_k)z+\left(\frac{13}{6}-\frac32x\right)p_{x,k}.
\]
Both the terminal linear predictor and the deformed continuation are block-wise linear, with $p_n^{(2)}=77+22q$, $s_n^{(2)}=15$, $p_n^{(1)}=44+11q$, and $s_n^{(1)}=32$. The continuation value for the expected terminal outcome falls outside the corresponding block-wise linear and logistic classes. The behavior probabilities, transition probabilities $p_{x,k}$, action increments $d_k$, and derivations of the continuation values are provided in Appendices~\ref{app:sim-dgp}--\ref{app:sim-continuation}.

We consider $(n,q)=(5000,500),(10000,500)$, with corresponding dimensions $(p_n^{(1)},p_n^{(2)})=(5544,11077),(5544,11077)$ and independent evaluation sample sizes $N=2000,3000$, respectively.

\subsection{Estimation and Inference}
We estimate the coefficients at both stages using Deformed Q-learning with the SCAD penalty. The tuning parameters previously selected using the Bayesian information criterion (BIC) are held fixed across the present 1000 replications. All estimation and inference results are averaged over 1000 independent replications. Table~\ref{tab:sim-coef} reports the $\ell_1$ and $\ell_2$ estimation errors for the full coefficient vectors and the mean counts of false positives (FP) and false negatives (FN), and average empirical coverage of 95\% confidence intervals for the true nonzero coefficients.

\begin{table}[!htbp]

\centering
\caption{Summary of coefficient estimation performance for $\bm\beta^{(2)}$ and $\bm\alpha^{(1)}$}\label{tab:sim-coef}

\begin{tabular*}{\linewidth}{@{\extracolsep{\fill}}llccccc}
\toprule
Stage & $n$ & $\ell_1$ & $\ell_2$ & FP & FN & Coverage \\
\midrule
Stage-2 & $5000$ & 1.9305 & 0.6724 & 0.0580 & 0.0400 & 0.935 \\
 & $10000$ & 1.2417 & 0.4231 & 0.0220 & 0.0000 & 0.950 \\
\midrule
Stage-1 & $5000$ & 3.5555 & 0.7772 & 0.0100 & 0.0000 & 0.937 \\
 & $10000$ & 2.4404 & 0.5351 & 0.0070 & 0.0000 & 0.948 \\
\bottomrule
\end{tabular*}

\end{table}

The $\ell_1$ and $\ell_2$ estimation errors decrease with sample size at both stages. The first-stage estimates are less precise, primarily because uncertainty from terminal-stage estimation accumulates in the upstream regression.

We next estimate the terminal optimal Q-value and the deformed optimal Q-values at the terminal and first stages and the corresponding value differences relative to the behavior policy $\pi_{\mathrm b}$. An independent evaluation sample of size $N$ is generated under the behavior policy. Table~\ref{tab:sim-values} reports the mean estimate, average estimated standard error (SE), and empirical coverage, using the joint covariance estimator to account for downstream estimation uncertainty.

\begin{table}[!htbp]

\centering
\caption{Optimal Q-value and deformed optimal Q-value estimates}\label{tab:sim-values}

\begin{tabular*}{\linewidth}{@{\extracolsep{\fill}}lcccccc}
\toprule
$(n,N)$ & \multicolumn{3}{c}{Value} & \multicolumn{3}{c}{Value Difference} \\
\cmidrule(lr){2-4}\cmidrule(lr){5-7}
 & Mean & SE & Coverage & Mean & SE & Coverage \\
\midrule
 & \multicolumn{3}{c}{$\mathbb E[Q_*^{(2)}]=0.8568$} & \multicolumn{3}{c}{$\mathbb E[Q_*^{(2)}]-\mathbb E[Q_{\pi_{\rm b}}^{(2)}]=0.4302$} \\
$(5000,2000)$ & 0.8559 & 0.0093 & 0.942 & 0.4294 & 0.0105 & 0.944 \\
$(10000,3000)$ & 0.8570 & 0.0068 & 0.943 & 0.4306 & 0.0080 & 0.939 \\
\midrule
 & \multicolumn{3}{c}{$\mathbb E[\widetilde Q_*^{(2)}]=3.3167$} & \multicolumn{3}{c}{$\mathbb E[\widetilde Q_*^{(2)}]-\mathbb E[\widetilde Q_{\pi_{\rm b}}^{(2)}]=3.9062$} \\
$(5000,2000)$ & 3.3214 & 0.1516 & 0.930 & 3.9124 & 0.1566 & 0.934 \\
$(10000,3000)$ & 3.3281 & 0.1102 & 0.955 & 3.9220 & 0.1138 & 0.944 \\
\midrule
 & \multicolumn{3}{c}{$\mathbb E[Q_{\mathrm{dfm},*}^{(1)}]=4.2160$} & \multicolumn{3}{c}{$\mathbb E[Q_{\mathrm{dfm},*}^{(1)}]-\mathbb E[Q_{\mathrm{dfm},\pi_{\rm b}}^{(1)}]=0.8993$} \\
$(5000,2000)$ & 4.2361 & 0.1746 & 0.929 & 0.9138 & 0.0540 & 0.957 \\
$(10000,3000)$ & 4.2359 & 0.1254 & 0.942 & 0.9071 & 0.0393 & 0.958 \\
\bottomrule
\end{tabular*}

\end{table}

The first two panels use the terminal probability and linear-predictor scales, respectively; the third uses the first-stage deformed scale. All quantities are averaged over the behavior-history distribution. The empirical coverage rates are generally close to the nominal 95\% level.

\subsection{Policy Comparison}
We compare Deformed Q-learning (DFM) with hard-threshold Q-learning (HQ) of \citet{Zhu2018} and two recursive adaptations of DROVE of \citet{Ding2024}. HQ selects between the endpoint actions $0$ and $1$. DROVE-G and DROVE-B retain all 11 actions and use Gaussian and logistic terminal regressions, respectively; both fit a Gaussian upstream regression of the optimized terminal mean. Details are given in Appendix~\ref{app:sim-benchmarks}.

Table~\ref{tab:sim-policy} reports the expected terminal outcome $V(\widehat\pi)=\mathbb E_{\widehat\pi}[Y^{(2)}]$ and regret $R(\widehat\pi)=V(\pi_*)-V(\widehat\pi)$. Evaluation uses histories generated under each learned regime. MCSE denotes the Monte Carlo standard error of mean regret.

\begin{table}[!htbp]

\centering
\caption{Expected terminal outcomes and regrets of learned regimes}\label{tab:sim-policy}

\begin{tabular*}{\linewidth}{@{\extracolsep{\fill}}llccc}
\toprule
$n$ & Method & Value & Regret & MCSE \\
\midrule
$5000$ & DFM & 0.913722 & 0.001812 & 0.000043 \\
 & DROVE-G & 0.818088 & 0.097446 & 0.000835 \\
 & DROVE-B & 0.659451 & 0.256083 & 0.002693 \\
 & HQ & 0.825260 & 0.090274 & 0.000081 \\
\midrule
$10000$ & DFM & 0.914359 & 0.001187 & 0.000019 \\
 & DROVE-G & 0.899328 & 0.016218 & 0.000440 \\
 & DROVE-B & 0.837977 & 0.077569 & 0.000334 \\
 & HQ & 0.823809 & 0.091737 & 0.000076 \\
\bottomrule
\end{tabular*}

\end{table}

DFM attains the lowest mean regret among the compared procedures, with regret decreasing as $n$ grows. HQ's larger loss partly reflects its endpoint-action restriction. The advantage persists against DROVE-B, which uses the same 11-action space and a logistic terminal model. In this design, the deformed continuation has an exact sparse block-wise linear representation, supporting upstream estimation under the nonlinear terminal mean model. The population comparison also favors DFM, whose small positive regret quantifies the decision cost of changing the recursive criterion (Table~\ref{tab:sim-population} in Appendix~\ref{app:sim-continuation}). Together with the estimation and coverage results, these findings illustrate the benefit of combining tractable recursive inference with a small loss in the expected terminal outcome.

\section{Empirical Study}\label{real}
The MIMIC-III database provides intensive care unit (ICU) data for sepsis research. Following \citet{Komorowski2018}, we discretize each patient's trajectory into 18 four-hour steps over a 72-hour window around sepsis onset. We focus on a 48-hour window from 12 hours before to 36 hours after the estimated sepsis onset (steps 4--15), which approximately captures the early intervention period commonly referred to as the golden hours in sepsis care \citep{rivers2001}. The earliest and latest 12-hour periods are excluded to reduce variability from remote pre-onset measurements and later downstream clinical dynamics.

To ensure data quality, we retain patients with at least eight observed steps, yielding approximately 13{,}000 samples. The window is divided into three stages: early (4--6), mid (7--9), and late (10--15). Stage-wise covariates are built by averaging continuous variables and taking the first occurrence of categorical ones. The dataset is randomly split into 80\% for training and 20\% for evaluation.

At each stage $i$, the covariate vector $\mathbf{X}^{(i)}$ includes 47 clinical variables, covering demographics, vital signs, laboratory results, and organ function scores. A complete variable list is provided in Appendix~E. Two treatment variables are considered: input-4hourly and max-dose-vaso, jointly forming the action vector \(\mathbf{A}^{(t)} = (A_1^{(t)}, A_2^{(t)}) \in \mathcal{A}=[0,1]^2\), where each coordinate is partitioned into four disjoint intervals determined by empirical quantiles. Specifically, $$
\mathcal{I}_i = \{[0, q_{0.05}),\ [q_{0.05}, q_{0.4}),\ [q_{0.4}, q_{0.7}),\ [q_{0.7}, 1.0]\},\quad i = 1, 2,$$ and the Cartesian products of intervals in \(\mathcal I_1\) and \(\mathcal I_2\) define the 16 action blocks. The final survival outcome is defined as the sole reward \( Y^{(3)} \in \{0,1\} \) with \( 1 \) indicating survival, and \( Y^{(1)} = Y^{(2)} = 0 \). To retain covariate effects, we include the intercept term \( c^{(i)} \) in Q-function estimation. The covariate dimensions are
\[
p^{(1)} = 753,\qquad p^{(2)} = 1537,\qquad p^{(3)} = 2321.
\] The final-stage Q-function is estimated via logistic regression, 
\[
Q^{(3)}(\mathbf{H}^{(3)}, \mathbf{A}^{(3)}) = \sigma\left((\bm{\beta}^{(3)})^\top\mathbf{S}^{(3)}\right),
\]
where \(\mathbf{H}^{(3)}\) is the history. The mid-stage deformed pseudo-outcome is defined as
\[
\widehat{Y}^{(2)}_{\mathrm{dfm}} = \max_{\mathbf a\in\mathcal A} \left\{ (\widehat{\bm{\beta}}^{(3)})^\top \mathbf{S}^{(3)}(\mathbf H^{(3)},\mathbf a) - (\widehat{\bm{\eta}}^{(3)}_{0,2})^\top \mathbf{H}^{(2)} \right\},
\]
which is used to estimate \( Q_{\mathrm{dfm}}^{(2)} \). Similarly, \(\widehat{Y}^{(1)}_{\mathrm{dfm}}\) estimates the upstream deformed Q-function \(Q_{\mathrm{dfm}}^{(1)}\).
Regularization parameters \( \lambda^{(i)} \) (\( i=1,2,3 \)) are chosen via cross-validation. 

\subsection{Value Estimation and Inference}
We first report stage-specific value estimates and confidence intervals in Table~\ref{tab:qvalue}. Let \(\pi_{\rm b}=(\pi_{\rm b}^{(1)},\pi_{\rm b}^{(2)},\pi_{\rm b}^{(3)})\) denote the behavior policy represented by the observed stage-specific treatment assignments. The terminal optimal Q-value uses the survival-probability scale; terminal and upstream deformed optimal Q-values use the log-odds scale.
The confidence interval for $\mathbb{E}[Q^{(3)}_{\pi_{\rm b}}]$ covers the empirical survival rate of 0.7923, supporting calibration of the final-stage model. Furthermore, the values of $\mathbb{E}[\widetilde{Q}^{(3)}_*]$ and $\mathbb{E}[Q_{{\mathrm{dfm}},*}^{(2)}]$ are consistently close to those of $\mathbb{E}[Q_{{\mathrm{dfm}},\pi_{\rm b}}^{(2)}]$ and $\mathbb{E}[Q_{{\mathrm{dfm}},\pi_{\rm b}}^{(1)}]$, suggesting that the estimated models successfully capture the continuation value from downstream stages. In addition, the ordering $\mathbb{E}[Q_{{\mathrm{dfm}},*}^{(1)}] > \mathbb{E}[Q_{{\mathrm{dfm}},*}^{(2)}] > \mathbb{E}[\widetilde{Q}^{(3)}_*]$ aligns with the recursive deformed-scale construction. The confidence intervals across all estimates are stable, reflecting strong statistical precision. Together, these findings support the internal consistency of the stage-wise framework and its potential for generating clinically interpretable treatment recommendations.

\begin{table}[!htbp]
	\centering
	\caption{Estimates of Values and Value Differences with 95\% Confidence Intervals}

	\begin{tabular*}{\linewidth}{@{\extracolsep{\fill}}lclc}
		\toprule
		\textbf{Value estimate} & \textbf{95\% CI} & \textbf{Difference estimate} & \textbf{95\% CI} \\
		\midrule
		$\mathbb{E}[Q^{(3)}_{\pi_{\rm b}}]\colon 0.7920$ & [0.7741, 0.8100] & $\mathbb{E}[Q^{(3)}_* - Q^{(3)}_{\pi_{\rm b}}]\colon 0.0429$ & [0.0285, 0.0572] \\
		$\mathbb{E}[Q^{(3)}_*]\colon 0.8349$ & [0.8202, 0.8496] & & \\
		
		$\mathbb{E}[\widetilde{Q}^{(3)}_{\pi_{\rm b}}]\colon 1.7706$ & [1.6461, 1.8951] & $\mathbb{E}[\widetilde{Q}^{(3)}_* - \widetilde{Q}^{(3)}_{\pi_{\rm b}}]\colon 0.3270$ & [0.2426, 0.4115] \\
		$\mathbb{E}[\widetilde{Q}^{(3)}_*]\colon 2.0976$ & [1.9545, 2.2407] & & \\
		
		$\mathbb{E}[Q^{(2)}_{{\mathrm{dfm}},\pi_{\rm b}}]\colon 2.0998$ & [1.9957, 2.2039] & $\mathbb{E}[Q^{(2)}_{{\mathrm{dfm}},*} - Q^{(2)}_{{\mathrm{dfm}},\pi_{\rm b}}]\colon 0.4444$ & [0.4035, 0.4853] \\
		$\mathbb{E}[Q^{(2)}_{{\mathrm{dfm}},*}]\colon 2.5443$ & [2.4482, 2.6403] & & \\
		
		$\mathbb{E}[Q^{(1)}_{{\mathrm{dfm}},\pi_{\rm b}}]\colon 2.5334$ & [2.4835, 2.5834] & $\mathbb{E}[Q^{(1)}_{{\mathrm{dfm}},*} - Q^{(1)}_{{\mathrm{dfm}},\pi_{\rm b}}]\colon 0.2104$ & [0.1905, 0.2304] \\
		$\mathbb{E}[Q^{(1)}_{{\mathrm{dfm}},*}]\colon 2.7439$ & [2.6927, 2.7951] & & \\
		\bottomrule
	\end{tabular*}
	\label{tab:qvalue}
\end{table}

\subsection{Interpretation of Model Decisions}
We next examine the selected Stage-3 Q-value coefficients to interpret the learned policy (Table~\ref{table:stage3_selected}). The "--$i$" suffix denotes the $i$-th time step. 

In Block (0,0), coefficients align with clinical intuition: higher SIRS, age, and SOFA reflect more evident systemic inflammatory response, lower physiologic reserve, and more severe organ dysfunction, and are therefore associated with lower estimated survival value; higher SpO2 usually indicates better oxygenation status and is associated with higher estimated survival value. In Block (1,0), the positive coefficient for SGPT-1 captures treatment heterogeneity under a moderate intervention block. In Block (3,0), the positive coefficients for SIRS and SOFA indicate that, when inflammatory response or organ dysfunction is more severe, a more intensive treatment block is estimated to yield higher survival value. Aggressive actions such as (3,3) are rarely selected in Stage 3, while action distributions across stages show clinically plausible variation. See Appendix~E for more details.

\begin{table}[!htbp]
	\centering
	\caption{Selected coefficient estimates from Stage 3}
	
	\begin{tabular*}{\textwidth}{@{\extracolsep{\fill}}cccc}
		\toprule
		\textbf{Value} & \textbf{Action\_Blocks} & \textbf{SubFeature} & \textbf{CI} \\
		\midrule
		-0.1642 & (0,0) & SIRS-1 & [-0.2311, -0.0973] \\
		-0.5918 & (0,0) & age-1 & [-0.6636, -0.5201] \\
		-0.3717 & (0,0) & SOFA-3 & [-0.4605, -0.2830] \\
		0.1291  & (0,0) & SpO2-3 & [0.0624, 0.1958] \\
		0.1160  & (1,0) & SGPT-1 & [0.0159, 0.2162] \\
		0.2349  & (3,0) & SIRS-1 & [0.0585, 0.4113] \\
		0.2581  & (3,0) & SOFA-2 & [0.0788, 0.4374] \\
		\bottomrule
	\end{tabular*}
	\label{table:stage3_selected}
\end{table}

\subsection{Auxiliary Action-Alignment Evaluation}\label{sec:action_alignment}
As a complementary assessment, we conduct a stage-wise action-alignment evaluation on the test set. Because actions are represented by treatment blocks, agreement between the observed action block and the learned recommendation is a discrete analogue of dose alignment. At each stage, we estimate the propensity score \( \widehat{e}(\mathbf{X}_i) = \mathbb{P}(T_i = 1 \mid \mathbf{X}_i) \) and the outcome model \( \widehat{m}_t(\mathbf{X}_i) = \mathbb{E}(Y_i \mid T_i = t, \mathbf{X}_i) \) via logistic regression, where \( T_i = 1 \) if the learned action matches the observed treatment block. The binary outcome \(Y_i\in\{0,1\}\) is defined on the survival scale, with larger values corresponding to more favorable outcomes. Final survival is used as a unified downstream endpoint for assessing the stage-specific action-alignment estimands.

We estimate the stage-wise action-alignment average treatment effect (ATE) via inverse probability weighting (IPW) and its doubly robust (DR) version:
\begin{align*}
	\widehat{\mathrm{ATE}}_{\mathrm{IPW}} &= \frac{1}{n} \sum_{i=1}^n \left( \frac{T_i Y_i}{\widehat{e}(\mathbf{X}_i)} - \frac{(1 - T_i) Y_i}{1 - \widehat{e}(\mathbf{X}_i)} \right), \\
	\widehat{\mathrm{ATE}}_{\mathrm{DR}} &= \frac{1}{n} \sum_{i=1}^n \left[ \frac{T_i - \widehat{e}(\mathbf{X}_i)}{\widehat{e}(\mathbf{X}_i)(1 - \widehat{e}(\mathbf{X}_i))} (Y_i - \widehat{m}_{T_i}(\mathbf{X}_i)) + \widehat{m}_1(\mathbf{X}_i) - \widehat{m}_0(\mathbf{X}_i) \right].
\end{align*}

To assess statistical significance, we compute 95\% confidence intervals for DR-ATE using the nonparametric bootstrap. Covariate balance is evaluated via the standardized mean difference (SMD) and variance ratio (VR) after weighting. For $j$-th covariate, $\mathrm{SMD}_j = (\bar{X}_{j,T=1}^{(w)} - \bar{X}_{j,T=0}^{(w)}) / \sqrt{0.5 \cdot (s_{j,T=1}^{2(w)} + s_{j,T=0}^{2(w)})}$, 
where $\bar{X}_{j,T=t}^{(w)}$ and $s_{j,T=t}^{2(w)}$ denote the weighted mean and variance in group $T = t$. We also compute the effective sample size (ESS) as 
$\mathrm{ESS} = (\sum_{i=1}^n w_i)^2 / \sum_{i=1}^n w_i^2$, and examine extreme weights.

In addition to our proposed three-stage model with 16 discrete actions, we evaluate several simplified alternatives: a fixed frequent-action policy, a two-stage and a single-stage design both using the same 16-action space, and a three-stage model with binary actions. Detailed configurations and results are provided in Appendix~\ref{app:alignment_alternatives}.

Within the proposed three-stage, 16-action model, only the Stage-1 DR-adjusted action-alignment contrast has a 95\% bootstrap confidence interval excluding zero. This stage also exhibits favorable covariate-balance and weighting diagnostics: all absolute weighted SMDs are below 0.1, including clinically important variables such as SOFA and SIRS; the VRs mostly fall within the conventional \(2/3\)--\(1.5\) range; no covariate has both a large SMD and an extreme VR; and the ESS remains adequate.

\begin{figure}[!htbp]
	\centering
	\begin{minipage}[t]{0.45\textwidth}
		\centering
		\includegraphics[width=\textwidth, trim=10 22 10 10, clip]{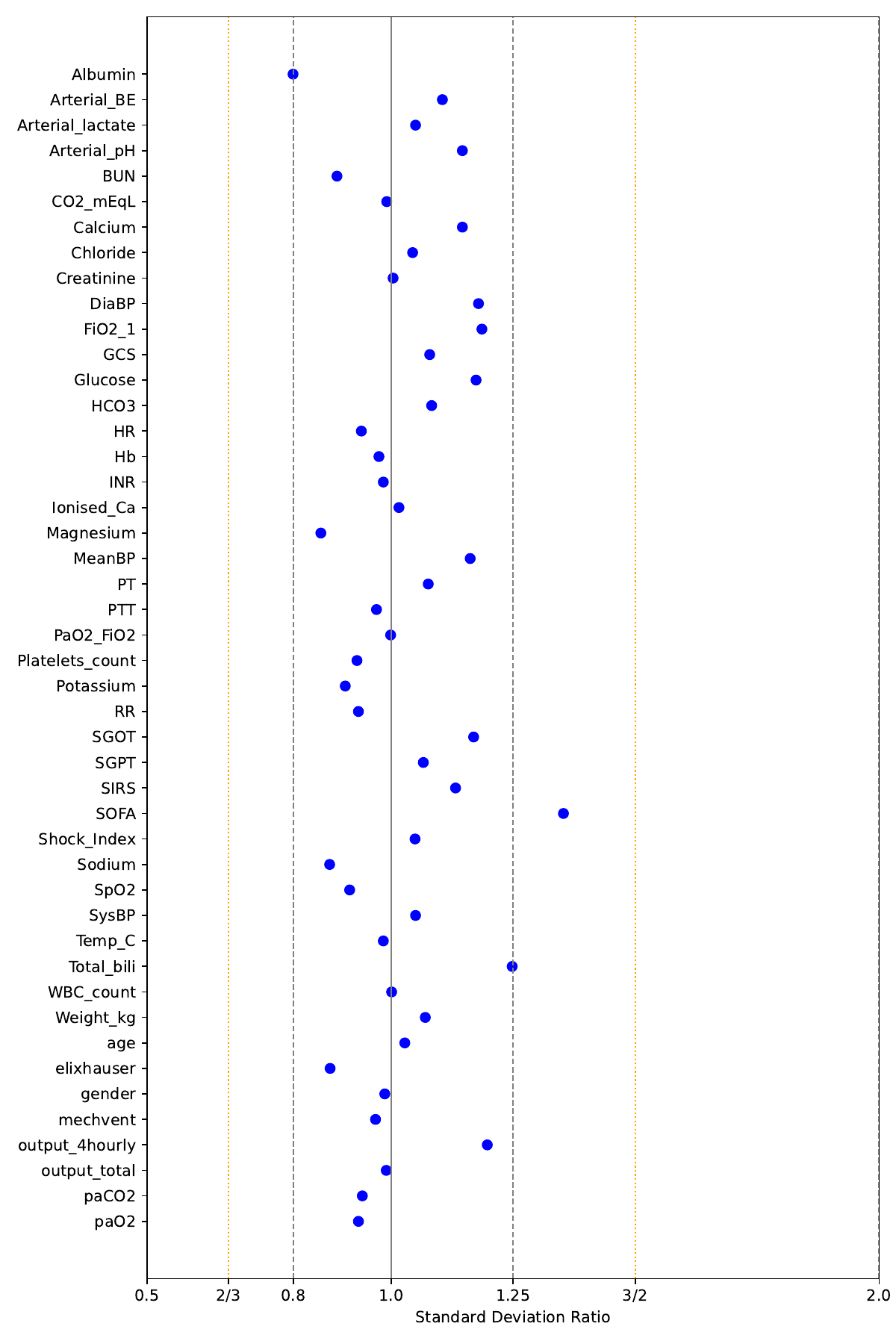}
	\end{minipage}
	\hfill
	\begin{minipage}[t]{0.45\textwidth}
		\centering
		\includegraphics[width=\textwidth, trim=10 22 10 10, clip]{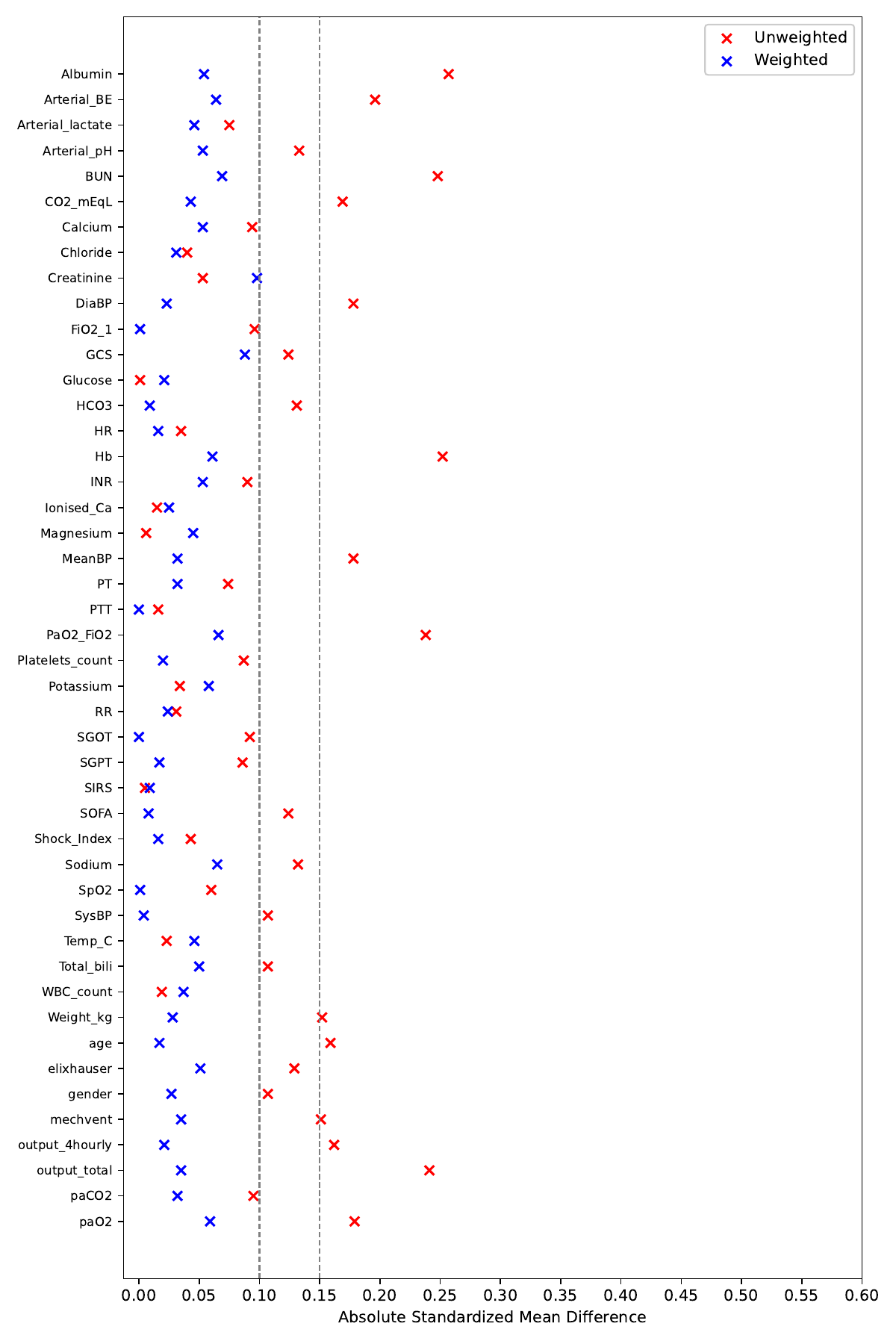}
	\end{minipage}
	\caption{Covariate balance diagnostics for Stage~1: VR (left) and SMD (right)}
	\label{fig:balance_stage1}
\end{figure}

\begin{table}[!htbp]
	\centering
	\caption{Action-alignment estimates and weighting diagnostics across stages}

		\begin{tabularx}{\textwidth}{@{}>{\raggedright\arraybackslash}Xccc@{}}
			\toprule
			\textbf{Metric} & \textbf{Stage 1} & \textbf{Stage 2} & \textbf{Stage 3}\\
			\midrule
				ATE (DR / IPW) & 0.069 / 0.095 & -0.0336 / 0.074 & 0.0217 / 0.043 \\
				95\% CI (DR) & [0.0184, 0.1146] & [-0.1498, 0.0763] & [-0.0163, 0.0600] \\
			Max SMD & \multirow{2}{*}{0.098 / 0.257} & \multirow{2}{*}{0.397 / 0.657} & \multirow{2}{*}{0.107 / 0.223} \\
            \mbox{(Weighted / Unweighted)} & & & \\
			ESS / Extreme Weights (\%) & 1079 / 1\% & 604 / 1\% & 1715 / 1\% \\
			\bottomrule
	\end{tabularx}
	\label{tab:stage_eval}
\end{table}

Together, these diagnostics support the adjusted comparison underlying the Stage-1 DR estimand. The positive Stage-1 DR-adjusted alignment contrast provides an exploratory signal that concordance with the learned recommendation during the early treatment window may be associated with improved survival.

\clearpage
\phantomsection
\addcontentsline{toc}{section}{References}
\bibliography{references}

@article{Laber2014IQ,
    author  = {Laber, Eric B. and Linn, Kristin A. and Stefanski, Leonard A.},
    title   = {Interactive model building for {Q}-learning},
    journal = {Biometrika},
    year    = {2014},
    volume  = {101},
    number  = {4},
    pages   = {831--847},
    doi     = {10.1093/biomet/asu043}
}

@article{Linn2017IQ,
    author  = {Linn, Kristin A. and Laber, Eric B. and Stefanski, Leonard A.},
    title   = {Interactive {Q}-Learning for Quantiles},
    journal = {Journal of the American Statistical Association},
    year    = {2017},
    volume  = {112},
    number  = {518},
    pages   = {638--649},
    doi     = {10.1080/01621459.2016.1155993}
}

@article{Li2023RobustHTE,
    author  = {Li, Ruohong and Wang, Honglang and Zhao, Yi and Su, Jing and Tu, Wanzhu},
    title   = {Robust Estimation of Heterogeneous Treatment Effects: An Algorithm-based Approach},
    journal = {Communications in Statistics---Simulation and Computation},
    year    = {2023},
    volume  = {52},
    number  = {10},
    pages   = {4981--4998},
    doi     = {10.1080/03610918.2021.1974883}
}

@article{Komorowski2018,
	author  = {Komorowski, M. and Celi, L. A. and Badawi, O.},
	title   = {The Artificial Intelligence Clinician learns optimal treatment strategies for sepsis in intensive care},
	journal = {Nature Medicine},
	volume  = {24},
	pages   = {1716--1720},
	year    = {2018}
}

@incollection{Robins2004,
	author    = {Robins, James M.},
	editor    = {Lin, D. Y. and Heagerty, P. J.},
	title     = {Optimal Structural Nested Models for Optimal Sequential Decisions},
	booktitle = {Proceedings of the Second Seattle Symposium in Biostatistics: Analysis of Correlated Data},
	year      = {2004},
	publisher = {Springer New York},
	address   = {New York, NY},
	pages     = {189--326}
}

@article{Murphy2003,
	author  = {Murphy, S. A.},
	title   = {Optimal Dynamic Treatment Regimes},
	journal = {Journal of the Royal Statistical Society: Series B (Statistical Methodology)},
	volume  = {65},
	number  = {2},
	pages   = {331--355},
	year    = {2003},
	month   = apr
}

@phdthesis{Watkins1989,
	author = {Watkins, C. J. C. H.},
	title  = {Learning From Delayed Rewards},
	school = {University of Cambridge},
	year   = {1989}
}

@article{Watkins1992,
	author  = {Watkins, C. J. C. H. and Dayan, P.},
	title   = {Q-learning},
	journal = {Machine Learning},
	volume  = {8},
	pages   = {279--292},
	year    = {1992}
}

@article{Chakraborty2010,
	author  = {Chakraborty, B. and Murphy, S. A. and Strecher, V.},
	title   = {Inference for non-regular parameters in optimal dynamic treatment regimes},
	journal = {Statistical Methods in Medical Research},
	year    = {2010},
	volume  = {19},
	number  = {3},
	pages   = {317--343}
}

@article{Song2015,
	author  = {Song, Rui and Kosorok, Michael R. and Zeng, Donglin},
	title   = {Penalized Q-Learning for Dynamic Treatment Regimens},
	journal = {Statistica Sinica},
	volume  = {25},
	number  = {3},
	pages   = {901--920},
	year    = {2015}
}

@article{Shi2018,
	author  = {Shi, Chenguang and Fan, Alden and Song, Rui and Lu, Wenbin},
	title   = {High-Dimensional A-Learning for Optimal Dynamic Treatment Regimes},
	journal = {Annals of Statistics},
	volume  = {46},
	number  = {3},
	pages   = {925--957},
	year    = {2018},
	month   = jun
}

@article{Zhu2018,
	author  = {Zhu, Wenjing and Zeng, Donglin and Song, Rui},
	title   = {Proper Inference for Value Function in High-Dimensional Q-Learning for Dynamic Treatment Regimes},
	journal = {Journal of the American Statistical Association},
	volume  = {114},
	number  = {527},
	pages   = {1404--1417},
	year    = {2019}
}

@article{Chen2016,
	author  = {Chen, Gongjun and Zeng, Donglin and Kosorok, Michael R.},
	title   = {Personalized Dose Finding Using Outcome Weighted Learning},
	journal = {Journal of the American Statistical Association},
	volume  = {111},
	number  = {516},
	pages   = {1509--1521},
	year    = {2016}
}

@InProceedings{Kallus2018,
	title = 	 {Policy Evaluation and Optimization with Continuous Treatments},
	author = 	 {Kallus, Nathan and Zhou, Angela},
	booktitle = 	 {Proceedings of the Twenty-First International Conference on Artificial Intelligence and Statistics},
	pages = 	 {1243--1251},
	year = 	 {2018},
	editor = 	 {Storkey, Amos and Perez-Cruz, Fernando},
	volume = 	 {84},
	series = 	 {Proceedings of Machine Learning Research},
	month = 	 {09--11 Apr},
	publisher =    {PMLR},
}

@inproceedings{Chernozhukov2019,
	author = {Chernozhukov, Victor and Demirer, Mert and Lewis, Greg and Syrgkanis, Vasilis},
	title = {Semi-Parametric Efficient Policy Learning with Continuous Actions},
	booktitle = {Advances in Neural Information Processing Systems},
	editor = {Wallach, H. and Larochelle, H. and Beygelzimer, A. and d'Alch\'{e}-Buc, F. and Fox, E. and Garnett, R.},
	publisher = {Curran Associates, Inc.},
	year = {2019},
	volume = {32}
}

@article{Zhang2010,
	author  = {Zhang, Cun-Hui},
	title   = {Nearly Unbiased Variable Selection under Minimax Concave Penalty},
	journal = {The Annals of Statistics},
	volume  = {38},
	number  = {2},
	pages   = {894--942},
	year    = {2010}
}

@article{Fan2001,
	author  = {Fan, Jianqing and Li, Runze},
	title   = {Variable Selection via Nonconcave Penalized Likelihood and Its Oracle Properties},
	journal = {Journal of the American Statistical Association},
	volume  = {96},
	number  = {456},
	pages   = {1348--1360},
	year    = {2001}
}

@article{Fan2011,
	author  = {Fan, Jianqing and Lv, Jinchi},
	title   = {Nonconcave Penalized Likelihood with NP-Dimensionality},
	journal = {IEEE Transactions on Information Theory},
	volume  = {57},
	number  = {8},
	pages   = {5467--5484},
	year    = {2011},
	month   = aug
}

@article{Cai2023,
	author  = {Cai, Hengrui and Shi, Chengchun and Song, Rui and Lu, Wenbin},
	title   = {Jump Interval-Learning for Individualized Decision Making with Continuous Treatments},
	journal = {Journal of Machine Learning Research},
	volume  = {24},
	pages   = {1--92},
	year    = {2023}
}

@article{rivers2001,
	title={Early goal-directed therapy in the treatment of severe sepsis and septic shock},
	author={Rivers, Emanuel and Nguyen, Bryant and Havstad, Suzanne and Ressler, Julie and Muzzin, Alexandria and Knoblich, Bernhard and Peterson, Edward and Tomlanovich, Michael},
	journal={New England Journal of Medicine},
	volume={345},
	number={19},
	pages={1368--1377},
	year={2001},
	publisher={Massachusetts Medical Society}
}

@article{Lv2009,
	author = {Jinchi Lv and Yingying Fan},
	journal = {The Annals of Statistics},
	number = {6A},
	pages = {3498--3528},
	publisher = {Institute of Mathematical Statistics},
	title = {A UNIFIED APPROACH TO MODEL SELECTION AND SPARSE RECOVERY USING REGULARIZED LEAST SQUARES},
	volume = {37},
	year = {2009}
}

@article{Fan2004,
	author = {Jianqing Fan and Heng Peng},
	title = {{Nonconcave penalized likelihood with a diverging number of parameters}},
	volume = {32},
	journal = {The Annals of Statistics},
	number = {3},
	publisher = {Institute of Mathematical Statistics},
	pages = {928 -- 961},
	year = {2004},
}

@article{Shi2019-highd,
	author = {Chengchun Shi and Rui Song and Zhao Chen and Runze Li},
	title = {{Linear hypothesis testing for high dimensional generalized linear models}},
	volume = {47},
	journal = {The Annals of Statistics},
	number = {5},
	publisher = {Institute of Mathematical Statistics},
	pages = {2671 -- 2703},
	year = {2019}
}

@book{vander2023,
	author    = {van der Vaart, A. W. and Wellner, Jon A.},
	title     = {Weak Convergence and Empirical Processes},
	subtitle  = {With Applications to Statistics},
	edition   = {2},
	series    = {Springer Series in Statistics},
	publisher = {Springer Cham},
	year      = {2023},
	isbn      = {978-3-031-29038-1},
	pages     = {XVII, 679},
}

@book{Vershynin2018,
	author       = {Roman Vershynin},
	title        = {High-Dimensional Probability: An Introduction with Applications in Data Science},
	year         = {2018},
	publisher    = {Cambridge University Press},
	location     = {Cambridge},
	series       = {Cambridge Series in Statistical and Probabilistic Mathematics},
	collection   = {Cambridge Series in Statistical and Probabilistic Mathematics}
}

@book{shao2003,
	author    = {Jun Shao},
	title     = {Mathematical Statistics},
	series    = {Springer Texts in Statistics},
	publisher = {Springer New York, NY},
	year      = {2003},
	edition   = {2},
	isbn      = {978-0-387-95382-3},
	pages     = {XVI, 592},
	note      = {Springer-Verlag New York}
}

@article{Candes2007,
	author    = {Emmanuel Candès and Terence Tao},
	title     = {The Dantzig selector: Statistical estimation when \( p \) is much larger than \( n \)},
	journal   = {Annals of Statistics},
	year      = {2007},
	volume    = {35},
	number    = {6},
	pages     = {2313--2351},
	doi       = {10.1214/009053606000001523}
}

@article{choi2024,
	title = {Deep reinforcement learning extracts the optimal sepsis treatment policy from treatment records},
	volume = {4},
	issn = {2730-664X},
	number = {1},
	journal = {Communications Medicine},
	author = {Choi, Yunho and Oh, Songmi and Huh, Jin Won and Joo, Ho-Taek and Lee, Hosu and You, Wonsang and Bae, Cheng-mok and Choi, Jae-Hun and Kim, Kyung-Joong},
	month = nov,
	year = {2024},
	pages = {245},
}

@article{Zhao2015,
	ISSN = {01621459},
	author = {Ying-Qi Zhao and Donglin Zeng and Eric B. Laber and Michael R. Kosorok},
	journal = {Journal of the American Statistical Association},
	number = {510},
	pages = {583--598},
	publisher = {[American Statistical Association, Taylor & Francis, Ltd.]},
	title = {New Statistical Learning Methods for Estimating Optimal Dynamic Treatment Regimes},
	urldate = {2025-06-19},
	volume = {110},
	year = {2015}
}

@article{Shi2022,
	title={Statistical Inference of the Value Function for Reinforcement Learning in Infinite-Horizon Settings},
	author={Shi, Chengchun and Zhang, Sheng and Lu, Wenbin and Song, Rui},
	journal={Journal of the Royal Statistical Society Series B: Statistical Methodology},
	volume={84},
	number={3},
	pages={765--793},
	year={2022},
	month={July},
	publisher={Wiley Online Library},
	doi={10.1111/rssb.12465}
}

@article{Shi20241,
	author = {Chengchun Shi and Shikai Luo and Yuan Le and Hongtu Zhu and Rui Song},
	title = {Statistically Efficient Advantage Learning for Offline Reinforcement Learning in Infinite Horizons},
	journal = {Journal of the American Statistical Association},
	volume = {119},
	number = {545},
	pages = {232--245},
	year = {2024},
	publisher = {ASA Website},
}

@article{Shi2018maximin,
	title={Maximin Projection Learning for Optimal Treatment Decision with Heterogeneous Individualized Treatment Effects},
	author={Shi, Chengchun and Song, Rui and Lu, Wenbin and Fu, Bo},
	journal={Journal of the Royal Statistical Society Series B: Statistical Methodology},
	volume={80},
	number={4},
	pages={681--702},
	year={2018},
	month={September},
	publisher={Wiley Online Library},
}

@article{Shi20243,
	author = {Chengchun Shi and Jin Zhu and Ye Shen and Shikai Luo and Hongtu Zhu and Rui Song},
	title = {Off-Policy Confidence Interval Estimation with Confounded Markov Decision Process},
	journal = {Journal of the American Statistical Association},
	volume = {119},
	number = {545},
	pages = {273--284},
	year = {2024},
	publisher = {ASA Website},
}

@article{Ding2024,
	author       = {Yi Ding and Yingying Li and Rui Song},
	title        = {Statistical Learning for Individualized Asset Allocation},
	journal      = {Journal of the American Statistical Association},
	year         = {2024},
	volume       = {119},
	number       = {545},
	pages        = {639--649},
	doi          = {10.1080/01621459.2022.2139265},
	publisher    = {Taylor \& Francis},
}

@article{ZouLi2008,
		author  = {Zou, Hui and Li, Runze},
		title   = {One-Step Sparse Estimates in Nonconcave Penalized Likelihood Models},
		journal = {The Annals of Statistics},
		volume  = {36},
		number  = {4},
		pages   = {1509--1533},
		year    = {2008},
		doi     = {10.1214/009053607000000802}
}

@article{FanXueZou2014,
		author  = {Fan, Jianqing and Xue, Lingzhou and Zou, Hui},
		title   = {Strong Oracle Optimality of Folded Concave Penalized Estimation},
		journal = {The Annals of Statistics},
		volume  = {42},
		number  = {3},
		pages   = {819--849},
		year    = {2014},
		doi     = {10.1214/13-AOS1198}
}

\clearpage
\appendix
\section*{Appendices}
\addcontentsline{toc}{section}{Appendices}
\allowdisplaybreaks
\spacingset{1.8}
\counterwithin{table}{section}
\counterwithin{figure}{section}

\counterwithin{equation}{section}

\textbf{Appendices Overview.} Appendix~\ref{A-thm} gives the assumptions, Appendix~\ref{omitted} supplementary theory, Appendix~\ref{Proofs} proofs, Appendix~\ref{A-simu} simulations, and Appendix~\ref{A-empi} empirical results.

\section{Technical Conditions}\label{A-thm}

We begin by discussing the classical conditions, drawing on frameworks established in the literature, such as \citet{Fan2011}. Although the main text focuses on $M_n=2$, all conditions are presented here in the general stage-$M_n$ form. 

First, we introduce some notations. Following the framework of \citet{Lv2009}, the local concavity of \(\rho\) at a given vector $\mathbf{v} = (v_1, \dots, v_q)^\top$ with $\|\mathbf{v}\|_0 = q,$ is defined as:
\begin{equation}\label{kapa}
	\kappa(\rho; \mathbf{v}) = \lim_{\epsilon \to 0^+}\max_{1 \leq j \leq q}\sup_{t_1<t_2 \in (|v_j|-\epsilon, |v_j|+\epsilon)} \frac{\rho'(t_2)-\rho'(t_1)}{t_2 - t_1}.
\end{equation}
Local concavity characterizes folded concave penalties, a widely studied class of nonconvex regularizers \citep{Fan2011}. 

\subsection{\texorpdfstring{Conditions for the $M_n$-th Stage}{Conditions for the Mn-th Stage}}
Let $p_n^{(M_n)}$ denote the dimension of $\bm{\beta}^{(M_n)}$, and $s_n^{(M_n)}$ denote the number of nonzero components in the corresponding true coefficient vector. Assume $a\leq K_n :=\max_i k_i$, then the dimension $d^{(M_n)}\asymp\sum_{m=1}^{M_n}k_m+(M_n-1)\,a$, and
$$p_n^{(M_n)}\lesssim L_n\,d^{(M_n)}\lesssim L_n\,M_n\,K_n.$$ Recall that
\begin{align*}
&\mathbf{H}^{(M_n)}=\left((\mathbf{X}^{(1)})^\top,(\mathbf{A}^{(1)})^\top,\cdots,(\mathbf{X}^{(M_n-1)})^\top,(\mathbf{A}^{(M_n-1)})^\top,(\mathbf{X}^{(M_n)})^\top\right)^\top,\\
&\mathbf{S}^{(M_n)} = \Big(1,(\mathbf{H}^{(M_n)})^\top,  \big((\varphi_k(\mathbf{A}^{(M_n)}))_{k=2}^{L_n}\otimes\mathbf{H}^{(M_n)}\big)^\top\Big)^\top.
\end{align*}
Let \(\mathbf{S}^{(M_n)}_1\) denote the subvector of \(\mathbf{S}^{(M_n)}\) consisting of the entries indexed by \(\operatorname{supp}(\bm{\beta}^{(M_n)}_*)\), and \(\mathbf{S}^{(M_n)}_2\) denote the subvector consisting of the entries indexed by the complement of \(\operatorname{supp}(\bm{\beta}^{(M_n)}_*)\). 

Let the design matrix $\bm{\Psi}^{(M_n)}=(\mathbf{h}_1^{(M_n)},\cdots,\mathbf{h}^{(M_n)}_{p_n^{(M_n)}})$ and $\bm{\Psi}_c^{(M_n)}=(\mathbf{h}_{c,1}^{(M_n)},\cdots) $ for $c=1,2$. For notational simplicity, we write $b^{(M_n)}$ as $b$ throughout. 
\subsubsection{\texorpdfstring{Stage-$M_n$ Conditions for Coefficient Estimation}{Stage-Mn Conditions for Coefficient Estimation}}\label{stageMn-coef}

\begin{ass}\label{design2}
	With probability at least $1-O(f(n))$ where $f(n)\to0$, the design matrix \(\bm{\Psi}^{(M_n)}\) satisfies
	\begin{align*}
		&\max_{1\leq j \leq p^{(M_n)}_n}\|\mathbf{h}_j^{(M_n)}\|_\infty=O\left(\sqrt{n/\log p_n^{(M_n)}}\right),\\&\max_{1\leq j \leq p^{(M_n)}_n}\|\mathbf{h}_j^{(M_n)}\|_2=O\left(\sqrt{n}\right),
		\\& \min_{\bm{\delta} \in \mathcal{N}^{(M_n)}_0} \lambda_{\min} \left[ (\bm{\Psi}_1^{(M_n)})^\top \bm{\Sigma}^{(M_n)} (\bm{\Psi}_1^{(M_n)} \bm{\delta}) \, \bm{\Psi}_1^{(M_n)} \right] \geq c n,\\& 
		\operatorname{tr} \left[ (\bm{\Psi}_1^{(M_n)})^\top \bm{\Sigma}^{(M_n)} (\bm{\theta}_*^{(M_n)})  \, \bm{\Psi}_1^{(M_n)} \right] = O(ns_n^{(M_n)}),\\&
		\left\| (\bm{\Psi}^{(M_n)}_2)^\top \bm{\Sigma}^{(M_n)} (\bm{\theta}_*^{(M_n)}) \, \bm{\Psi}_1^{(M_n)}\right\|_{2, \infty} = O(n),\\&	\max_{\bm{\delta} \in \mathcal{N}_0^{(M_n)}} \max_{j=1}^{p_n^{(M_n)}} \lambda_{\max} 
		\left[ (\bm{\Psi}_1^{(M_n)})^\top \operatorname{diag} \{ |\mathbf{h}^{(M_n)}_j| \} \circ | \bm{\mu}^{(M_n)''} (\bm{\Psi}_1^{(M_n)} \bm{\delta}) | \, \bm{\Psi}_1^{(M_n)} \right] = O(n),
	\end{align*}
	where \( \mathcal{N}^{(M_n)}_0 = \{ \bm{\delta} \in \mathbb{R}^{s_n^{(M_n)}} : \|\bm{\delta} - \bm{\beta}^{(M_n)}_{*,1}\|_{\infty} \leq g^{(M_n)}_n \} \), $\bm{\mu}^{(M_n)}(\bm{\theta})=(b^{'}(\theta_1),\cdots,b^{'}(\theta_n))^\top$, $ \bm{\Sigma}^{(M_n)}(\bm{\theta})=\operatorname{diag}\{b^{''}(\theta_1),\cdots,b^{''}(\theta_n)\}$ for $\bm{\theta}=(\theta_1,\cdots,\theta_n)^\top$. \( c \) is some positive constant, and $
	\|\mathbf{B}\|_{2, \infty} = \max_{\|\mathbf{v}\|_2 = 1} \|\mathbf{B} \mathbf{v}\|_{\infty}.$ 
\end{ass}

\begin{ass}\label{compare2}
	Assume that $
	g^{(M_n)}_n \gg \lambda^{(M_n)}_n \gg \max \left\{ (s_n^{(M_n)}/n)^{1/2}, (\log p_n^{(M_n)}/n)^{1/2}\right\}$, $ 
	\quad p'_{\lambda^{(M_n)}_n} (g^{(M_n)}_n) = o((ns_n^{(M_n)})^{-1/2})$, and 
	$	\lambda^{(M_n)}_n \kappa^{(M_n)}_0 = o(1),
	$ where $
	\kappa_0^{(M_n)} = \max_{\bm{\delta} \in \mathcal{N}^{(M_n)}_0} \kappa(\rho; \bm{\delta})
	$.
\end{ass} 

\begin{ass}\label{subG_mom}
	Assume $\max_j\mathbb{E}[b''(\theta_*)(h^{(M_n)}_{j,1})^2]<\infty$, and for $i\neq j$
	$$\mathbb{E}[\varepsilon_i^{(M_n)}\varepsilon_j^{(M_n)}\mid \{{\mathbf{H}}_i^{(M_n)},{\mathbf{A}}_i^{(M_n)}\}_{i=1}^n]=0.$$ Moreover, there exist some \( M, v_0 \in (0, \infty) \) such that
	\begin{align}
		\mathbb{ E }\Big\{ \exp \left[ \frac{\, |Y_i^{(M_n)}- b' (\theta^{(M_n)}_{*,i})|}{M} \right] - 1 - \frac{|Y_i^{(M_n)} - b' (\theta^{(M_n)}_{*,i})|}{M} \mid \{{\mathbf{H}}_i^{(M_n)},{\mathbf{A}}_i^{(M_n)}\}_{i=1}^n \Big\} M^2 
		\leq \frac{v_0}{2}\label{subG_mom1}
	\end{align}
	with \( (\theta_{*,1}, \dots, \theta_{*,n})^\top = \bm{\theta}_* \). 
\end{ass} 

\begin{ass}\label{normal2}
	\item (a) Assume $\max_j\mathbb{E}[|\varepsilon^{(M_n)}h^{(M_n)}_{j,1}|^3]<\infty$, and the following holds:
	\begin{align}
		&\lambda_{\min} \left[\mathbf{F}^{(M_n)}\right]\geq c >0,\label{normal20}
	\end{align}
	Here
	\( (\theta^{(M_n)}_{*,1}, \dots, \theta^{(M_n)}_{*,n})^\top = \bm{\theta}^{(M_n)}_* \) and 
	\( \mathbf{F}^{(M_n)} = \mathbb{E}\left[b''({\theta}_*^{(M_n)}) \mathbf{S}_{1}^{(M_n)}(\mathbf{S}_{1}^{(M_n)})^\top\right] \).
	
	\item (b) Let $\mathbf{F}^{(M_n)}_n=(\bm{\Psi}_1^{(M_n)})^\top \bm{\Sigma}^{(M_n)} (\bm{\theta}_*^{(M_n)}) \, \bm{\Psi}_1^{(M_n)} $, assume with probability at least $1-O(f(n))$ where $f(n)\to0$,
	\begin{align}
		\|\mathbf{F}_n^{(M_n)}-n \, \mathbf{F}^{(M_n)} \|_2\leq o(n/\sqrt{s_n^{(M_n)}}).\label{normal22}
	\end{align}
\end{ass}

\begin{remark}
	Assumptions~\ref{design2}--\ref{normal2} are modified versions of conditions widely adopted in the literature on high-dimensional inference and generalized linear models. Due to the randomness of the design matrix in this context, we require slightly different conditions. Specifically, Assumption~\ref{design2} imposes standard constraints on the design matrix: it requires that each column $\mathbf{h}_j^{(M_n)}$ is subject to constraints on its $\ell_2$ and $\ell_\infty$ norms, and that the submatrix  \(\bm{\Psi}_1^{(M_n)}\) exhibits relatively low multicollinearity to ensure identifiability and invertibility. Assumption~\ref{compare2} places mild restrictions on the minimum signal strength and the choice of penalty function; for instance, it excludes Lasso due to the violation of certain curvature conditions, but accommodates folded-concave penalties such as SCAD. Assumption~\ref{subG_mom} ensures conditional Bernstein tail behavior, enabling sharp concentration bounds via Bernstein-type inequalities. It is slightly stronger than standard assumptions in fixed design settings. Finally, Assumption~\ref{normal2} (a) is used to verify the Lyapunov condition required for establishing the central limit theorem, and (b) guarantees that the estimated covariance matrix can be consistently replaced by the population covariance matrix. For a detailed discussion of such assumptions and their roles in high-dimensional theory, see, for example, \citet{Fan2011}.
\end{remark}

We next examine the conditions under which the above assumptions hold, with particular emphasis on the validity of Assumption~\ref{design2} in the context of our stage-$M_n$ model. This type of condition has been studied in prior works (e.g., \citet{Shi2019-highd}), and here we provide a detailed analysis in the context of our structured design. Now, for Gaussian linear regression, Assumption \ref{design2} degenerates to
\begin{ass}\label{design2'}
	With probability at least $1-O(f(n))$,
	\begin{align}
		&\max_{1\le j\le  p_n^{(M_n)}} \| \mathbf h^{(M_n)}_j\|_{\infty}= O\left(\sqrt{n/\log p_n^{(M_n)}}\right), \label{g1}\\&
		\max_{1\le j\le  p_n^{(M_n)}} \| \mathbf h^{(M_n)}_j\|_{2}= O\left(\sqrt{n}\right),\label{g2} \\&
		\lambda_{\min}\left[
		(	\bm{\Psi}^{(M_n)}_{1})^{T}\bm{\Psi}^{(M_n)}_{1}
		\right] \ge c\,n, \label{g3}\\&
		\operatorname{tr} \left[
		(\bm{\Psi}^{(M_n)}_{1})^{T}\bm{\Psi}^{(M_n)}_{1}
		\right]= O(ns_n^{(M_n)}),\label{g4} \\&
		\|
		(	\bm{\Psi}^{(M_n)}_{2})^{T}
		\bm{\Psi}^{(M_n)}_{1}
		\|_{2,\infty}= O(n),\label{g5}
	\end{align}
	for some constant $c>0$.
\end{ass}
Let the natural–parameter space be $\Theta=(c,d)$ with $c,d\in\mathbb{R}\cup\{\pm\infty\}$.  
For a general GLM, Assumption~\ref{design2} is implied by Assumption~\ref{design2'} together with the additional Assumption~\ref{design2''} stated below.

\begin{ass}\label{design2''}
	With probability at least $1-O(f(n))$,
	\begin{align} 
		&\max_{1\le i\le n}
		(\mathbf{S}^{(M_n)}_{i})^{T}\boldsymbol{\beta}^{(M_n)}_{*}
		\leq D<d ,\text{if} \ b''(d)=\infty \ \operatorname{or} \ 0 \  \operatorname{or} \ b'''(d)=\pm \infty ,\label{max1}\\& \min_{1\le i\le n}
		(\mathbf{S}^{(M_n)}_{i})^{T}\boldsymbol{\beta}^{(M_n)}_{*}
		\geq C>c,\text{if} \ b''(c)=\infty \ \operatorname{or} \ 0 \ \operatorname{or}\ b'''(c)=\pm \infty, \label{min1}\\&
		\max_{1\le j\le p_n^{(M_n)}}\left\lVert\mathbf h^{(M_n)}_j\right\rVert_{\infty}= O \Biggl(
		\frac{n^{1/2}}
		{(s_n^{(M_n)})^{1/2}\,\log^{1/2} n}
		\Biggr),\label{g6}\\&
		\max_{1\le j\le  p_n^{(M_n)}}
		\lambda_{\max}\left[
		(	\bm{\Psi}^{(M_n)}_{1})^{T}
		\operatorname{diag} \{ | \mathbf {h}^{(M_n)}_j | \}
		\, \bm{\Psi}_{1}^{(M_n)}
		\right]= O(n).\label{g7}
	\end{align}
	For poisson or logistic model, \eqref{max1} and \eqref{min1} reduces to
	\begin{equation}
		\max_{1\le i\le n}
		|(\mathbf{S}^{(M_n)}_{i})^{T}\boldsymbol{\beta}^{(M_n)}_{*}|=O(1).
	\end{equation}
\end{ass}

Following \citet{Shi2019-highd}, we examine sufficient conditions for Assumptions~\ref{design2'} and~\ref{design2''} under sub-Gaussian and uniformly bounded designs.

Write \(\mathbf{S}^{(M_n)}_1 = \left(\mathbf{r}_1^\top, \left((\varphi_k(\mathbf{A}^{(M_n)}))_{k=2}^{L_n}\otimes\mathbf{r}_{k}\right)^\top \right)^\top\), where for each $k = 1, \dots, L_n$, $\mathbf{r}_k$ denotes the subvector of the corresponding block of $\mathbf{S}^{(M_n)}$ consisting only of the entries whose coordinates belong to $\operatorname{supp}(\bm{\beta}_*^{(M_n)})$.

If none of the entries in the k-th block belong to $\operatorname{supp}(\bm{\beta}_*^{(M_n)})$, we set \(\mathbf{r}_k:= \varnothing\). Then 
\begin{equation}
	\mathbf{S}^{(M_n)}_1 = \left(\mathbf{r}_{k_1}^\top, \left((\varphi_{k_j}(\mathbf{A}^{(M_n)}))_{j=2}^{L_n}\otimes\mathbf{r}_{k_j}\right)^\top \right)^\top\label{r1}
\end{equation}
with $1=k_1<k_2<\cdots<k_r\leq L_n.$ Based on the above construction, we further define a new vector $\mathbf{r}$ as a subvector of $\mathbf{H}^{(M_n)}$, formed by collecting all the entries of $\mathbf{H}^{(M_n)}$ that appear in the subvectors $\mathbf{r}_{k_1}, \mathbf{r}_{k_2}, \dots, \mathbf{r}_{k_r}$ in \eqref{r1}. In other words, $\mathbf{r}$ is the minimal subvector of $\mathbf{H }^{(M_n)}$ that contains all the effective components involved in $\mathbf{S}^{(M_n)}_1.$ 

Using notations in \eqref{r1}, let $\bm{\Lambda}=\mathbb{E}[\mathbf{S}^{(M_n)}_1 (\mathbf{S}^{(M_n)}_1)^\top]$, $p_{k_i}=\mathbb{P}[\mathbf{A}^{(M_n)}\in\mathcal{A}_{(k_i)}]$ and $\mathbf{C}_{k_i}=\mathbb{E}[\varphi_{k_i}(\mathbf{A}^{(M_n)})\mathbf{r}_{k_i}\mathbf{r}^\top_{k_i}\mid \mathbf{A}^{(M_n)}\in\mathcal{A}_{(k_i)}]$ for $i\geq 2$. Besides, define
\begin{align*}
	\mathbf{B}&=\mathbb{E}\left(\mathbf{r}_{k_1}\mathbf{r}^\top_{k_2}\cdot \varphi_{k_1}(\mathbf{A}^{(M_n)}),\cdots,\mathbf{r}_{k_1}\mathbf{r}^\top_{k_r}\cdot  \varphi_{k_r}(\mathbf{A}^{(M_n)}) \right)\\
	&=\left(p_{k_2}\mathbf{D}_{k_{12}},\cdots, p_{k_r}\mathbf{D}_{k_{1r}}\right),
\end{align*}
with $\mathbf{D}_{k_{1j}}=\mathbb{E}[\varphi_{k_j}(\mathbf{A}^{(M_n)})\mathbf{r}_{k_1}\mathbf{r}^\top_{k_j}\mid \mathbf{A}^{(M_n)}\in\mathcal{A}_{(k_j)}]$.
Let $m_r=\min_i \{ p_{k_i}\lambda_{\min}[\mathbf{C}_{k_i}]\}$, $M_r=\|\mathbf{B}\|_2$. Recall that $\varphi_k(a)\le c_{0}$.

The following two lemmas establish sufficient conditions under each case.

\begin{lemma}\label{lemmaA.1}
	For $f(n)=O(1/n)$, assume $\log K_n=O(n^{u})$, $\log M_n=O(n^{v})$, $\log L_n=O(n^{w})$, then $\log p_n^{(M_n)}=O(n^{\max{\{u,v,w\}}})$. If $u,v,w$ satisfies $\max\{u,v\}+\max\{u,v,w\} \leq 1$, $s_n^{(M_n)}\log n=O(n^{1-{\max \{u,v\}}})$ and
	\begin{align}
		&\max\{1,c_0\} \cdot \max_{1\leq m \leq M_n,1\leq j \leq k_m}\| X_{j}^{(m)}\|_{\psi_2} \leq C_0,\, \max\{1,c_0\}\cdot\max_{1\leq m <M_n,1\leq j\leq a}\| {A}_j^{(m)}\|_{\psi_2}\leq  C_0,\label{maxxa}
	\end{align} 
	then \eqref{g1}, \eqref{g2}, \eqref{g6} hold.

	Further assume $(s_n^{(M_n)})^2\log n\log s_n^{(M_n)}=o(n)$ and the following conditions hold:
	\item (a) $\lambda_{\min} \left[\mathbb{E}\left[\mathbf{r}_1 \mathbf{r}_1^\top\right]\right]\geq c_1$ for some constant $c_1>0$.
	\item (b) $m_r$ has a positive lower bound.
	\item (c) $c_1-{M^2_r}/{m_r}$ has a positive lower bound.
	
	Then \eqref{g3} and \eqref{g4} hold.

	If further assume $s_n^{(M_n)}=O(n^{1-\max\{u,v,w\}})$ and the following holds:
	\item (d) For each \(i\), \(\mathbf{r}_{k_i}\) is sub-Gaussian with its \(\psi_2\)-norm uniformly bounded by a constant. In particular, \(\|\mathbf{r}\|_{\psi_2} \leq C\).
	
	Then \eqref{g5} holds; if $(s_n^{(M_n)})^2\log^4 n=O(n^{1-\max\{u,v\}})$, so does \eqref{g7}.
	
	Finally, if $b''$ is bounded (e.g., for logistic or Gaussian regression), and $(s_n^{(M_n)})^3\allowbreak\log n \allowbreak\log s_n^{(M_n)}=o(n)$, then \eqref{normal22} holds. 
\end{lemma}

\begin{lemma}\label{lemmaA.2}
	For $f(n)=O(1/n)$, assume $\log K_n=O(n^{u})$, $\log M_n=O(n^{v})$, $\log L_n=O(n^{w})$. If 
	\begin{align}
		&\max\{1,c_0\} \cdot\max_{1\leq m \leq M_n}\| \mathbf{X}^{(m)}\|_\infty\leq C_0,\, \max\{1,c_0\} \cdot \max_{1\leq m <M_n}\| \mathbf{A}^{(m)}\|_\infty\leq C_0, \quad \text{a.s.},\label{maxxa1}
	\end{align} 
	then \eqref{g1}, \eqref{g2}, \eqref{g6} hold.

	Further assume $(s_n^{(M_n)})^2\log n\log s_n^{(M_n)}=o(n)$ and the following conditions hold:
	\item\text{(a)} $\lambda_{\min} \left[\mathbb{E}\left[\mathbf{r}_1 \mathbf{r}_1^\top\right]\right]\geq c_1$ for some constant $c_1>0$.
	\item \text{(b)} $m_r$ has a positive lower bound.
	\item \text{(c)} $c_1-{M^2_r}/{m_r}$ has a positive lower bound.
	
	Then \eqref{g3} and \eqref{g4} hold with probability tending to 1.

	If further assume $s_n^{(M_n)}=O(n^{1-\max\{u,v,w\}})$, and the following holds: 
	\item \text{(d$\,^{\prime}$)} $\max_i\lambda_{\max}\left[\mathbb{E}[\mathbf{r}_{k_i}\mathbf{r}^\top_{k_i}]\right]\leq C_1.$
	
	Then \eqref{g5} and \eqref{g7} hold.
	
	Finally, if $b''$ is bounded, and $(s_n^{(M_n)})^3\log n$ $\log s_n^{(M_n)}=o(n)$, then \eqref{normal22} holds. 
\end{lemma}

\begin{remark}
	\textbf{On the rate conditions.} All the rate conditions in the lemma are satisfied if either of the following holds:
	\begin{itemize}
		\item \(s_n^{(M_n)} = O(n^{\ell^{(M_n)}_2})\) with \(\ell^{(M_n)}_2 < \tfrac{1}{3}\), and \((u,v,w) = (\tfrac{1}{3}, \tfrac{1}{3}, w)\) for some \(w < \tfrac{2}{3}\);
		\item \(s_n^{(M_n)} = O(n^{\ell^{(M_n)}_2})\) with \(\ell^{(M_n)}_2  \leq \tfrac{1}{4}\), and \(u = v = w < \tfrac{1}{2}\).
	\end{itemize}
	For further inference (e.g., concerning optimal Q-values and deformed optimal Q-values and inference for earlier stages), stronger restrictions on $L_n$ and $M_n$ may be necessary. Specifically, the condition may shift to $ L_n=O(n^{l_1}),M_n=O(n^{l_6})$.
	
	\textbf{On the sub-Gaussian assumptions.} 
	We do not impose sub-Gaussianity directly on composite vectors such as $\mathbf{S}^{(M_n)}$ or $\mathbf{S}_1^{(M_n)}$. Instead, our assumptions are stated on components like ${X}^{(m)}_j$ and ${A}_j^{(m)}$. For instance, Condition~\eqref{maxxa} requires that the feature and action components at each stage are sub-Gaussian with uniformly bounded $\psi_2$-norms. This is a strictly weaker requirement than assuming sub-Gaussianity of $\mathbf{S}^{(M_n)}$, whose complex dependence structure may violate such bounds.
	
	\textbf{On assumptions (a)--(c).} 
	Conditions~(a) and (b) require the per-block covariance matrix to have a uniformly bounded minimum eigenvalue and \(m_r\) to be bounded below, respectively. Condition~(c) places an upper bound on \(\|\mathbf{B}\|_2\), which effectively controls the cross-action conditional correlation among \(\mathbf{r}_{k_1}\) and \(\mathbf{r}_{k_j}\) under $ \{\mathbf{A}^{(M_n)}\in\mathcal{A}_{k_j}\}$. 
	
	\textbf{On condition (d).} 
	Since both \(\mathbf{r}_{k_i}\) and \(\mathbf{r}\) are sub-vectors of \(\mathbf{H}^{(M_n)}\), the sub-Gaussian Assumption in (d) also only imposes conditions on the components of \(\mathbf{H}^{(M_n)}\). \
	
\end{remark}

Finally, we give a useful lemma.
\begin{lemma}\label{FboundMn}
	Under Assumption~\ref{subG_mom} and \ref{normal2}, 
	\begin{align*}
		&\mathbb{E}\left[b''({\theta}_*^{(M_n)})(\mathbf{S}_1^{(M_n)})^\top\mathbf{S}_1^{(M_n)}\right]\leq C_1s_n^{(M_n)},
		\\&\frac{1}{\sqrt{n}}\mathbb{E}\left|(\varepsilon_i^{(M_n)})^2 (\mathbf{S}_{i,1}^{(M_n)})^\top(\mathbf{F}^{(M_n)})^{-1}\mathbf{S}_{i,1}^{(M_n)}\right|^{3/2}\to 0.
	\end{align*}
	for an absolute constant $C_1$.
\end{lemma}

For detailed proofs, see Appendix~\ref{Proofs}.

\subsubsection{\texorpdfstring{Stage-$M_n$ Conditions for Value Estimation}{Stage-Mn Conditions for Value Estimation}}
In this subsection, we discuss several conditions for value estimation in the $M_n$-th stage. We first introduce some notations.
Let the vector $\mathcal{R}_i=(\mathcal{R}_{i,j})_j$ be the $i$-th testing sample of $\mathbf{r}$ defined in \eqref{r1}. For any policy $\pi^{(M_n)}$, write $Q^{(M_n)}(\mathcal{H}^{(M_n)},\pi^{(M_n)})=Q_{\pi^{(M_n)}}(\mathcal{H}^{(M_n)})$ for simplicity.
Moreover, define the optimal policy
\begin{align*}
	&{\Pi}_{*}^{(M_n)}(\mathcal{H}^{(M_n)})=
	\left\{
	\mathcal{A}^{(M_n)} \in \mathcal{A}
	:
	{Q}^{(M_n)}\bigl(\mathcal{H}^{(M_n)},\mathcal{A}^{(M_n)}\bigr)
	= 
	\max_{{\mathcal{A}}^{(M_n)}\in\mathcal{A}}
	{Q}^{(M_n)}\bigl(\mathcal{H}^{(M_n)},{\mathcal{A}}^{(M_n)}\bigr)
	\right\},\\&\widehat{\Pi}_{*}^{(M_n)}(\mathcal{H}^{(M_n)})=
	\left\{
	\mathcal{A}^{(M_n)} \in \mathcal{A}
	:
	\widehat{Q}^{(M_n)}\bigl(\mathcal{H}^{(M_n)},\mathcal{A}^{(M_n)}\bigr)
	= 
	\max_{{\mathcal{A}}^{(M_n)}\in\mathcal{A}}
	\widehat{Q}^{(M_n)}\bigl(\mathcal{H}^{(M_n)},{\mathcal{A}}^{(M_n)}\bigr)
	\right\}.
\end{align*}

\begin{ass}\label{maxmincd}
	Recall the space $\Theta=(c,d)$. Assume $\mathbb{P}[|\mathcal{R}_j|>t]\leq c_1\exp({-c_2t^r})$ for each $j$ with $c_1,c_2,r>0$, and there exists $c<C<D<d$ such that
	\begin{align*}
		&\max_i\max_k \left\{(\bm{\eta}^{(M_n)}_{*,0}+\bm{\eta}^{(M_n)}_{*,k})^\top\mathcal{H}_i^{(M_n)}\right\}\leq D+o_p(1) , \, \text{if} \, \, b''(d)=\infty,\\ &\min_i\min_k \left\{(\bm{\eta}^{(M_n)}_{*,0}+\bm{\eta}^{(M_n)}_{*,k})^\top\mathcal{H}_i^{(M_n)}\right\}\geq C+o_p(1),  \, \text{if} \, \, b''(c)=\infty.
	\end{align*}
\end{ass}
\begin{ass}\label{Eb''}
	Assume 
	\begin{align*}
		\mathbb{E}\left|b'\left(\mathcal{S}^\top_{\pi_*^{(M_n)}}\bm{\beta}_* ^{(M_n)}\right)\right|<\infty.
	\end{align*}
	Moreover, there exists $\varepsilon>0, M>0$, for all 
	$$\|\bm{\beta}^{(M_n)}-\bm{\beta}_*^{(M_n)}\|_2<\varepsilon,$$ we have 
	$$\max_j \mathbb{E}\left|b''\left(\mathcal{S}^\top_{\pi_*^{(M_n)}}\bm{\beta}^{(M_n)}\right)r_j\right|^k\leq M$$
	for $k=1,2$, where $\mathbf{r}=(r_j)_j$ are defined in \eqref{r1}.
\end{ass}
\begin{remark}
	Assumption \ref{maxmincd} is a subtle refinement of \eqref{max1} and \eqref{min1}, which ensures that with high probability, if \( b'' \) is unbounded, its argument is kept sufficiently away from divergence points. The first part of Assumption \ref{Eb''} requires the tail bound of $r_j$. These guarantee that the subsequent inference under any policy is well-defined, which is naturally satisfied for logistic regression.

	The second part requires that within a small neighborhood of the true value, the moments of \( b''\left(\mathcal{S}^\top_{\pi_*^{(M_n)}}\bm{\beta}^{(M_n)}\right)r_j \) are uniformly controlled. In fact, this can be ensured by the following classical condition (similar conditions in \cite{shao2003}, Section 4): there exists a function \( M(\cdot) \) such that
	\[
	\sup_{\|\bm{\beta}^{(M_n)}-\bm{\beta}_*^{(M_n)}\|_2<\varepsilon}\left|b''\left(\mathcal{S}^\top_{\pi_*^{(M_n)}}\bm{\beta}^{(M_n)}\right)\right| \leq M(\mathcal{S}),
	\]
	and 
	\[
	\sup_j \mathbb{E}\left[ M(\mathcal{S}) r_j \right]^2 < \infty.
	\]
\end{remark}

\begin{ass}\label{difflb}
	Define
	\begin{align}
		D_{k,j}=-D_{j,k}= (\bm{\eta}^{(M_n)}_{*,k})^\top\mathcal{H}^{(M_n)} -(\bm{\eta}^{(M_n)}_{*,j})^\top\mathcal{H}^{(M_n)}\label{dkj},
	\end{align}
	assume the following probability bound holds: 
	$$\mathbb{P}\left[\, \exists k, \,\pi_*^{(M_n)}(\mathcal{H}^{(M_n)})\in\mathcal{A}_{(k)}, \inf_{j\neq k}\left\{D_{k,j}\right\}\leq \frac{\varepsilon}{L_n}	\,  \right]=O(\varepsilon).$$
	Moreover, define $\mathcal{U}_n=\left\{ |\Pi_*^{(M_n)}|=1 \right\}$, assume that for some $\delta>\ell^{(M_n)}_2$, 
	$$\mathbb{P}\left[ \, \mathcal{U}_n  \, \right]\geq 1-O(n^{-\delta}).$$
\end{ass}

\begin{remark}
	Assumption \ref{difflb} consists of two key parts. 
	
	The first part requires that the difference between the optimal Q-value and the sub-optimal Q-value has a probabilistic lower bound. This ensures that, with high probability, the optimal Q-value is sufficiently distinct from other values. The second part asserts that with high probability, the optimal policy \( \pi_*^{(M_n)} \) is unique. These assumptions are a modification of similar conditions in \citet{Ding2024}.
\end{remark}

\subsection{\texorpdfstring{Conditions for the $(M_n \! -\!1)$-th Stage}{Conditions for the (Mn - 1)-th Stage}}\label{A-thm1}
\counterwithin{equation}{section}

Let \(\mathbf{S}^{(M_n-1)}_{\mathrm{dfm},1}\) denote the subvector of \(\mathbf{S}^{(M_n-1)}_{\mathrm{dfm}}\) consisting of the entries indexed by \(\operatorname{supp}(\bm{\alpha}^{(M_n-1)}_*)\), and \(\mathbf{S}^{(M_n-1)}_{\mathrm{dfm},2}\) denote the subvector consisting of the entries indexed by the complement of \(\operatorname{supp}(\bm{\alpha}^{(M_n-1)}_*)\). Similarly, we write the vector as 
\begin{equation}
	\mathbf{S}^{(M_n-1)}_{\mathrm{dfm},1} = \left((\varphi_k(\mathbf{A}^{(M_n -1)})_{k=1}^{L_n})\otimes \mathbf{H}^{(M_n  - 1)}\right)^\top\label{r11}
\end{equation}
with $1\leq k_1<k_2<\cdots<k_r\leq L_n,$ $\mathbf{r}_{k_i}=\mathbf{r}^{(M_n-1)}_{k_i}$ and $\mathbf{r}=\mathbf{r}^{(M_n-1)}$ is the minimal subvector of $\mathbf{H }^{(M_n-1)}$ that contains all the effective components involved in $\mathbf{S}^{(M_n-1)}_{\mathrm{dfm},1}.$

Now let $\bm{\Lambda}^{(M_n-1)}=\mathbb{E}[\mathbf{S}^{(M_n-1)}_{\mathrm{dfm},1} (\mathbf{S}^{(M_n-1)}_{\mathrm{dfm},1})^\top]$, $p_{k_i}=\mathbb{P}[\mathbf{A}^{(M_n-1)}\in\mathcal{A}_{(k_i)}]$ and $\mathbf{C}_{k_i}=\mathbb{E}[\varphi_{k_i}^2(\mathbf{A}^{(M_n-1)})\mathbf{r}_{k_i}\mathbf{r}^\top_{k_i}\mid \mathbf{A}^{(M_n-1)}\in\mathcal{A}_{(k_i)}]$ for $i\geq 1$. Let $m_r=\min_i \{ p_{k_i}\lambda_{\min}[\mathbf{C}_{k_i}]\}$.

Let the design matrix $\bm{\Psi}_{\mathrm{dfm}}^{(M_n-1)}=(\mathbf{h}_1^{(M_n-1)},\cdots,\mathbf{h}^{(M_n-1)}_{p_n^{(M_n-1)}})$ and $\bm{\Psi}_{\mathrm{dfm},c}^{(M_n-1)}=(\mathbf{h}_{c,1}^{(M_n-1)},\cdots) $ for $c=1,2$. 

\subsubsection{\texorpdfstring{Stage-$(M_n \!-\!1 )$ Conditions for Coefficient Estimation}{Stage-(Mn - 1) Conditions for Coefficient Estimation}}\label{stageMn-1-coef}

\begin{ass}\label{design1}
	With probability at least $1-O(f(n))$ with $f(n)\to 0$, for any policy $\pi^{(M_n)}$ in the $M_n$-th stage, the design matrix \(\bm{\Psi}_{\mathrm{dfm}}^{(M_n-1)}\) satisfies
	\begin{align}
		&\max_{1\leq j \leq p^{(M_n-1)}_n}\|\mathbf{h}_j^{(M_n-1)}\|_\infty=O\left(\sqrt{n/\log p_n^{(M_n-1)}}\right)\label{g1Mn-1},\\&\max_{1\leq j \leq p^{(M_n-1)}_n}\|\mathbf{h}_j^{(M_n-1)}\|_2=O\left(\sqrt{n}\right),\label{g2Mn-1}
		\\& \lambda_{\min} \left[ (\bm{\Psi}_{\mathrm{dfm},1}^{(M_n-1)})^\top \, \bm{\Psi}_{\mathrm{dfm},1}^{(M_n-1)} \right] \geq c n,\label{g3Mn-1}\\& 
		\left\| (\bm{\Psi}_{\mathrm{dfm},1}^{(M_n-1)})^\top  \, \bm{\Psi}_{\mathrm{dfm},1}^{(M_n-1)} \right\|_2 = O(n),\label{g4Mn-1}\\&\left\| (\bm{\Psi}^{(M_n-1)}_{\mathrm{dfm},2})^\top \, \bm{\Psi}_{\mathrm{dfm},1}^{(M_n-1)}\right\|_{2, \infty} = O(n)\label{g5Mn-1},\\& \left\| (\bm{\Psi}_{\mathrm{dfm},\pi^{(M_n)},1})^\top  \, \bm{\Psi}_{\mathrm{dfm},\pi^{(M_n)},1} \right\|_{2} = O(n),\label{g6M_n-1}\\& \left\| (\bm{\Psi}_{\mathrm{dfm},2}^{(M_n-1)})^\top  \, \bm{\Psi}_{\mathrm{dfm},\pi^{(M_n)},1} \right\|_{2,\infty} = O(n),\label{g7M_n-1}\\& \left\| |\bm{\Psi}_{\mathrm{dfm},2}^{(M_n-1)}|^\top  \, \bm{\Psi}_{\mathrm{dfm},\pi^{(M_n)},1} \right\|_{2,\infty} = O(n).\label{g8M_n-1}
	\end{align}
	where \( |\cdot| \) denotes the element-wise absolute value, \( \mathbf{S}_{\mathrm{dfm},\pi^{(M_n)},1} \) denotes the sub-vector of \( \mathbf{S}_{\pi^{(M_n)}} \) (see Section~3) corresponding to \( \bm{\alpha}^{(M_n)}_{*,1} \), and \( \bm{\Psi}_{\mathrm{dfm},\pi^{(M_n)},1} \) denotes the associated design matrix.
\end{ass}

Notably, conditions \eqref{g1Mn-1} and \eqref{g2Mn-1} are directly satisfied under conditions \eqref{g1} and \eqref{g2}.

\begin{ass}\label{compare1}
	Assume $
	g^{(M_n-1)}_n \gg\lambda^{(M_n-1)}_n \gg \max \big\{ ((s_n^{(M_n-1)}+s_n^{(M_n)})/n)^{1/2},$ $(\log p_n^{(M_n-1)}/n)^{1/2}\big\}$, $p'_{\lambda^{(M_n-1)}_n} (g^{(M_n-1)}_n) = o((ns_n^{(M_n-1)})^{-1/2})$, and 
	$	\lambda^{(M_n-1)}_n \kappa^{(M_n-1)}_0 = o(1),
	$ where $
	\kappa_0^{(M_n-1)} = \max_{\bm{\delta} \in \mathcal{N}^{(M_n-1)}_0} \kappa(\rho; \bm{\delta})
	$. Here $ \mathcal{N}^{(M_n-1)}_0 = \{ \bm{\delta} \in \mathbb{R}^{s_n^{(M_n-1)}} : \|\bm{\delta} - \bm{\alpha}^{(M_n-1)}_{*,1}\|_{\infty} \leq g^{(M_n-1)}_n \}.$
\end{ass} 

\begin{ass}\label{subG_Mn-1}
	Assume $\max_j\mathbb{E}[(\varepsilon^{(M_n-1)}h^{(M_n-1)}_{j,1})^2]<\infty$, and for $i\neq j$
	$$\mathbb{E}[\varepsilon_i^{(M_n-1)}\varepsilon_j^{(M_n-1)}\mid \{{\mathbf{H}}_i^{(M_n-1)},{\mathbf{A}}_i^{(M_n-1)}\}_{i=1}^n]=0.$$ Moreover, there exist some \( M, v_0 \in (0, \infty) \) such that
	\begin{align}
		\mathbb{ E }\Big\{ \exp \left[ \frac{\, |\tilde{Y}_{\mathrm{dfm},i}^{(M_n-1)}- \theta^{(M_n-1)}_{*,i}|}{M} \right] - 1 - \frac{|\tilde{Y}_{\mathrm{dfm},i}^{(M_n-1)} - \theta^{(M_n-1)}_{*,i}|}{M}&\\ \mid \{{\mathbf{H}}_i^{(M_n-1)},{\mathbf{A}}_i^{(M_n-1)}\}_{i=1}^n \Big\} M^2 
		\leq \frac{v_0}{2}.&\label{subG_mom1_Mn-1}
	\end{align}
\end{ass}
The diagonal entries of $\bm\Sigma^{(M_n-1)}$ below are $\operatorname{Var}(\varepsilon_i^{(M_n-1)}\mid\{\mathbf H_l^{(M_n-1)},\mathbf A_l^{(M_n-1)}\}_{l=1}^n)$.

Here we introduce some notations. Let \( \bm{\Sigma}^{(M_n-1)} \) be a diagonal matrix whose diagonal entries are the conditional variances of \( \varepsilon_i^{(M_n-1)} =\tilde{Y}^{(M_n-1)}_{\mathrm{dfm},i}-\theta_{*,i}^{(M_n-1)}\), and define the matrices as follows:

\begin{align*}
	&\mathbf{D}^{(M_n-1)}=\mathbb{E}[\mathbf{S}^{(M_n-1)}_{\mathrm{dfm},1}(\mathbf{S}^{(M_n-1)}_{\mathrm{dfm},1})^\top]\\&\mathbf{D}_n^{(M_n-1)}={(\bm{\Psi}^{(M_n-1)}_{\mathrm{dfm},1})^\top\bm{\Psi}^{(M_n-1)}_{\mathrm{dfm},1}} \\	&\mathbf{F}^{(M_n-1)}=\mathbb{E}[(\varepsilon^{(M_n-1)})^2\mathbf{S}^{(M_n-1)}_{\mathrm{dfm},1}(\mathbf{S}^{(M_n-1)}_{\mathrm{dfm},1})^\top]\\&\mathbf{F}_n^{(M_n-1)}={(\bm{\Psi}^{(M_n-1)}_{\mathrm{dfm},1})^\top\bm{\Sigma}^{(M_n-1)}\bm{\Psi}^{(M_n-1)}_{\mathrm{dfm},1}} \\&\mathbf{J}^{(M_n-1)}=\mathbb{E}[\mathbf{S}^{(M_n-1)}_{\mathrm{dfm},1}(\mathbf{S}_{\mathrm{dfm},\pi_*^{(M_n)},1})^\top]\\&\mathbf{J}_n^{(M_n-1)}=(\bm{\Psi}^{(M_n-1)}_{\mathrm{dfm},1})^\top\bm{\Psi}_{\mathrm{dfm},\pi_*^{(M_n)},1} \\&\mathbf{K}^{(M_n-1)}=\mathbf{J}^{(M_n-1)}\mathbf{P}^{(M_n)}{\mathbf{V}}^{(M_n)}(\mathbf{P}^{(M_n)})^\top(\mathbf{J}^{(M_n-1)})^\top+\mathbf{F}^{(M_n-1)}\\&\mathbf{K}_n^{(M_n-1)}=n^{-1}\mathbf J_n^{(M_n-1)}\mathbf P^{(M_n)}\mathbf V^{(M_n)}(\mathbf P^{(M_n)})^\top(\mathbf J_n^{(M_n-1)})^\top+\mathbf F_n^{(M_n-1)}.
\end{align*}

where \( \mathbf{P}^{(M_n)} \) is the linear operator that extracts \( \bm{\alpha}^{(M_n)}_{*,1} \) from \( \bm{\beta}^{(M_n)}_{*,1} \).
\begin{ass}\label{normal1}
	\item (a) Assume $\max_j\mathbb{E}[|\varepsilon^{(M_n-1)}h^{(M_n-1)}_{j,1}|^3]<\infty$, $\delta$ in Assumption~\ref{difflb} satisfies $\delta>2l^{(M_n)}_2$, and the following conditions hold:
	\begin{align}
		&\lambda_{\min} \left[\mathbf{F}^{(M_n-1)}\right]\geq c >0,\\&
		\left\|\mathbf{J}^{(M_n-1)}\right\|_2=O(1)\label{normal200}.
	\end{align}
	
	\item (b) Assume with probability at least $1-O(f(n))$ where $f(n)\to0$,
	\begin{align}
		&	\|\mathbf{D}_n^{(M_n-1)}-n \, \mathbf{D}^{(M_n-1)} \|_2\leq o\left(n/\sqrt{(s_n^{(M_n)}+s_n^{(M_n-1)})}\right),\label{normal221}\\&
		\|\mathbf{J}_n^{(M_n-1)}-n \, \mathbf{J}^{(M_n-1)} \|_2\leq o\left(n/\sqrt{(s_n^{(M_n)}+s_n^{(M_n-1)})}\right).\label{normal222}
	\end{align}
	
\end{ass}

\begin{ass}\label{B5}
	Define the filtration $\mathcal{F}_m=\sigma\left(\mathbf{X}^{(1)},\mathbf{A}^{(1)}, \cdots,\mathbf{X}^{(m)},\mathbf{A}^{(m)}\right)$ for $1\leq m \leq M_n$. Assume that \begin{equation}
		\operatorname{Cov}(Y^{(M_n-1)},Y^{(M_n)}\mid\mathcal{F}_{M_n})=0.\label{cov0}
	\end{equation}
\end{ass}

\begin{remark}
	Assumption~\ref{design1} imposes norm constraints on the cross-stage coupling matrices. As will be shown later, this condition holds with high probability under either sub-Gaussian or bounded cases. Assumption~\ref{compare1} requires a rate of order $\lambda_n^{(M_n-1)}\gg \sqrt{s^{(M_n)}_n + s^{(M_n-1)}_n/n}$, which arises from the joint estimation variability across both stages. Assumption~\ref{subG_Mn-1} parallels previously stated moment conditions. Regarding Assumption~\ref{normal1}, part (a) ensures the Lyapunov condition for asymptotic normality, while part (b) guarantees the uniform replacement of empirical matrices (e.g., sample covariance) with their population counterparts. We will show that condition (b) also holds with high probability under sub-Gaussian or bounded designs.

	Assumption~\ref{B5} requires that $Y^{(M_n-1)}$ and $Y^{(M_n)}$ are uncorrelated conditional on $\mathcal{F}_{M_n}$. Define
	$
	\varepsilon_{1}^{(M_n-1)} = Y^{(M_n-1)} - \mathbb{E}\bigl[Y^{(M_n-1)} \mid \mathcal{F}_{M_n-1}\bigr],
	$
	then it is equivalent to
	$$
	\mathbb{E}\Bigl[\varepsilon_{1}^{(M_n-1)} \mathbb{E}\bigl[\varepsilon^{(M_n)} \mid \mathcal{F}_{M_n} , \varepsilon_{1}^{(M_n-1)}\bigr] \mid \mathcal{F}_{M_n}\Bigr] = 0.
	$$
	Thus it holds if $\varepsilon_{1}^{(M_n-1)}$ cannot provide additional information once $\mathcal{F}_{M_n}$ is known, i.e.
	$$
	\mathbb{E}\bigl[\varepsilon^{(M_n)} \mid \mathcal{F}_{M_n},\varepsilon_{1}^{(M_n-1)}\bigr] \in \mathcal{F}_{M_n},
	$$
	which is a mild condition. Note that in many clinical situations, the reward is simply “survival” or “death” at the final stage, so $Y^{(M_n-1)}=0$.  In that case, Assumption~\ref{B5} holds trivially. 
	
	It is worth noting that \( \varepsilon_{1}^{(M_n-1)} \) only constitutes part of the stage-\( (M_n\!-\!1) \) residual—the other component $\varepsilon^{(M_n-1)} - \varepsilon_{1}^{(M_n-1)}$ arises from the approximation error in \( \max Q^{(M_n)} \). Assumption~\ref{B5} requires uncorrelatedness only for the first part, while no assumption is imposed on the second. However, in the proof, we show that the second component is also asymptotically uncorrelated.
\end{remark}

The following two lemmas establish sufficient conditions under each case, which are parallel to Lemmas~\ref{lemmaA.1} and \ref{lemmaA.2}.

\begin{lemma}\label{lemmaA.1-Mn-1}
	Under assumptions in Lemma~\ref{lemmaA.1},  \eqref{g1Mn-1}, \eqref{g2Mn-1}, and \eqref{g6M_n-1} hold with probability tending to1. 
	
	Further assume $(s_n^{(M_n-1)})^2\log s_n^{(M_n-1)}=o(n)$ and $m_r$ has a positive lower bound, then \eqref{g3Mn-1} and \eqref{g4Mn-1} hold with probability tending to 1.

	If further assume $s_n^{(M_n-1)}=O(n^{1-\max\{u,v,w\}})$ and the following holds:
	\item For each \(i\), \(\mathbf{r}^{(M_n-1)}_{k_i}\) is sub-Gaussian with its \(\psi_2\)-norm uniformly bounded by a constant. In particular, \(\|\mathbf{r}^{(M_n-1)}\|_{\psi_2} \leq C\).
	
	Then \eqref{g5Mn-1}, \eqref{g7M_n-1}, \eqref{g8M_n-1} and \eqref{normal200} hold with probability tending to 1.
	
	If $t_n^3\log t_n=o(n)$ for $t_n=s_n^{(M_n)}+s_n^{(M_n-1)}$, then \eqref{normal221} and \eqref{normal222} hold.
\end{lemma}

\begin{lemma}\label{lemmaA.2-Mn-1}
	Under assumptions in Lemma~\ref{lemmaA.2},  \eqref{g1Mn-1}, \eqref{g2Mn-1}, and \eqref{g6M_n-1} hold with probability tending to1. 

	Further assume $(s_n^{(M_n-1)})^2\log s_n^{(M_n-1)}=o(n)$ and $m_r$ has a positive lower bound, then \eqref{g3Mn-1} and \eqref{g4Mn-1} hold with probability tending to 1.

	If further assume $s_n^{(M_n-1)}=O(n^{1-\max\{u,v,w\}})$, and the following holds: 
	\item $\max_i\lambda_{\max}\left[\mathbb{E}[\mathbf{r}^{(M_n-1)}_{k_i}(\mathbf{r}^{(M_n-1)}_{k_i})^\top]\right]\leq C_1.$
	
	Then \eqref{g5Mn-1}, \eqref{g7M_n-1}, \eqref{g8M_n-1} and \eqref{normal200} hold with probability tending to 1.
	
	If $t_n^3\log t_n=o(n)$ for $t_n=s_n^{(M_n)}+s_n^{(M_n-1)}$, then \eqref{normal221} and \eqref{normal222} hold.
\end{lemma}

The final Lemma is parallel to Lemma~\ref{FboundMn}.

\begin{lemma}\label{Kbound}
	Under Assumption \ref{subG_mom}, \ref{normal2}, \ref{subG_Mn-1} and \ref{normal1},
	\begin{align*}
		&\mathbb{E}\left[(\varepsilon_i^{(M_n-1)})^2(\mathbf{S}^{(M_n-1)}_{\mathrm{dfm},1})^\top\mathbf{S}^{(M_n-1)}_{\mathrm{dfm},1}\right]=O(s_n^{(M_n-1)}),\\&\frac{1}{\sqrt{n}}\mathbb{E}\left|(\varepsilon_i^{(M_n-1)})^2 (\mathbf{S}_{\mathrm{dfm},i,1}^{(M_n-1)})^\top(\mathbf{K}^{(M_n-1)})^{-1}\mathbf{S}_{\mathrm{dfm},i,1}^{(M_n-1)}\right|^{3/2}\to 0,\\&\frac{1}{\sqrt{n}} \mathbb{E}\Big|(\varepsilon_i^{(M_n)})^2(\mathbf{J}^{(M_n-1)}\mathbf{P}^{(M_n)}\{\mathbf V^{(M_n)}/\phi^{(M_n)}\}{\mathbf{S}}_{i,1}^{(M_n)})^\top \,\\&(\mathbf{K}^{(M_n-1)})^{-1}\mathbf{J}^{(M_n-1)}\mathbf{P}^{(M_n)}\{\mathbf V^{(M_n)}/\phi^{(M_n)}\}{\mathbf{S}}_{i,1}^{(M_n)}\Big|^{3/2}\to 0.
	\end{align*}
\end{lemma}

For detailed proofs, see Appendix~\ref{Proofs}.

\counterwithin{equation}{section}

\section{Supplementary Theoretical Results} \label{omitted}

For the likelihood calculations below, write \(\mathbf b(\bm\theta)=(b(\theta_1),\ldots,b(\theta_n))^\top\) for the coordinatewise cumulant function.

The following result parallels those in \citet{Fan2011} or \citet{Ding2024}, but is established under a random design.
\begin{pro}\label{convergence2}
	Under Assumptions~3.1, A.1, A.2, A.3, there exists a strict local maximizer $
	\bm{\widehat{\beta}}^{(2)} = \left((\bm{\widehat{\beta}}^{(2)}_1)^\top, (\bm{\widehat{\beta}}^{(2)}_2)^\top\right)^\top$
	of the penalized likelihood \( L^{(2)}_n(\bm{\beta}^{(2)}) \), 	where \( \bm{\widehat{\beta}}^{(2)}_1 \) is the subvector of \( \bm{\widehat{\beta}}^{(2)} \) formed by components in \( \operatorname{supp}(\bm{\beta}^{(2)}_*) \), such that the following properties hold with probability tending to 1 as \( n \to \infty \):
	\item[a)] (Sparsity). \( \bm{\widehat{\beta}}^{(2)}_2 = \mathbf{0} \).
	\item[b)] (Estimation consistency). $
	\|\bm{\widehat{\beta}}^{(2)} - \bm{\beta}^{(2)}_*\|_2 = O_p(\sqrt{s_n^{(2)}/n}).
	$
	
	\item[c)] (Asymptotic normality). If $s_n^{(2)}=O(n^{\ell^{(2)}_2})$ with $\ell^{(2)}_2<1/3$, and Assumption~A.4 (a) additionally holds, then
	\[
	{n}^{-1/2}\, \mathbf{C}^{(2)}( \mathbf{F}^{(2)})^{-1/2}  \mathbf{F}_n^{(2)}
	( \bm{\widehat{\beta}}^{(2)}_1 - \bm{\beta}^{(2)}_{*,1} ) \xrightarrow{d} \mathcal{N}(0, \phi^{(2)} \mathbf{G}^{(2)}).
	\]
	where $\mathbf{F}^{(2)} = \mathbb{E}[b^{(2)''}(\theta^{(2)}_*)\mathbf{S}^{(2)}_1(\mathbf{S}^{(2)}_1)^\top]$,
	$\mathbf{F}_n^{(2)}=(\bm{\Psi}_1^{(2)})^\top\bm{\Sigma}^{(2)}(\bm{\theta}^{(2)}_*)\bm{\Psi}_1^{(2)}$, 
	$\mathbf{C}^{(2)}\in \mathbb{R}^{q\times s_n^{(2)}}$, $\mathbf{C}^{(2)}(\mathbf{C}^{(2)})^\top\to \mathbf{G}^{(2)}$ is positive definite. If Assumption~A.4 (b) further holds, then
	\[
	{\sqrt{n}}\mathbf{C}^{(2)}( \mathbf{F}^{(2)})^{1/2} 
	( \bm{\widehat{\beta}}^{(2)}_1 - \bm{\beta}^{(2)}_{*,1} ) \xrightarrow{d} \mathcal{N}(0, \phi^{(2)} \mathbf{G}^{(2)}).
	\]
\end{pro}
If all assumptions in Proposition \ref{convergence2} hold, the asymptotic covariance matrix of $\widehat{\bm{\beta}}^{(2)}_1$ can be estimated by the known sandwich formula \citep{Fan2004}:
\begin{align}
	&\mathbf{V}^{(2)}=\phi^{(2)} (\mathbf{F}^{(2)})^{-1},\quad
	\widehat{\mathbf{V}}^{(2)}=n \, (\widehat{\mathbf{F}}_n^{(2)})^{-1}\left((\widehat{\bm{\Psi}}_1^{(2)})^\top\widehat{\bm{\Sigma}}_\varepsilon^{(2)}\widehat{\bm{\Psi}}_1^{(2)}\right)(\widehat{\mathbf{F}}_n^{(2)})^{-1},
\end{align}
where $\widehat{\bm{\Psi}}_1^{(2)}$ is the sub-matrix of $\bm{\Psi}^{(2)}$ with columns supported in $\operatorname{supp}(\widehat{\bm{\beta}}^{(2)})$, and $\widehat{\mathbf{F}}^{(2)}_n = (\widehat{\bm{\Psi}}_1^{(2)})^\top \bm{\Sigma}^{(2)} (\widehat{\bm{\theta}}^{(2)}_0) \widehat{\bm{\Psi}}_1^{(2)}$. Here  $\widehat{\bm{\Sigma}}_\varepsilon^{(2)}=\operatorname{diag}\{(\widehat{\varepsilon}^{(2)}_i)^2\}^n_{i=1}$, $\widehat{\varepsilon}^{(2)}_i$ is the $i$-th residual.

\begin{remark}\label{remark:lla-oracle}
For either the terminal penalized GLM or an upstream penalized least-squares problem, let $\mathcal{L}^{(m)}_n(\bm\vartheta)$ denote the corresponding unpenalized loss (the negative normalized log-likelihood at the terminal stage), and apply the local linear approximation (LLA) to the folded-concave penalty. Given an initializer $\widetilde{\bm\vartheta}^{(m)}$, one update solves the weighted $\ell_1$ problem
\[
\bm\vartheta^{(m),[1]}
=\arg\min_{\bm\vartheta}
\left\{\mathcal{L}^{(m)}_n(\bm\vartheta)
+\sum_j p'_{\lambda_n^{(m)}}\bigl(|\widetilde\vartheta^{(m)}_j|\bigr)|\vartheta_j|\right\},
\]
over the penalized coordinates. Let $S_m$ be the true support and let $\bm\vartheta^{(m),\mathrm{o}}$ be the unique oracle estimator obtained by minimizing $\mathcal{L}^{(m)}_n$ subject to $\bm\vartheta_{S_m^c}=0$. Under the minimum-signal conditions in Assumptions~\ref{compare2}, \ref{compare1}, or \ref{comparem}, the active LLA weights vanish for SCAD or MCP whenever the initializer lies in the required local neighborhood. If, in addition, the inactive LLA weights strictly dominate the inactive oracle score, the KKT conditions of the weighted $\ell_1$ problem coincide with those of the oracle restricted problem. Hence $\bm\vartheta^{(m),[1]}=\bm\vartheta^{(m),\mathrm{o}}$ on this event. At the oracle solution the same KKT conditions continue to hold, so the next LLA update leaves the solution unchanged. Consequently, for an initializer for which the localization and score events hold with probability tending to one, one-step LLA reaches the oracle local solution with probability tending to one \citep{ZouLi2008,FanXueZou2014}.
\end{remark}

For \(M_n\) stages, estimation proceeds backward from the terminal stage.
The details are as follows. Let the trajectory be $(\mathbf{X}^{(1)},\mathbf{A}^{(1)},\ldots,\mathbf{X}^{(M_n)},\mathbf{A}^{(M_n)},Y^{(M_n)})$.
Under our primary setting, the regime is evaluated solely by \(Y^{(M_n)}\).
The stage-$M_n$ Q-function is modeled as
\begin{align*}
	&Q^{(M_n)}(\mathbf{H}^{(M_n)},\mathbf{A}^{(M_n)})\\=&\mathbb{E}\left[Y^{(M_n)}\mid \mathbf{H}^{(M_n)},\mathbf{A}^{(M_n)}\right]\\=&b^{(M_n)'}\left(c^{(M_n)}+(\bm{\eta}_0^{(M_n)})^\top \mathbf{H}^{(M_n)}+\sum_{k=2}^{L_n}(\bm{\eta}_k^{(M_n)})^\top \mathbf{H}^{(M_n)}\cdot I_{\{\mathbf{A}^{(M_n)}\in  \mathcal{A}_{(k)}\}}\right)\\=&b^{(M_n)'}\left((\bm{\beta}^{(M_n)})^\top\mathbf{S}^{(M_n)}\right).
\end{align*}
Choosing the deformed function 	$$d\left(\mathbf{H}^{(M_n)},\mathbf{A}^{(M_n)}\right)=(b^{(M_n)'})^{-1}\left(\max_{\mathbf{A}^{(M_n)}}Q^{(M_n)}\left(\mathbf{H}^{(M_n)},\mathbf{A}^{(M_n)}\right)\right),$$
then the stage-$(M_n\!-\!1)$ deformed Q-function is modeled as
\begin{align*}
	&Q_{{\mathrm{dfm}}}^{(M_n\!-\!1)}(\mathbf{H}^{(M_n\!-\!1)},\mathbf{A}^{(M_n\!-\!1)})\\=&\mathbb{E}\left[\max_{\mathbf{A}^{(M_n)}}(b^{(M_n)'})^{-1}\left(Q^{(M_n)}(\mathbf{H}^{(M_n)},\mathbf{A}^{(M_n)})\right)\mid \mathbf{H}^{(M_n\!-\!1)},\mathbf{A}^{(M_n\!-\!1)}\right]\\=&c^{(M_n\!-\!1)}+(\bm{\eta}_0^{(M_n\!-\!1)})^\top \mathbf{H}^{(M_n\!-\!1)}+\sum_{k=1}^{L_n}(\bm{\eta}_k^{(M_n\!-\!1)})^\top \mathbf{H}^{(M_n\!-\!1)}\cdot I_{\{\mathbf{A}^{(M_n\!-\!1)}\in  \mathcal{A}_{(k)}\}}\\=&(\bm{\beta}^{(M_n\!-\!1)})^\top\mathbf{S}^{(M_n\!-\!1)}.
\end{align*}
For $1\leq m \leq M_n-2$,
\begin{align*}
	&Q_{{\mathrm{dfm}}}^{(m)}(\mathbf{H}^{(m)},\mathbf{A}^{(m)})\\=&\mathbb{E}\left[\max_{\mathbf{A}^{(m+1)}} 	Q_{{\mathrm{dfm}}}^{(m+1)}(\mathbf{H}^{(m+1)},\mathbf{A}^{(m+1)})\mid \mathbf{H}^{(m)},\mathbf{A}^{(m)}\right]\\=&c^{(m)}+(\bm{\eta}_0^{(m)})^\top \mathbf{H}^{(m)}+\sum_{k=1}^{L_n}(\bm{\eta}_k^{(m)})^\top \mathbf{H}^{(m)}\cdot I_{\{\mathbf{A}^{(m)}\in \mathcal{A}_{(k)}\}}=(\bm{\beta}^{(m)})^\top\mathbf{S}^{(m)}.
\end{align*}

The estimation procedures are as follows :

Stage-$M_n$: Get the estimator $\widehat{\bm{\beta}}^{(M_n)}$ and the estimated asymptotic covariance matrix $\widehat{\mathbf{V}}^{(M_n)}$. Let $\widehat{\bm{\beta}}^{(M_n)}=(\widehat{c}^{(M_n)},(\widehat{\bm{\eta}}^{(M_n)}_0)^\top,(\widehat{\bm{\eta}}^{(M_n)}_2)^\top,\cdots,(\widehat{\bm{\eta}}^{(M_n)}_{L_n})^\top)$, then fix the first $d^{(m)} = \sum_{i=1}^m k_i+(m-1)\cdot a$ components of $\widehat{\bm{\eta}}^{(M_n)}_0$ to $\widehat{\bm{\eta}}^{(m)}_0$. 

Stage-$(M_n\!-\!1)$: Use 
$$\widehat{Y}^{(M_n\!-\!1)}_{\mathrm{dfm}}=\widehat{c}^{(M_n)} +(\widehat{\bm{\eta}}_{0,2}^{(M_n)})^\top \mathbf{A}^{(M_n\!-\!1)}+(\widehat{\bm{\eta}}_{0,3}^{(M_n)})^\top\mathbf{X}^{(M_n)}+\max_{k\geq2}\{0,(\widehat{\bm{\eta}}_k^{(M_n)})^\top \mathbf{H}^{(M_n)}\}$$
to estimate \begin{align*}
	&Q^{(M_n\!-\!1)}_{\mathrm{dfm}}(\mathbf{H}^{(M_n\!-\!1)},\mathbf{A}^{(M_n\!-\!1)})\\=&c^{(M_n\!-\!1)}+\sum_{k=1}^{L_n}(\bm{\eta}_k^{(M_n\!-\!1)})^\top \mathbf{H}^{(M_n\!-\!1)}\cdot I_{\{\mathbf{A}^{(M_n\!-\!1)}\in  \mathcal{A}_{(k)}\}}\\=&(\bm{\alpha}^{(M_n\!-\!1)})^\top\mathbf{S}^{(M_n\!-\!1)}_{\mathrm{dfm}}
\end{align*}
by penalized least squares, where $\widehat{\bm{\eta}}_{0,2}^{(M_n)}$ and $\widehat{\bm{\eta}}_{0,3}^{(M_n)}$ denote the components of $\bm{\widehat{\eta}}^{(M_n)}_0$ with respect to $\mathbf{A}^{(M_n)}$ and $\mathbf{X}^{(M_n)}$, respectively.

Then we get the estimator $\widehat{\bm{\alpha}}^{(M_n\!-\!1)}=\left(\widehat{c}^{(M_n\!-\!1)}, (\widehat{\bm{\eta}}_1^{(M_n\!-\!1)})^\top,\cdots,(\widehat{\bm{\eta}}_{L_n}^{(M_n\!-\!1)})^\top\right)^\top$ and the estimated covariance matrix $\widehat{\mathbf{V}}^{(M_n\!-\!1)}$. Now $\widehat{\bm{\beta}}^{(M_n\!-\!1)}$ is a permutation of  $\widehat{\bm{\xi}}^{(M_n\!-\!1)}=((\bm{\widehat{\eta}}^{(M_n\!-\!1)}_0)^\top,(\widehat{\bm{\alpha}}^{(M_n\!-\!1)})^\top)$, whose estimated covariance matrix $\widehat{\tilde{\mathbf{V}}}^{(M_n\!-\!1)}$ can be calculated.

$\cdots$

Stage-$m$ for $1\leq m \leq M_n-2$: Use
$$\widehat{Y}^{(m)}_{\mathrm{dfm}}=\widehat{c}^{(m+1)}+(\widehat{\bm{\eta}}_{0,2}^{(m+1)})^\top \mathbf{A}^{(m)}+(\widehat{\bm{\eta}}_{0,3}^{(m+1)})^\top\mathbf{X}^{(m+1)}+\max_{k\geq1}\{(\widehat{\bm{\eta}}_k^{(m+1)})^\top\mathbf{H}^{(m+1)}\}, $$
to estimate $$
Q^{(m)}_{\mathrm{dfm}}(\mathbf{H}^{(m)},\mathbf{A}^{(m)})=	c^{(m)}+\sum_{k=1}^{L_n}(\bm{\eta}_k^{(m)})^\top \mathbf{H}^{(m)}\cdot I_{\{\mathbf{A}^{(m)}\in \mathcal{A}_{(k)}\}}=(\bm{\alpha}^{(m)})^\top\mathbf{S}^{(m)}_{\mathrm{dfm}},$$
where $\widehat{\bm{\eta}}_{0,2}^{(m+1)}$ and $\widehat{\bm{\eta}}_{0,3}^{(m+1)}$ denote the components of $\bm{\widehat{\eta}}^{(m+1)}_0$ with respect to $\mathbf{A}^{(m)}$ and $\mathbf{X}^{(m+1)}$, respectively. Then $\widehat{\bm{\alpha}}^{(m)}$, $\widehat{\bm{\xi}}^{(m)}$, $\widehat{\mathbf{V}}^{(m)}$ and $\widehat{\tilde{\mathbf{V}}}^{(m)}$ can be calculated using similar methods.

Finally, each stage's deformed optimal Q-value and policy value differences can be calculated as in the two-stage case, using $\widehat{\mathbf{V}}^{(M_n)}$ and $\{\widehat{\tilde{\mathbf{V}}}^{(m)}\}_{1\leq m \leq M_n\!-\!1}$. When \(M_n\) is fixed, the theoretical arguments follow by a finite backward recursion: the terminal-stage GLM result initializes the induction, and each upstream stage applies the same deformed pseudo-outcome construction with a finite number of additional covariance blocks.

For nonzero intermediate rewards, \(Y^{(m)}\) is added to the stage-\(m\) pseudo-outcome; the estimation results continue to hold under the assumptions stated below. The CE and regret results are stated for \(Y^{(m)}=0\), \(m<M_n\).

\subsection{Additional Results for Terminal Stage}
\subsubsection*{Comparison Between Optimal and Candidate Policies}

For a given policy \(\pi^{(M_n)}(\cdot) \), the associated pseudo and true Q-functions are given by
\begin{align*}
\tilde{Q}^{(M_n)}(\mathcal{H}^{(M_n)}, \pi^{(M_n)})
&=c^{(M_n)}_*+(\bm{\eta}_{*,0}^{(M_n)})^\top\mathcal{H}^{(M_n)}\\
&\quad+\sum_{k=2}^{L_n}(\bm{\eta}_{*,k}^{(M_n)})^\top\mathcal{H}^{(M_n)}\cdot I_{\{\pi^{(M_n)}(\mathcal{H}^{(M_n)})\in\mathcal{A}_{(k)}\}},\\
Q^{(M_n)}(\mathcal{H}^{(M_n)}, \pi^{(M_n)})
&=b'\Bigl[c^{(M_n)}_*+(\bm{\eta}_{*,0}^{(M_n)})^\top\mathcal{H}^{(M_n)}\\
&\qquad+\sum_{k=2}^{L_n}(\bm{\eta}_{*,k}^{(M_n)})^\top\mathcal{H}^{(M_n)}\cdot I_{\{\pi^{(M_n)}(\mathcal{H}^{(M_n)})\in\mathcal{A}_{(k)}\}}\Bigr],
\end{align*}
and the corresponding true value is defined as
\begin{align*}\mathbb{E}[Q^{(M_n)}(\mathcal{H}^{(M_n)}, \pi^{(M_n)})]= \int Q^{(M_n)}(\mathcal{H}^{(M_n)}, \pi^{(M_n)}) dF(\mathcal{H}^{(M_n)}) =\mathbb{E}\left[b'\left((\bm{\beta}_*^{(M_n)})^\top \mathcal{S}_{\pi^{(M_n)}}\right)\right].
\end{align*}
The value difference between $\pi_*^{(M_n)}$ and $\pi^{(M_n)}$ is defined as
\begin{align*}
	\mathbb{E}[Q_*^{(M_n)}(\mathcal{H}^{(M_n)})] - \mathbb{E}[Q^{(M_n)}(\mathcal{H}^{(M_n)}, \pi^{(M_n)})].
\end{align*}
To this end, the values associated with \(\pi^{(M_n)}(\cdot)\) is estimated by
\begin{align*}
	\mathbb{P}_N \widehat{Q}^{(M_n)}(\mathcal{H}^{(M_n)}, \pi^{(M_n)}) = \frac{1}{N} \sum_{i=1}^{N} \widehat{Q}^{(M_n)} \left(\mathcal{H}^{(M_n)}_{i}, \pi^{(M_n)}_{i} \right) =  \mathbb{P}_N\left[b'\left((\widehat{\beta}^{(M_n)})^\top\mathcal{S}_{\pi^{(M_n)}}\right)\right].
\end{align*}
Accordingly, the value difference is estimated by \(\mathbb{P}_N \widehat{Q}^{(M_n)}_*(\mathcal{H}^{(M_n)}) - \mathbb{P}_N \widehat{Q}^{(M_n)}(\mathcal{H}^{(M_n)}, \pi^{(M_n)})\).

\begin{thm}\label{2-diff}
	Suppose all conditions in Proposition~\ref{convergence2} and Assumptions~\ref{maxmincd}, \ref{Eb''} hold. Also assume $L_n=O(n^{\ell_1}), s_n^{(M_n)}=O(n^{\ell^{(M_n)}_2}), n^{\ell_3}=O(N),N=O(n^{\ell_4})$ with $\ell^{(M_n)}_2<1/3$.
	\item[a)] (Consistency). As \(n, N \to \infty\),
	\[
	\mathbb{P}_N \widehat{Q}_*^{(M_n)}(\mathcal{H}^{(M_n)}) - \mathbb{P}_N \widehat{Q}^{(M_n)}(\mathcal{H}^{(M_n)}, \pi^{(M_n)}) 
	\xrightarrow{p} \mathbb{E}[Q_*^{(M_n)}(\mathcal{H}^{(M_n)})] - \mathbb{E}[Q^{(M_n)}(\mathcal{H}^{(M_n)}, \pi^{(M_n)})].
	\]
	
	\item[b)] (Asymptotic normality). Further suppose that $3\ell^{(M_n)}_2 < \ell_3$, $2\ell^{(M_n)}_2 + \ell_1 < 1/2$, and Assumption~\ref{difflb} hold. Moreover, let
	\begin{align*}
		v_n^{(M_n)}= \mathbb{E}\left[{b}''\left(\tilde{Q}_*^{(M_n)}(\mathcal{H}^{(M_n)})\right)\mathcal{S}_{\pi_*^{(M_n)}} - {b}''\left(\tilde{Q}^{(M_n)}(\mathcal{H}^{(M_n)},\pi^{(M_n)})\right)\mathcal{S}_{\pi^{(M_n)}}\right],
	\end{align*}
	suppose
	\vspace{-1em}
	\begin{align*}
		&\lim_{n \to \infty}  (v_n^{(M_n)})^\top \mathbf{V}^{(M_n)} v_n^{(M_n)} = (\sigma_1^{(M_n)})^2,\\&
		\lim_{n,N \to \infty} \operatorname{Var} \left(Q_*^{(M_n)}(\mathcal{H}^{(M_n)}) - Q^{(M_n)}(\mathcal{H}^{(M_n)}, \pi^{(M_n)})\right) \, n/{N} = (\sigma_2^{(M_n)})^2,
	\end{align*}
	for constants \( \sigma_1^{(M_n)} > 0 \), \( \sigma_2^{(M_n)} \geq 0 \). Define \( (\sigma^{(M_n)}_{*,\pi})^2 = (\sigma_1^{(M_n)})^2 + (\sigma_2^{(M_n)})^2 \), then
	\begin{align*}
		&\sqrt{n} \Big( \mathbb{P}_N \widehat{Q}_*^{(M_n)}(\mathcal{H}^{(M_n)}) - \mathbb{P}_N \widehat{Q}^{(M_n)}(\mathcal{H}^{(M_n)}, \pi^{(M_n)})  \\ -&\left\{ \mathbb{E}[Q_*^{(M_n)}(\mathcal{H}^{(M_n)})] - \mathbb{E}[Q^{(M_n)}(\mathcal{H}^{(M_n)}, \pi^{(M_n)})] \right\} \Big)
		\xrightarrow{d} \mathcal{N}(0, (\sigma^{(M_n)}_{*,\pi})^2).
	\end{align*}
\end{thm}
Similarly, the asymptotic variance can be consistently estimated by
\begin{align*}
	(\widehat{\sigma}_*^{(M_n)})^2 = &   \left(\widehat{v}_n^{(M_n)}\right)^\top \widehat{\mathbf{V}}^{(M_n)} \, \widehat{v}_n^{(M_n)}
	+ \widehat{\operatorname{Var}}\left( \widehat{Q}_*^{(M_n)}(\mathcal{H}^{(M_n)})- \widehat{Q}^{(M_n)}(\mathcal{H}^{(M_n)},\pi^{(M_n)}) \right) \, n/N,
\end{align*}
where $\widehat{v}_n^{(M_n)}=\mathbb{P}_N\left[{b}''\left(\widehat{\tilde{Q}}_*^{(M_n)}(\mathcal{H}^{(M_n)})\right)\mathcal{S}_{\widehat{\pi}_*^{(M_n)}} - {b}''\left(\widehat{\tilde{Q}}^{(M_n)}(\mathcal{H}^{(M_n)},\pi^{(M_n)})\right)\mathcal{S}_{\pi^{(M_n)}}\right]$.

The following lemma concerns the analysis of the terminal deformed optimal Q-value and parallels a known result in \cite{Ding2024}. We include it here for completeness, as it supports subsequent analysis. The structure is similar, and the assumptions and derivations differ slightly.

\begin{lemma}\label{2-opt-ps}
	Suppose all conditions in Proposition~\ref{convergence2} and Assumption~\ref{maxmincd} hold. Also assume $L_n=O(n^{\ell_1}), s_n^{(M_n)}=O(n^{\ell^{(M_n)}_2}), n^{\ell_3}=O(N),N=O(n^{\ell_4})$ with $\ell^{(M_n)}_2<1/3$.
	
	\item (i) For the deformed optimal Q-value associated with \(\tilde{Q}_*^{(M_n)}\):
	
	\item[a)] (Consistency). As \( n, N \to \infty \),
	\[
	\mathbb{P}_N \widehat{\tilde{Q}}_*^{(M_n)}(\mathcal{H}^{(M_n)}) \xrightarrow{p} \mathbb{E}[\tilde{Q}_*^{(M_n)}(\mathcal{H}^{(M_n)})].
	\]
	
	\item[b)] (Asymptotic normality). Further suppose that $3\ell^{(M_n)}_2 < \ell_3$, $2\ell^{(M_n)}_2 + \ell_1 < 1/2$, and Assumption~\ref{difflb} hold. Moreover, 
	\begin{align*}
		&\lim_{n \to \infty}  \mathbb{E} [\mathcal{S}_{\pi_*^{(M_n)}}^\top] \mathbf{V}^{(M_n)} \, \mathbb{E} [\mathcal{S}_{\pi_*^{(M_n)}}] = (\tilde{\sigma}_1^{(M_n)})^2,\\
		&\lim_{n,N \to \infty} \operatorname{Var}( \tilde{Q}_*^{(M_n)}(\mathcal{H}^{(M_n)})) \, n/N= (\tilde{\sigma}_2^{(M_n)})^2,
	\end{align*}
	for some constants \( \tilde{\sigma}_1^{(M_n)} > 0 \), \( \tilde{\sigma}_2^{(M_n)} \geq 0 \). Define 
	\[
	(\tilde{\sigma}_*^{(M_n)})^2 = (\tilde{\sigma}_1^{(M_n)})^2 + (\tilde{\sigma}_2^{(M_n)})^2,
	\]
	then
	\[
	\sqrt{n} \left( \mathbb{P}_N \widehat{\tilde{Q}}_*^{(M_n)}(\mathcal{H}^{(M_n)}) - \mathbb{E}[\tilde{Q}_*^{(M_n)}(\mathcal{H}^{(M_n)})] \right) \xrightarrow{d} \mathcal{N}(0, (\tilde{\sigma}_*^{(M_n)})^2).
	\]
	
\end{lemma}
The asymptotic variances can be consistently estimated by
\[
(\widehat{\tilde{\sigma}}_*^{(M_n)})^2 =   \mathbb{P}_N\left(\mathcal{S}_{\widehat{\pi}_*^{(M_n)}}^\top\right) \widehat{\mathbf{V}}^{(M_n)} \mathbb{P}_N\left(\mathcal{S}_{\widehat{\pi}_*^{(M_n)}}\right)
+ \widehat{\operatorname{Var}}\left( \widehat{\tilde{Q}}_*^{(M_n)}(\mathcal{H}^{(M_n)}) \right) \, n/N.
\]
Moreover, the inference for the deformed value differences can be established similarly.

\begin{remark}\label{remark:fixed-rate-feasible}
The exponent restrictions used for fixed-stage value and upstream inference are joint sufficient inequalities, not competing equalities. Writing $\ell_2$ for the relevant sparsity exponent, the restrictions
\[
3\ell_2<\ell_3,\qquad
2\ell_2+\ell_1<\tfrac12,\qquad
3\ell_2+\ell_1<\tfrac12
\]
have a nonempty feasible region. For example, each row below satisfies all three inequalities and $\ell_3\leq\ell_4$, as required by $n^{\ell_3}=O(N)$ and $N=O(n^{\ell_4})$:
\[
\begin{array}{c|cccc}
&\ell_1&\ell_2&\ell_3&\ell_4\\ \hline
\mathrm{F1}&0.05&0.10&0.35&1.00\\
\mathrm{F2}&0.10&0.05&0.20&0.50\\
\mathrm{F3}&0.15&0.05&0.25&0.75
\end{array}
\]
Thus the growth conditions on the number of action blocks, sparsity, and testing-sample size can hold simultaneously.
\end{remark}

\subsection{Additional Results for Upstream Deformation}

\subsubsection*{Evaluation of Stage-$(M_n\!-\!1)$ Deformed Optimal Q-value}

Let \( \mathcal{H}^{(M_n\!-\!1)} \) follow the testing distribution of \(\mathbf{H}^{(M_n\!-\!1)}\). Throughout this subsection, \(Q^{(M_n-1)}\), \(\pi_*^{(M_n-1)}\), and the associated values refer to the deformed criterion.
Define
\begin{align*}
	\pi^{(M_n\!-\!1)}_*\left(\mathcal{H}^{(M_n\!-\!1)}\right) 
	&:= \min_{\mathcal{A}^{(M_n\!-\!1)}} 
	\left\{
	\mathcal{A}^{(M_n\!-\!1)} : 
	\right. \\
	&\qquad \left.
	Q^{(M_n\!-\!1)}\left(\mathcal{H}^{(M_n\!-\!1)}, \mathcal{A}^{(M_n\!-\!1)}\right) = 
	\max_{\mathcal{A}^{(M_n\!-\!1)}} 
	Q^{(M_n\!-\!1)}\left(\mathcal{H}^{(M_n\!-\!1)}, \mathcal{A}^{(M_n\!-\!1)}\right)
	\right\}, \\
	Q^{(M_n\!-\!1)}_*\left(\mathcal{H}^{(M_n\!-\!1)}\right) 
	&= c_*^{(M_n\!-\!1)} 
	+ \left(\bm{\eta}_{*,0}^{(M_n\!-\!1)}\right)^\top \mathcal{H}^{(M_n\!-\!1)} 
	+ \max_{k \geq 1} 
	\left\{ \left(\bm{\eta}_{*,k}^{(M_n\!-\!1)}\right)^\top \mathcal{H}^{(M_n\!-\!1)} \right\}.
\end{align*}
Let
\begin{align*}
	\mathcal{S}_{\pi^{(M_n\!-\!1)}} = \Big(1, (\mathcal{H}^{(M_n\!-\!1)})^\top, (\mathcal{H}^{(M_n\!-\!1)})^\top \cdot I_{\{\pi^{(M_n\!-\!1)}(\mathcal{H}^{(M_n\!-\!1)}) \in \mathcal{A}_{(1)}\}} \dots, &\\(\mathcal{H}^{(M_n\!-\!1)})^\top \cdot I_{\{\pi^{(M_n\!-\!1)}(\mathcal{H}^{(M_n\!-\!1)}) \in\mathcal{A}_{(L_n)}\}} \Big)^\top.&
\end{align*}
Using the notation above, the deformed optimal Q-function is \( Q^{(M_n\!-\!1)}_* =( \bm{\beta}^{(M_n\!-\!1)}_*)^\top \mathcal{S}_{\pi^{(M_n\!-\!1)}_*} \), the deformed optimal Q-value \( \mathbb{E}[Q^{(M_n\!-\!1)}_*(\mathcal{H}^{(M_n\!-\!1)})] = (\bm{\beta}^{(M_n\!-\!1)}_*)^\top \mathbb{E}( \mathcal{S}_{\pi^{(M_n\!-\!1)}_*}) \), and the estimator is
\[
\mathbb{P}_N \widehat{Q}_*^{(M_n\!-\!1)}(\mathcal{H}^{(M_n\!-\!1)}) = \frac{1}{N} \sum_{i=1}^N \widehat{Q}^{(M_n\!-\!1)}_*(\mathcal{H}^{(M_n\!-\!1)}_{i})= (\widehat{\bm{\beta}}^{(M_n\!-\!1)})^\top \mathbb{P}_N (\mathcal{H}_{\widehat{\pi}^{(M_n\!-\!1)}_*}),
\]
where
\begin{align*}
	\widehat{\pi}_*^{(M_n\!-\!1)}\left(\mathcal{H}^{(M_n\!-\!1)}\right) 
	&:= \min_{\mathcal{A}^{(M_n\!-\!1)}}
	\left\{
	\mathcal{A}^{(M_n\!-\!1)} :
	\right. \\
	&\qquad \left.
	\widehat{Q}^{(M_n\!-\!1)}\left(\mathcal{H}^{(M_n\!-\!1)}, \mathcal{A}^{(M_n\!-\!1)}\right) = 
	\max_{\mathcal{A}^{(M_n\!-\!1)}} 
	\widehat{Q}^{(M_n\!-\!1)}\left(\mathcal{H}^{(M_n\!-\!1)}, \mathcal{A}^{(M_n\!-\!1)}\right)
	\right\}, \\
	\widehat{Q}_*^{(M_n\!-\!1)}\left(\mathcal{H}^{(M_n\!-\!1)}\right) 
	&= \widehat{c}^{(M_n\!-\!1)} 
	+ \left(\widehat{\bm{\eta}}_{0}^{(M_n\!-\!1)}\right)^\top \mathcal{H}^{(M_n\!-\!1)} 
	+ \max_{k \geq 1} 
	\left\{ \left(\widehat{\bm{\eta}}_{k}^{(M_n\!-\!1)}\right)^\top \mathcal{H}^{(M_n\!-\!1)} \right\}.
\end{align*}

\begin{cor}\label{1-opt}Assume all the assumptions in Theorem~\ref{convergence1} hold, and conditions of Theorem~\ref{2-opt} similarly hold for the $(M_n \!-\!1)$-th stage.
	\item[a)] (Consistency). As \( n, N \to \infty \),
	\begin{align*}
		\mathbb{P}_N \widehat{Q}_*^{(M_n\!-\!1)}(\mathcal{H}^{(M_n\!-\!1)}) \xrightarrow{p} \mathbb{E}[Q^{(M_n\!-\!1)}_*(\mathcal{H}^{(M_n\!-\!1)})].
	\end{align*}

	\item[b)] (Asymptotic normality). If
	\begin{align*}
		&\lim_{n \to \infty}  \mathbb{E} [\mathcal{S}_{\pi^{(M_n\!-\!1)}_*}^\top] \tilde{\mathbf{V}}^{(M_n\!-\!1)}\mathbb{E} [{\mathcal{S}}_{\pi_*^{(M_n\!-\!1)}}]= (\sigma^{(M_n\!-\!1)}_1)^2 ,\\&\lim_{n,N \to \infty} \operatorname{Var}( Q_*^{(M_n\!-\!1)}(\mathcal{H}_1))\, n/N = (\sigma_2^{(M_n\!-\!1)})^2
	\end{align*}
	for some constants \( \sigma^{(M_n\!-\!1)}_1 > 0, \sigma^{(M_n\!-\!1)}_2\geq 0 \). Let \( (\sigma^{(M_n\!-\!1)}_* )^2= (\sigma^{(M_n\!-\!1)}_1)^2 + (\sigma^{(M_n\!-\!1)}_2)^2 \), then
	\[
	\sqrt{n} \left( \mathbb{P}_N \widehat{Q}_*^{(M_n\!-\!1)}(\mathcal{H}^{(M_n\!-\!1)}) - \mathbb{E}[Q^{(M_n\!-\!1)}_*(\mathcal{H}^{(M_n\!-\!1)})] \right) \xrightarrow{d} \mathcal{N}(0, (\sigma_*^{(M_n\!-\!1)})^2).
	\]
\end{cor}
The estimated variance is 
$$(\widehat{\sigma}^{(M_n\!-\!1)}_*)^2 =  \mathbb{P}_N(\mathcal{S}_{\widehat{\pi}^{(M_n\!-\!1)}_*}^{T}) \widehat{\tilde{\mathbf{V}}}^{(M_n\!-\!1)} \mathbb{P}_N(\mathcal{S}_{\widehat{\pi}^{(M_n\!-\!1)}_*})+ \widehat{\operatorname{Var}}\left( \widehat{Q}_*^{(M_n\!-\!1)}(\mathcal{H}^{(M_n\!-\!1)}) \right) \, n / N $$

\subsubsection*{Comparison Between Stage-$(M_n\!-\!1)$ Deformed-Optimal and Candidate Policies}
For a given decision rule $\pi^{(M_n\!-\!1)}$, the following corollary parallels Theorem~\ref{2-diff}.
\begin{cor}\label{1-diff}Assume all conditions of Theorem~\ref{convergence1} and the analogous conditions of Theorem~\ref{2-opt} for stage $(M_n \!-\!1)$.
	\item[a)] (Consistency). As \(n, N \to \infty\),
	\begin{align*}
		&\mathbb{P}_N \widehat{Q}^{(M_n\!-\!1)}_*(\mathcal{H}^{(M_n\!-\!1)}) - \mathbb{P}_N \widehat{Q}^{(M_n\!-\!1)}(\mathcal{H}^{(M_n\!-\!1)}, \pi^{(M_n\!-\!1)}) \\&
		\xrightarrow{p} \mathbb{E}[Q_*^{(M_n\!-\!1)}(\mathcal{H}^{(M_n\!-\!1)})] - \mathbb{E}[Q^{(M_n\!-\!1)}(\mathcal{H}^{(M_n\!-\!1)}, \pi^{(M_n\!-\!1)})].
	\end{align*}
	\item[b)] (Asymptotic normality). If 
	\begin{align*}
		&\lim_{n \to \infty} \mathbb{E}[\mathcal{S}_{\pi_*^{(M_n\!-\!1)}} - \mathcal{S}_{\pi^{(M_n\!-\!1)}}]^\top \tilde{\mathbf{V}}^{(M_n\!-\!1)} \mathbb{E}[\mathcal{S}_{\pi_*^{(M_n\!-\!1)}} - \mathcal{S}_{\pi^{(M_n\!-\!1)}}] = (\sigma_1^{(M_n\!-\!1)})^2, \\&
		\lim_{n,N \to \infty} \operatorname{Var} \left(Q_*^{(M_n\!-\!1)}(\mathcal{H}^{(M_n\!-\!1)}) - Q^{(M_n\!-\!1)}(\mathcal{H}^{(M_n\!-\!1)}, \pi^{(M_n\!-\!1)})\right)\, n/N = (\sigma_2^{(M_n\!-\!1)})^2
	\end{align*}
	for some constants \(\sigma^{(M_n\!-\!1)}_1 > 0\), and \(\sigma^{(M_n\!-\!1)}_2 \geq 0\). Let \((\sigma^{(M_n\!-\!1)}_{*, \pi})^2 = (\sigma_1^{(M_n\!-\!1)} )^2+ (\sigma_2^{(M_n\!-\!1)})^2\), then
	\begin{align}
		&\sqrt{n} \Big( \mathbb{P}_N \widehat{Q}^{(M_n\!-\!1)}_*(\mathcal{H}^{(M_n\!-\!1)}) - \mathbb{P}_N \widehat{Q}^{(M_n\!-\!1)}(\mathcal{H}^{(M_n\!-\!1)}, \pi^{(M_n\!-\!1)}) \nonumber \\ 
		& -(\mathbb{E}[Q_*^{(M_n\!-\!1)}(\mathcal{H}^{(M_n\!-\!1)})] - \mathbb{E}[Q^{(M_n\!-\!1)}(\mathcal{H}^{(M_n\!-\!1)}, \pi^{(M_n\!-\!1)})]) \Big)\xrightarrow{d} \mathcal{N}(0, (\sigma^{(M_n\!-\!1)}_{*, \pi})^2).\label{1-diff-an}
	\end{align}
\end{cor}

Similarly, we can estimate \(\sigma^{(M_n\!-\!1)}_{*, \pi}\) using the following sandwich formula
\begin{align*}
	(\widehat{\sigma}^{(M_n\!-\!1)}_{*, \pi})^2 &=  \mathbb{P}_N (\mathcal{S}_{\widehat{\pi}_*^{(M_n\!-\!1)}} - \mathcal{S}_{\pi^{(M_n\!-\!1)}})^\top \widehat{\tilde{\mathbf{V}}}^{(M_n\!-\!1)} \mathbb{P}_N (\mathcal{S}_{\widehat{\pi}_*^{(M_n\!-\!1)}} - \mathcal{S}_{\pi^{(M_n\!-\!1)}})
	\\&+ \widehat{\operatorname{Var}}( \widehat{Q}^{(M_n\!-\!1)}_*(\mathcal{H}^{(M_n\!-\!1)}) - \widehat{Q}^{(M_n\!-\!1)}(\mathcal{H}^{(M_n\!-\!1)}, \pi^{(M_n\!-\!1)})) \, n/N
\end{align*}

\subsection{\texorpdfstring{Theoretical Results for Diverging $M_n$}{Theoretical Results for Diverging Mn}}\label{diverge}
For diverging \(M_n\), impose the following conditions uniformly over stages. All constants in $O$ and $o$ notation are uniform in $m$.

\begin{ass}\label{designm}
	Assume that the matrix norm conditions in Assumptions~\ref{design2} and~\ref{design1}, and that
	\begin{equation}
		\left\| \left( \bm{\Psi}_1^{(m)} \right)^\top \bm{\Psi}_1^{(m)} \right\|_2 = O(n) \label{newMncondition}
	\end{equation}
	hold. These conditions are required to hold with probability at least \( 1 - O(1/n) \) separately at each stage \( m \).
\end{ass}

\begin{ass}\label{comparem}
	Assume that $S_n: =\max_m \{s_n^{(m)}\}=O(n^{\ell_5}), M_n=O(n^{\ell_6})$ with $\ell_5+\ell_6<1$. Moreover, $
	g^{(m)}_n \gg \lambda^{(m)}_n \gg \max \Big\{ (\sum_{j\geq m}s_n^{(j)}\log n /n)^{1/2}$, $(\log p_n^{(m)}/n)^{1/2}\Big\}$, $ 
	\quad p'_{\lambda^{(m)}_n} (g^{(m)}_n) = o((ns_n^{(m)})^{-1/2})$, and 
	$	\lambda^{(m)}_n \kappa^{(m)}_0 = o(1).$

For diverging $M_n$, also require
\begin{equation}
\max_{m\leq M_n}\frac{a_{m,n}}{r_m}\longrightarrow0,
\label{eq:consistency-accumulation}
\end{equation}
where $r_m$ and $a_{m,n}$ are defined in \eqref{eq:local-remainder-rates}--\eqref{eq:stage-accumulation}. Lemma~\ref{lemma:accumulated-verification} verifies this condition under the sub-Gaussian or bounded designs considered above.

\end{ass}

\begin{ass}\label{subgm}
	Assume $\max_{m,j}\mathbb{E}[(\varepsilon^{(m)}h^{(m)}_{j,1})^2]<\infty$, and for $i\neq j$
	$$\mathbb{E}[\varepsilon_i^{(m)}\varepsilon_j^{(m)}\mid \{{\mathbf{H}}_i^{(m)},{\mathbf{A}}_i^{(m)}\}_{i=1}^n]=0.$$ Moreover, there exists \( v_0 \in (0, \infty) \) such that for any $\lambda\in\mathbb{R}$,
	\begin{align}
		\mathbb{ E }\Big\{ \exp \left[ {\lambda \, \{\tilde{Y}_{\mathrm{dfm},i}^{(m)}- b' (\theta^{(m)}_{*,i})\}} \right]\mid \{{\mathbf{H}}_i^{(m)},{\mathbf{A}}_i^{(m)}\}_{i=1}^n \Big\} 
		\leq \exp\left(\frac{\lambda^2 v_0^2}{2}\right)\label{subG_momm}.
	\end{align}
	where $b'=b^{(M_n)'}$ if $m=M_n$, and $b'(x)=x$ otherwise.
\end{ass}

\begin{ass}\label{difflbm}
	Define
	\begin{align}
		D^{(m)}_{k,j}=-D^{(m)}_{j,k}= (\bm{\eta}^{(m)}_{*,k})^\top\mathcal{H}^{(m)} -(\bm{\eta}^{(m)}_{*,j})^\top\mathcal{H}^{(m)}\label{dkjm},
	\end{align}
	assume the following probability bound holds: 
	$$\mathbb{P}\left[\, \exists k, \,\pi_*^{(m)}(\mathcal{H}^{(m)})\in\mathcal{A}_{(k)}, \inf_{j\neq k}\left\{D_{k,j}\right\}\leq \frac{\varepsilon}{L_n}	\,  \right]=O(\varepsilon).$$
	Moreover, define $\mathcal{U}^{(m)}_n=\left\{ |\Pi_*^{(m)}|=1 \right\}$, assume that for some $\delta^{(m)}>\sum_{j\geq m}\ell^{(j)}_2$, 
	$$\mathbb{P}\left[ \mathcal{U}^{(m)}_n \right]\geq 1-O(n^{-\delta^{(m)}}).$$
\end{ass}

Let $\boldsymbol\vartheta_m$ contain the active inherited coefficients $\boldsymbol\gamma_m$ and the active coefficients $\boldsymbol\alpha^{(m)}_1$ estimated at stage $m$, each coordinate appearing once. At the terminal stage, $\boldsymbol\vartheta_{M_n}=\boldsymbol\beta^{(M_n)}_{*,1}$. Let $\mathbf E_m$ select $\boldsymbol\alpha^{(m)}_1$ and let $\mathbf R_m$ select the coefficients inherited from $\boldsymbol\vartheta_{m+1}$.

For $m<M_n$, let $\mathbf u_{mi}=\mathbf S^{(m)}_{\mathrm{dfm},i,1}\varepsilon_i^{(m)}$. Let $\mathbf W_{mi}$ be the derivative, with respect to the full vector $\boldsymbol\vartheta_{m+1}$, of the stage-$m$ adjusted pseudo-outcome evaluated at the true maximizing action, including the subtraction of the inherited history term. Define
\[
\begin{aligned}
\mathbf D^{(m)}&=\mathbb E\{\mathbf S^{(m)}_{\mathrm{dfm},i,1}(\mathbf S^{(m)}_{\mathrm{dfm},i,1})^{\top}\},\\
\boldsymbol{\mathcal J}_m&=\mathbb E(\mathbf S^{(m)}_{\mathrm{dfm},i,1}\mathbf W_{mi}^{\top}),\qquad
\boldsymbol\Omega_m=\mathbb E(\mathbf u_{mi}\mathbf u_{mi}^{\top})=\mathbf F^{(m)}.
\end{aligned}
\]
Define
\begin{equation}
\mathbf B_m=\begin{pmatrix}\mathbf R_m\\ (\mathbf D^{(m)})^{-1}\boldsymbol{\mathcal J}_m\end{pmatrix},\qquad
\mathbf A_m=\begin{pmatrix}\mathbf 0\\ (\mathbf D^{(m)})^{-1}\end{pmatrix},\qquad m<M_n.
\label{eq:full-transfer}
\end{equation}
At stage $M_n$, set $\mathbf u_{M_n i}=\mathbf S^{(M_n)}_{i,1}\varepsilon_i^{(M_n)}$, $\mathbf H_{M_n}=\mathbb E[b^{(M_n)''}(\theta_{*,i}^{(M_n)})\mathbf S^{(M_n)}_{i,1}(\mathbf S^{(M_n)}_{i,1})^\top]$, and $\mathbf A_{M_n}=\mathbf H_{M_n}^{-1}$. Thus $\boldsymbol\Omega_{M_n}=\phi^{(M_n)}\mathbf H_{M_n}$ and $\mathbf A_{M_n}\boldsymbol\Omega_{M_n}\mathbf A_{M_n}^\top=\mathbf V^{(M_n)}$.

Define
\begin{equation}
\begin{split}
\boldsymbol\Gamma_{mj}&=\mathbf B_m\cdots\mathbf B_{j-1},\qquad \boldsymbol\Gamma_{mm}=\mathbf I,\\
\boldsymbol\psi_{mi}&=\sum_{j=m}^{M_n}\boldsymbol\Gamma_{mj}\mathbf A_j\mathbf u_{ji},\qquad
\widetilde{\mathbf V}_m=\sum_{j=m}^{M_n}\boldsymbol\Gamma_{mj}\mathbf A_j\boldsymbol\Omega_j\mathbf A_j^\top\boldsymbol\Gamma_{mj}^\top.
\end{split}
\label{eq:full-influence}
\end{equation}
The covariance expression follows because the scores from different stages are uncorrelated, as shown below. For an upstream stage, define
\begin{equation}
\begin{split}
\mathbf L_{mj}&=\mathbf D^{(m)}\mathbf E_m\boldsymbol\Gamma_{mj}\mathbf A_j,\\
\mathbf K^{(m)}&=\mathbf D^{(m)}\mathbf E_m\widetilde{\mathbf V}_m\mathbf E_m^\top\mathbf D^{(m)}
=\sum_{j=m}^{M_n}\mathbf L_{mj}\boldsymbol\Omega_j\mathbf L_{mj}^\top.
\end{split}
\label{eq:full-covariance}
\end{equation}
Equivalently, $\widetilde{\mathbf V}_m=\mathbf B_m\widetilde{\mathbf V}_{m+1}\mathbf B_m^\top+\mathbf A_m\boldsymbol\Omega_m\mathbf A_m^\top$. At $m=M_n-1$, $\boldsymbol{\mathcal J}_m=\mathbf J^{(m)}\mathbf P^{(M_n)}$, and it reduces to the two-stage covariance in Theorem~\ref{convergence1}.

Let $G_m$ be the gap between the largest and second-largest action-block values, with $G_m=0$ at a tie. For the rate conditions, put $T_m=\sum_{j=m}^{M_n}s_n^{(j)}$, $r_m=(T_m\log n/n)^{1/2}$, and $x_n=\log(nM_n)$. Let $b_n\geq1$ be a $(1-(nM_n)^{-2})$ upper quantile of the maximum absolute sample coordinate among the active designs, the derivatives evaluated at the true maximizing actions, and the active history coordinates in action contrasts. Let $d_{m,n}$ be a $(1-(nM_n)^{-2})$ upper quantile of
\[
\max\{\|\widehat{\mathbf D}^{(m)}-\mathbf D^{(m)}\|_2,
\|\widehat{\boldsymbol{\mathcal J}}_m-\boldsymbol{\mathcal J}_m\|_2\},\quad m<M_n,
\]
and use $\|\widehat{\mathbf H}_{M_n}-\mathbf H_{M_n}\|_2$ at $m=M_n$. The hats denote empirical averages with divisor $n$, evaluated at the true parameters and true winning actions. The quantities $b_n$ and $d_{m,n}$ are deterministic. For any sufficiently large fixed $C_0$, define
\begin{equation}
\begin{split}
t_{m,n}&=C_0b_n\sqrt{s_n^{(m)}}\,r_m,\qquad
h_{m,n}=\sqrt{s_n^{(m)}}\,p'_{\lambda_n^{(m)}}(g_n^{(m)}),\\
q_{m,n}&=d_{m,n}(r_m+r_{m+1})+h_{m,n}\\
&\quad+b_n\sqrt{s_n^{(m)}}\,t_{m+1,n}
\{L_nt_{m+1,n}+x_n/n\},\qquad m<M_n,\\
q_{M_n,n}&=d_{M_n,n}r_{M_n}+h_{M_n,n}+\sqrt{s_n^{(M_n)}}\,r_{M_n}^2.
\end{split}
\label{eq:local-remainder-rates}
\end{equation}
Set
\begin{equation}
a_{m,n}=\sum_{j=m}^{M_n}\|\boldsymbol\Gamma_{mj}\|_2q_{j,n}.
\label{eq:stage-accumulation}
\end{equation}
The design conditions in Assumption~\ref{designm} also apply to $\mathbf W_m$.

\begin{ass}\label{normalm}
\begin{enumerate}
\item[(a)] Assume $\ell_1+3\ell_5+2\ell_6<1/2$, $\ell_5+\ell_6<1/3$, and $\delta^{(m)}$ in Assumption~\ref{difflbm} satisfies $\delta^{(m)}>\ell_2^{(m)}+\sum_{j\geq m}\ell_2^{(j)}$. Assume $\max_{m,k}\mathbb E|u_{mi,k}|^3<\infty$, $\lambda_{\min}(\boldsymbol\Omega_m)\geq c>0$, and $\|\boldsymbol{\mathcal J}_m\|_2=O(1)$, uniformly over stages. The terminal dispersion is a fixed positive constant.
\item[(b)] With $q_{j,n}$ defined in \eqref{eq:local-remainder-rates}, require
\begin{equation}
\sqrt n\max_{m<M_n}\sum_{j=m}^{M_n}
\big\|(\mathbf K^{(m)})^{-1/2}\mathbf D^{(m)}\mathbf E_m
\boldsymbol\Gamma_{mj}\big\|_2\,q_{j,n}\longrightarrow0.
\label{eq:normalized-accumulation}
\end{equation}
\end{enumerate}
The rate conditions are required for each sufficiently large fixed $C_0$ in \eqref{eq:local-remainder-rates}.
\end{ass}

\begin{ass}\label{B5m}
Use the increasing filtration $\mathcal F_m=\sigma(\mathbf X^{(1)},\mathbf A^{(1)},\ldots,\mathbf X^{(m)},\mathbf A^{(m)})$. For the extension with nonzero intermediate utilities, assume, for $i<j$, that $\mathbb E(Y^{(i)}\mid\mathcal F_j)$ is $\mathcal F_i$-measurable and $\operatorname{Cov}(Y^{(i)},Y^{(j)}\mid\mathcal F_j)=0$. In the primary terminal-outcome setting, $Y^{(i)}=0$ for $i<M_n$, so these utility conditions hold automatically.
\end{ass}
The concentration argument below uses the terminal-outcome setting, in which the stage residuals are martingale differences: $\varepsilon^{(m)}$ is $\mathcal F_{m+1}$-measurable and has conditional mean zero given $\mathcal F_m$, for $m<M_n$. The terminal residual has conditional mean zero given $\mathcal F_{M_n}$. To apply the same growing-stage result to nonzero intermediate utilities, additionally require this martingale-difference representation, with the conditional exponential-moment bound in Assumption~\ref{subgm}, for the score innovations.

\begin{ass}\label{propagationm}
Assume
\begin{equation}
\sup_{1\leq m\leq j\leq M_n}\|\boldsymbol\Gamma_{mj}\|_2
\leq C_\Gamma<\infty.
\label{eq:product-stability}
\end{equation}
\end{ass}

\begin{remark}\label{convergencemremark}
A sufficient condition for \eqref{eq:product-stability} is the existence of positive definite matrices $\mathbf G_m$ with $c\mathbf I\preceq\mathbf G_m\preceq C\mathbf I$ and $\mathbf B_m^\top\mathbf G_m\mathbf B_m\preceq\mathbf G_{m+1}$. Iteration gives $\boldsymbol\Gamma_{mj}^\top\mathbf G_m\boldsymbol\Gamma_{mj}\preceq\mathbf G_j$, hence $\|\boldsymbol\Gamma_{mj}\|_2\leq\sqrt{C/c}$.
\end{remark}

\begin{lemma}\label{lemma:accumulated-verification}
Suppose the sub-Gaussian or bounded sufficient conditions in Lemmas~\ref{lemmaA.1}--\ref{lemmaA.2} and \ref{lemmaA.1-Mn-1}--\ref{lemmaA.2-Mn-1} hold with constants uniform over stages. Retain the design, conditional moment, margin, and minimum-signal conditions above, and Assumption~\ref{propagationm}. Then, writing $x=\log(nM_nS_n)$, the quantiles can be chosen so that
\begin{equation}
b_n=O(\sqrt{x}),\qquad
\max_m d_{m,n}=O\{S_n\sqrt{x/n}\}.
\label{eq:verified-matrix-rates}
\end{equation}
For bounded designs, $b_n=O(1)$ suffices. Under Assumption~\ref{normalm}(a), and SCAD or MCP with the stated uniform signal separation, both \eqref{eq:consistency-accumulation} and \eqref{eq:normalized-accumulation} follow. For another penalty, it suffices additionally that $\sqrt n\sum_{j=1}^{M_n}h_{j,n}\to0$.
\end{lemma}

\subsection{Certainty-Equivalent Interpretation}\label{sec:ce-interpretation}

Let \(\psi=(b^{(2)\prime})^{-1}\). Deformation yields a conditional certainty-equivalent representation:
\begin{align}
\operatorname{CE}_{\psi}\!\left\{
\Gamma_0(\mathbf H^{(2)})
\,\middle|\,
\mathbf H^{(1)},\mathbf A^{(1)}
\right\}
&:=
\psi^{-1}\!\left[
\mathbb E\!\left[
\psi\{\Gamma_0(\mathbf H^{(2)})\}
\mid\mathbf H^{(1)},\mathbf A^{(1)}
\right]
\right]
\nonumber\\*
&=
b^{(2)'}\!\left\{
Q_{\mathrm{dfm}}^{(1)}(\mathbf H^{(1)},\mathbf A^{(1)})
\right\}.
\label{eq:CElink}
\end{align}
Because the terminal mean function is increasing, this representation induces the same action ordering as the deformed Q-function. For a logistic terminal model, \(Q_{\mathrm{dfm}}^{(1)}\) is the conditional mean of the optimized terminal log-odds. This corresponds to the upstream regression on the log-odds scale for discrete Q-learning discussed by \citet[Remark~2.4]{Linn2017IQ}.

\counterwithin{equation}{section}
\section{Proofs}\label{Proofs}

\subsection{Proofs for the Terminal Stage}

\begin{proof}[\textbf{Proof of Lemma}~\ref{lemmaA.1}]
	\item \textbf{First Part.} 
	Note that for any $\ell$, 
	\begin{equation}
		\max_{1\leq i \leq n} \max_j  \left| h_{i,j}^{(M_n)}\cdot \varphi_\ell(\mathbf{A}^{(M_n)})\right| \leq c_0 \max_{1\leq i \leq n} \max_j \left| h_{i,j}^{(M_n)} \right| ,\label{inftygood}
	\end{equation}
	thus 
	$$\max_{1\le j\le  p_n^{(M_n)}} \| \mathbf h^{(M_n)}_j\|_{\infty}\leq \max\{1,c_0\}  \max\left\{\max_{1\leq m\leq M_n} \max_{1\leq k\leq k_m} \|\mathbf{x}_k^{(m)}\|_\infty,\max_{1\leq m\leq M_n-1} \max_{1\leq k\leq a} \|\mathbf{a}_k^{(m)}\|_\infty\right\}. $$
	By Bernstein's inequality and Hoeffding's inequality (see Section 2 in \citet{vander2023}), 
	\begin{align*}
		&\mathbb{P}	\left(\max_{1\le j\le  p_n^{(M_n)}} \| \mathbf h^{(M_n)}_j\|_{\infty}>2 \, C_0\sqrt{\log(n\, K_nM_n)}\right)\\\lesssim  & \, n K_n M_n \max_{i,j} \mathbb{P}	\left(|  h^{(M_n)}_{i,j}|>2\, C_0\sqrt{\log(n\, K_nM_n)}\right)\\\leq&\, 2\exp\left\{-\frac{4C_0^2 \log(n \, K_n M_n)}{2C_0^2}+\log(n \, K_n M_n)\right\}\\\leq&\, O(\frac{1}{n K_nM_n}),
	\end{align*}
	which proves \eqref{g1} in Assumption~\ref{design2'}. Moreover, $n/\left(
	{s_n^{(M_n)}\,\log n}\right)\gtrsim \log(n \, M_n \, K_n)$ under the settings, which proves \eqref{g6}. 
	
	Next, note that
	\begin{equation}
		\left\|\left( h_{i,j}^{(M_n)}\cdot  \varphi_k(\mathbf{A}^{(M_n)}) \right)_{i=1}^n\right\|_2\leq c_0 \left\|\left( h_{i,j}^{(M_n)}\right)_{i=1}^n \right\|_2,\label{2good}
	\end{equation}
	thus $$\max_{1\leq j \leq p^{(M_n)}_n}\|\mathbf{h}_j^{(M_n)}\|_2 \leq \max\{1,c_0\} \max_{1\leq j \leq d^{(M_n)}} \|\mathbf{h}_j^{(M_n)}\|_2,$$
	which means we just need to bound the last term. Now
	\begin{align*}
		\max_{1\le j\le  d^{(M_n)}}\mathbb{E} [\| \mathbf h^{(M_n)}_j\|^2_{2}]&=n \,\max_{1\le j\le  d^{(M_n)}} \mathbb{E}[h^{(M_n)^2}_{1,j}]\\&\lesssim n \,\max_{1\le j\le d^{(M_n)}} \|h^{(M_n)}_{1,j}\|^2_{\psi_2}\\&\lesssim n.
	\end{align*}
	Moreover, by Bernstein's inequality and Bonferroni's inequality, 
	\begin{align*}
		&\mathbb{P}\left(\max_{1\leq j\leq d^{(M_n)}}\left|\sum_{i=1}^{n}h^{(M_n)^2}_{j,i}-\mathbb{E} [\| \mathbf h^{(M_n)}_j\|^2_{2}]\right|\geq 2 (C_0+\sqrt{C_0})\sqrt{n\log d^{(M_n)}}\right)\\\lesssim&\exp\left(\log d^{(M_n)}-c\min\left\{\frac{4C_0^2n\log d^{(M_n)}}{n \, C_0^2},\frac{4C_0\sqrt{n\log d^{(M_n)}}}{C_0 } \right\}\right)\\\leq&O\left(\frac{1}{K_nM_n}\right).
	\end{align*}
	But $\sqrt{n\log d^{(M_n)}}=O(n)$, which in turn ensures that condition \eqref{g2} is satisfied. 
	\item \textbf{Second Part.}  For \eqref{g3}, note that by~\eqref{r1}
	\begin{align*}
		\bm{\Lambda}^{(M_n)}= &\, \mathbb{E}\left[\mathbf{S}_1^{(M_n)}(\mathbf{S}_1^{(M_n)})^\top\right]\\=&\,\mathbb{E}
		\begin{bmatrix}
			\mathbf{r}_{k_1}\mathbf{r}^\top_{k_1}& \mathbf{r}_{k_1}\mathbf{r}^\top_{k_2}\cdot \varphi_{k_2}(\mathbf{A}^{(M_n)})  & \cdots & \mathbf{r}_{k_1}\mathbf{r}^\top_{k_r}\cdot  \varphi_{k_r}(\mathbf{A}^{(M_n)})  \\
			\mathbf{r}_{k_2}\mathbf{r}^\top_{k_1}\cdot \varphi_{k_2}(\mathbf{A}^{(M_n)})  & \mathbf{r}_{k_2}\mathbf{r}^\top_{k_2}\cdot  \varphi^2_{k_2}(\mathbf{A}^{(M_n)})  & \cdots & 0\\
			\vdots & \vdots & \ddots & \vdots \\
			\mathbf{r}_{k_r}\mathbf{r}^\top_{k_1}\cdot \varphi_{k_r}(\mathbf{A}^{(M_n)}) & 0 & \cdots & \mathbf{r}_{k_r}\mathbf{r}^\top_{k_r}\cdot  \varphi^2_{k_r}(\mathbf{A}^{(M_n)}) 
		\end{bmatrix}
		\\:=&\begin{bmatrix}
			\bm{\Sigma}_{1}&\mathbf{B}\\
			\mathbf{B}^\top&\bm{\Sigma}_2
		\end{bmatrix}
	\end{align*}
	where $\bm{\Sigma}_1=\mathbb{E}[\mathbf{r}_{k_1}\mathbf{r}^\top_{k_1}]$. Then by (a)(b)(c) and Schur complement, we have
	$$\lambda_{\min}\left[\bm{\Lambda}^{(M_n)}\right]=\min\{m_r, \lambda_{\min}[\bm{\Sigma}_1-\mathbf{B}\bm{\Sigma}^{-1}_2\mathbf{B}^\top]\}\geq \min\{m_r,c_1-\frac{M_r^2}{m_r}\}\geq\bar{c}.$$
	Thus
	\begin{align*}
		&\frac{1}{n}\lambda_{\min}\left[(\bm{\Psi}_1^{(M_n)})^\top\bm{\Psi}_1^{(M_n)}\right]\\=&\,\frac{1}{n}\inf_{\|\mathbf{a}\|_2=1}\mathbf{a}^\top(\bm{\Psi}_1^{(M_n)})^\top\bm{\Psi}_1^{(M_n)}\mathbf{a}\\\geq& \inf_{\|\mathbf{a}\|_2=1}\mathbf{a}^\top\bm{\Lambda}^{(M_n)}\mathbf{a}-\|\frac{1}{n}\sum_{i=1}^{n}\mathbf{S}_{i,1}^{(M_n)}(\mathbf{S}_{i,1}^{(M_n)})^\top-\bm{\Lambda}^{(M_n)}\|_2\\\geq&\, \bar{c}-\|\frac{1}{n}\sum_{i=1}^{n}\mathbf{S}_{i,1}^{(M_n)}(\mathbf{S}_{i,1}^{(M_n)})^\top-\bm{\Lambda}^{(M_n)}\|_2.
	\end{align*}
	However, we can prove that with probability at least $1-O(1/n^2)$, 
	\begin{equation}
		\sup_{\|a\|_2=1}\left|\frac{1}{n}\sum_{i=1}^{n}(\mathbf{a}^\top\mathbf{S}^{(M_n)}_{i,1})^2-\mathbf{a}^\top\bm{\Lambda}^{(M_n)}\mathbf{a}\right|=O\left(\sqrt{\frac{(s_n^{(M_n)})^2 \log n \log s_n^{(M_n) }}{n}}\right)=o(1),\label{barc}
	\end{equation}
	which implies
	$$\mathbb{P}\left[\frac{1}{n}\lambda_{\min}\left[(\bm{\Psi}_1^{(M_n)})^\top\bm{\Psi}_1^{(M_n)}\right]\geq \frac{\bar{c}}{2}\right]\to 1,$$
	and thus \eqref{g3} holds.
	
	In fact, $h_{i,k,1}h_{j,l,1}$ is sub-exponential with $\psi_1$ norm bounded by $C_0^2$ for $i,j,k,l$, see Section 2 in \citet{Vershynin2018}. Thus, by Bernstein's inequality and Bonferroni's inequality, 
	\begin{align*}
		&\mathbb{P}\left(	\|(\bm{\Psi}_1^{(M_n)})^\top\bm{\Psi}_1^{(M_n)}-n \bm{\Lambda}^{(M_n)}\|_{\max}>2(C_0+C_0^2)\sqrt{n \log n\log s_n^{(M_n)}}\right)\\=& \mathbb{P}\left(\max_{k,l}(\sum_{i=1}^{n}h_{k,i,1}h_{l,i,1}-n \mathbb{E}[h_{k,i,1}h_{l,i,1}])>2(C_0+C_0^2)\sqrt{n\log n \log s_n^{(M_n)}}\right)\\\leq &s_n^{(M_n)} \max_{k,l}\mathbb{P}\left((\sum_{i=1}^{n}h_{k,i,1}h_{l,i,1}-n \mathbb{E}[h_{k,i,1}h_{l,i,1}])>2(C_0+C_0^2)\sqrt{n\log n\log s_n^{(M_n)}}\right)\\\leq&2\exp (-\min\left\{\frac{4C_0^4 n\log n\log s_n^{(M_n)}}{n C_0^4},\frac{4C_0^2\sqrt{ n\log n \log s_n^{(M_n)}}}{C_0^2}\right\}+\log s_n^{(M_n)}) \\\leq &O(\frac{1}{n^2}),
	\end{align*}
	which implies with probability at least $1-O(1/n^2)$,
	\begin{align*}
		\|(\bm{\Psi}_1^{(M_n)})^\top\bm{\Psi}_1^{(M_n)}-n \, \bm{\Lambda}^{(M_n)}\|_{2}&\leq s_n^{(M_n)}\|(\bm{\Psi}_1^{(M_n)})^\top\bm{\Psi}_1^{(M_n)}-n \, \bm{\Lambda}^{(M_n)}\|_{\max}\\&\leq 2(C_0+C_0^2)\sqrt{n \,\log n (s_n^{(M_n)})^2  \log s_n^{(M_n) }}. 
	\end{align*}
	For \eqref{g4}, first notice that $
	\mathbb{E}[\mathbf{r}^\top\mathbf{r}]\leq s_n^{(M_n)}\max_j\mathbb{E}[r_j^2]\leq C_0^2 s_n^{(M_n)}$ by \eqref{maxxa}, where $\mathbf{r}=(r_1,\cdots,r_i,\cdots)^\top$ with $\operatorname{dim}\{\mathbf{r}\}\leq s_n^{(M_n)}$. Thus 
	\begin{align*}
		&\operatorname{tr}\left[\frac{1}{n}\sum_{i=1}^{n}\mathbf{S}_{i,1}^{(M_n)}(\mathbf{S}_{i,1}^{(M_n)})^\top\right]\\=&\frac{1}{n}\sum_{i=1}^{n}(\mathbf{S}_{i,1}^{(M_n)})^\top\mathbf{S}_{i,1}^{(M_n)}\\=&\frac{1}{n}\sum_{i=1}^{n}\left(\mathbf{r}_{i,1}^\top\mathbf{r}_{i,1}+\sum_{j\geq 2} \mathbf{r}_{i,k_j}^\top\mathbf{r}_{i,k_j}\cdot  \varphi^2_{k_j}(\mathbf{A}^{(M_n)}) \right)\\\leq&\frac{1}{n}\sum_{i=1}^{n}\left(\mathbf{r}_i^\top\mathbf{r}_i+\sum_{j\geq 2} \mathbf{r}_i^\top\mathbf{r}_i\cdot\varphi^2_{k_j}(\mathbf{A}^{(M_n)}) \right)\\\leq&\left(\frac{1+c^2_0}{n}\sum_{i=1}^{n}\mathbf{r}_i^\top\mathbf{r}_i-2\mathbb{E}[\mathbf{r}^\top\mathbf{r}]\right)+2C_0^2s_n^{(M_n)}.
	\end{align*}
	where $\mathbf{r}_i$ is the $i$-th sample from the population $\mathbf{r}$. Then, by Markov's inequality, Marcinkiewicz--Zygmund inequality, and Minkowski's inequality, we obtain
	\begin{align*}
		&\mathbb{P}\left(\operatorname{tr}\left[\frac{1}{n}\sum_{i=1}^{n}\mathbf{S}_{i,1}^{(M_n)}(\mathbf{S}_{i,1}^{(M_n)})^\top\right]>4C_0^2s_n^{(M_n)} \right)\\\leq & \, \mathbb{P}\left[\left|\sum_{i=1}^{n}\mathbf{r}_i^\top\mathbf{r}_i-n\mathbb{E}[\mathbf{r}^\top\mathbf{r}]\right|>nC_0^2s_n^{(M_n)}\right]\\\leq& \, \frac{ 1}{C_0^8 n^4 (s_n^{(M_n)})^4} \mathbb{E}\left|\sum_{i=1}^{n}\mathbf{r}_i^\top\mathbf{r}_i-n\mathbb{E}[\mathbf{r}^\top\mathbf{r}]\right|^4\\\lesssim& \, \frac{n^2\mathbb{E}[\mathbf{r}^\top\mathbf{r}-\mathbb{E}[\mathbf{r}^\top\mathbf{r}]]^4}{C_0^8 n^4(s_n^{(M_n)})^4}\\\lesssim& \,  \frac{n^2}{C_0^8 n^4(s_n^{(M_n)})^4}\left(\sum_{j=1}^{s_n^{(M_n)}}\|r^2_j\|_{L_4}\right)^4\\=& \, O\left(1/n^2\right),
	\end{align*}
	which implies \eqref{g4} with $f(n)=O(1/n^2)$.
	
	\item \textbf{Third Part.}
	If condition (d) holds, first note that
	\begin{align}
		\lambda_{\max}[\bm{\Lambda}^{(M_n)}]&=\sup_{\|\mathbf{a}\|_2=1}\mathbb{E}\left[\mathbf{a}^\top\mathbf{S}_1^{(M_n)}\right]^2\nonumber\\&=\sup_{\|\mathbf{a}\|_2=1}\mathbb{E}\left[\mathbf{a}_1^\top\mathbf{r}_1+\sum_{i=2}^{r}\mathbf{a}_i^\top\mathbf{r}_{k_i}\cdot  \varphi_{k_i}(\mathbf{A}^{(M_n)})\right]^2\nonumber\\&\lesssim\sup_{\|\mathbf{a}\|_2=1}\mathbb{E}\left[\mathbf{a}_1^\top\mathbf{r}_1\right]^2+\mathbb{E}\left[\sum_{i=2}^{r}\mathbf{a}_i^\top\mathbf{r}_{k_i}\cdot  \varphi_{k_i}(\mathbf{A}^{(M_n)})\right]^2\nonumber\\&=\sup_{\|\mathbf{a}\|_2=1}\mathbb{E}\left[\mathbf{a}_1^\top\mathbf{r}_1\right]^2+\sum_{i=2}^{r}\mathbb{E}\left[\left(\mathbf{a}_{k_i}^\top\mathbf{r}_{k_i}\right)^2\cdot  \varphi^2_{k_i}(\mathbf{A}^{(M_n)})\right]\nonumber\\&\leq \sup_{\|\mathbf{a}\|_2=1}\mathbb{E}\left[\mathbf{a}_1^\top\mathbf{r}_1\right]^2+c^2_0\sum_{i=2}^{r}\mathbb{E}\left[\mathbf{a}_i^\top\mathbf{r}_{k_i}\right]^2\nonumber\\&\leq \max\{1,c^2_0\}\max_i \|\mathbf{r}_{k_i}\|^2_{\psi_2} \sup_{\|\mathbf{a}\|_2=1} \,(\|\mathbf{a}_1\|_2^2+\sum_{i=2}^{r}\|\mathbf{a}_i\|_2^2)\nonumber\\&\leq \max\{1,c^2_0\} \max_i \|\mathbf{r}_{k_i}\|^2_{\psi_2}=O(1).\label{lammax1}
	\end{align}
	For \eqref{g5}, now it holds that
	\begin{align*}
		\sup_{\|a\|_2=1}\left|\mathbb{E}h^{(M_n)}_j\mathbf{a}^\top\mathbf{S}_1^{(M_n)}\right|^2&\leq \mathbb{E}\left|h^{(M_n)}_j\right|^2\sup_{\|a\|_2=1}\mathbb{ E }\left|\mathbf{a}^\top\mathbf{S}_1^{(M_n)}\right|^2\\&\leq\max_i {\mathbb{E}[r_i^2]}\, \lambda_{\max}[\bm{\Lambda}^{(M_n)}]\\&=O(1).
	\end{align*}
	This implies
	$$\|\mathbb{E}\,\mathbf{S}^{(M_n)}_2(\mathbf{S}^{(M_n)}_1)^\top\|_{2, \infty}=O(1).$$
	Similar to $\eqref{barc}$, with probability at least $1-O(1/p_n^{(M_n)})$
	$$\|\,\sum_{i=1}^{n}\mathbf{S}^{(M_n)}_{i,2}(\mathbf{S}^{(M_n)}_{i,1})^\top-n\, \mathbb{E}\,\mathbf{S}^{(M_n)}_2(\mathbf{S}^{(M_n)}_1)^\top\|_{2, \infty}\lesssim\sqrt{s_n^{(M_n)}n\log p_n^{(M_n)}}=O(n),$$
	which proves \eqref{g5}. For \eqref{g7}, using similar arguments in \eqref{inftygood}, \eqref{2good}, we just need to show 
	$$	\max_{1\le j\le  d^{(M_n)}}
	\lambda_{\max}\left[
	(	\bm{\Psi}^{(M_n)}_{1})^{T}
	\operatorname{diag} \{ | \mathbf {h}^{(M_n)}_j | \}
	\, \bm{\Psi}_{1}^{(M_n)}
	\right]= O(n).$$ Note that
	\begin{align*}
		&\mathbb{P}\left(\max_i\|\mathbf{S}_{i,1}^{(M_n)}\|_\infty>C_0{\log n} \right)\\\leq& 2\exp\left(-\frac{C_0^2\log^2n}{2 C_0^2}+\log (ns_n^{(M_n)})\right)\\\leq &O(\frac{1}{n^2}),
	\end{align*}
	thus define $\mathbf{Z}_{i,1}=\mathbf{S}_{i,1}^{(M_n)}\, I_{\{\|\mathbf{S}_{i,1}^{(M_n)}\|_\infty>C_0\log n\}}$, it suffices to show :
	\begin{align}
		&\max_j \lambda_{\max}(\sum_{i=1}^{n}\mathbf{Z}_{i,1}\mathbf{Z}_{i,1}^\top h^{(M_n)}_{i,j}-n \, \mathbb{E}[\mathbf{Z}_{i,1}\mathbf{Z}_{i,1}^\top h^{(M_n)}_{i,j}])=O_p(n),\label{g71}\\&\max_j \lambda_{\max}( \mathbb{E}[\mathbf{Z}_{i,1}\mathbf{Z}_{i,1}^\top h^{(M_n)}_{i,j}])=O(1).\label{g72}
	\end{align}
	In fact, 
	\begin{align*}
		&\mathbb{P}\left[\max_j\lambda_{\max}(\sum_{i=1}^{n}\mathbf{Z}_{i,1}\mathbf{Z}_{i,1}^\top h^{(M_n)}_{i,j}-n \, \mathbb{E}[\mathbf{Z}_{i,1}\mathbf{Z}_{i,1}^\top h^{(M_n)}_{i,j}])>Cn\right]\\\leq &\mathbb{P}\left[\max_j s_n^{(M_n)}\|\sum_{i=1}^{n}\mathbf{Z}_{i,1}\mathbf{Z}_{i,1}^\top h^{(M_n)}_{i,j}-n \, \mathbb{E}[\mathbf{Z}_{i,1}\mathbf{Z}_{i,1}^\top h^{(M_n)}_{i,j}]\|_{\max}>Cn\right]\\\leq &d^{(M_n)}\max_j\mathbb{P}\left[ \|\sum_{i=1}^{n}\mathbf{Z}_{i,1}\mathbf{Z}_{i,1}^\top h^{(M_n)}_{i,j}-n \, \mathbb{E}[\mathbf{Z}_{i,1}\mathbf{Z}_{i,1}^\top h^{(M_n)}_{i,j}]\|_{\max}>Cn/s_n^{(M_n)}\right],
	\end{align*}
	Now $|z_{k,i,1}z_{l,i,1}h^{(M_n)}_{i,j}|\leq C_0^2\log^2n|h^{(M_n)}_{i,j}|$ thus is sub-Gaussian with $\psi_2$ norm bounded by $c_1\log^2n$ for some $c_1$, thus the term above can be further bounded by
	\begin{align*}
		&d^{(M_n)}(s_n^{(M_n)})^2\exp\left(-\frac{C^2n^2}{2n(s_n^{(M_n)})^2c^2_1\log^4n}\right)\\\leq& \exp\left(-\frac{C^2n}{2c^2_1(s_n^{(M_n)})^2\log^4n}+3\log d^{(M_n)}\right)\\\leq&O(1/K_nM_n).
	\end{align*}
	for large $C$ in that $(s_n^{(M_n)})^2\log^4 n=O(n^{1-\max\{u,v\}})$, impling \eqref{g71}. Moreover,
	\begin{align*}
		&\max_j\sup_{\|\mathbf{a}\|_2=1}\mathbb{E}\left[h_{i,j}^{(M_n)}(\mathbf{a}^\top\mathbf{Z}_{i,1})^2\right]\\\leq&\max_j\sup_{\|\mathbf{a}\|_2=1}\sqrt{\mathbb{E}\left[h_{i,j}^{(M_n)}\right]^2\mathbb{E}\left[\mathbf{a}^\top\mathbf{Z}_{i,1}\right]^4}\\\lesssim&\sup_{\|\mathbf{a}\|_2=1}\sqrt{\mathbb{E}\left[\mathbf{a}^\top\mathbf{S}^{(M_n)}_{i,1}\right]^4}.
	\end{align*}
	Using similar arguments in \eqref{lammax1}, we can prove the last term is $O(1)$, thus \eqref{g72} holds. Finally, for Gaussian or logistic regression, $b''$ is bounded, thus using similar methods, with probability at least $1-O(1/n^2)$
	\begin{align*}
		&\|\mathbf{F}_n^{(M_n)}- n \, \mathbf{F}^{(M_n)}\|_2\\\leq& s_n^{(M_n)} \|\mathbf{F}_n^{(M_n)}- n \, \mathbf{F}^{(M_n)}\|_{\max}\\=&O\left(s_n^{(M_n)} \sqrt{n \log n \log s_n^{(M_n)}}\right)\\=&o(n/\sqrt{s_n^{(M_n)}}).
	\end{align*}
	which completes our proof.
\end{proof}

\begin{proof}[\textbf{Proof of Lemma}~\ref{lemmaA.2}]
	\item \textbf{First Part.} By applying similar arguments as in $\eqref{inftygood}$ and $\eqref{2good}$, and under condition \eqref{maxxa1}, it is straightforward to verify that \eqref{g1}, \eqref{g2}, and \eqref{g6} hold almost surely.
	
	\item \textbf{Second Part.}  From the proof of Lemma~\ref{lemmaA.1}, we know that $\lambda_{\min}\left[\bm{\Lambda}^{(M_n)}\right]\geq\bar{c},$ and
	\begin{align*}
		&\frac{1}{n}\lambda_{\min}\left[(\bm{\Psi}_1^{(M_n)})^\top\bm{\Psi}_1^{(M_n)}\right]\\\geq&\, \bar{c}-\|\frac{1}{n}\sum_{i=1}^{n}\mathbf{S}_{i,1}^{(M_n)}(\mathbf{S}_{i,1}^{(M_n)})^\top-\bm{\Lambda}^{(M_n)}\|_2.
	\end{align*}
	Using the same arguments in sub-Gaussian cases, with probability at least $1-O(1/n^2)$,
	\begin{align}
		&\|(\bm{\Psi}_1^{(M_n)})^\top\bm{\Psi}_1^{(M_n)}-\bm{\Lambda}^{(M_n)}\|_{2}\nonumber\\\leq& s_n^{(M_n)}\|(\bm{\Psi}_1^{(M_n)})^\top\bm{\Psi}_1^{(M_n)}-\bm{\Lambda}^{(M_n)}\|_{\max}\nonumber\nonumber\\=&O\left(\sqrt{n \log n \, (s_n^{(M_n)})^2  \log s_n^{(M_n) }}\right)\nonumber\\=&o(n)\label{lammaxdiff}. 
	\end{align}
	This implies \eqref{g3}.
	For \eqref{g4}, now
	\begin{align*}
		&\operatorname{tr}\left[\frac{1}{n}\sum_{i=1}^{n}\mathbf{S}_{i,1}^{(M_n)}(\mathbf{S}_{i,1}^{(M_n)})^\top\right]\\=&\frac{1}{n}\sum_{i=1}^{n}(\mathbf{S}_{i,1}^{(M_n)})^\top\mathbf{S}_{i,1}^{(M_n)}\\=&\frac{1}{n}\sum_{i=1}^{n}\left(\mathbf{r}_{i,1}^\top\mathbf{r}_{i,1}+\sum_{j\geq 2} \mathbf{r}_{i,k_j}^\top\mathbf{r}_{i,k_j}\cdot\varphi^2_{k_j}(\mathbf{A}^{(M_n)})\right)\\\leq&\frac{1}{n}\sum_{i=1}^{n}\left(\mathbf{r}_i^\top\mathbf{r}_i+\sum_{j\geq 2} \mathbf{r}_i^\top\mathbf{r}_i\cdot \varphi^2_{k_j}(\mathbf{A}^{(M_n)})\right)\\\leq&\frac{1+c_0^2}{n}\sum_{i=1}^{n}\mathbf{r}_i^\top\mathbf{r}_i\\\leq& 2C_0^2\,s_n^{(M_n)}.
	\end{align*}
	where $\mathbf{r}_i$ is the $i$-th sample from the population $\mathbf{r}$. 
	
	\item \textbf{Third Part.}
	If condition (d) holds, similarly
	\begin{align}
		\lambda_{\max}[\bm{\Lambda}^{(M_n)}]&=\sup_{\|\mathbf{a}\|_2=1}\mathbb{E}\left[\mathbf{a}^\top\mathbf{S}_1^{(M_n)}\right]^2\nonumber\\&\lesssim \sup_{\|\mathbf{a}\|_2=1}\mathbb{E}\left[\mathbf{a}_1^\top\mathbf{r}_1\right]^2+\sum_{i=2}^{r}\mathbb{E}\left[\mathbf{a}_i^\top\mathbf{r}_{k_i}\right]^2\nonumber\\&\leq \max_i\lambda_{\max}\left[\mathbb{E}[\mathbf{r}_{k_i}\mathbf{r}^\top_{k_i}]\right]\sup_{\|\mathbf{a}\|_2=1} \,(\|\mathbf{a}_1\|_2^2+\sum_{i=2}^{r}\|\mathbf{a}_i\|_2^2)\nonumber\\&\leq C_1.\label{lammax11}
	\end{align}
	For \eqref{g5}, now it holds that
	\begin{align*}
		\sup_{\|a\|_2=1}\left|\mathbb{E}h^{(M_n)}_j\mathbf{a}^\top\mathbf{S}_1^{(M_n)}\right|^2&\leq C_0^2\sup_{\|a\|_2=1}\mathbb{ E }\left|\mathbf{a}^\top\mathbf{S}_1^{(M_n)}\right|^2=O(1).
	\end{align*}
	Following the same methods in sub-Gaussian cases, we can get \eqref{g5}. For \eqref{g7}, note that
	\begin{align*}
		&\max_{j}
		\lambda_{\max}\left[
		(	\bm{\Psi}^{(M_n)}_{1})^{T}
		\operatorname{diag} \{ | \mathbf {h}^{(M_n)}_j | \}
		\, \bm{\Psi}_{1}^{(M_n)}
		\right]\\\leq&C_0
		\lambda_{\max}\left[
		(	\bm{\Psi}^{(M_n)}_{1})^{T}
		\, \bm{\Psi}_{1}^{(M_n)}
		\right].
	\end{align*}	
	However, combining \eqref{lammaxdiff} and \eqref{lammax11}, the last term is $O_p(n)$. The asserted bound follows by substituting this estimate into the preceding display, completing the proof.
\end{proof}

\begin{proof}[\textbf{Proof of Lemma}~\ref{FboundMn}]
	In fact, 
	\begin{align*}
		&\mathbb{E}\left[b''({\theta}_*^{(M_n)})(\mathbf{S}_1^{(M_n)})^\top\mathbf{S}_1^{(M_n)}\right]\\=&\sum_{j=1}^{s_n^{(M_n)}}\mathbb{E}\left[b''({\theta}_*^{(M_n)})(h_{j,1}^{(M_n)})^2\right]\\\lesssim&O(s_n^{(M_n)}).
	\end{align*}
	Moreover, if Assumption~\ref{normal2} further holds, then by Minkowski's inequality,
	\begin{align*}
		&\frac{1}{\sqrt{n}}\mathbb{E}\left|(\varepsilon_i^{(M_n)})^2 (\mathbf{S}_{i,1}^{(M_n)})^\top(\mathbf{F}^{(M_n)})^{-1}\mathbf{S}_{i,1}^{(M_n)}\right|^{3/2}\\\leq &\frac{1}{c\sqrt{n}}\mathbb{E}\left|(\varepsilon_i^{(M_n)})^2 (\mathbf{S}_{i,1}^{(M_n)})^\top\mathbf{S}_{i,1}^{(M_n)}\right|^{3/2}\\=&\frac{1}{c\sqrt{n}}\mathbb{E}\left|\sum_{j=1}^{s_n^{(M_n)}}(\varepsilon_i^{(M_n)}h^{(M_n)}_{i,j,1})^2 \right|^{3/2}\\\leq &\frac{1}{c\sqrt{n}}\left(\sum_{j=1}^{s_n^{(M_n)}}\left(\mathbb{E}\left|\varepsilon_i^{(M_n)}h^{(M_n)}_{i,j,1}\right|^3 \right)^{2/3}\right)^{3/2}\\\lesssim&\sqrt{\frac{(s_n^{(M_n)})^3}{n}}\\=&o(1).
	\end{align*}
	This completes the proof.
\end{proof}

\begin{proof}[\textbf{Proof of Proposition}~\ref{convergence2}]
	In the following analysis, we restrict our attention to the event $\mathcal{W}_n$ where all matrix norms in Assumption~\ref{design2} are uniformly bounded by $\bar{C}$. We have $\mathbb{P}[\mathcal{W}_n]=1-f(n)$.
	\item \textbf{First Part (Consistency).}

	To prove the conclusions, it suffices to show that under the given regularity conditions, there exists a strict local maximizer \( \bm{\widehat{\beta}}^{(M_n)} \) of the penalized likelihood \( {L}^{(M_n)} (\bm{\beta}^{(M_n)}) \) such that 
	\item (a) \( \bm{\widehat{\beta}}^{(M_n)}_2 = 0 \) with probability tending to 1 as \( n \to \infty \) , and 
	\item (b) \( \|\widehat{\bm{\beta}}^{(M_n)}_1 - \bm{\beta}^{(M_n)}_{*,1}\|_2 = O_p(\sqrt{s_n^{(M_n)}/n}) \) .
	
	\subsubsection*{Step 1: Consistency in the \( s_n^{(M_n)} \)-Dimensional Subspace}
	
	We first constrain \( {L}^{(M_n)} (\bm{\beta}^{(M_n)})  \) on the \( s_n^{(M_n)} \)-dimensional subspace \( \{ \bm{\beta} \in \mathbb{R}^{p_n^{(M_n)}} : \bm{\beta}_{2} = 0 \} \). This constrained penalized loss is given by
	\begin{equation}
		\overline{L} (\bm{\delta}) =  \overline{\ell} (\bm{\delta}) - \sum_{j=1}^{s_n^{(M_n)}} p_{\lambda^{(M_n)}} (|\delta_j|) 
	\end{equation}
	where $
	\overline{\ell} (\bm{\delta}) = n^{-1} [({\mathbf{Y}}^{(M_n)})^\top \bm{\Psi}_1 ^{(M_n)}\bm{\delta} -\mathbf{1}^\top\mathbf{b}( \bm{\Psi}^{(M_n)}_1 \bm{\delta})]$
	and $
	\bm{\delta} = (\delta_1, \dots, \delta_{s_n^{(M_n)}})^\top.
	$ We now show that there exists a strict local maximizer \( \widehat{\bm{\beta}}^{(M_n)}_1 \) of \( \overline{L} (\bm{\delta}) \) such that \( \|\widehat{\bm{\beta}}^{(M_n)}_1 - \bm{\beta}^{(M_n)}_{*,1}\|_2 = O_p(\sqrt{s_n^{(M_n)}/n}) \). To this end, we define an event
	\[
	H_n = \left\{ \overline{L} (\bm{\beta}^{(M_n)}_{*,1}) > \max_{\bm{\delta} \in \partial N_{\tau}} \overline{L} (\bm{\delta}) \right\}
	\]
	where \( \partial N_{\tau} \) denotes the boundary of the closed set 
	\[
	N_{\tau} = \{ \bm{\delta} \in \mathbb{R}^{s_n^{(M_n)}} : \|\bm{\delta} - \bm{\beta}^{(M_n)}_{*,1}\|_2 \leq \sqrt{s_n^{(M_n)}/n} \tau \}
	\]
	and \( \tau \in (0, \infty) \). Clearly, under the event \( H_n \), there exists a local maximizer \( \widehat{\bm{\beta}}^{(M_n)}_1 \) of \( \overline{L} (\bm{\delta}) \) in \( N_{\tau} \). Thus it suffices to show that \( \mathbb{P}(H_n)\to 1 \) as \( n \to \infty \) for sufficient large \( \tau \). 
	
	Let \( n \) be sufficiently large such that \( \sqrt{s_n^{(M_n)}/n} \tau \leq g^{(M_n)}_n \). Since \( g^{(M_n)}_n \gg \sqrt{s_n^{(M_n)}/n}  \) by Assumption~\ref{compare2}, it is easy to see that $
	\bm{\delta} = (\delta_1, \dots, \delta_{s_n^{(M_n)}})^\top \in N_{\tau} \text{ entails } \text{sgn} (\bm{\delta}) = \text{sgn} (\bm{\beta}^{(M_n)}_{*,1}), \|\bm{\delta} - \bm{\beta}^{(M_n)}_{*,1}\|_{\infty} \leq g^{(M_n)}_n, \text{ and } \min_j |\delta_j| \geq g_n^{(M_n)}.
	$
	
	Define $\bar{\rho}(t)=\operatorname{sgn}(t)\rho'(|t|)$, and $\bar{\rho}(\mathbf{v})=(\bar{\rho}(v_1),\cdots,\bar{\rho}(v_q))^\top$ for $\mathbf{v}=(v_1,\cdots,v_q)^\top$. Then by Taylor’s expansion,we have for any \( \bm{\delta} \in N_{\tau} \),
	\begin{equation}
		\overline{L} (\bm{\delta}) - \overline{L} (\bm{\beta}^{(M_n)}_{*,1}) = (\bm{\delta} - \bm{\beta}^{(M_n)}_{*,1})^\top \mathbf{v} - \frac{1}{2} (\bm{\delta} -\bm{\beta}^{(M_n)}_{*,1})^\top \mathbf{D} (\bm{\delta} - \bm{\beta}^{(M_n)}_{*,1})
	\end{equation}
	where 
	\begin{align*}
		&	\mathbf{v} = n^{-1} (\bm{\Psi}^{(M_n)}_1)^\top [{\mathbf{Y}}^{(M_n)}-  \bm{\mu}^{(M_n)}(\bm{\theta}^{(M_n)}_*)] - \bar{\rho}_{\lambda_n} (\bm{\beta}^{(M_n)}_{*,1}),\\&  \bm{\theta}^{(M_n)}_* = \bm{\Psi}^{(M_n)} \bm{\beta}^{(M_n)}_{*} = \bm{\Psi}^{(M_n)}_1\bm{\beta}^{(M_n)}_{*,1},
	\end{align*}
	and
	\[
	\mathbf{D} = n^{-1} (\bm{\Psi}_1)^\top\bm{\Sigma}^{(M_n)}(\bar{\bm{\theta}})\bm{\Psi}^{(M_n)}_1 + \text{diag} \left\{ p''_{\lambda_n} (|\bar{\bm{\beta}}|) \right\},
	\]
	$\bar{\bm{\theta}}=\bm{\Psi}^{(M_n)}_1\bar{\bm{\beta}}$, where $\bar{\bm{\beta}}$ lies on the line segment joining $\bm{\delta} \text{ and } \bm{\beta}^{(M_n)}_{*,1}.
	$ More generally, when the second derivative of the penalty function \( p_{\lambda}^{(M_n)} \) does not necessarily exist, the second part of the matrix \( \mathbf{D} \) can be replaced by a diagonal matrix with maximum absolute element bounded by \( \lambda^{(M_n)} \kappa_0 \). Define
	\[
	\mathcal{N}_0 = \{ \mathbf{b} \in \mathbb{R}^{s_n^{(M_n)}} : \|\mathbf{b} - \bm{\beta}^{(M_n)}_{*,1}\|_{\infty} \leq g^{(M_n)}_n \}
	\]
	and \( \kappa_0 = \max_{\mathbf{b} \in \mathcal{N}_0} \kappa(\rho; \mathbf{b}) \), where \( \kappa(\rho; \mathbf{b}) \) is given by (5). For any \( \bm{\delta} \in \partial N_{\tau} \), we have \( \|\bm{\delta} - \bm{\beta}^{(M_n)}_{*,1}\|_2 = \sqrt{s_n^{(M_n)}/n} \tau \) and \( \bar{\bm{\beta}} \in \mathcal{N}_0 \). Then for sufficiently large \( n \), by Assumption~\ref{design2} and \( \lambda_n \kappa_0 = o(1) \) in Assumption~\ref{compare2} we have
	\[
	\lambda_{\min} (\mathbf{D}) \geq c - \lambda_n \kappa_0 \geq \frac{c}{2}.
	\]
	Thus, 
	\[
	\max_{\bm{\delta} \in \partial N_{\tau}} \left\{\overline{L} (\bm{\delta}) - \overline{L} (\bm{\beta}^{(M_n)}_{*,1})\right\} \leq \sqrt{s_n^{(M_n)}/n} \tau \left( \|\mathbf{v}\|_2 - c \sqrt{s_n^{(M_n)}/n} \tau / 4 \right),
	\]
	which entails that
	\[
	P(H_n) \geq P \left( \|\mathbf{v}\|_2^2 < \frac{c^2 s_n^{(M_n)} \tau^2}{16n} \right).
	\]
	Now 
	\begin{align}
		\|\mathbf{v}\|_2^2 =& \left\|n^{-1} (\bm{\Psi}^{(M_n)}_1)^\top [{\mathbf{Y}}^{(M_n)}-  \bm{\mu}(\bm{\theta}^{(M_n)}_*)] - \bar{\rho}_{\lambda_n} (\bm{\beta}^{(M_n)}_{*,1})\right\|_2^2\nonumber\\\lesssim&\left\|n^{-1} (\bm{\Psi}^{(M_n)}_1)^\top [{\mathbf{Y}}^{(M_n)}-\bm{\mu}(\bm{\theta}^{(M_n)}_*)] \right\|_2^2+\left\|\bar{\rho}_{\lambda_n} (\bm{\beta}^{(M_n)}_{*,1})\right\|_2^2\nonumber\\:=&I_1+I_2.\label{I1I2}
	\end{align}
	By Lemma~\ref{FboundMn}, 
	\begin{align*}
		\mathbb{E}[I_1]=&\mathbb{E}\left\|n^{-1} (\bm{\Psi}^{(M_n)}_1)^\top [{\mathbf{Y}}^{(M_n)}-\bm{\mu}^{(M_n)}(\bm{\theta}^{(M_n)}_*)] \right\|_2^2\\=&n^{-2}\mathbb{E}\left\{\mathbb{E}\left[(\bm{\varepsilon}^{(M_n)})^\top\bm{\Psi}^{(M_n)}_1(\bm{\Psi}^{(M_n)}_1)^\top\bm{\varepsilon}^{(M_n)}\mid \bm{\Psi}^{(M_n)}_1 \right]\right\}\\=&n^{-2}\operatorname{tr}\left\{\mathbb{E}\left[(\bm{\Psi}^{(M_n)}_1)^\top\bm{\Sigma}^{(M_n)}\bm{\Psi}^{(M_n)}_1\right]\right\}\, \phi^{(M_n)}\\=&n^{-2}\mathbb{E}\left[\sum_{i=1}^{n}b''({\theta}_{*,i}^{(M_n)})(\mathbf{S}^{(M_n)}_{i,1})^\top\mathbf{S}^{(M_n)}_{i,1}\right]\, \phi^{(M_n)}\\=&O(\frac{s_n^{(M_n)}}{n}).
	\end{align*}
	
	Moreover, $I_2=O_p(s_n^{(M_n)}/n)$ since \( p'_{\lambda_n} (t) \) is decreasing in \( t \in [0, \infty) \). Hence, $P(H_n) \to 1$ for large $\tau$. This proves \( \|\widehat{\bm{\beta}}^{(M_n)}_1 - \bm{\beta}^{(M_n)}_{*,1}\|_2 = O_P(\sqrt{s_n^{(M_n)}/n}) \).
	
	\subsubsection*{Step 2: Sparsity}
	
	Let \( \widehat{\bm{\beta}} \in \mathbb{R}^{p_n^{(M_n)}} \) with \( \widehat{\bm{\beta}}^{(M_n)}_1 \in N_{\tau} \subset \mathcal{N}_0 \) a strict local maximizer of \( \overline{L}(\bm{\delta}) \) and \( \widehat{\bm{\beta}}^{(M_n)}_2=0 \), and \( \widehat{\bm{\theta}}^{(M_n)} = \bm{\Psi}^{(M_n)} \widehat{\bm{\beta}}^{(M_n)} \). It remains to prove that the vector \( \widehat{\bm{\beta}}^{(M_n)} \) is indeed a strict local maximizer of \( \widehat{L} (\bm{\beta}) \) on the space \( \mathbb{R}^{p_n^{(M_n)}} \). From the proof in \citet{Fan2011}, it suffices to show
	$$\|\mathbf{w}^{(M_n)}_1\|_\infty<\rho'(0+),$$ where 
	\begin{equation}
		\mathbf{w}^{(M_n)}_1=(n\lambda^{(M_n)})^{-1}(\bm{\Psi}^{(M_n)}_2)^\top({\mathbf{Y}}^{(M_n)}-\bm{\mu}(\widehat{\bm{\theta}}^{(M_n)})).\label{w1Mn}
	\end{equation}
	Now
	\begin{align*}
		\|\mathbf{w}^{(M_n)}_1\|_\infty\leq &\left\|(n\lambda^{(M_n)})^{-1}(\bm{\Psi}^{(M_n)}_2)^\top({\mathbf{Y}}^{(M_n)}-\bm{\mu}^{(M_n)}({\bm{\theta}}_*^{(M_n)}))\right\|_\infty\\&+\left\|(n\lambda^{(M_n)})^{-1}(\bm{\Psi}^{(M_n)}_2)^\top(\bm{\mu}({\bm{\theta}}_*^{(M_n)})-\bm{\mu}(\widehat{\bm{\theta}}^{(M_n)}))\right\|_\infty\\=&I'_1+I'_2.
	\end{align*}
	Let \begin{align}
		\bm{\xi}^{(M_n)} =&( (\bm{\xi}^{(M_n)}_1)^\top,(\bm{\xi}_2^{(M_n)})^\top )^\top=(\xi^{(M_n)}_1, \dots, \xi^{(M_n)}_{p_n^{(M_n)}})^\top \nonumber\\=& (\bm{\Psi}^{(M_n)})^\top {\mathbf{Y}}^{(M_n)} - (\bm{\Psi}^{(M_n)})^\top \bm{\mu}^{(M_n)}(\bm{\theta}_*^{(M_n)}),\label{xiMn}
	\end{align}
	and consider the event
	\[
	\mathcal{E}_2 = \left\{ \|\bm{\xi}^{(M_n)}_2 \|_{\infty} \leq u_n \sqrt{n} \right\}
	\]
	where \( u_n = C\sqrt{\log p_n^{(M_n)} } \). 
	Under assumption~\ref{subG_mom},
	\begin{align*}
		&\max_j	\mathbb{P} \Big( \left| ({\mathbf{h}}_j^{(M_n)})^\top \mathbf{Y}^{(M_n)} - ({\mathbf{h}}_j^{(M_n)})^\top \bm{\mu}^{(M_n)} (\bm{\theta}_*) \right| > \varepsilon \mid \{{\mathbf{H}}_i^{(M_n)},{\mathbf{A}}_i^{(M_n)}\}_{i=1}^n \Big) \\\leq& 2\, \exp\left(-\frac{1}{2}\frac{\varepsilon^2}{\|\mathbf{h}_j\|^2_2 \, v_0+\|\mathbf{h}_j\|_\infty M\varepsilon}\right)\\\leq & 2\, \exp\left(-\frac{1}{2}\frac{\varepsilon^2}{n \bar{C}^2 \, v_0+\sqrt{n/\log p_n^{(M_n)}}\bar{C} M\varepsilon}\right).
	\end{align*}
	Then for large $C$, 
	\begin{align*}
		P (\mathcal{E}_2) \geq& 1 - (p_n^{(M_n)} - s_n^{(M_n)}) \mathbb{P}(|\xi^{(M_n)}_j|>u_n\sqrt{n})\\\gtrsim& 1-2\, \, \exp\left(-\frac{1}{2}\frac{C^2\log p_n^{(M_n)}}{ \bar{C}^2 \, v_0+C\bar{C} M}+\log p_n^{(M_n)}\right)\\\gtrsim&1-O({1}/{p_n^{(M_n)}}).
	\end{align*}
	which implies $$I_1'=O_p\left(\frac{\sqrt{n \log p_n^{(M_n)} }}{n\lambda^{(M_n)}}\right)=o_p(1).$$

	Next, by Taylor's expansion, under Assumption~\ref{design2},
	\begin{align*}
		&(\bm{\Psi}^{(M_n)}_2)^\top(\bm{\mu}(\widehat{\bm{\theta}}^{(M_n)})-\bm{\mu}({\bm{\theta}}_*^{(M_n)}))\\=&(\bm{\Psi}^{(M_n)}_2)^\top\bm{\Sigma}^{(M_n)}(\bm{\theta}_*^{(M_n)})\bm{\Psi}^{(M_n)}_1(\widehat{\bm{\beta}}^{(M_n)}_1 - \bm{\beta}^{(M_n)}_{*,1})+\bm{\eta}
	\end{align*}
	with 
	$$\|\bm{\eta}\|_\infty\leq (\bar{C}n)\, O_p(\|\widehat{\bm{\beta}}^{(M_n)}_1 - \bm{\beta}^{(M_n)}_{*,1}\|^2_2)=O_p(s_n^{(M_n)}).$$
	Thus,
	\begin{align*}
		I_2'=&\left\|(n\lambda^{(M_n)})^{-1}(\bm{\Psi}^{(M_n)}_2)^\top(\bm{\mu}({\bm{\theta}}_*^{(M_n)})-\bm{\mu}(\widehat{\bm{\theta}}^{(M_n)}))\right\|_\infty\\\leq &(n\lambda^{(M_n)})^{-1}\|(\bm{\Psi}^{(M_n)}_2)^\top\bm{\Sigma}^{(M_n)}(\bm{\theta}_*^{(M_n)})\bm{\Psi}^{(M_n)}_1(\widehat{\bm{\beta}}^{(M_n)}_1 - \bm{\beta}^{(M_n)}_{*,1})\|_\infty+O_p\!\left(\frac{s_n^{(M_n)}}{n \lambda^{(M_n)}}\right)\\\leq &(n\lambda^{(M_n)})^{-1} \, (\bar{C}n)\, \|(\widehat{\bm{\beta}}^{(M_n)}_1 - \bm{\beta}^{(M_n)}_{*,1})\|_2+O_p\!\left(\frac{s_n^{(M_n)}}{n \lambda^{(M_n)}}\right)\\=&O_p\!\left(\frac{\sqrt{s_n^{(M_n)}}}{\sqrt{n} \lambda^{(M_n)}}\right)+O_p\!\left(\frac{s_n^{(M_n)}}{n \lambda^{(M_n)}}\right)\\=&o_p(1).
	\end{align*}
	Thus, we complete the first part.

	\item \textbf{Second Part (Asymptotic normality).}

	We only need to prove the asymptotic normality of \( \widehat{\bm{\beta}}^{(M_n)}_1 \). Now on the event \( H_n \), \( \widehat{\bm{\beta}}^{(M_n)}_1 \in N_{\tau} \subset \mathcal{N}_0 \) is a strict local maximizer of \( \overline{L} (\bm{\delta}) \) and \( \widehat{\bm{\beta}}^{(M_n)}_2 = 0 \). It follows easily that \( \nabla \overline{L} (\widehat{\bm{\beta}}^{(M_n)}_1) = 0 \), thus by Taylor's expansion, 
	\begin{align*}
		0 =& \nabla \overline{L} (\widehat{\bm{\beta}}^{(M_n)}_1) \\=&n^{-1} [(\bm{\Psi}_1 ^{(M_n)})^\top{\mathbf{Y}}^{(M_n)} -(\bm{\Psi}_1^{(M_n)})^\top\bm{\mu}^{(M_n)}(\bm{\theta}_*^{(M_n)})]\\&-n^{-1}[(\bm{\Psi}_1^{(M_n)})^\top\bm{\Sigma}^{(M_n)}(\bm{\theta}_*^{(M_n)}) \bm{\Psi}^{(M_n)}_1](\widehat{\bm{\beta}}_1^{(M_n)}- {\bm{\beta}}_{*,1}^{(M_n)})\\ &+\bar{p}_{\lambda^{(M_n)}}(\widehat{\bm{\beta}}_1^{(M_n)})\\&+O_p((s_n^{(M_n)})^3/n)\\:=&I''_1-I''_2+I''_3+o_p(1).
	\end{align*}
	It follows from \( \widehat{\bm{\beta}}_1 \in \mathcal{N}_0 \), and \( p'_{\lambda^{(M_n)}} (g^{(M_n)}_n) = o(1/\sqrt{n s_n^{(M_n)}}) \)  that
	\[
	\left\| \bar{p}_{\lambda_n} (\widehat{\bm{\beta}}^{(M_n)}_1) \right\|_2 \leq O_p(\sqrt{s_n^{(M_n)}} ) \, p'_{\lambda_n} (g^{(M_n)}_n) = o_p(1/\sqrt{n}),
	\]
	which means $\|I''_3\|_2=o_p(1/\sqrt{n})$. Now for any $\mathbf{C}^{(M_n)}\in \mathbb{R}^{q\times s_n^{(M_n)}}$, let
	\begin{align*}
		&\mathbf{F}^{(M_n)}=\mathbb{E}[b''(\theta_*^{(M_n)})\mathbf{S}^{(M_n)}_1(\mathbf{S}^{(M_n)}_1)^\top]\\&\mathbf{F}_n^{(M_n)}={(\bm{\Psi}^{(M_n)}_1)^\top\bm{\Sigma}^{(M_n)}\bm{\Psi}^{(M_n)}_1} 
	\end{align*}  
	Then by Assumption~\ref{normal2} (a), 
	\begin{align*}
		&\frac{1}{\sqrt{n}}(\mathbf{F}^{(M_n)})^{-1/2}\mathbf{F}_n^{(M_n)}(\widehat{\bm{\beta}}_1^{(M_n)}- {\bm{\beta}}_{*,1}^{(M_n)})\\=&\frac{1}{\sqrt{n}}(\mathbf{F}^{(M_n)})^{-1/2}[(\bm{\Psi}_1 ^{(M_n)})^\top\bm{\varepsilon}^{(M_n)}]+o_p(1).
	\end{align*}
	Then for any unit vector $\mathbf{a}$, we just need to focus on
	$$\frac{1}{\sqrt{n}}\mathbf{a}^\top\mathbf{C}^{(M_n)}(\mathbf{D}^{(M_n)})^{-1/2}[(\bm{\Psi}_1 ^{(M_n)})^\top\bm{\varepsilon}^{(M_n)}]:=\sum_{i=1}^{n}\xi_i$$
	with independent components
	$$\xi_i=\frac{1}{\sqrt{n}}\mathbf{a}^\top\mathbf{C}^{(M_n)}(\mathbf{F}^{(M_n)})^{-1/2}[\mathbf{S}_{i,1}^{(M_n)}{\varepsilon}_i^{(M_n)}].$$  Thus
	\begin{align*}
		&\sum_{i=1}^{n}\operatorname{Var}\xi_i\\=&\frac{1}{n}\sum_{i=1}^{n} \phi^{(M_n)}\, \mathbf{a}^\top \,\mathbf{C}^{(M_n)}(\mathbf{F}^{(M_n)})^{-1/2}\, \mathbf{F}^{(M_n)} (\mathbf{F}^{(M_n)})^{-1/2}(\mathbf{C}^{(M_n)})^\top\mathbf{a}\\\to &\phi^{(M_n)} \, \mathbf{a}^\top\mathbf{G}^{(M_n)}\mathbf{a}.
	\end{align*}
	Moreover, by Lemma~\ref{FboundMn},
	\begin{align*}
		\sum_{i=1}^{n}\mathbb{E}|\xi_i|^3=&n^{-3/2}\sum_{i=1}^{n}\mathbb{E}\left|\mathbf{a}^\top\mathbf{C}^{(M_n)}(\mathbf{F}^{(M_n)})^{-1/2}[\mathbf{S}_{i,1}^{(M_n)}{\varepsilon}_i^{(M_n)}]\right|^3\\\leq &n^{-3/2}\sum_{i=1}^{n}\|\mathbf{a}^\top\mathbf{C}^{(M_n)}\|_2^3\,  \mathbb{E}\left\|(\mathbf{F}^{(M_n)})^{-1/2}\mathbf{S}_{i,1}^{(M_n)}{\varepsilon}_i^{(M_n)}\right\|_2^3\\\lesssim& n^{-3/2}\sum_{i=1}^{n}\,  \mathbb{E}\left|({\varepsilon}_i^{(M_n)})^2 (\mathbf{S}_{i,1}^{(M_n)})^\top(\mathbf{F}^{(M_n)})^{-1}\mathbf{S}_{i,1}^{(M_n)}\right|^{3/2}\\\to&0.
	\end{align*}
	Thus, by Lyapunov's Theorem, we complete the part. If further assume $$\|{\mathbf{F}^{(M_n)}_n}-n \, \mathbf{F}^{(M_n)}\|_2=o_p(n / \sqrt{s_n^{(M_n)}}),$$ then
	\begin{align*}
		&(	{\mathbf{F}^{(M_n)}_n}-n \, \mathbf{F}^{(M_n)})(\widehat{\bm{\beta}}^{(M_n)}_1 - \bm{\beta}^{(M_n)}_{*,1})\\\leq &\|	{\mathbf{F}^{(M_n)}_n}-n \, \mathbf{F}^{(M_n)}\|_2\|\widehat{\bm{\beta}}^{(M_n)}_1 - \bm{\beta}^{(M_n)}_{*,1}\|_2\\\leq &o_p(n / \sqrt{s_n^{(M_n)}})\, O_p(\sqrt{s_n^{(M_n)}/n})\\=&o_p(\sqrt{n}),
	\end{align*}
	which means we can replace $\mathbf{F}_n^{(M_n)}$ with $n \, \mathbf{F}^{(M_n)}$ in $I''_2$, all subsequent results then follow immediately.
\end{proof}

\begin{proof}[\textbf{Proof of Theorem}~\ref{2-opt}]
	\item \textbf{First Part-I (Consistency of optimal Q-value estimation).}

	In the two-stage setting, under the polynomial growth rates stated in Theorem~\ref{2-opt}, consistency \(\widehat\mu_*^{(2)}-\mu_*^{(2)}\xrightarrow{p}0\) already follows from Proposition~\ref{convergence2}, Assumptions~\ref{maxmincd}--\ref{Eb''}, and \(\ell_2^{(2)}<1/3\).

	By Proposition~B.1, if we define 
	$$\mathcal{S}_n=\left\{\|\widehat{\bm{\beta}}^{(M_n)}-\bm{\beta}_*^{(M_n)} \|_2\leq\log n \sqrt{s_n^{(M_n)}/n}\right\}\cap\left\{\widehat{\bm{\beta}}^{(M_n)}_2=\mathbf{0}\right\}$$
	then $\mathbb{P}[\mathcal{S}_n]\to 1$. Now note that
	\begin{align*}
		&\mathbb{P}_N \widehat{{Q}}_*^{(M_n)}(\mathcal{H}^{(M_n)}) - \mathbb{E}[{Q}_*^{(M_n)}(\mathcal{H}^{(M_n)})]\\=&\mathbb{P}_N \left\{b'\left[(\widehat{\bm{\beta}}^{(M_n)})^\top\mathcal{S}_{\widehat{\pi}_*^{(M_n)}}\right]\right\}-\mathbb{E} \left\{b'\left[({\bm{\beta}}_*^{(M_n)})^\top\mathcal{S}_{\pi_*^{(M_n)}}\right] \right\}\\=&\mathbb{P}_N \left\{b'\left[(\widehat{\bm{\beta}}^{(M_n)})^\top\mathcal{S}_{\widehat{\pi}_*^{(M_n)}}\right]-b'\left[(\widehat{\bm{\beta}}^{(M_n)})^\top\mathcal{S}_{{\pi}_*^{(M_n)}}\right]\right\}\\&+ \mathbb{E} \left\{b'\left[({\widehat{\bm{\beta}}}^{(M_n)})^\top\mathcal{S}_{\pi_*^{(M_n)}}\right] \right\} -\mathbb{E} \left\{b'\left[({{\bm{\beta}}}_*^{(M_n)})^\top\mathcal{S}_{\pi_*^{(M_n)}}\right] \right\} \\&+\mathbb{P}_N \left\{b'\left[({\bm{\beta}}_*^{(M_n)})^\top\mathcal{S}_{\pi_*^{(M_n)}}\right]\right\}-\mathbb{E} \left\{b'\left[({{\bm{\beta}}}_*^{(M_n)})^\top\mathcal{S}_{\pi_*^{(M_n)}}\right] \right\}\\&+(\mathbb{P}_N-\mathbb{E}) \left\{b'\left[(\widehat{\bm{\beta}}^{(M_n)})^\top\mathcal{S}_{\pi_*^{(M_n)}}\right]-b'\left[({\bm{\beta}}_*^{(M_n)})^\top\mathcal{S}_{\pi_*^{(M_n)}}\right]\right\}  \\:=&S_1+S_2+S_3+S_4.
	\end{align*}
	For $S_1$, by the definition of $\widehat{\pi}_*^{(M_n)}$ and $\pi_*^{(M_n)}$, under event $\mathcal{S}_n$ we have 
	\begin{align*}
		0\leq& S_1=\mathbb{P}_N \left\{b'\left[(\widehat{\bm{\beta}}^{(M_n)})^\top\mathcal{S}_{\widehat{\pi}_*^{(M_n)}}\right]-b'\left[(\widehat{\bm{\beta}}^{(M_n)})^\top\mathcal{S}_{{\pi}_*^{(M_n)}}\right]\right\}\\\leq &\mathbb{P}_N \left\{b'\left[(\widehat{\bm{\beta}}^{(M_n)})^\top\mathcal{S}_{\widehat{\pi}_*^{(M_n)}}\right]-b'\left[(\widehat{\bm{\beta}}^{(M_n)})^\top\mathcal{S}_{{\pi}_*^{(M_n)}}\right]\right\}\\&-\left(\mathbb{P}_N \left\{b'\left[({\bm{\beta}}_*^{(M_n)})^\top\mathcal{S}_{\widehat{\pi}_*^{(M_n)}}\right]-b'\left[({\bm{\beta}}_*^{(M_n)})^\top\mathcal{S}_{{\pi}_*^{(M_n)}}\right]\right\}\right)\\=&\mathbb{P}_N \left\{b'\left[(\widehat{\bm{\beta}}^{(M_n)})^\top\mathcal{S}_{\widehat{\pi}_*^{(M_n)}}\right]-b'\left[({\bm{\beta}}_*^{(M_n)})^\top\mathcal{S}_{\widehat{\pi}_*^{(M_n)}}\right]\right\}\\&-\left(\mathbb{P}_N \left\{b'\left[(\widehat{\bm{\beta}}^{(M_n)})^\top\mathcal{S}_{{\pi}_*^{(M_n)}}\right]-b'\left[({\bm{\beta}}_*^{(M_n)})^\top\mathcal{S}_{{\pi}_*^{(M_n)}}\right]\right\}\right)\\=&\mathbb{P}_N \left\{b''(\xi_n)\mathcal{S}_{\widehat{\pi}_*^{(M_n)}}^\top\right\}\left(\widehat{\bm{\beta}}^{(M_n)}-{\bm{\beta}}_*^{(M_n)}\right)\\&-\mathbb{P}_N \left\{b''(\zeta_n)\mathcal{S}_{{\pi}_*^{(M_n)}}^\top\right\}\left(\widehat{\bm{\beta}}^{(M_n)}-{\bm{\beta}}_*^{(M_n)}\right)\\=&\mathbb{P}_N \left\{b''(\xi_n)\mathcal{S}_{\widehat{\pi}_*^{(M_n)}}^\top-b''(\zeta_n)\mathcal{S}_{{\pi}_*^{(M_n)}}^\top\right\}\left(\widehat{\bm{\beta}}^{(M_n)}-{\bm{\beta}}_*^{(M_n)}\right) \\\leq& \| \widehat{\bm{\beta}}^{(M_n)}-\bm{\beta}_*^{(M_n)}\|_1 \, \left\| \mathbb{P}_N \left\{b''(\xi_n)\mathcal{S}_{\widehat{\pi}_*^{(M_n)}}^\top-b''(\zeta_n)\mathcal{S}_{{\pi}_*^{(M_n)}}^\top\right\}_{\{\operatorname{supp}(\bm{\beta}_*^{(M_n)})\}}\right\|_\infty\\\lesssim &\sqrt{s_n^{(M_n)}} \,  \log n\sqrt{ \, s_n^{(M_n)}/n} \, \max_j \max_{1\leq i \leq N}|\mathcal{R}_{i,j}|\, C_n\\\lesssim&\log n \, s_n^{(M_n)}/\sqrt{n} \, O_p\left(\log^{1/r}(Ns_n^{(M_n)})\right)C_n.
	\end{align*}
	where $C_n=\max_{i}\{b''(\xi_{n,i}), b''(\zeta_{n,i})\}$ with $\xi_{n}=\bm{\xi}_n^\top\mathcal{S}_{\widehat{\pi}_*^{(M_n)}}$, $\zeta_{n}=\bm{\zeta}_n^\top\mathcal{S}_{{\pi}_*^{(M_n)}}$. Here $\bm{\xi}_n$ and $\bm{\zeta}_{n}$ are vectors between $\widehat{\bm{\beta}}^{(M_n)}$ and ${\bm{\beta}}_*^{(M_n)}$. Now under Assumption~\ref{maxmincd}, if $b''$ is unbounded, then 
	$$\max _i \left\{({\bm{\beta}}_*^{(M_n)})^\top\mathcal{S}_{{\pi}_*^{(M_n)},i} \right\}\leq D+o_p(1),$$ thus
	\begin{align*}
		&\max_i \left\{(\widehat{\bm{\beta}}^{(M_n)})^\top\mathcal{S}_{{\pi}_*^{(M_n)},i} \right\}\\\leq&\max_i \left|(\widehat{\bm{\beta}}^{(M_n)}-{\bm{\beta}}_*^{(M_n)})^\top\mathcal{S}_{{\pi}_*^{(M_n)},i} \right|+D+o_p(1)\\\leq&\log n \, s_n^{(M_n)}/\sqrt{n}\,  O_p\left(\log^{1/r}(Ns_n^{(M_n)})\right)+D+o_p(1)\\=&D+o_p(1),
	\end{align*}
	similarly $\min_i \left\{(\widehat{\bm{\beta}}^{(M_n)})^\top\mathcal{S}_{{\pi}_*^{(M_n)},i}\right\} \geq C+o_p(1)$, and similar bounds can be established for $\xi_{n,i}$. Thus $C_n=O_p(1)$, and then $S_1=o_p(1)$.

	For $S_2$, 
	\begin{align*}
		S_2&=\mathbb{E} \left\{b'\left[({\widehat{\bm{\beta}}}^{(M_n)})^\top\mathcal{S}_{\pi_*^{(M_n)}}\right] -b'\left[({{\bm{\beta}}}_*^{(M_n)})^\top\mathcal{S}_{\pi_*^{(M_n)}}\right] \right\} \\&=\mathbb{E}\left\{b''(\zeta_n)\mathcal{S}_{\pi_*^{(M_n)}}^\top\right\}(\widehat{\bm{\beta}}^{(M_n)}-{\bm{\beta}}_*^{(M_n)})\\&\lesssim \left\|\mathbb{E}\left\{b''(\zeta_n)\mathcal{S}_{\pi_*^{(M_n)}}\right\}\right\|_\infty O_p(s_n^{(M_n)}\log n/\sqrt{n})
	\end{align*}
	However, under Assumption~\ref{Eb''},  $\left\|\mathbb{E}\left\{b''(\zeta_n)\mathcal{S}_{\pi_*^{(M_n)}}\right\}\right\|_\infty =O_p(1)$. Thus $S_2=o_p(1)$.
	
	For \( S_3 \), note that under Assumption~\ref{Eb''},
	$$ \mathbb{E} \left|b'\left[({{\bm{\beta}}}_*^{(M_n)})^\top\mathcal{S}_{\pi_*^{(M_n)}}\right] \right|<\infty ,$$ thus
	$$S_3=\mathbb{P}_N \left\{b'\left[({\bm{\beta}}_*^{(M_n)})^\top\mathcal{S}_{\pi_*^{(M_n)}}\right]\right\}-\mathbb{E} \left\{b'\left[({{\bm{\beta}}}_*^{(M_n)})^\top\mathcal{S}_{\pi_*^{(M_n)}}\right] \right\} $$
	is a sum of independent and identically distributed (i.i.d.) variables with a finite mean. Consequently, by the Law of Large Numbers (LLN), it immediately follows that \( S_3 = o_p(1) \).
	
	For $S_4$, using similar arguments in $S_1$ and $S_2$, 
	\begin{align*}
		S_4=&(\mathbb{P}_N-\mathbb{E}) \left\{b'\left[(\widehat{\bm{\beta}}^{(M_n)})^\top\mathcal{S}_{\pi_*^{(M_n)}}\right]-b'\left[({\bm{\beta}}_*^{(M_n)})^\top\mathcal{S}_{\pi_*^{(M_n)}}\right]\right\} \\=&(\mathbb{P}_N-\mathbb{E}) \left\{b''(\zeta_n)\mathcal{S}_{\pi_*^{(M_n)}}^\top\right\}(\widehat{\bm{\beta}}^{(M_n)}-\bm{\beta}_*^{(M_n)}) \\\leq&\| \widehat{\bm{\beta}}^{(M_n)}-\bm{\beta}_*^{(M_n)}\|_1 \, \left\|((\mathbb{P}_N-\mathbb{E}) \left\{b''(\zeta_n)\mathcal{S}_{\pi_*^{(M_n)}}^\top\right\})_{\{\operatorname{supp}(\bm{\beta}_*^{(M_n)})\}}\right\|_\infty\\\lesssim&\sqrt{s_n^{(M_n)}} \, \log n\sqrt{ \, s_n^{(M_n)}/n} \max_j \, |\frac{1}{N}\sum_{i=1}^{N}\mathcal{R}_{i,j}b''(\zeta_{n,i})-\mathbb{E}[\mathcal{R}_{i,j}b''(\zeta_{n,i})]|.
	\end{align*}
	However,
	\begin{align*}
		&\max_j \, |\frac{1}{N}\sum_{i=1}^{N}\mathcal{R}_{j,i}b''(\zeta_{n,i})-\mathbb{E}[\mathcal{R}_{j,i}b''(\zeta_{n,i})]|\\=&O_p\left(\sqrt{\mathbb{E}\left[\max_j \, |\frac{1}{N}\sum_{i=1}^{N}\mathcal{R}_{j,i}b''(\zeta_{n,i})-\mathbb{E}[\mathcal{R}_{j,i}b''(\zeta_{n,i})]|\right]^2}\right)\\\leq& O_p\left( \sqrt{s_n^{(M_n)}/N} \, \max_j\sqrt{\mathbb{E}\left[\mathcal{R}_{j}b''(\zeta_{n})\right]^2}\right)\\=&O_p(\sqrt{s_n^{(M_n)}/N})
	\end{align*}
	Thus $S_4=o_p(1)$, which completes the proof of the first part-I.

	\item \textbf{First Part-II (Asymptotic normality of optimal Q-value estimation).}
	
	Now $S_4=o_p(1/\sqrt{n})$. By Taylor's expansion and variance conditions,
	\begin{align*}
		S_2&=\mathbb{E} \left\{b'\left[({\widehat{\bm{\beta}}}^{(M_n)})^\top\mathcal{S}_{\pi_*^{(M_n)}}\right]\right\} -\mathbb{E}\left\{b'\left[({{\bm{\beta}}}_*^{(M_n)})^\top\mathcal{S}_{\pi_*^{(M_n)}}\right] \right\}\\&=\mathbb{E}\left\{b''\left[({{\bm{\beta}}}_*^{(M_n)})^\top\mathcal{S}_{\pi_*^{(M_n)}}\right]\mathcal{S}_{\pi_*^{(M_n)}}^\top \right\}(\widehat{\bm{\beta}}^{(M_n)}-\bm{\beta}_*^{(M_n)})+o_p(n^{-1/2}).
	\end{align*}
	Thus, the asymptotic distribution of $S_2$ follows from Proposition~B.1. Furthermore, the asymptotic distribution of $S_3$ can be calculated by classical CLT under the conditions. Moreover, they are independent by definition. If either limiting variance is zero, the corresponding centered influence term converges to zero in probability by Chebyshev's inequality. The normal limit therefore also includes zero total variance, with \(\mathcal N(0,0)\) interpreted as a point mass at zero. 
	
	Now it suffices to show that $S_1=o_p(1/\sqrt{n})$. Define the event
	\begin{align}
		&\mathcal{T}=\left\{\widehat{\pi}_*^{(M_n)}(\mathcal{H}^{(M_n)})\neq \pi_*^{(M_n)}(\mathcal{H}^{(M_n)}) \right\},\label{mathcalt}
	\end{align}
	then
	\begin{align}
		S_1\leq&\mathbb{P}_N \left\{b'\left[(\widehat{\bm{\beta}}^{(M_n)})^\top\mathcal{S}_{\widehat{\pi}_*^{(M_n)}}\right]-b'\left[({\bm{\beta}}_*^{(M_n)})^\top\mathcal{S}_{\widehat{\pi}_*^{(M_n)}}\right]\right\}\nonumber\\&-\left(\mathbb{P}_N \left\{b'\left[(\widehat{\bm{\beta}}^{(M_n)})^\top\mathcal{S}_{{\pi}_*^{(M_n)}}\right]-b'\left[({\bm{\beta}}_*^{(M_n)})^\top\mathcal{S}_{{\pi}_*^{(M_n)}}\right]\right\}\right)\nonumber \\ =&\mathbb{P}_N \left\{\left(b''(\xi_n)\mathcal{S}_{\widehat{\pi}_*^{(M_n)}}^\top-b''(\zeta_n)\mathcal{S}_{{\pi}_*^{(M_n)}}^\top\right)\cdot I_{\mathcal{T}}\right\}\left(\widehat{\bm{\beta}}^{(M_n)}-{\bm{\beta}}_*^{(M_n)}\right) \nonumber \\\leq& \| \widehat{\bm{\beta}}^{(M_n)}-\bm{\beta}_*^{(M_n)}\|_1 \, \left\| \mathbb{P}_N \left\{\left(b''(\xi_n)\mathcal{S}_{\widehat{\pi}_*^{(M_n)}}^\top-b''(\zeta_n)\mathcal{S}_{{\pi}_*^{(M_n)}}^\top\right)\cdot I_{\mathcal{T}}\right\}_{\{\operatorname{supp}(\bm{\beta}_*^{(M_n)})\}}\right\|_\infty\nonumber\\\lesssim &\sqrt{s_n^{(M_n)}} \,  \log n\sqrt{ \, s_n^{(M_n)}/n} \, \max_j \max_{1\leq i \leq N}|\mathcal{R}_{j,i}|\, C_n\, \mathbb{P}_N( I_{\mathcal{T}})\nonumber\\\lesssim&\log n \, s_n^{(M_n)}/\sqrt{n} \, O_p\left(\log^{1/r}(Ns_n^{(M_n)})\right)O_p(1) \, \mathbb{P}_N(1_{\mathcal{T}}), \label{pnT1}
	\end{align}
	where the second equality holds in that, if $\pi_{*}^{(M_n)}=\widehat{\pi}_{*}^{(M_n)}$, then $\xi_{n}=\zeta_n$ and $\mathcal{S}_{\widehat{\pi}_*^{(M_n)}}=\mathcal{S}_{{\pi}_*^{(M_n)}}$.
	Define 
	\begin{align*}
		&C_{k,j}=-C_{j,k}=(\widehat{\bm{\eta}}^{(M_n)}_k-\bm{\eta}^{(M_n)}_{*,k}-\widehat{\bm{\eta}}^{(M_n)}_j+\bm{\eta}^{(M_n)}_{*,j})^\top\mathcal{H}^{(M_n)},\\&D_{k,j}=-D_{j,k}= (\bm{\eta}^{(M_n)}_{*,k})^\top\mathcal{H}^{(M_n)} -(\bm{\eta}^{(M_n)}_{*,j})^\top\mathcal{H}^{(M_n)},
	\end{align*}
	then under event $\mathcal{S}_n$
	\begin{align*}
		|C_{k,j}|\leq&|(\widehat{\bm{\eta}}^{(M_n)}_k-\bm{\eta}^{(M_n)}_{*,k})^\top\mathcal{H}^{(M_n)}|+|(\widehat{\bm{\eta}}^{(M_n)}_j-\bm{\eta}^{(M_n)}_{*,j})^\top\mathcal{H}^{(M_n)}|\\\leq &2\max_k|(\widehat{\bm{\eta}}^{(M_n)}_k-\bm{\eta}^{(M_n)}_{*,k})^\top\mathcal{H}^{(M_n)}|\\\leq& 2\sqrt{s_n^{(M_n)}}\|\widehat{\bm{\beta}}^{(M_n)}-\bm{\beta}^{(M_n)}_{*}\|_2\, \max_j|\mathcal{R}_{j}|\\\leq&2\, s_n^{(M_n)}\log n/\sqrt{n} \max_j|\mathcal{R}_{j}|.
	\end{align*}
	Now under Assumption~\ref{Eb''}, for large $C$,
	\begin{align}
		&\mathbb{P}\left[\max_j|\mathcal{R}_{j}|>C \, \log^{1/r}(ns_n^{(M_n)} )\right]\nonumber\\\lesssim&s_n^{(M_n)}\exp(-\frac{\, C^r \log(ns_n^{(M_n)})}{O(1)})\nonumber\\=&n^{-{C^r}/{O(1)}}\, (s_n^{(M_n)})^{1-{C^r}/{O(1)}}\label{log1r},
	\end{align}
	thus choosing large $C_1>C/2^r$, \begin{equation}
		\mathbb{P}\left[\max_{k,j}|C_{k,j}|\geq \frac{2 C_1 s_n^{(M_n)}\log^{1+1/r} n}{  \sqrt{n}}\right]\leq O(\frac{1}{n^2}).
	\end{equation}
	Note that
	$$(\mathcal{T} \cap \, \mathcal{U}_n)\subseteq\left\{\,  \exists j\neq k, \, (\widehat{\bm{\eta}}^{(M_n)}_j)^\top\mathcal{H}^{(M_n)}\geq (\widehat{\bm{\eta}}^{(M_n)}_{k})^\top\mathcal{H}^{(M_n)} ,\,  \pi_*^{(M_n)}(\mathcal{H}^{(M_n)})\in\mathcal{A}_{(k)} \right\},$$
	As a result, under Assumption~\ref{difflb}, let $k_*$ be the index of $\pi_*^{(M_n)}(\mathcal{H}^{(M_n)})$, 
	\begin{align*}
		\mathbb{P}[\mathcal{T} \cap \, \mathcal{U}_n]\leq&\mathbb{P}\left[ \, \exists j\neq k_*, C_{j,k_*}\geq (\bm{\eta}^{(M_n)}_{*,k_*})^\top\mathcal{H}^{(M_n)} -(\bm{\eta}^{(M_n)}_{*,j})^\top\mathcal{H}^{(M_n)}	\, \right]\\\leq&\mathbb{P}\left[ \, \exists j\neq k_*,   C_{j,k_*}\geq \frac{2 C_1 s_n^{(M_n)}\log^{1+1/r} n}{  \sqrt{n}}\, \right]\\&+\mathbb{P}\left[\, \inf_{j\neq k_*}\left\{D_{k_*,j}\right\}\leq\frac{2 C_1 s_n^{(M_n)}\log^{1+1/r} n}{  \sqrt{n}}	\,   \right]\\\leq &O\left(\frac{1}{n^2s_n^{(M_n)}}\right)+O\left(\frac{2 C_1 s_n^{(M_n)}L_n\log^{1+1/r} n}{  \sqrt{n}}\right).
	\end{align*}
	Thus \begin{align*}
		\mathbb{P}[\mathcal{T}]&\leq\mathbb{P}[\, \mathcal{U}_n^c]+\mathbb{P}[\mathcal{T}\cap \mathcal{U}_n]\\&\leq O\left(\frac{1}{n^{\delta}}\right)+O\left(\frac{1}{n^2s_n^{(M_n)}}\right)+O\left(\frac{2 C_1 s_n^{(M_n)}L_n\log^{1+1/r} n}{  \sqrt{n}}\right),
	\end{align*}
	combining it with \eqref{pnT1}, $|S_1|=o_p(1/\sqrt{n})$, which completes the proof.
	
	\item \textbf{Second Part (Consistency and Asymptotic normality of value-difference estimation).}
	
	For a given policy $\pi^{(M_n)}$, note that
	\begin{align*}
		&\mathbb{P}_N \widehat{{Q}}_*^{(M_n)}(\mathcal{H}^{(M_n)})-\mathbb{P}_N \widehat{{Q}}_{\pi^{(M_n)}}^{(M_n)}(\mathcal{H}^{(M_n)}) -\mathbb{E}[{Q}_*^{(M_n)}(\mathcal{H}^{(M_n)})]+ \mathbb{E}[{Q}_{\pi^{(M_n)}}^{(M_n)}(\mathcal{H}^{(M_n)})]\\=&\mathbb{P}_N \widehat{{Q}}_*^{(M_n)}(\mathcal{H}^{(M_n)})-\mathbb{P}_N \widehat{{Q}}^{(M_n)}(\mathcal{H}^{(M_n)},\pi_{*}^{(M_n)})\\&+\mathbb{E}\left\{ \widehat{{Q}}_{\pi_{*}^{(M_n)}}^{(M_n)}(\mathcal{H}^{(M_n)})-\widehat{{Q}}_{\pi^{(M_n)}}^{(M_n)}(\mathcal{H}^{(M_n)})- {{Q}}_{\pi_{*}^{(M_n)}}^{(M_n)}(\mathcal{H}^{(M_n)})+{{Q}}_{\pi^{(M_n)}}^{(M_n)}(\mathcal{H}^{(M_n)})\right\}\\&+(\mathbb{P}_N-\mathbb{E})\left\{ {{Q}}_{\pi_{*}^{(M_n)}}^{(M_n)}(\mathcal{H}^{(M_n)})-{{Q}}_{\pi^{(M_n)}}^{(M_n)}(\mathcal{H}^{(M_n)})\right\}\\&+(\mathbb{P}_N-\mathbb{E})\left\{ \widehat{{Q}}_{\pi_{*}^{(M_n)}}^{(M_n)}(\mathcal{H}^{(M_n)})-\widehat{{Q}}_{\pi^{(M_n)}}^{(M_n)}(\mathcal{H}^{(M_n)})- {{Q}}_{\pi_{*}^{(M_n)}}^{(M_n)}(\mathcal{H}^{(M_n)})+{{Q}}_{\pi^{(M_n)}}^{(M_n)}(\mathcal{H}^{(M_n)})\right\}\\:=&S'_1+S'_2+S'_3+S'_4.
	\end{align*}
	The consistency and asymptotic normality arguments follow by applying the decomposition used in \textbf{First Part-I} and \textbf{First Part-II} to \(S'_1,\ldots,S'_4\), with the candidate policy terms replacing the optimal-policy terms in the linear expansion.
\end{proof}

\subsection{\texorpdfstring{Proofs for $(M_n \! -\!1)$-th Stage}{Proofs for the (Mn - 1)-th Stage}}

\counterwithin{equation}{section}

\begin{proof}[\textbf{Proof of Lemma}~\ref{lemmaA.1-Mn-1} and \ref{lemmaA.2-Mn-1}]
	Since $p_n^{(M_n)}>p_n^{(M_n-1)}, d_n^{(M_n)}>d_n^{(M_n-1)}$, now
	\begin{align*}
		\bm{\Lambda}^{(M_n-1)}= &\, \mathbb{E}\left[\mathbf{S}_{\mathrm{dfm},1}^{(M_n-1)}(\mathbf{S}_{\mathrm{dfm},1}^{(M_n-1)})^\top\right]\\=&\,\mathbb{E}
		\begin{bmatrix}
			\mathbf{r}_{k_1}\mathbf{r}^\top_{k_1}\cdot\varphi_{k_1}(\mathbf{A}^{(M_n-1)})& 0& \cdots & 0 \\
			0 & \mathbf{r}_{k_2}\mathbf{r}^\top_{k_2}\cdot \varphi_{k_2}(\mathbf{A}^{(M_n-1)}) & \cdots & 0\\
			\vdots & \vdots & \ddots & \vdots \\
			0& 0 & \cdots & \mathbf{r}_{k_r}\mathbf{r}^\top_{k_r}\cdot \varphi_{k_r}(\mathbf{A}^{(M_n-1)})
		\end{bmatrix},
	\end{align*}
	which is block diagonal. Therefore, ensuring a lower bound for the minimum eigenvalue only requires a lower bound for $m_r.$ The arguments for \eqref{g1Mn-1}, \eqref{g2Mn-1}, \eqref{g3Mn-1}, \eqref{g4Mn-1}, \eqref{g5Mn-1} and \eqref{normal200} follow the corresponding terminal-stage bounds after replacing the terminal design by the deformed upstream design. For \eqref{g6M_n-1}, by the proof in Lemma~\ref{lemmaA.1} or \ref{lemmaA.2} (see the third part), it naturally holds in that the indicator-structure remains the same for any policy $\pi^{(M_n)}$. The verification of \eqref{g7M_n-1} and \eqref{g8M_n-1} follows from the same cross-block concentration argument used for \eqref{g5}, together with the identity \(\|X\|_{\psi_2}=\||X|\|_{\psi_2}\) for sub-Gaussian random variables. Finally for \eqref{normal221} and \eqref{normal222}, the proofs are similar to \eqref{normal22}.
\end{proof}

\begin{proof}[\textbf{Proof of Lemma}~\ref{Kbound}]
	
The first two assertions follow from $\mathbf F^{(M_n-1)}\preceq\mathbf K^{(M_n-1)}$ and the moment bounds as in Lemma~\ref{FboundMn}. For the terminal contribution, let
\[
\mathbf T=\mathbf J^{(M_n-1)}\mathbf P^{(M_n)}\mathbf V^{(M_n)}/\phi^{(M_n)}.
\]
Then $\mathbf T\boldsymbol\Omega_{M_n}\mathbf T^\top=\mathbf J^{(M_n-1)}\mathbf P^{(M_n)}\mathbf V^{(M_n)}(\mathbf P^{(M_n)})^\top(\mathbf J^{(M_n-1)})^\top\preceq\mathbf K^{(M_n-1)}$. Consequently,
\[
\mathbf T^\top(\mathbf K^{(M_n-1)})^{-1}\mathbf T\preceq\boldsymbol\Omega_{M_n}^{-1}.
\]
The terminal covariance lower bound and the coordinate third moments therefore give the required terminal Lyapunov bound of order $(s_n^{(M_n)})^{3/2}/\sqrt n=o(1)$.

\end{proof}

\begin{proof}[\textbf{Proof of Theorem}~\ref{convergence1}]
	
	\item \textbf{First Part (Consistency).}
	Now 
	\begin{align*}
		\widehat{\bm{\alpha}}^{(M_n-1)}&:=\arg \min_{\bm{\alpha}^{(M_n-1)}} \widehat{L}^{(M_n-1)}(\bm{\alpha}^{(M_n-1)})\\&= \arg \min_{\bm{\alpha}^{(M_n-1)}} \frac{1}{2}\, \mathbb{P}_n \left(\widehat{Y}^{(M_n-1)}_{\mathrm{dfm}}-Q^{(M_n-1)}_{\mathrm{dfm}}(\mathbf{H}^{(M_n-1)},\mathbf{A}^{(M_n-1)})\right)^2+p_{\lambda^{(M_n-1)}_n}(\bm{\alpha}^{(M_n-1)}).
	\end{align*}
	Here we define
	\begin{align*}
		&\widehat{Y}^{(M_n-1)}_{\mathrm{dfm}}
		=
		Y^{(M_n-1)}
		+
		\widehat{\tilde{Q}}^{(M_n)}_{\mathrm{dfm},*}
		=
		Y^{(M_n-1)}
		+
		(\widehat{\bm{\alpha}}^{(M_n)})^\top\mathbf{S}^{(M_n)}_{\mathrm{dfm},\widehat{\pi}_*^{(M_n)}},\\
		&\tilde{Y}^{(M_n-1)}_{\mathrm{dfm}}
		=
		Y^{(M_n-1)}
		+
		{\tilde{Q}}^{(M_n)}_{\mathrm{dfm},*}
		=
		Y^{(M_n-1)}
		+
		({\bm{\alpha}}_*^{(M_n)})^\top\mathbf{S}^{(M_n)}_{\mathrm{dfm},{\pi}_*^{(M_n)}}.
	\end{align*}

	To prove the conclusions, it suffices to show that under the given regularity conditions, there exists a strict local minimizer \( \bm{\widehat{\alpha}}^{(M_n-1)} \) of the penalized loss \( \widehat{L}^{(M_n-1)} (\bm{\alpha}^{(M_n-1)}) \) such that 
	\item (a) \( \bm{\widehat{\alpha}}^{(M_n-1)}_2 = 0 \) with probability tending to 1 as \( n \to \infty \) , and 
	\item (b) \( \|\widehat{\bm{\alpha}}^{(M_n-1)}_1 - \bm{\alpha}^{(M_n-1)}_{*,1}\|_2 = O_p(\sqrt{(s_n^{(M_n-1)}+s_n^{(M_n)})/n}) \) .
	
	\subsubsection*{Step 1: Consistency in the \( s_n^{(M_n-1)} \)-Dimensional Subspace}
	
	We first constrain \( \widehat{L}^{(M_n-1)} (\bm{\alpha}^{(M_n-1)})  \) on the \( s_n^{(M_n-1)} \)-dimensional subspace \( \{ \bm{\alpha} \in \mathbb{R}^{p_n^{(M_n-1)}} : \bm{\alpha}_{2} = 0 \} \). This constrained penalized loss is given by
	\begin{equation}
		\overline{L} (\bm{\delta}) = - \overline{\ell} (\bm{\delta}) + \sum_{j=1}^{s_n^{(M_n-1)}} p_{\lambda^{(M_n-1)}} (|\delta_j|) 
	\end{equation}
	where
\[
\begin{aligned}
\overline{\ell}(\bm{\delta})
&=n^{-1}\Bigl[(\widehat{\mathbf{Y}}^{(M_n-1)}_{\mathrm{dfm}})^\top
\bm{\Psi}_{\mathrm{dfm},1}^{(M_n-1)}\bm{\delta}\\
&\qquad-\frac{1}{2}\,\bm{\delta}^\top
(\bm{\Psi}_{\mathrm{dfm},1}^{(M_n-1)})^\top
\bm{\Psi}^{(M_n-1)}_{\mathrm{dfm},1}\bm{\delta}\Bigr]
\end{aligned}
\]
and $\bm{\delta}=(\delta_1,\dots,\delta_{s_n^{(M_n-1)}})^\top.$ We now show that there exists a strict local minimizer \( \widehat{\bm{\alpha}}^{(M_n-1)}_1 \) of \( \overline{L} (\bm{\delta}) \) such that \( \|\widehat{\bm{\alpha}}^{(M_n-1)}_1 - \bm{\alpha}^{(M_n-1)}_{*,1}\|_2 = O_p(\sqrt{(s_n^{(M_n-1)}+s_n^{(M_n)})/n}) \). Define an event
	\[
	H_n = \left\{ \overline{L} (\bm{\alpha}^{(M_n-1)}_{*,1}) < \min_{\bm{\delta} \in \partial N_{\tau}} \overline{L} (\bm{\delta}) \right\}
	\]
	where \( \partial N_{\tau} \) denotes the boundary of 
	\[
	N_{\tau} = \{ \bm{\delta} \in \mathbb{R}^{s_n^{(M_n-1)}} : \|\bm{\delta} - \bm{\alpha}^{(M_n-1)}_{*,1}\|_2 \leq \sqrt{(s_n^{(M_n-1)}+s_n^{(M_n)})/n} \tau \}
	\]
	and \( \tau \in (0, \infty) \). Clearly, under the event \( H_n \), there exists a local minimizer \( \widehat{\bm{\alpha}}^{(M_n-1)}_1 \) of \( \overline{L} (\bm{\delta}) \) in \( N_{\tau} \). Let \( n \) be sufficiently large such that \( \sqrt{(s_n^{(M_n-1)}+s_n^{(M_n)})/n} \tau \leq g^{(M_n-1) }_n \). Since \( g^{(M_n-1)}_n \gg \sqrt{ (s_n^{(M_n-1)}+s_n^{(M_n)})/n}  \) by Assumption~\ref{compare1}, it is easy to see that $
	\bm{\delta} = (\delta_1, \dots, \delta_{s_n^{(M_n-1)}})^\top \in N_{\tau} \text{ entails } \text{sgn} (\bm{\delta}) = \text{sgn} (\bm{\alpha}^{(M_n-1)}_{*,1}), \|\bm{\delta} - \bm{\alpha}^{(M_n-1)}_{*,1}\|_{\infty} \leq g^{(M_n-1)}_n, \text{ and } \min_j |\delta_j| \geq g_n^{(M_n-1)}.
	$
	
	Define $\bar{\rho}(t)=\operatorname{sgn}(t)\rho'(|t|)$, and $\bar{\rho}(\mathbf{v})=(\bar{\rho}(v_1),\cdots,\bar{\rho}(v_q))^\top$ for $\mathbf{v}=(v_1,\cdots,v_q)^\top$. Then by Taylor’s expansion,we have for any \( \bm{\delta} \in N_{\tau} \),
	\begin{equation}
		\overline{L} (\bm{\delta}) - \overline{L} (\bm{\alpha}^{(M_n-1)}_{*,1}) = -(\bm{\delta} - \bm{\alpha}^{(M_n-1)}_{*,1})^\top \mathbf{v} + \frac{1}{2} (\bm{\delta} -\bm{\alpha}^{(M_n-1)}_{*,1})^\top \mathbf{D} (\bm{\delta} - \bm{\alpha}^{(M_n-1)}_{*,1})
	\end{equation}
	where 
	\begin{align*}
		&	\mathbf{v} = n^{-1} (\bm{\Psi}^{(M_n-1)}_{\mathrm{dfm},1})^\top [\widehat{\mathbf{Y}}^{(M_n-1)}_{\mathrm{dfm}}-  \bm{\theta}^{(M_n-1)}_*] - \bar{\rho}_{\lambda_n} (\bm{\alpha}^{(M_n-1)}_{*,1}),\\&  \bm{\theta}^{(M_n-1)}_* = \bm{\Psi}_{\mathrm{dfm}}^{(M_n-1)} \bm{\alpha}^{(M_n-1)}_{*} = \bm{\Psi}^{(M_n-1)}_{\mathrm{dfm},1}\bm{\alpha}^{(M_n-1)}_{*,1},
	\end{align*}
	and
	\[
	\mathbf{D} = n^{-1} (\bm{\Psi}^{(M_n-1)}_{\mathrm{dfm},1})^\top\bm{\Psi}^{(M_n-1)}_{\mathrm{dfm},1} + \text{diag} \left\{ p''_{\lambda_n} (|\bar{\bm{\alpha}}|) \right\}
	\]
	where $\bar{\bm{\alpha} }\text{ lies on the line segment joining } \bm{\delta} \text{ and } \bm{\alpha}^{(M_n-1)}_{*,1}.
	$ More generally, \( \mathbf{D} \) can be replaced by a diagonal matrix with maximum absolute element bounded by \( \lambda^{(M_n-1)} \kappa_0 \). Define
	\[
	\mathcal{N}_0 = \{ \mathbf{b} \in \mathbb{R}^{s_n^{(M_n-1)}} : \|\mathbf{b} - \bm{\alpha}^{(M_n-1)}_{*,1}\|_{\infty} \leq g^{(M_n-1)}_n \}
	\]
	and \( \kappa_0 = \max_{\mathbf{b} \in \mathcal{N}_0} \kappa(\rho; \mathbf{b}) \). For any \( \bm{\delta} \in \partial N_{\tau} \), we have \( \|\bm{\delta} - \bm{\alpha}^{(M_n-1)}_{*,1}\|_2 = \sqrt{(s_n^{(M_n-1)}+s_n^{(M_n)})/n} \tau \) and \( \bar{\bm{\alpha}} \in \mathcal{N}_0 \). Then for sufficiently large \( n \), by Assumption~\ref{design1} and \( \lambda_n \kappa_0 = o(1) \) in Assumption~\ref{compare1} we have
	\[
	\lambda_{\min} (\mathbf{D}) \geq c - \lambda_n \kappa_0 \geq \frac{c}{2}.
	\]
	Thus, 
	\[
	\min_{\bm{\delta} \in \partial N_{\tau}} \overline{L} (\bm{\delta}) - \overline{L} (\bm{\alpha}^{(M_n-1)}_{*,1}) \geq \sqrt{(s_n^{(M_n-1)}+s_n^{(M_n)})/n} \tau \left( -\|\mathbf{v}\|_2 + c \sqrt{(s_n^{(M_n-1)}+s_n^{(M_n)})/n} \tau / 4 \right),
	\]
	which entails that
	\[
	P(H_n) \geq P \left( \|\mathbf{v}\|_2^2 < \frac{c^2 (s_n^{(M_n-1)}+s_n^{(M_n)}) \tau^2}{16n} \right).
	\]
	
	Now 
	\begin{align}
		\|\mathbf{v}\|_2^2 
		=& \left\|n^{-1} (\bm{\Psi}^{(M_n-1)}_{\mathrm{dfm},1})^\top [\widehat{\mathbf{Y}}^{(M_n-1)}_{\mathrm{dfm}}-  \bm{\theta}^{(M_n-1)}_*] - \bar{\rho}_{\lambda_n} (\bm{\alpha}^{(M_n-1)}_{*,1})\right\|_2^2 \nonumber \\
		\lesssim&\left\|n^{-1} (\bm{\Psi}^{(M_n-1)}_{\mathrm{dfm},1})^\top [\widehat{\mathbf{Y}}^{(M_n-1)}_{\mathrm{dfm}}-\tilde{\mathbf{Y}}^{(M_n-1)}_{\mathrm{dfm}}]\right\|_2^2 \nonumber \\
		&+\left\|n^{-1} (\bm{\Psi}^{(M_n-1)}_{\mathrm{dfm},1})^\top [\tilde{\mathbf{Y}}^{(M_n-1)}_{\mathrm{dfm}}-\bm{\theta}^{(M_n-1)}_*] \right\|_2^2 \nonumber \\
		&+\left\|\bar{\rho}_{\lambda_n} (\bm{\alpha}^{(M_n-1)}_{*,1})\right\|_2^2 \nonumber \\
		:=&I_1+I_2+I_3 \label{I1Mn-1}
	\end{align}
	and
	\begin{align*}
		I_1=&\left\|n^{-1} (\bm{\Psi}^{(M_n-1)}_{\mathrm{dfm},1})^\top [\widehat{\mathbf{Y}}^{(M_n-1)}_{\mathrm{dfm}}-\tilde{\mathbf{Y}}^{(M_n-1)}_{\mathrm{dfm}}]\right\|_2^2\\=&\left\|n^{-1} (\bm{\Psi}^{(M_n-1)}_{\mathrm{dfm},1})^\top [\bm{\Psi}_{\mathrm{dfm},\widehat{\pi}^{(M_n)}_*}\widehat{\bm{\alpha}}^{(M_n)}-\bm{\Psi}_{\mathrm{dfm},\pi_*^{(M_n)}}{\bm{\alpha}}^{(M_n)}_*]\right\|_2^2\\\lesssim &\left\|n^{-1} (\bm{\Psi}^{(M_n-1)}_{\mathrm{dfm},1})^\top [\bm{\Psi}_{\mathrm{dfm},\widehat{\pi}^{(M_n)}_*}-\bm{\Psi}_{\mathrm{dfm},\pi_*^{(M_n)}}]\widehat{\bm{\alpha}}^{(M_n)}\right\|_2^2 \\&+\left\|n^{-1} (\bm{\Psi}^{(M_n-1)}_{\mathrm{dfm},1})^\top \bm{\Psi}_{\mathrm{dfm},\pi_*^{(M_n)}}[\widehat{\bm{\alpha}}^{(M_n)}-{\bm{\alpha}}^{(M_n)}_*]\right\|_2^2\\:=&II_1+II_2.
	\end{align*}
	Now  if the assumptions in Theorem~\ref{convergence1} hold for the estimation sample, then for each $1\leq i \leq n$,
	\begin{align*}
		0\leq&(\mathbf{S}_{\mathrm{dfm},\widehat{\pi}^{(M_n)}_*,i})^\top\widehat{\bm{\alpha}}^{(M_n)}-(\mathbf{S}_{\mathrm{dfm},\pi_*^{(M_n)},i})^\top\widehat{\bm{\alpha}}^{(M_n)}\\=&\left([\bm{\Psi}_{\mathrm{dfm},\widehat{\pi}^{(M_n)}_*}-\bm{\Psi}_{\mathrm{dfm},\pi_*^{(M_n)}}]\widehat{\bm{\alpha}}^{(M_n)}\right)_i\\\leq&\left([\bm{\Psi}_{\mathrm{dfm},\widehat{\pi}^{(M_n)}_*}-\bm{\Psi}_{\mathrm{dfm},\pi_*^{(M_n)}}](\widehat{\bm{\alpha}}^{(M_n)}-{\bm{\alpha}}_*^{(M_n)})\right)_i,
	\end{align*}
	thus
	\begin{align*}
		II_1=&\left\|n^{-1} (\bm{\Psi}^{(M_n-1)}_{\mathrm{dfm},1})^\top [\bm{\Psi}_{\mathrm{dfm},\widehat{\pi}^{(M_n)}_*}-\bm{\Psi}_{\mathrm{dfm},\pi_*^{(M_n)}}]\widehat{\bm{\alpha}}^{(M_n)}\right\|_2^2\\\leq &\left\|n^{-2} (\bm{\Psi}^{(M_n-1)}_{\mathrm{dfm},1})^\top\bm{\Psi}^{(M_n-1)}_{\mathrm{dfm},1}\right\|_2 \, \left\|[\bm{\Psi}_{\mathrm{dfm},\widehat{\pi}^{(M_n)}_*}-\bm{\Psi}_{\mathrm{dfm},\pi_*^{(M_n)}}]\widehat{\bm{\alpha}}^{(M_n)}\right\|_2^2\\=&O_p({1}/{n}) \, \left\|[\bm{\Psi}_{\mathrm{dfm},\widehat{\pi}^{(M_n)}_*}-\bm{\Psi}_{\mathrm{dfm},\pi_*^{(M_n)}}]\widehat{\bm{\alpha}}^{(M_n)}\right\|_2^2\\\leq& O_p({1}/{n})\left\|[\bm{\Psi}_{\mathrm{dfm},\widehat{\pi}^{(M_n)}_*}-\bm{\Psi}_{\mathrm{dfm},\pi_*^{(M_n)}}](\widehat{\bm{\alpha}}^{(M_n)}-\bm{\alpha}_*^{(M_n)})\right\|_2^2\\\leq& O_p(1) \, \mathbb{P}_n\left\{ (\mathbf{S}_{\mathrm{dfm},\widehat{\pi}^{(M_n)}_*}-\mathbf{S}_{\mathrm{dfm},\pi_*^{(M_n)}})^\top(\widehat{\bm{\alpha}}^{(M_n)}-\bm{\alpha}_*^{(M_n)})\right\}^2\\\leq & O_p((s_n^{(M_n)})^2/n) \, \mathbb{P}_n\left\| (\mathbf{S}_{\mathrm{dfm},\widehat{\pi}^{(M_n)}_*}-\mathbf{S}_{\mathrm{dfm},\pi_*^{(M_n)}})_{\operatorname{supp}(\bm{\alpha}_*^{(M_n)})}\right\|_\infty^2\\\lesssim&  O_p((s_n^{(M_n)})^2/n) \, \max_{i,j} r^2_{i,j} \mathbb{P}_n(I_{\mathcal{T}})
	\end{align*}
	
To bound the empirical frequency of the set $\mathcal T$ in \eqref{mathcalt}, fix a large constant $C$ and restrict to $\|\widehat{\boldsymbol\alpha}^{(M_n)}-\boldsymbol\alpha_*^{(M_n)}\|_2\leq C\sqrt{s_n^{(M_n)}/n}$. Let $b_n$ be the coordinate bound supplied by the tail condition and $t_n=C b_n s_n^{(M_n)}/\sqrt n$. When the sample coordinates are bounded by $b_n$, $\mathcal T$ on the training observations is contained in $\{G_{M_n}\leq2t_n\}$, where $G_{M_n}$ is the true action-block gap (zero at a tie). The indicators of this latter set are independent across trajectories. Writing $\nu_n=\mathbb P\{(\mathcal U_n^{(M_n)})^c\}$, Bernstein's inequality gives
\[
\mathbb P_n I_{\mathcal T}\leq O_p\{L_nt_n+\nu_n+(\log n)/n\}.
\]
Under the consistency exponent and tie conditions in Theorem~\ref{convergence1}, the bound above implies $II_1=o_p(s_n^{(M_n)}/n)$. The constant $C$ can then be taken arbitrarily large. Similarly,

\begin{align*}
		II_2=&\left\|n^{-1} (\bm{\Psi}^{(M_n-1)}_{\mathrm{dfm},1})^\top \bm{\Psi}_{\mathrm{dfm},\pi_*^{(M_n)}}[\widehat{\bm{\alpha}}^{(M_n)}-{\bm{\alpha}}^{(M_n)}_*]\right\|_2^2\\\leq &n^{-2}\left\| (\bm{\Psi}^{(M_n-1)}_{\mathrm{dfm},1})^\top\bm{\Psi}^{(M_n-1)}_{\mathrm{dfm},1}\right\|_2\left\|(\bm{\Psi}_{\mathrm{dfm},\pi_*^{(M_n)},1})^\top\bm{\Psi}_{\mathrm{dfm},\pi_*^{(M_n)},1}\right\|_2\left\|[\widehat{\bm{\alpha}}_1^{(M_n)}-{\bm{\alpha}}^{(M_n)}_{*,1}]\right\|_2^2\\\leq& O_p(\frac{s^{(M_n)}_n}{n})
	\end{align*}
	For $I_2$,
	\begin{align*}
		\mathbb{E}[I_2]=&\mathbb{E}\left\|n^{-1} (\bm{\Psi}^{(M_n-1)}_{\mathrm{dfm},1})^\top [\tilde{\mathbf{Y}}^{(M_n-1)}_{\mathrm{dfm}}-\bm{\theta}_*] \right\|_2^2\\=&n^{-2}\mathbb{E}\left\{\mathbb{E}\left[(\bm{\varepsilon}^{(M_n-1)})^\top\bm{\Psi}^{(M_n-1)}_{\mathrm{dfm},1}(\bm{\Psi}^{(M_n-1)}_{\mathrm{dfm},1})^\top\bm{\varepsilon}^{(M_n-1)}\mid \bm{\Psi}^{(M_n-1)}_{\mathrm{dfm},1} \right]\right\}\\=&n^{-2}\operatorname{tr}\left\{\mathbb{E}\left[(\bm{\Psi}^{(M_n-1)}_{\mathrm{dfm},1})^\top\bm{\Sigma}^{(M_n-1)}\bm{\Psi}^{(M_n-1)}_{\mathrm{dfm},1}\right]\right\}\\=&n^{-2}\mathbb{E}\left[\sum_{i=1}^{n}(\varepsilon_i^{(M_n-1)})^2(\mathbf{S}^{(M_n-1)}_{\mathrm{dfm},i,1})^\top\mathbf{S}^{(M_n-1)}_{\mathrm{dfm},i,1}\right]\\=&O(\frac{s_n^{(M_n-1)}}{n}).
	\end{align*}
	
	Moreover, $I_3=O_p(s^{(M_n-1)}_n/n)$ since \( p'_{\lambda_n} (t) \) is decreasing in \( t \in [0, \infty) \). Hence, $P(H_n) \to 1$ for large $\tau$. This proves \( \|\widehat{\bm{\alpha}}^{(M_n-1)}_1 - \bm{\alpha}^{(M_n-1)}_{*,1}\|_2 = O_P(\sqrt{(s_n^{(M_n-1)}+s_n^{(M_n)})/n}) \).
	
	\subsubsection*{Step 2: Sparsity}
	Let \( \widehat{\bm{\alpha}} \in \mathbb{R}^{p_n^{(M_n-1)}} \) with \( \widehat{\bm{\alpha}}^{(M_n-1)}_1 \in N_{\tau} \subset \mathcal{N}_0 \) a strict local minimizer of \( \overline{L}(\bm{\delta}) \) and \( \widehat{\bm{\alpha}}^{(M_n-1)}_2=0 \), and \( \widehat{\bm{\theta}}^{(M_n-1)} = \bm{\Psi}_{\mathrm{dfm}}^{(M_n-1)} \widehat{\bm{\alpha}}^{(M_n-1)} \). It remains to prove that the vector \( \widehat{\bm{\alpha}}^{(M_n-1)} \) is indeed a strict local minimizer of \( \widehat{L} (\bm{\alpha}) \) on the space \( \mathbb{R}^{p_n^{(M_n-1)}} \), which suffices to show
	$$\|\mathbf{w}_1\|_\infty<\rho'(0+),$$ where $$\mathbf{w}_1=(n\lambda^{(M_n-1)})^{-1}(\bm{\Psi}^{(M_n-1)}_{\mathrm{dfm},2})^\top(\widehat{\mathbf{Y}}^{(M_n-1)}_{\mathrm{dfm}}-\widehat{\bm{\theta}}^{(M_n-1)}).$$
	Now
	\begin{align*}
		\|\mathbf{w}_1\|_\infty=&\left\|(n\lambda^{(M_n-1)})^{-1}(\bm{\Psi}^{(M_n-1)}_{\mathrm{dfm},2})^\top(\widehat{\mathbf{Y}}^{(M_n-1)}_{\mathrm{dfm}}-\widehat{\bm{\theta}}^{(M_n-1)})\right\|_\infty\\\leq &\left\|(n\lambda^{(M_n-1)})^{-1}(\bm{\Psi}^{(M_n-1)}_{\mathrm{dfm},2})^\top(\widehat{\mathbf{Y}}^{(M_n-1)}_{\mathrm{dfm}}-\tilde{\mathbf{Y}}^{(M_n-1)}_{\mathrm{dfm}})\right\|_\infty\\&+\left\|(n\lambda^{(M_n-1)})^{-1}(\bm{\Psi}^{(M_n-1)}_{\mathrm{dfm},2})^\top(\tilde{\mathbf{Y}}^{(M_n-1)}_{\mathrm{dfm}}-{\bm{\theta}}_*^{(M_n-1)})\right\|_\infty\\&+\left\|(n\lambda^{(M_n-1)})^{-1}(\bm{\Psi}^{(M_n-1)}_{\mathrm{dfm},2})^\top({\bm{\theta}}_*^{(M_n-1)}-\widehat{\bm{\theta}}^{(M_n-1)})\right\|_\infty\\=&I'_1+I'_2+I'_3.
	\end{align*}
	By similar arguments in step 1
	\begin{align*}
		I'_1\leq &\left\|(n\lambda^{(M_n-1)})^{-1} (\bm{\Psi}^{(M_n-1)}_{\mathrm{dfm},2})^\top [\bm{\Psi}_{\mathrm{dfm},\widehat{\pi}^{(M_n)}_*}-\bm{\Psi}_{\mathrm{dfm},\pi_*^{(M_n)}}]\widehat{{\bm{\alpha}}}^{(M_n)}\right\|_\infty\\&+  \left\|(n\lambda^{(M_n-1)})^{-1} (\bm{\Psi}^{(M_n-1)}_{\mathrm{dfm},2})^\top [\bm{\Psi}_{\mathrm{dfm},{\pi}_*^{(M_n)}}](\widehat{\bm{\alpha}}^{(M_n)}-{\bm{\alpha}}_*^{(M_n)})\right\|_\infty\\:=&II'_1+II'_2,
	\end{align*}
	and	
	\begin{align*}
		II'_1\leq &\left\|(n\lambda^{(M_n-1)})^{-1} |\bm{\Psi}^{(M_n-1)}_{\mathrm{dfm},2}|^\top [\bm{\Psi}_{\mathrm{dfm},\widehat{\pi}^{(M_n)}_*}-\bm{\Psi}_{\mathrm{dfm},\pi_*^{(M_n)}}]\widehat{{\bm{\alpha}}}^{(M_n)}\right\|_\infty\\\leq &(n\lambda^{(M_n-1)})^{-1}\left\| |\bm{\Psi}^{(M_n-1)}_{\mathrm{dfm},2}|^\top [\bm{\Psi}_{\mathrm{dfm},\widehat{\pi}^{(M_n)}_*}-\bm{\Psi}_{\mathrm{dfm},\pi_*^{(M_n)}}](\widehat{{\bm{\alpha}}}^{(M_n)}-{{\bm{\alpha}}}_*^{(M_n)})\right\|_\infty\\\lesssim&(n\lambda^{(M_n-1)})^{-1}\left\| |\bm{\Psi}^{(M_n-1)}_{\mathrm{dfm},2}|^\top [\bm{\Psi}_{\mathrm{dfm},\widehat{\pi}^{(M_n)}_*}-\bm{\Psi}_{\mathrm{dfm},\pi_*^{(M_n)}}]\right\|_{2,\infty}\left\|(\widehat{{\bm{\alpha}}}^{(M_n)}-{{\bm{\alpha}}}_*^{(M_n)})\right\|_2\\\leq &O_p\left(\frac{\sqrt{s_n^{(M_n)}}}{\sqrt{n}\lambda^{(M_n-1)}}\right)\\=&o_p(1).
	\end{align*}
	Moreover
	\begin{align*}
		II'_2\leq &(n\lambda^{(M_n-1)})^{-1}  \left\|(\bm{\Psi}^{(M_n-1)}_{\mathrm{dfm},2})^\top \bm{\Psi}_{\mathrm{dfm},\widehat{\pi}_*^{(M_n)},1}\right\|_{2,\infty}\left\|(\widehat{\bm{\alpha}}^{(M_n)}-{\bm{\alpha}}_*^{(M_n)})\right\|_2\\\leq &O_p(\frac{\sqrt{s_n^{(M_n)}}}{\lambda^{(M_n-1)} \sqrt{n}})\\=&o_p(1).
	\end{align*}
	Next, let $$\bm{\xi} =( \bm{\xi}^\top_{1},\bm{\xi}^\top_2 )^\top=(\xi_1, \dots, \xi_{p_n^{(M_n-1)}})^\top = (\bm{\Psi}_{\mathrm{dfm}}^{(M_n-1)})^\top \tilde{\mathbf{Y}}^{(M_n-1)}_{\mathrm{dfm}} - (\bm{\Psi}_{\mathrm{dfm}}^{(M_n-1)})^\top \bm{\theta}_*^{(M_n-1)},$$
	and consider the event
	\[
	\mathcal{E}_2 = \left\{ \|\bm{\xi}_2 \|_{\infty} \leq u_n \sqrt{n} \right\}
	\]
	where \( u_n = 	C\sqrt{\log p_n^{(M_n-1)} } \). Then similarly in the proof of the $M_n-$th stage, we can prove that
	$$\mathbb{P}(\mathcal{E}_2)\gtrsim 1- O(1/p_n^{(M_n-1)}),$$
	which implies $$I_2'=O_p\left(\frac{\sqrt{n \log p_n^{(M_n-1)} }}{n\lambda^{(M_n-1)}}\right)=o_p(1).$$
	Finally
	\begin{align*}
		I_3'=&\left\|(n\lambda^{(M_n-1)})^{-1}(\bm{\Psi}^{(M_n-1)}_{\mathrm{dfm},2})^\top({\bm{\theta}}_*^{(M_n-1)}-\widehat{\bm{\theta}}^{(M_n-1)})\right\|_\infty\\=&(n\lambda^{(M_n-1)})^{-1}\left\|(\bm{\Psi}^{(M_n-1)}_{\mathrm{dfm},2})^\top\bm{\Psi}_{\mathrm{dfm},1}^{(M_n-1)}(\widehat{\bm{\alpha}}^{(M_n-1)}-{\bm{\alpha}}_*^{(M_n-1)})\right\|_\infty \\\leq &(n\lambda^{(M_n-1)})^{-1}\left\|(\bm{\Psi}^{(M_n-1)}_{\mathrm{dfm},2})^\top\bm{\Psi}_{\mathrm{dfm},1}^{(M_n-1)}\right\|_{2,\infty}\left\|(\widehat{\bm{\alpha}}^{(M_n-1)}-{\bm{\alpha}}_*^{(M_n-1)})\right\|_2\\=&O_p\left(\frac{n \sqrt{ (s_n^{(M_n-1)}+ s_n^{(M_n)})}}{n\sqrt{n}\lambda^{(M_n-1)}}\right)\\=&o_p(1).
	\end{align*}
	Thus, we complete the first part.

	\item \textbf{Second Part (Asymptotic normality).}

	We only need to prove the asymptotic normality of \( \widehat{\bm{\alpha}}^{(M_n-1)}_1 \). Now on the event \( H_n \), \( \widehat{\bm{\alpha}}^{(M_n-1)}_1 \in N_{\tau} \subset \mathcal{N}_0 \) is a strict local minimizer of \( \overline{L} (\bm{\delta}) \) and \( \widehat{\bm{\alpha}}^{(M_n)}_2 = 0 \). It follows easily that \( \nabla \overline{L} (\widehat{\bm{\alpha}}^{(M_n-1)}_1) = 0 \). Thus, by Taylor's expansion, 
	\begin{align*}
		0 =& \nabla \overline{L} (\widehat{\bm{\alpha}}^{(M_n-1)}_1) \\=&-n^{-1} [(\bm{\Psi}_{\mathrm{dfm},1} ^{(M_n-1)})^\top\widehat{\mathbf{Y}}^{(M_n-1)}_{\mathrm{dfm}} -(\bm{\Psi}_{\mathrm{dfm},1}^{(M_n-1)})^\top \bm{\Psi}^{(M_n-1)}_{\mathrm{dfm},1} \widehat{\bm{\alpha}}_1^{(M_n-1)}]+\bar{p}_{\lambda^{(M_n-1)}}(\widehat{\bm{\alpha}}_1^{(M_n-1)})\\=&-n^{-1} [(\bm{\Psi}_{\mathrm{dfm},1} ^{(M_n-1)})^\top(\widehat{\mathbf{Y}}^{(M_n-1)}_{\mathrm{dfm}}-\tilde{\mathbf{Y}}^{(M_n-1)}_{\mathrm{dfm}}) \\&-n^{-1} [(\bm{\Psi}_{\mathrm{dfm},1} ^{(M_n-1)})^\top\tilde{\mathbf{Y}}^{(M_n-1)}_{\mathrm{dfm}} -(\bm{\Psi}_{\mathrm{dfm},1}^{(M_n-1)})^\top \bm{\Psi}^{(M_n-1)}_{\mathrm{dfm},1} {\bm{\alpha}}_{*,1}^{(M_n-1)}]\\&+n^{-1}[(\bm{\Psi}_{\mathrm{dfm},1}^{(M_n-1)})^\top \bm{\Psi}^{(M_n-1)}_{\mathrm{dfm},1}](\widehat{\bm{\alpha}}_1^{(M_n-1)}- {\bm{\alpha}}_{*,1}^{(M_n-1)})\\ &+\bar{p}_{\lambda^{(M_n-1)}}(\widehat{\bm{\alpha}}_1^{(M_n-1)})\\:=&-I''_1-I''_2+I''_3+I''_4,
	\end{align*}
	and 
	\begin{align*}
		I''_1=&n^{-1} (\bm{\Psi}^{(M_n-1)}_{\mathrm{dfm},1})^\top [\bm{\Psi}_{\mathrm{dfm},\widehat{\pi}^{(M_n)}_*}-\bm{\Psi}_{\mathrm{dfm},\pi_*^{(M_n)}}]\widehat{\bm{\alpha}}^{(M_n)} \\&+n^{-1} (\bm{\Psi}^{(M_n-1)}_{\mathrm{dfm},1})^\top \bm{\Psi}_{\mathrm{dfm},\pi_*^{(M_n)}}[\widehat{\bm{\alpha}}^{(M_n)}-{\bm{\alpha}}^{(M_n)}_*]\\:=&II''_1+II''_2.
	\end{align*}
	It follows from \( \widehat{\bm{\alpha}}_1 \in \mathcal{N}_0 \), and \( p'_{\lambda^{(M_n-1)}} (g^{(M_n-1)}_n) = o(1/\sqrt{n s_n^{(M_n-1)}}) \)  that
	\[
	\left\| \bar{p}_{\lambda_n} (\widehat{\bm{\alpha}}^{(M_n-1)}_1) \right\|_2 \leq \sqrt{s_n^{(M_n-1)}} p'_{\lambda_n} (d_n) = o_p(1/\sqrt{n}),
	\]
	which means $\|I''_4\|_2=o_p(1/\sqrt{n})$. Now for any $\mathbf{C}^{(M_n-1)}\in \mathbb{R}^{q\times s_n^{(M_n-1)}}$, let
	
\begin{align*}
		&\mathbf{D}^{(M_n-1)}=\mathbb{E}[\mathbf{S}^{(M_n-1)}_{\mathrm{dfm},1}(\mathbf{S}^{(M_n-1)}_{\mathrm{dfm},1})^\top]\\&\mathbf{D}_n^{(M_n-1)}={(\bm{\Psi}^{(M_n-1)}_{\mathrm{dfm},1})^\top\bm{\Psi}^{(M_n-1)}_{\mathrm{dfm},1}} \\	&\mathbf{F}^{(M_n-1)}=\mathbb{E}[(\varepsilon^{(M_n-1)})^2\mathbf{S}^{(M_n-1)}_{\mathrm{dfm},1}(\mathbf{S}^{(M_n-1)}_{\mathrm{dfm},1})^\top]\\&\mathbf{F}_n^{(M_n-1)}={(\bm{\Psi}^{(M_n-1)}_{\mathrm{dfm},1})^\top\bm{\Sigma}^{(M_n-1)}\bm{\Psi}^{(M_n-1)}_{\mathrm{dfm},1}} \\&\mathbf{J}^{(M_n-1)}=\mathbb{E}[\mathbf{S}^{(M_n-1)}_{\mathrm{dfm},1}(\mathbf{S}_{\mathrm{dfm},\pi_*^{(M_n)},1})^\top]\\&\mathbf{J}_n^{(M_n-1)}=(\bm{\Psi}^{(M_n-1)}_{\mathrm{dfm},1})^\top\bm{\Psi}_{\mathrm{dfm},\pi_*^{(M_n)},1} \\&\mathbf{K}^{(M_n-1)}=\mathbf{J}^{(M_n-1)}\mathbf{P}^{(M_n)}{\mathbf{V}}^{(M_n)}(\mathbf{P}^{(M_n)})^\top(\mathbf{J}^{(M_n-1)})^\top+\mathbf{F}^{(M_n-1)}\\&\mathbf{K}_n^{(M_n-1)}=n^{-1}\mathbf J_n^{(M_n-1)}\mathbf P^{(M_n)}\mathbf V^{(M_n)}(\mathbf P^{(M_n)})^\top(\mathbf J_n^{(M_n-1)})^\top+\mathbf F_n^{(M_n-1)},
\end{align*}

	where \( \mathbf{P}^{(M_n)} \) is the linear operator that extracts \( \bm{\alpha}^{(M_n)}_{*,1} \) from \( \bm{\beta}^{(M_n)}_{*,1} \).
	Note that under Assumption~\ref{normal1},
	\begin{align*}
		&\|(\mathbf{D}_n^{(M_n-1)}-n \, \mathbf{D}^{(M_n-1)})(\widehat{\bm{\alpha}}_1^{(M_n-1)}- {\bm{\alpha}}_{*,1}^{(M_n-1)})\|_2\\\leq &\|(\mathbf{D}_n^{(M_n-1)}-n \, \mathbf{D}^{(M_n-1)})\|_2\|(\widehat{\bm{\alpha}}_1^{(M_n-1)}- {\bm{\alpha}}_{*,1}^{(M_n-1)})\|_2\\=&o_p\left(n/\sqrt{(s_n^{(M_n)}+s_n^{(M_n-1)})}\right)O_p\left(\sqrt{(s_n^{(M_n)}+s_n^{(M_n-1)})/n}\right)\\=&o_p(\sqrt{n}),
	\end{align*}
	which means we can replace $\mathbf{D}_n^{(M_n-1)}$ with $n \,\mathbf{D}^{(M_n-1)}$ in $I''_3$. Similarly, we can replace $\mathbf{J}_n^{(M_n-1)}$ with $n \, \mathbf{J}^{(M_n-1)}$ in $II_2''$. Moreover, define
	$$\mathbf{O}^{(M_n-1)}=\mathbf{C}^{(M_n-1)}(\mathbf{K}^{(M_n-1)})^{-1/2},$$ then under regular conditions in Proposition~\ref{convergence2}, 
	\begin{align}
		&{\sqrt{n}} \, \mathbf{O}^{(M_n-1)} \mathbf{J}^{(M_n-1)}(\widehat{\bm{\alpha}}^{(M_n)}-{\bm{\alpha}}^{(M_n)}_*)\nonumber\\=&\frac{1}{\sqrt{n}{\phi^{(M_n)}}} \,  \mathbf{O}^{(M_n-1)} \mathbf{J}^{(M_n-1)} \mathbf{P}^{(M_n)}{\mathbf{V}}^{(M_n)}(\bm{\Psi}_1^{(M_n)})^\top(\mathbf{Y}^{(M_n)}-\bm{\mu}^{(M_n)}(\bm{\theta}_*^{(M_n)}))+o_p(1).\label{alphaop1}
	\end{align}
	Thus define $\bm{\varepsilon}^{(M_n)}=\mathbf{Y}^{(M_n)}-\bm{\mu}^{(M_n)}(\bm{\theta}_*^{(M_n)})$, we have
	\begin{align}
		&\sqrt{n}\mathbf{C}^{(M_n-1)}(\mathbf{K}^{(M_n-1)})^{-1/2}\mathbf{D}^{(M_n-1)}(\widehat{\bm{\alpha}}_1^{(M_n-1)}- {\bm{\alpha}}_{*,1}^{(M_n-1)})\nonumber\\= &\frac{1}{\sqrt{n}} \,\mathbf{C}^{(M_n-1)}(\mathbf{K}^{(M_n-1)})^{-1/2}\Big[(\bm{\Psi}^{(M_n-1)}_{\mathrm{dfm},1})^\top\bm{\varepsilon}^{(M_n-1)}\nonumber\\&+\frac{1}{\phi^{(M_n)}} \mathbf{J}^{(M_n-1)}\mathbf{P}^{(M_n)}{\mathbf{V}}^{(M_n)}(\bm{\Psi}_1^{(M_n)})^\top\bm{\varepsilon}^{(M_n)} \nonumber\\&+(\bm{\Psi}^{(M_n-1)}_{\mathrm{dfm},1})^\top [\bm{\Psi}_{\mathrm{dfm},\widehat{\pi}^{(M_n)}_*}-\bm{\Psi}_{\mathrm{dfm},\pi_*^{(M_n)}}]\widehat{\bm{\alpha}}^{(M_n)}\Big]\nonumber\\&+o_p(1).\label{un1}
	\end{align}
	
	To show the asymptotic normality, we break the proof into two parts:
	\item \textbf{Part I}
	For any unit vector $\mathbf{a}$,
	\begin{equation}
		\frac{1}{\sqrt{n}} \,\mathbf{a}^\top\mathbf{C}^{(M_n-1)}(\mathbf{K}^{(M_n-1)})^{-1/2} (\bm{\Psi}_{\mathrm{dfm},1}^{(M_n-1)})^\top[\bm{\Psi}_{\mathrm{dfm},\widehat{\pi}^{(M_n)}_*}-\bm{\Psi}_{\mathrm{dfm},\pi_*^{(M_n)}}]\widehat{\bm{\alpha}}^{(M_n)}\to 0.\label{amn-1}
	\end{equation}
	
Using the bound on the empirical frequency established above, the square of the left-hand side of \eqref{amn-1} is at most
\[
O_p\bigl[(s_n^{(M)})^2b_n^2\{L_n b_n s_n^{(M)}/\sqrt n+\nu_n+(\log n)/n\}\bigr]=o_p(1).
\]
Indeed, the first term vanishes under $\ell_1+3\ell_2^{(M)}<1/2$, the second under $\delta>2\ell_2^{(M)}$, and the last under the same strict sparsity exponent.

	\item \textbf{Part II}
	Then by \eqref{un1}, we just need to focus on 
	\begin{align*}
		\mathbf{u}_n=&\frac{1}{\sqrt{n}} \,\mathbf{C}^{(M_n-1)}(\mathbf{K}^{(M_n-1)})^{-1/2}\Big[(\bm{\Psi}^{(M_n-1)}_{\mathrm{dfm},1})^\top\bm{\varepsilon}^{(M_n-1)}\\&+ \frac{1}{\phi^{(M_n)}} \mathbf{J}^{(M_n-1)}\mathbf{P}^{(M_n)}{\mathbf{V}}^{(M_n)}(\bm{\Psi}_1^{(M_n)})^\top\bm{\varepsilon}^{(M_n)} \Big]
	\end{align*}
	For any unit vector $\mathbf{a}$, we have $\mathbf{a}^\top\mathbf{u}_n=\sum_{i=1}^{n}\xi_i$ with 
	\begin{align*}
		\xi_i=&\frac{1}{\sqrt{n}} \,\mathbf{a}^\top\mathbf{C}^{(M_n-1)}(\mathbf{K}^{(M_n-1)})^{-1/2}\Big[{\mathbf{S}}^{(M_n-1)}_{\mathrm{dfm},i,1}{\varepsilon}_i^{(M_n-1)}\\&+\frac{1}{\phi^{(M_n)}} \mathbf{J}^{(M_n-1)}\mathbf{P}^{(M_n)}{\mathbf{V}}^{(M_n)}{\mathbf{S}}_{i,1}^{(M_n)}{\varepsilon}_i^{(M_n)} \Big]\\:=&a_i+b_i.
	\end{align*}
	However, note that
	\begin{align*}
		&\mathbb{E}[{\mathbf{S}}^{(M_n)}_{i,1}{\varepsilon}_i^{(M_n)}]=\mathbb{E}\left[{\mathbf{S}}^{(M_n)}_{i,1}\mathbb{E}[{\varepsilon}_i^{(M_n)}\mid{\mathbf{S}}^{(M_n)}_{i,1}]\right]=0,\\	&\mathbb{E}[{\mathbf{S}}^{(M_n-1)}_{\mathrm{dfm},i,1}{\varepsilon}_i^{(M_n-1)}]=\mathbb{E}\left[{\mathbf{S}}^{(M_n-1)}_{\mathrm{dfm},i,1}\mathbb{E}[{\varepsilon}_i^{(M_n-1)}\mid{\mathbf{S}}^{(M_n-1)}_{\mathrm{dfm},i,1}]\right]=0,\\	
	\end{align*}
	and 
	\begin{align*}
		&\operatorname{Cov}[{\mathbf{S}}^{(M_n)}_{i,1}{\varepsilon}_i^{(M_n)}]\\=&\mathbb{E}\left[({\varepsilon}_i^{(M_n)})^2{\mathbf{S}}^{(M_n)}_{i,1}({\mathbf{S}}^{(M_n)}_{i,1})^\top\right]\\=&\mathbb{E}\left[{\mathbf{S}}^{(M_n)}_{i,1}({\mathbf{S}}^{(M_n)}_{i,1})^\top\mathbb{E}\left[({\varepsilon}_i^{(M_n)})^2\mid{\mathbf{S}}^{(M_n)}_{i,1} \right]\right]\\=& \phi^{(M_n)}\mathbf{F}^{(M_n)}=({\phi^{(M_n)}})^2(\mathbf{V}^{(M_n)})^{-1}
	\end{align*}
	and similarly, $\operatorname{Cov}[{\mathbf{S}}^{(M_n-1)}_{i,1}{\varepsilon}_i^{(M_n-1)}]=\mathbf{F}^{(M_n-1)}.$
	
	Moreover, define the filtration 
	\begin{align}
		\mathcal{G}_m=\sigma\left(\mathbf{X}_i^{(1)},\mathbf{A}_i^{(1)}, \cdots,\mathbf{X}_i^{(m)},\mathbf{A}_i^{(m)},\ 1\leq i\leq n \right)\label{gmn}
	\end{align}
	for $1\leq m \leq M_n$, then  $\tilde{Y}_{\mathrm{dfm},i}^{(M_n-1)}=Y_i^{(M_n-1)}+Z_i^{(M_n-1)}$
	with $Z_i^{(M_n-1)}\in \mathcal{G}_{M_n}$. Thus we can decompose $\varepsilon_i^{(M_n-1)}=	\varepsilon^{(M_n-1)}_{i,1}+\varepsilon^{(M_n-1)}_{i,2}$ with $\varepsilon_{i,2}^{(M_n-1)}\in \mathcal{G}_{M_n}$, 
	where
	\begin{align*}
		&\varepsilon^{(M_n-1)}_{i,1}=Y_i^{(M_n-1)}-\mathbb{E}[Y_i^{(M_n-1)}\mid \mathbf{H}_i^{(M_n-1)},\mathbf{A}_i^{(M_n-1)}],\\
		&\varepsilon^{(M_n-1)}_{i,2}=Z_i^{(M_n-1)}-\mathbb{E}[Z_i^{(M_n-1)}\mid \mathbf{H}_i^{(M_n-1)},\mathbf{A}_i^{(M_n-1)}].
	\end{align*}
	Moreover, notice that $\mathbb{E}\left[{\varepsilon}_i^{(M_n)}\mid \mathcal{G}_{M_n} \right]=0$, and the samples are independent for $i\neq j$ by definition, thus by Assumption~\ref{B5}, 
	\begin{align*}
		&\operatorname{Cov}[{\mathbf{S}}^{(M_n)}_{i,1}{\varepsilon}_i^{(M_n)},{\mathbf{S}}^{(M_n-1)}_{\mathrm{dfm},i,1}{\varepsilon}_i^{(M_n-1)}]\\=&\mathbb{E}\left[{\mathbf{S}}^{(M_n)}_{i,1}{\varepsilon}_i^{(M_n)}{\varepsilon}_i^{(M_n-1)}({\mathbf{S}}^{(M_n-1)}_{\mathrm{dfm},i,1})^\top\right]\\=&\mathbb{E}\left[{\mathbf{S}}^{(M_n)}_{i,1} \, \mathbb{E}\left[{\varepsilon}_i^{(M_n)}{\varepsilon}_{i,1}^{(M_n-1)}\mid  \mathcal{G}_{M_n}\right] \,({\mathbf{S}}^{(M_n-1)}_{\mathrm{dfm},i,1})^\top \right]\\&+\mathbb{E}\left[{\mathbf{S}}^{(M_n)}_{i,1} \, \mathbb{E}\left[{\varepsilon}_i^{(M_n)}{\varepsilon}_{i,2}^{(M_n-1)}\mid \mathcal{G}_{M_n} \right] \,({\mathbf{S}}^{(M_n-1)}_{\mathrm{dfm},i,1})^\top \right]\\=&0+\mathbb{E}\left[{\mathbf{S}}^{(M_n)}_{i,1} \, \mathbb{E}\left[{\varepsilon}_i^{(M_n)}\mid \mathcal{G}_{M_n} \right] {\varepsilon}_{i,2}^{(M_n-1)} \,({\mathbf{S}}^{(M_n-1)}_{\mathrm{dfm},i,1})^\top \right]\\=&0.
	\end{align*}
	Thus
	\begin{align*}
		&\sum_{i=1}^{n}\operatorname{Var}\xi_i\\=&\frac{1}{n}\sum_{i=1}^{n}\mathbf{a}^\top \,\mathbf{C}^{(M_n-1)}(\mathbf{K}^{(M_n-1)})^{-1/2}\Big[\mathbf{F}^{(M_n-1)}\\&+ \mathbf{J}^{(M_n-1)}\mathbf{P}^{(M_n)}{\mathbf{V}}^{(M_n)}(\mathbf{P}^{(M_n)})^\top(\mathbf{J}^{(M_n-1)})^\top \Big]((\mathbf{K}^{(M_n-1)})^{-1/2})^\top(\mathbf{C}^{(M_n-1)})^\top\mathbf{a}\\=&\mathbf{a}^\top\mathbf{C}^{(M_n-1)}(\mathbf{C}^{(M_n-1)})^\top\mathbf{a}\\ \to &\mathbf{a}^\top\mathbf{G}^{(M_n-1)}\mathbf{a}.
	\end{align*}
	Moreover,
	\begin{align*}
		\sum_{i=1}^{n}\mathbb{E}|\xi_i|^3\lesssim&\sum_{i=1}^{n}\mathbb{E}|a_i|^3+\sum_{i=1}^{n}\mathbb{E}|b_i|^3:=III_1+III_2.
	\end{align*}
	Now by Lemma~\ref{Kbound},
	\begin{align*}
		III_1=&n^{-3/2}\sum_{i=1}^{n}\mathbb{E}\left|\mathbf{a}^\top\mathbf{C}^{(M_n-1)}(\mathbf{K}^{(M_n-1)})^{-1/2}{\mathbf{S}}^{(M_n-1)}_{\mathrm{dfm},i,1}{\varepsilon}_i^{(M_n-1)}\right|^3\\\leq &n^{-3/2}\sum_{i=1}^{n}\|\mathbf{a}^\top\mathbf{C}^{(M_n-1)}\|_2^3\,  \mathbb{E}\left\|(\mathbf{K}^{(M_n-1)})^{-1/2}{\mathbf{S}}^{(M_n-1)}_{\mathrm{dfm},i,1}{\varepsilon}_i^{(M_n-1)}\right\|_2^3\\\lesssim& n^{-3/2}\sum_{i=1}^{n}\,  \mathbb{E}\left|({\varepsilon}_i^{(M_n-1)})^2 ({\mathbf{S}}^{(M_n-1)}_{\mathrm{dfm},i,1})^\top(\mathbf{K}^{(M_n-1)})^{-1}{\mathbf{S}}^{(M_n-1)}_{\mathrm{dfm},i,1}\right|^{3/2}\\\to&0.
	\end{align*}
	Similarly,
	
Using $\mathbf T$ from the proof of Lemma~\ref{Kbound},
\begin{align*}
III_2&\leq\frac C{\sqrt n}\mathbb E\left|
\mathbf a^\top\mathbf C^{(M_n-1)}(\mathbf K^{(M_n-1)})^{-1/2}\mathbf T\mathbf u_{M_n i}\right|^3\\
&\leq\frac C{\sqrt n}\mathbb E\left\|
(\mathbf K^{(M_n-1)})^{-1/2}\mathbf T\mathbf u_{M_n i}\right\|_2^3\longrightarrow0.
\end{align*}

	Thus, by Lyapunov's theorem, we complete the proofs.
\end{proof}

\begin{proof}[Proof of Corollary~\ref{convergence1full}]
For two stages, the joint influence representation is
\[
\sqrt n\begin{pmatrix}\widehat{\boldsymbol\gamma}_1-\boldsymbol\gamma_{*,1}\\
\widehat{\boldsymbol\alpha}^{(1)}_1-\boldsymbol\alpha^{(1)}_{*,1}\end{pmatrix}
=\frac1{\sqrt n}\sum_{i=1}^n
\begin{pmatrix}
\mathbf P_{\gamma}^{(2)}\mathbf V^{(2)}\mathbf u_{2i}/\phi^{(2)}\\
(\mathbf D^{(1)})^{-1}\{\mathbf u_{1i}+\mathbf J^{(1)}\mathbf P^{(2)}\mathbf V^{(2)}\mathbf u_{2i}/\phi^{(2)}\}
\end{pmatrix}+o_p(1).
\]
Since $\operatorname{Var}\{\mathbf V^{(2)}\mathbf u_{2i}/\phi^{(2)}\}=\mathbf V^{(2)}$ and the scores from the two stages are uncorrelated, the covariance is the matrix displayed in Corollary~\ref{convergence1full}. The same triangular-array argument applied to this joint influence vector proves the corollary.
\end{proof}

\begin{proof}[Proofs of Corollaries~\ref{1-opt} and \ref{1-diff}]
	If all conditions in the theorems before hold, note that $\mathbf{H}^{(M_n-1)}$ is a sub-vector of $\mathbf{H}^{(M_n)}$, the tail behaviour of each component can be bounded using the first part of Assumption~\ref{Eb''}. The value inference proof then follows the same decomposition as the terminal-stage value proof, with Corollary~\ref{convergence1full} supplying the asymptotic distribution of \(\widehat{\bm{\beta}}_{*,1}^{(M_n-1)}\).
\end{proof}

\subsection{\texorpdfstring{Proofs for Diverging $M_n$}{Proofs for Diverging Mn}}

\begin{proof}[Proof of Lemma~\ref{lemma:accumulated-verification}]
Bounded action indicators preserve coordinate sub-Gaussian tails, and the derivatives evaluated at the true maximizing actions are differences of the same history coordinates multiplied by bounded indicators. There are at most polynomially many such design components across stages. A union bound therefore gives $b_n=O(\sqrt{x})$ with the required quantile probability. Products of two such coordinates have uniformly bounded sub-exponential norms. Applying Bernstein's inequality to the independent trajectories, followed by a union bound over the $O(M_nS_n^2)$ relevant matrix entries, bounds the maximum entrywise error by $C\{\sqrt{x/n}+x/n\}$. The inequality $\|\mathbf U\|_2\leq\sqrt{ab}\|\mathbf U\|_{\max}$ for an $a\times b$ matrix yields \eqref{eq:verified-matrix-rates}. The matrix $\boldsymbol{\mathcal J}_m$ has $O(S_n)$ rows and columns, including those for the inherited coefficients. At the true terminal predictor, the conditional sub-Gaussian bound in Assumption~\ref{subgm} and the fixed dispersion imply $b''\leq v_0^2/\phi^{(M_n)}$, so the same argument applies to the weighted terminal information matrix.

For the terminal Taylor term, use the last matrix condition of Assumption~\ref{design2}, already included in Assumption~\ref{designm}. The integral Taylor remainder in each active score coordinate is bounded by $C\|\mathbf e_{M_n}\|_2^2$. Consequently,
\[
\|\boldsymbol\Delta_{M_n}\|_2
\leq C\sqrt{s_n^{(M_n)}}\,\|\mathbf e_{M_n}\|_2^2,
\]
which gives the last term of $q_{M_n,n}$.

Within the coefficient neighborhoods used in the proof, SCAD and MCP have zero active penalty derivatives eventually, because $g_n^{(m)}/\lambda_n^{(m)}\to\infty$ uniformly. Thus $h_{m,n}=0$. The unconditional gap bound displayed in Assumption~\ref{difflbm} gives $\mathbb P(G_m\leq u)\leq CL_nu$ at the relevant radii. Applying Bernstein's inequality to these fixed gap events controls the maximization term in \eqref{eq:local-remainder-rates}.

The preceding bounds give
\begin{equation}
\begin{split}
\sqrt n\sum_{j=1}^{M_n}q_{j,n}
\leq C\bigg\{&\frac{(M_nS_n)^{3/2}x}{\sqrt n}
+\frac{L_nM_n^2S_n^{5/2}x^{5/2}}{\sqrt n}\\
&+\frac{(M_nS_n)^{3/2}x^{5/2}}{n}
+\frac{(s_n^{(M_n)})^{3/2}\log n}{\sqrt n}\bigg\}=o(1).
\end{split}
\label{eq:verified-accumulation}
\end{equation}
Indeed, $\ell_5+\ell_6<1/3$ controls the first, third, and terminal terms, while $\ell_1+3\ell_5+2\ell_6<1/2$ controls the second. Under Assumption~\ref{propagationm}, $a_{m,n}\leq C\sum_{j=m}^{M_n}q_{j,n}$, and $r_m\geq\sqrt{\log n/n}$ for a nonempty active vector, proving \eqref{eq:consistency-accumulation}. The existing Gram bounds and \eqref{eq:verified-matrix-rates} imply $\sup_m\|\mathbf D^{(m)}\|_2<\infty$; moreover, $\mathbf K^{(m)}\succeq\boldsymbol\Omega_m\succeq c\mathbf I$. Equation~\eqref{eq:verified-accumulation} therefore also proves \eqref{eq:normalized-accumulation}. For a general penalty, add $\sqrt n\sum_jh_{j,n}$ to the right-hand side.
\end{proof}

\begin{proof}[Proof of Theorem~\ref{convergencem}]
Write $\mathbf e_m=\widehat{\boldsymbol\vartheta}_m-\boldsymbol\vartheta_m$. Constants are uniform over stages. We work on the joint design events in Assumption~\ref{designm} and the events on which the quantities used to define $b_n$ and $d_{m,n}$ do not exceed these bounds. A union bound gives probability $1-o(1)$ for their intersection. Equation~\eqref{eq:consistency-accumulation} implies $\max_m d_{m,n}=o(1)$, because $a_{m,n}\geq q_{m,n}\geq d_{m,n}r_m$. Together with the existing empirical Gram and information bounds, this gives uniform bounds on $\mathbf D^{(m)}$, $(\mathbf D^{(m)})^{-1}$, and $\mathbf H_{M_n}^{-1}$.

For all active coefficient vectors in a fixed ball of radius $C r_j$, the fitted and true action-dependent values differ by at most $t_{j,n}$ when the relevant sample coordinates are bounded by $b_n$. The action-independent inherited term cancels when comparing actions. Thus, simultaneously over that entire ball,
\begin{equation}
\{\widehat\pi_j(H_i)\ne\pi_j(H_i)\}\subseteq\{G_j(H_i)\leq 2t_{j,n}\},\qquad
|\omega_{j-1,i}|\leq 2t_{j,n}\,1\{G_j(H_i)\leq2t_{j,n}\},
\label{eq:training-envelope}
\end{equation}
where $\omega_{j-1,i}$ is the remainder after linearizing the adjusted maximized pseudo-outcome at a true maximizing action. Selection is understood at the action-block level, with a fixed tie-breaking rule. The set on the right depends on the true parameters and a deterministic radius.

The unconditional gap bound in Assumption~\ref{difflbm} bounds its probability by $CL_nt_{j,n}$. Applying Bernstein's inequality to these fixed Bernoulli indicators and taking a union bound over stages gives, with probability $1-o(1)$,
\begin{equation}
\mathbb P_n\{G_j\leq2t_{j,n}\}
\leq C\{L_nt_{j,n}+x_n/n\},\qquad j\leq M_n.
\label{eq:training-bernstein}
\end{equation}
Indeed, the usual deviation $C\{\sqrt{p x_n/n}+x_n/n\}$ is at most $C'(p+x_n/n)$; its failure probability can be made $O((nM_n)^{-2})$ at each stage. Consequently,
\begin{equation}
\|\mathbb P_n(\mathbf S^{(m)}_{\mathrm{dfm},1}\omega_m)\|_2
\leq C b_n\sqrt{s_n^{(m)}}\,t_{m+1,n}\{L_nt_{m+1,n}+x_n/n\}.
\label{eq:max-score-remainder}
\end{equation}

The oracle active estimating equations, linearized at the true parameters and true maximizing actions, give
\begin{equation}
\begin{split}
\mathbf e_m&=\mathbf B_m\mathbf e_{m+1}+\mathbf A_m\mathbb P_n\mathbf u_m+\mathbf q_m,\qquad m<M_n,\\
\mathbf e_{M_n}&=\mathbf A_{M_n}\mathbb P_n\mathbf u_{M_n}+\mathbf q_{M_n}.
\end{split}
\label{eq:local-full-recursion}
\end{equation}
On the coefficient balls, the terms from the errors in the empirical matrices are bounded by $Cd_{m,n}(r_m+r_{m+1})$. The active penalty gradient has norm at most $h_{m,n}$, because the signal condition keeps these coefficients above $g_n^{(m)}$. Equation~\eqref{eq:max-score-remainder} controls the nonsmooth maximization term. Therefore
\[
\|\mathbf q_m\|_2\leq Cq_{m,n},\qquad m<M_n.
\]
At stage $M_n$, denote the normalized score Taylor remainder by $\boldsymbol\Delta_{M_n}$. With $\boldsymbol\beta_t=\boldsymbol\beta_{*,1}^{(M_n)}+t\mathbf e_{M_n}$, its $k$th coordinate satisfies
\[\begin{aligned}
|\Delta_{M_n,k}|
&\leq\int_0^1(1-t)\,\mathbf e_{M_n}^\top\Big[n^{-1}(\bm\Psi_1^{(M_n)})^\top\\
&\qquad\operatorname{diag}\{|\mathbf h_{k,1}^{(M_n)}|\circ
|\bm\mu^{(M_n)''}(\bm\Psi_1^{(M_n)}\boldsymbol\beta_t)|\}\,
\bm\Psi_1^{(M_n)}\Big]\mathbf e_{M_n}\,dt\\
&\leq C\|\mathbf e_{M_n}\|_2^2.
\end{aligned}\]
The last inequality is exactly the last matrix condition in Assumption~\ref{design2}. Thus
\[
\|\boldsymbol\Delta_{M_n}\|_2\leq C\sqrt{s_n^{(M_n)}}\,\|\mathbf e_{M_n}\|_2^2,
\qquad \|\mathbf q_{M_n}\|_2\leq Cq_{M_n,n}.
\]
Iterating \eqref{eq:local-full-recursion} gives
\begin{equation}
\mathbf e_m=\mathbb P_n\boldsymbol\psi_m+\mathbf{Rem}_m,\qquad
\mathbf{Rem}_m=\sum_{j=m}^{M_n}\boldsymbol\Gamma_{mj}\mathbf q_j,\qquad
\|\mathbf{Rem}_m\|_2\leq C a_{m,n}.
\label{eq:accumulated-expansion}
\end{equation}

When only the terminal outcome is used, the scores have conditional mean zero given the history at the corresponding stage. Conditional on the design at that stage, Assumption~\ref{subgm} supplies a uniform sub-Gaussian bound. Revealing the $n$ independent trajectories stage by stage therefore gives the exponential-supermartingale bound for any deterministic linear combination of these scores. On the event where the Gram bounds hold its variance proxy for $\mathbb P_n\boldsymbol\psi_m$ is bounded by
\[
\mathbf Q_m=\frac{C}{n}\sum_{j=m}^{M_n}
\boldsymbol\Gamma_{mj}\mathbf A_j\mathbf A_j^\top\boldsymbol\Gamma_{mj}^\top,
\qquad \operatorname{tr}(\mathbf Q_m)\leq C T_m/n.
\]
Integrating the scalar exponential bound against an independent standard Gaussian vector gives, on this event and for $0<t<\|\mathbf Q_m\|_2^{-1}$,
\[
\mathbb E\left[1_{\mathrm{Gram}}\exp\{t\|\mathbb P_n\boldsymbol\psi_m\|_2^2/2\}\right]
\leq\det(\mathbf I-t\mathbf Q_m)^{-1/2}.
\]
The same inequality follows by stopping before the predictable variance exceeds its deterministic bound. Taking $t$ to be a fixed fraction of $\{\operatorname{tr}(\mathbf Q_m)\}^{-1}$ and using Markov's inequality shows that $\|\mathbb P_n\boldsymbol\psi_m\|_2\leq C r_m$ simultaneously for all stages, with probability tending to one. The union bound uses $M_n=O(n^{\ell_6})$.

The preceding remainder bounds hold uniformly over fixed coefficient balls. Scale the score perturbations, empirical-matrix perturbations, and active penalty gradients from zero to their observed values. The restricted objectives have positive definite local Hessians under the design and penalty-curvature conditions, and their local roots vary continuously with the downstream pseudo-outcomes. If the resulting sequence of roots first reached the boundary of a sufficiently large ball $\|\mathbf e_m\|_2\leq C r_m$, \eqref{eq:accumulated-expansion}, the score bound, and $\max_m a_{m,n}/r_m=o(1)$ would place it strictly inside every ball, a contradiction. This proves the uniform coefficient rates. The inactive-score and cross-design bounds in Assumptions~\ref{designm}--\ref{subgm}, together with $\lambda_n^{(m)}\gg r_m$ and $\lambda_n^{(m)}\gg\sqrt{\log p_n^{(m)}/n}$, give the strict inactive KKT inequalities as in the proof of Theorem~\ref{convergence1}. In taking the stage-wise union, the noise threshold may be chosen proportional to $\sqrt{\log p_n^{(m)}+\log(nM_n)}$; its extra $\sqrt{\log(nM_n)/n}$ term is dominated by $r_m$. Positive active curvature then gives strict local minimizers with the stated zero coordinates.

For $j<k$ in the primary setting, $\mathbf u_j$ is $\mathcal F_{j+1}$-measurable and $\mathbb E(\mathbf u_k\mid\mathcal F_k)=0$. Thus
\[
\mathbb E(\mathbf u_j\mathbf u_k^\top)
=\mathbb E\{\mathbf u_j\mathbb E(\mathbf u_k^\top\mid\mathcal F_k)\}=\mathbf0.
\]
The utility conditions in Assumption~\ref{B5m} give the same orthogonality for the stated extension. Equations~\eqref{eq:full-influence}--\eqref{eq:full-covariance} are therefore the covariance identities for the linear term. In particular, $\mathbf L_{mm}=\mathbf I$ and $\mathbf K^{(m)}\succeq\boldsymbol\Omega_m\succeq c\mathbf I$. For uniformly bounded contrast matrices,
\begin{align*}
&\max_{m<M_n}\sqrt n\,
\|\mathbf C^{(m)}(\mathbf K^{(m)})^{-1/2}\mathbf D^{(m)}\mathbf E_m\mathbf{Rem}_m\|_2\\*
&\quad\leq C\sqrt n\max_{m<M_n}\sum_{j=m}^{M_n}
\big\|(\mathbf K^{(m)})^{-1/2}\mathbf D^{(m)}\mathbf E_m\boldsymbol\Gamma_{mj}\big\|_2q_{j,n}
=o(1)
\end{align*}
by Assumption~\ref{normalm}(b). The distributional assertion applies to any sequence of stages and contrasts satisfying the theorem.

Fix such a stage and a vector $\mathbf a\in\mathbb R^q$, and define
\[
\mathbf v_j=\mathbf L_{mj}^\top(\mathbf K^{(m)})^{-1/2}(\mathbf C^{(m)})^\top\mathbf a,\qquad
X_{ni}=\sum_{j=m}^{M_n}\mathbf v_j^\top\mathbf u_{ji}.
\]
The covariance identity gives
\begin{equation}
\sum_{j=m}^{M_n}\mathbf v_j^\top\boldsymbol\Omega_j\mathbf v_j
=\mathbf a^\top\mathbf C^{(m)}(\mathbf C^{(m)})^\top\mathbf a=O(1),
\qquad \sum_{j=m}^{M_n}\|\mathbf v_j\|_2^2\leq C.
\label{eq:normalized-energy}
\end{equation}
The third-moment condition in Assumption~\ref{normalm}(a) and Minkowski's inequality imply
\[
\|\mathbf v_j^\top\mathbf u_j\|_{L^3}
\leq C\|\mathbf v_j\|_1\leq C\sqrt{s_n^{(j)}}\|\mathbf v_j\|_2.
\]
Applying Minkowski once more across stages, followed by Cauchy--Schwarz and \eqref{eq:normalized-energy}, yields
\begin{equation}
\frac{\mathbb E|X_{ni}|^3}{\sqrt n}
\leq \frac{C}{\sqrt n}\left(\sum_{j=m}^{M_n}\sqrt{s_n^{(j)}}\|\mathbf v_j\|_2\right)^3
\leq C\sqrt{\frac{T_m^3}{n}}
\leq C\sqrt{\frac{M_n^3S_n^3}{n}}=o(1).
\label{eq:lyapunov-energy}
\end{equation}
The observations indexed by $i$ are independent trajectories. Their variance converges to $\mathbf a^\top\mathbf G^{(m)}\mathbf a$, so the triangular-array Lyapunov theorem and Cram\'er--Wold device establish the claimed normal limit. Combining this limit with the bound on the sum of the remainders completes the proof. Terminal-stage inference uses the terminal GLM normalization in Proposition~\ref{convergence2}.
\end{proof}

\subsection{Proofs for Policy Regret}\label{proofs:regret}

\begin{proof}[Proof of Proposition~\ref{prop:regret-multistage}]
For brevity, write
\[
Z_*=\widetilde\Gamma(\mathbf H_{\pi_*}^{(M)}),
\qquad
Z_{\mathrm{dfm}}
=\widetilde\Gamma(\mathbf H_{\pi_{\mathrm{dfm}}}^{(M)}),
\]
and let \(F_*\) and \(F_{\mathrm{dfm}}\) be their distribution functions. The optimality of \(\pi_{\mathrm{dfm}}\) for \(V_{\mathrm{dfm}}\) gives
\[
\Delta
:=
\mathbb E(Z_{\mathrm{dfm}})-\mathbb E(Z_*)\geq0.
\]
Finite first moments imply
\[
\int_{\mathbb R}\{F_{\mathrm{dfm}}(t)-F_*(t)\}\,dt=-\Delta,
\qquad
\int_{\mathbb R}|F_{\mathrm{dfm}}(t)-F_*(t)|\,dt
=W_1\{\mathcal L(Z_*),\mathcal L(Z_{\mathrm{dfm}})\}.
\]
Consequently,
\begin{equation}
\int_{\mathbb R}
\{F_{\mathrm{dfm}}(t)-F_*(t)\}_+\,dt
=
\frac12\left[
W_1\{\mathcal L(Z_*),\mathcal L(Z_{\mathrm{dfm}})\}
-\Delta
\right].
\label{eq:positive-cdf-area}
\end{equation}
Let \(g=b^{(M)'}\). An increasing Lipschitz function is absolutely continuous on compact intervals, differentiable almost everywhere, and satisfies \(0\leq g'(t)\leq L_b\) almost everywhere. The integrated-CDF identity, first for truncated variables and then by passage to the limit, gives
\begin{align*}
\mathcal R_{\mathrm{dfm}}
&=
\mathbb E\{g(Z_*)\}-\mathbb E\{g(Z_{\mathrm{dfm}})\}\\
&=
\int_{\mathbb R}g'(t)
\{F_{\mathrm{dfm}}(t)-F_*(t)\}\,dt\\
&\leq
L_b
\int_{\mathbb R}
\{F_{\mathrm{dfm}}(t)-F_*(t)\}_+\,dt.
\end{align*}
Substitution of \eqref{eq:positive-cdf-area} proves the claimed directional Wasserstein bound. For the logistic mean function, \(\sup_t g'(t)=1/4\), yielding the coefficient \(1/8\).

The bound also implies the centered-distribution bound. Let \(\mu_*=\mathbb E(Z_*)\) and \(\mu_{\mathrm{dfm}}=\mathbb E(Z_{\mathrm{dfm}})\). Coupling the centered variables and using the triangle inequality yields
\[
W_1\{\mathcal L(Z_*),\mathcal L(Z_{\mathrm{dfm}})\}
-\Delta
\leq
W_1\{\mathcal L(Z_*-\mu_*),
\mathcal L(Z_{\mathrm{dfm}}-\mu_{\mathrm{dfm}})\}.
\]
Thus the directional result is never weaker than the centered Wasserstein bound.
\end{proof}

\begin{proof}[Proof of Corollary~\ref{cor:regret-fosd}]
The deformed Bellman recursion gives
\[
\mathbb E[Z_m\mid\mathbf H^{(m)}=\mathbf h,\mathbf A^{(m)}=\mathbf a]
=Q_{\mathrm{dfm}}^{(m)}(\mathbf h,\mathbf a),
\]
with subsequent actions chosen by \(\pi_{\mathrm{dfm}}\). Under conditional independence, let \(G_m(\cdot\mid\mathbf h)\) denote the common conditional CDF of \(U^{(m)}\). Then
\[
\mathbb P(Z_m\leq t\mid\mathbf H^{(m)}=\mathbf h,\mathbf A^{(m)}=\mathbf a)
=G_m\!\left(t-Q_{\mathrm{dfm}}^{(m)}(\mathbf h,\mathbf a)\mid\mathbf h\right).
\]
Thus \(\pi_{\mathrm{dfm}}^{(m)}(\mathbf h)\) minimizes this CDF for every \(t\). Fix a threshold \(t\). At stage \(M-1\), this gives the smallest conditional CDF of the terminal deformed optimal Q-function. If the downstream deformed regime minimizes that CDF at every subsequent history, replacing any competing downstream regime by it cannot increase the CDF for a fixed current action. Choosing \(\pi_{\mathrm{dfm}}^{(m)}\) then minimizes the CDF over current actions as well. Backward induction and integration over the baseline history yield
\[
\mathbb P\!\left\{\widetilde\Gamma(\mathbf H_{\pi_{\mathrm{dfm}}}^{(M)})\leq t\right\}
\leq
\mathbb P\!\left\{\widetilde\Gamma(\mathbf H_{\pi_*}^{(M)})\leq t\right\}.
\]
Since \(t\) is arbitrary, \(F_{\mathrm{dfm}}\leq F_*\). Under this ordering, the positive CDF difference in \eqref{eq:positive-cdf-area} vanishes, so Proposition~\ref{prop:regret-multistage} gives \(\mathcal R_{\mathrm{dfm}}=0\).
\end{proof}

For the two-stage results below, let \(g=b^{(2)'}\) and define
\[
Q^{(1)}(\mathbf h,\mathbf a)
=\mathbb E\!\left[g\{\widetilde\Gamma(\mathbf H^{(2)})\}
\mid\mathbf H^{(1)}=\mathbf h,\mathbf A^{(1)}=\mathbf a\right].
\]

\begin{proof}[Proof of Proposition~\ref{prop:regret-second-order}]
Conditionally on \((\mathbf H^{(1)},\mathbf A^{(1)})=(\mathbf h,\mathbf a)\), write
\[
\widetilde\Gamma(\mathbf H^{(2)})
=
Q_{\mathrm{dfm}}^{(1)}(\mathbf h,\mathbf a)
+U_0^{(1)}+R^{(1)}.
\]
Define
\[
\overline Q^{(1)}(\mathbf h,\mathbf a)
=
\mathbb E\!\left[
g\!\left\{
Q_{\mathrm{dfm}}^{(1)}(\mathbf h,\mathbf a)+U_0^{(1)}
\right\}
\mid\mathbf H^{(1)}=\mathbf h
\right].
\]
The conditional distribution of \(U_0^{(1)}\) is common across actions and \(g\) is increasing. Hence every action maximizing \(Q_{\mathrm{dfm}}^{(1)}(\mathbf h,\mathbf a)\) also maximizes \(\overline Q^{(1)}(\mathbf h,\mathbf a)\).

Taylor expansion in \(R^{(1)}\), together with
\[
\mathbb E\{R^{(1)}\mid
U_0^{(1)},\mathbf H^{(1)},\mathbf A^{(1)}\}=0,
\]
gives
\begin{align*}
&\left|
Q^{(1)}(\mathbf h,\mathbf a)
-
\overline Q^{(1)}(\mathbf h,\mathbf a)
\right|\\
&\qquad\leq
\frac{C_b}{2}
\mathbb E\!\left\{
(R^{(1)})^2
\mid
\mathbf H^{(1)}=\mathbf h,\mathbf A^{(1)}=\mathbf a
\right\},
\qquad
C_b=\sup_t|b^{(2)'''}(t)|.
\end{align*}
Comparing an action maximizing the expected terminal outcome with a deformed-optimal action, the difference in $\overline Q^{(1)}$ is nonpositive. The two Taylor remainders therefore give
\[
\mathcal R_{\mathrm{dfm}}
\leq
C_b\,
\mathbb E\!\left[
\sup_{\mathbf a\in\mathcal A}
\mathbb E\!\left\{
(R^{(1)})^2
\mid
\mathbf H^{(1)},
\mathbf A^{(1)}=\mathbf a
\right\}
\right].
\]
For the logistic mean function,
\(\sup_t|b^{(2)'''}(t)|=1/(6\sqrt3)\).
\end{proof}

The following example shows that conditional independence is not necessary for zero regret.

\begin{example}\label{ex:two-point}
Conditionally on \(\mathbf H^{(1)}=\mathbf h\), suppose that
\(\widetilde\Gamma(\mathbf H^{(2)})\) takes the same two values
\(z_-(\mathbf h)<z_+(\mathbf h)\) under every action, while
\[
p(\mathbf h,\mathbf a)
=
\mathbb P\!\left\{
\widetilde\Gamma(\mathbf H^{(2)})=z_+(\mathbf h)
\mid\mathbf H^{(1)}=\mathbf h,\mathbf A^{(1)}=\mathbf a
\right\}
\]
may depend arbitrarily on \(\mathbf a\). Then
\begin{align*}
Q_{\mathrm{dfm}}^{(1)}(\mathbf h,\mathbf a)
&=
z_-(\mathbf h)
+p(\mathbf h,\mathbf a)\{z_+(\mathbf h)-z_-(\mathbf h)\},\\*
Q^{(1)}(\mathbf h,\mathbf a)
&=
g\{z_-(\mathbf h)\}
+p(\mathbf h,\mathbf a)
\bigl[g\{z_+(\mathbf h)\}-g\{z_-(\mathbf h)\}\bigr].
\end{align*}
The deformed objective is strictly increasing and the expected terminal outcome is nondecreasing in the same mixing probability, so every deformed-optimal action is also optimal for the expected terminal outcome. This is also an instance of first-order stochastic dominance, although the centered residual distribution generally changes with the action.
\end{example}

\section{Supporting Materials for Simulation}\label{app:simulation}\label{A-simu}

\label{app:sim-dgp}
We specify the complete two-stage model used in Section~\ref{simu} of the main text. The action set is $\mathcal A_{11}=\{a_k=(k-1)/10:k=1,\ldots,11\}$. Independent trajectories are generated as follows.

The baseline history is $\mathbf H^{(1)}=(1,X,Z,V,\bm\xi_1^\top)^\top$, where $X$ takes values $-1$ and $1$ with equal probabilities, $V\sim N(0,1)$, and $Z=(U-2/7)/\sqrt{10/392}$ with $U\sim\operatorname{Beta}(2,5)$. These three variables are mutually independent. The added noise vectors $\bm\xi_1,\bm\xi_2\sim N_q(0,\Sigma_q)$, with $(\Sigma_q)_{ij}=0.5^{|i-j|}$, are mutually independent and independent of all other variables, with zero true coefficients. Conditional on $\mathbf H^{(1)}$, the first-stage behavior probabilities are
\[
b_k(\mathbf H^{(1)})=\frac{0.15}{11}+0.85\frac{\exp\{2\tanh(Z+0.5V)(2a_k-1)\}}{\sum_{j=1}^{11}\exp\{2\tanh(Z+0.5V)(2a_j-1)\}}.
\]
Thus every action has positive assignment probability. Define $p^0_{+,k}=\sigma(0.2k-1)$, $p^0_{-,k}=\sigma(-0.4k+0.9)$, and $\bar p^0_x=11^{-1}\sum_{k=1}^{11}p^0_{x,k}$. The binary transition satisfies $\mathbb P(X_{21}=1\mid\mathbf H^{(1)},A^{(1)}=a_k)=p_{X,k}$, where $p_{x,k}=(p^0_{x,k}+\bar p^0_x)/2$; otherwise $X_{21}=-1$.

Independently of the draw of $X_{21}$ given the baseline history and action, let
\[
W=g(A^{(1)})Z+\{1+0.8\tanh(V)\}\varepsilon,\qquad \varepsilon\sim N(0,1).
\]
The positive-action loadings interpolate linearly between $g(0.1)=0.8$ and $g(1)=3$; the loading at $a=0$ is set to zero.

\[
g(a)=\begin{cases}
0,&a=0,\\
\dfrac45+\dfrac{22}{9}\left(a-\dfrac1{10}\right),&a\in\{0.1,0.2,\ldots,1\}.
\end{cases}
\]

Set $\mathbf H^{(2)}=((\mathbf H^{(1)})^\top,A^{(1)},X_{21},W,\bm\xi_2^\top)^\top$. The second-stage action is uniform on $\mathcal A_{11}$ independently of $\mathbf H^{(2)}$. Set $Y^{(1)}=0$ and generate $Y^{(2)}\mid\mathbf H^{(2)},A^{(2)}=a_k\sim\operatorname{Bernoulli}\{\sigma(\eta_k)\}$, with
\[
\eta_k=-\frac32-\frac32X+\frac32 A^{(1)}-\frac32X_{21}+W+d_k(X,X_{21}).
\]
The action increments are
\[
(d_1,\ldots,d_{11})=\left(0,3X,-2,2X,\frac53X,\frac52X,-2,\frac{12}{5}X,
\frac95X_{21},\frac52X_{21},\frac{20}{3}X_{21}\right).
\]

The $(77+22q)$-dimensional terminal coefficient vector has 15 nonzero entries. The main study uses $(n,q)=(5000,250),(5000,500),(10000,500)$ and 1000 replications per setting.

\subsection{Continuation Values and Population Regret}\label{app:sim-continuation}
Let $B=I(X_{21}=1)$. Direct maximization over the 11 terminal actions gives
\[
T:=\max_k\eta_k=\frac32+\frac32 A^{(1)}+\left(\frac{13}{6}-\frac32X\right)B+W.
\]
The maximizing terminal action is $1$ when $X_{21}=1$, $0.1$ when $X_{21}=-1$ and $X=1$, and $0$ when $X_{21}=-1$ and $X=-1$. The terminal deformed optimal Q-function is separated from its closest competitor by at least $1/2$.

For $h=(1,x,z,v,\bm\xi_1^\top)^\top$ and $a=a_k$, write $c_x=13/6-3x/2$, $\ell(h,a)=3/2+(3/2)a+g(a)z$, and $s(v)=1+0.8\tanh(v)$. Then
\begin{equation}\label{eq:sim-mean}
\mu(h,a):=Q^{(1)}_{\mathrm{dfm}}(h,a)=\ell(h,a)+c_xp_{x,k}.
\end{equation}
The corresponding centered population residual is $c_X(B-p_{X,k})+s(V)\varepsilon$, with conditional variance $c_X^2p_{X,k}(1-p_{X,k})+s(V)^2$. Its conditional mean is zero, while its distribution depends on the history and action. In particular, varying $V$ changes the transition variance while preserving \eqref{eq:sim-mean}.

To match the recursive parameterization, let $t_0=(-3/2,-3/2,0,0,\mathbf 0_q^\top)^\top$ and write $\mu(h,a_k)=t_0^\top h+\alpha_k^\top h$. The first four entries of $\alpha_k$ are
\[
\alpha_{k,1:4}=\left(3+\frac32a_k+\frac13p_{+,k}+\frac{11}{6}p_{-,k},\quad
\frac32+\frac13p_{+,k}-\frac{11}{6}p_{-,k},\quad g(a_k),\quad 0\right)^\top.
\]
The remaining $q$ entries of each $\alpha_k$ are zero. There are 32 nonzero entries among the $44+11q$ action-block coefficients. The common vector $t_0$ is inherited from the terminal fit. Since the slopes $g(a_k)$ are distinct and $Z$ has a bounded density, first-stage score ties have probability zero and the probability of a score gap at most $u$ is $O(u)$ as $u\downarrow0$.

Let $F(m,s)=\int\sigma(m+se)\phi(e)\,de$, where $\phi$ is the standard normal density. The continuation value for the expected terminal outcome, with the optimal terminal action, equals
\begin{equation}\label{eq:sim-original}
q(h,a_k)=(1-p_{x,k})F\{\ell(h,a_k),s(v)\}+p_{x,k}F\{\ell(h,a_k)+c_x,s(v)\}.
\end{equation}
The deformed continuation $\mu$ is block-wise linear, whereas $q$ generally depends on $v$ through the conditional variance.

For example, at $X=-1$ and $a=0$, we have $q(h,0)=(1-p_{-,1})F\{3/2,s(v)\}+p_{-,1}F\{31/6,s(v)\}$. For $m,s>0$, pairing positive and negative Gaussian innovations shows that $\partial F(m,s)/\partial s<0$, because $\sigma'$ is even and strictly decreases with the absolute value of its argument. Hence $q(h,0)$ strictly decreases with $v$ and has distinct limits inside $(0,1)$ as $v\to-\infty$ and $v\to\infty$. These properties place $q$ outside the corresponding affine and logistic-affine classes in $v$, while \eqref{eq:sim-mean} remains exactly specified.

Define $\pi^{(1)}_{\mathrm{dfm}}(h)\in\arg\max_k\mu(h,a_k)$ and $\pi_*^{(1)}(h)\in\arg\max_kq(h,a_k)$, resolving numerical ties by the smallest index. Both use the same optimal terminal rule. The regret of the deformed regime is $\mathbb E[\max_kq(\mathbf H^{(1)},a_k)-q\{\mathbf H^{(1)},\pi^{(1)}_{\mathrm{dfm}}(\mathbf H^{(1)})\}]$, which is positive in this design.

For the population comparison in Table~\ref{tab:sim-population}, DFM uses the known $\mu$. The Gaussian and HQ comparators use unpenalized least-squares working-model projections; the binomial comparator uses the true terminal logistic predictor and an upstream least-squares projection of its optimized probability. Gaussian and Gauss--Jacobi quadrature integrate the projection moments; the refined rule uses 91, 51, and 181 nodes for $Z$, $V$, and the Gaussian innovation, respectively, followed by $2^{20}$ Sobol evaluation points. Irrelevant covariates independent of the core history have zero population projection slopes. The endpoint oracle separately optimizes both stages over $\{0,1\}$ while retaining full histories. The unrestricted and endpoint-restricted oracle values are 0.915562 and 0.835659, respectively, giving restriction regret 0.079903.

\begin{table}[!htbp]

\centering
\caption{Population regret for the expected terminal outcome}\label{tab:sim-population}
{\small\renewcommand{\baselinestretch}{1}\selectfont
\begin{tabular*}{\linewidth}{@{\extracolsep{\fill}}lc}
\toprule
Strategy & Regret \\
\midrule
DFM & 0.000735 \\
DROVE-G projection & 0.001059 \\
DROVE-B projection & 0.010253 \\
HQ projection & 0.093855 \\
\bottomrule
\end{tabular*}}

\end{table}

\subsection{Estimation, Comparison Procedures, and Policy Evaluation}\label{app:sim-estimation}\label{app:sim-benchmarks}
All methods reuse the previously selected tuning parameters, held fixed across 1000 replications per setting. DFM uses SCAD penalization with shape parameter 3.7, lasso-initialized LLA, and backward BIC calibration. Inference follows the theory in the main text; implementation details are provided in the simulation code.

HQ adapts the binary-action construction of \citet{Zhu2018}, coding doses below 0.5 as $-1$ and the remaining doses as $1$, while retaining the observed first-stage dose in the terminal history. Its calibration uses SCAD with BIC and thresholds chosen by bootstrap coverage. The learned rule selects dose 0 or 1 according to the fitted contrast sign.

DROVE uses the generalized folded-concave penalty of \citet{Ding2024} on coefficients and adjacent action-increment differences, with generalized-lasso initialization, generalized LLA, and holdout calibration. DROVE-G and DROVE-B propagate the optimized Gaussian terminal mean and logistic terminal probability, respectively. Both use Gaussian upstream regression and retain all 11 actions; DROVE-B extends the binomial GLM option to backward recursion.

Each replication evaluates all methods on 10000 shared independent baseline histories, generating transitions under each learned first-stage action with common random innovations. Regret decomposes into first-stage and terminal losses:
\[
R(\widehat\pi)=\mathbb E[\max_a q(H_1,a)-q(H_1,\widehat a_1)]
+\mathbb E_{\widehat\pi^{(1)}}[\max_a Q^{(2)}(H_2,a)-Q^{(2)}(H_2,\widehat a_2)].
\]
Table~\ref{tab:sim-loss-decomposition} reports both probability-scale losses, with the terminal loss evaluated under the learned first-stage rule. Table~\ref{tab:sim-paired} reports paired regret differences and Monte Carlo intervals.

\begin{table}[!htbp]

\centering
\caption{Decomposition of regret for the expected terminal outcome}\label{tab:sim-loss-decomposition}
{\small\renewcommand{\baselinestretch}{1}\selectfont
\begin{tabular*}{\linewidth}{@{\extracolsep{\fill}}llccc}
\toprule
$(n,q)$ & Method & Stage-1 loss & Stage-2 loss & Total regret \\
\midrule
$(5000,250)$ & DFM & 0.001690 & 0.000064 & 0.001754 \\
 & DROVE-G & 0.011725 & 0.010786 & 0.022511 \\
 & DROVE-B & 0.071259 & 0.170149 & 0.241408 \\
 & HQ & 0.008849 & 0.083403 & 0.092251 \\
\midrule
$(5000,500)$ & DFM & 0.001716 & 0.000095 & 0.001812 \\
 & DROVE-G & 0.013972 & 0.083474 & 0.097446 \\
 & DROVE-B & 0.075840 & 0.180243 & 0.256083 \\
 & HQ & 0.007595 & 0.082679 & 0.090274 \\
\midrule
$(10000,500)$ & DFM & 0.001187 & 0.000000 & 0.001187 \\
 & DROVE-G & 0.011030 & 0.005189 & 0.016218 \\
 & DROVE-B & 0.075861 & 0.001708 & 0.077569 \\
 & HQ & 0.008381 & 0.083356 & 0.091737 \\
\bottomrule
\end{tabular*}}

\end{table}

\begin{table}[!htbp]

\centering
\caption{Paired differences in mean regret: comparator minus DFM}\label{tab:sim-paired}
{\small\renewcommand{\baselinestretch}{1}\selectfont
\begin{tabular*}{\linewidth}{@{\extracolsep{\fill}}llccc}
\toprule
$(n,q)$ & Comparator & Difference & MCSE & 95\% MC interval \\
\midrule
$(5000,250)$ & DROVE-G & 0.020757 & 0.000708 & (0.019370, 0.022144) \\
 & DROVE-B & 0.239654 & 0.001320 & (0.237067, 0.242241) \\
 & HQ & 0.090498 & 0.000108 & (0.090286, 0.090709) \\
\midrule
$(5000,500)$ & DROVE-G & 0.095634 & 0.000831 & (0.094005, 0.097263) \\
 & DROVE-B & 0.254271 & 0.002696 & (0.248987, 0.259556) \\
 & HQ & 0.088462 & 0.000095 & (0.088276, 0.088648) \\
\midrule
$(10000,500)$ & DROVE-G & 0.015032 & 0.000440 & (0.014169, 0.015894) \\
 & DROVE-B & 0.076382 & 0.000334 & (0.075727, 0.077037) \\
 & HQ & 0.090550 & 0.000080 & (0.090393, 0.090707) \\
\bottomrule
\end{tabular*}}

\end{table}

\providecommand{\popgridappref}[1]{\ref{#1}}

\subsection{Population Regret}\label{subsec:sim-population-grid}
We extend the main simulation design in Appendix~\popgridappref{app:sim-dgp} to examine population regret across a broader range of data-generating mechanisms. We take $Z$ to be either standard normal or the standardized beta variable defined there, and the first-stage behavior policy to be either uniform or the specified history-dependent policy. The transitions become
\[
\begin{aligned}
\mathbb P(X_{21}=1\mid\mathbf H^{(1)},A^{(1)}=a_k)
 &=p_{X,k}(\tau):=(1-\tau)\bar p^0_X+\tau p^0_{X,k},\\
W&=\kappa g(A^{(1)})Z+s_0\{1+\gamma\tanh(V)\}\varepsilon,
\end{aligned}
\]
with $\tau\in\{0,0.1,0.2,0.3,0.4,0.5\}$, $\kappa\in\{0.5,1,2\}$, and six noise configurations $(s_0,\gamma)\in\{0.5,1,2\}\times\{0,0.8\}$. All remaining components follow the main simulation design. The grid contains 72 DGPs at each $\tau$, totaling 432. The main simulation corresponds to $(\tau,\kappa,s_0,\gamma)=(0.5,1,1,0.8)$ with the standardized beta covariate and history-dependent behavior policy.

The terminal deformed optimal Q-function has a block-wise linear conditional mean throughout the grid. At $\tau=0$, its centered conditional distribution is common across actions, yielding zero DFM regret. Positive $\tau$ introduces action-dependent changes in the binary mixture while retaining the block-wise linear deformed mean.

We use the population projections and regret criterion in Appendix~\popgridappref{app:sim-continuation}, with penalties and HQ thresholds set to zero. Gaussian and Gauss--Jacobi quadrature use 91, 51, and 181 nodes for $Z$, $V$, and the Gaussian innovation, respectively, and $2^{20}$ Sobol evaluation points approximate regime values. Table~\ref{tab:sim-popgrid-means} reports equally weighted mean regrets at each $\tau$, including the optimal regime within the two-endpoint action class.

\begin{table}[!htbp]
\centering
\caption{Mean population regret across transition structures}\label{tab:sim-popgrid-means}
{\small\renewcommand{\baselinestretch}{1}\selectfont
\begin{tabular*}{\linewidth}{@{\extracolsep{\fill}}lccccc}
\toprule
$\tau$ & DFM & DROVE-G & DROVE-B & HQ & Endpoint oracle \\
\midrule
0.0 & 0.000000 & 0.000799 & 0.013210 & 0.084141 & 0.079850 \\
0.1 & 0.000038 & 0.000783 & 0.013265 & 0.083828 & 0.079500 \\
0.2 & 0.000149 & 0.000770 & 0.013281 & 0.084188 & 0.078756 \\
0.3 & 0.000324 & 0.000766 & 0.013235 & 0.085232 & 0.077663 \\
0.4 & 0.000548 & 0.000777 & 0.013119 & 0.086961 & 0.076275 \\
0.5 & 0.000803 & 0.000807 & 0.012948 & 0.089376 & 0.074651 \\
\bottomrule
\end{tabular*}}
\end{table}

DFM has the smallest mean regret at all six evaluated $\tau$ values. Its regret is zero at $\tau=0$, where the centered continuation residual has a common conditional distribution across actions, and increases as the binary transition becomes action-dependent. Across individual DGPs, DFM has the lowest regret among the four procedures in 72, 72, 67, 52, 42, and 35 cases at the respective grid values. The endpoint-oracle results quantify the contribution of the action restriction to the HQ comparison.

\subsection{Additional Coefficient Results}\label{app:sim-coordinates}
Tables~\ref{tab:sim-beta-coordinates} and \ref{tab:sim-alpha-coordinates} report every core coefficient, including true zeros. SE and SD denote the mean estimated standard error and empirical standard deviation, respectively. Coverage is reported only for true nonzero coefficients; a dash indicates an unreported coverage rate for a true zero. Terminal core coordinates are ordered by the common vector and then increments for actions $a_2,\ldots,a_{11}$; each vector uses the seven core history variables in Appendix~\ref{app:sim-dgp}. Upstream core coordinates are ordered by action $a_1,\ldots,a_{11}$ and then $(1,X,Z,V)$.

\begingroup\renewcommand{\baselinestretch}{1}\fontsize{8.5}{10}\selectfont\setlength{\tabcolsep}{3pt}

\begin{longtable}{rr*{12}{r}}
\caption{Complete core coefficient results for $\bm\beta^{(2)}$}\label{tab:sim-beta-coordinates}\\
\toprule
Comp. & True & \multicolumn{4}{c}{$(5000,250)$} & \multicolumn{4}{c}{$(5000,500)$} & \multicolumn{4}{c}{$(10000,500)$} \\
\cmidrule(lr){3-6}\cmidrule(lr){7-10}\cmidrule(lr){11-14}
 &  & Mean & SE & SD & Cov. & Mean & SE & SD & Cov. & Mean & SE & SD & Cov. \\
\midrule\endfirsthead
\caption[]{Complete core coefficient results for $\bm\beta^{(2)}$ (continued)}\\
\toprule
Comp. & True & \multicolumn{4}{c}{$(5000,250)$} & \multicolumn{4}{c}{$(5000,500)$} & \multicolumn{4}{c}{$(10000,500)$} \\
\cmidrule(lr){3-6}\cmidrule(lr){7-10}\cmidrule(lr){11-14}
 &  & Mean & SE & SD & Cov. & Mean & SE & SD & Cov. & Mean & SE & SD & Cov. \\
\midrule\endhead
\midrule\multicolumn{14}{r}{Continued on next page}\\\endfoot
\bottomrule\endlastfoot
1 & -1.500 & -1.5103 & 0.0855 & 0.0862 & 0.949 & -1.5040 & 0.0853 & 0.0880 & 0.944 & -1.5060 & 0.0603 & 0.0600 & 0.954 \\
2 & -1.500 & -1.5063 & 0.0779 & 0.0839 & 0.932 & -1.4906 & 0.0775 & 0.1108 & 0.914 & -1.5027 & 0.0550 & 0.0543 & 0.957 \\
3 & 0.000 & 0.0000 & 0.0000 & 0.0000 & -- & 0.0000 & 0.0000 & 0.0000 & -- & 0.0000 & 0.0000 & 0.0000 & -- \\
4 & 0.000 & 0.0000 & 0.0000 & 0.0000 & -- & 0.0000 & 0.0000 & 0.0000 & -- & 0.0000 & 0.0000 & 0.0000 & -- \\
5 & 1.500 & 1.5102 & 0.1393 & 0.1427 & 0.946 & 1.5028 & 0.1391 & 0.1408 & 0.949 & 1.5097 & 0.0982 & 0.0985 & 0.943 \\
6 & -1.500 & -1.5090 & 0.0665 & 0.0655 & 0.947 & -1.5039 & 0.0663 & 0.0676 & 0.948 & -1.5048 & 0.0469 & 0.0468 & 0.956 \\
7 & 1.000 & 1.0065 & 0.0346 & 0.0367 & 0.933 & 1.0038 & 0.0344 & 0.0366 & 0.937 & 1.0024 & 0.0243 & 0.0243 & 0.949 \\
8 & 0.000 & 0.0000 & 0.0000 & 0.0000 & -- & 0.0000 & 0.0000 & 0.0000 & -- & 0.0000 & 0.0000 & 0.0000 & -- \\
9 & 3.000 & 3.0134 & 0.1709 & 0.1732 & 0.952 & 2.9998 & 0.1707 & 0.1928 & 0.914 & 3.0106 & 0.1207 & 0.1196 & 0.962 \\
10 & 0.000 & 0.0000 & 0.0000 & 0.0000 & -- & 0.0000 & 0.0000 & 0.0000 & -- & 0.0000 & 0.0000 & 0.0000 & -- \\
11 & 0.000 & 0.0000 & 0.0000 & 0.0000 & -- & 0.0000 & 0.0000 & 0.0000 & -- & 0.0000 & 0.0000 & 0.0000 & -- \\
12 & 0.000 & 0.0000 & 0.0000 & 0.0000 & -- & -0.0010 & 0.0003 & 0.0323 & -- & 0.0007 & 0.0002 & 0.0214 & -- \\
13 & 0.000 & 0.0000 & 0.0000 & 0.0000 & -- & 0.0000 & 0.0000 & 0.0000 & -- & 0.0000 & 0.0000 & 0.0000 & -- \\
14 & 0.000 & 0.0000 & 0.0000 & 0.0000 & -- & 0.0000 & 0.0000 & 0.0000 & -- & 0.0000 & 0.0000 & 0.0000 & -- \\
15 & -2.000 & -2.0079 & 0.1787 & 0.1838 & 0.946 & -2.0005 & 0.1783 & 0.1940 & 0.941 & -2.0114 & 0.1250 & 0.1254 & 0.943 \\
16 & 0.000 & 0.0000 & 0.0000 & 0.0000 & -- & -0.0009 & 0.0002 & 0.0284 & -- & 0.0000 & 0.0000 & 0.0000 & -- \\
17 & 0.000 & 0.0000 & 0.0000 & 0.0000 & -- & 0.0000 & 0.0000 & 0.0000 & -- & 0.0000 & 0.0000 & 0.0000 & -- \\
18 & 0.000 & 0.0000 & 0.0000 & 0.0000 & -- & 0.0000 & 0.0000 & 0.0000 & -- & 0.0000 & 0.0000 & 0.0000 & -- \\
19 & 0.000 & -0.0164 & 0.0073 & 0.1353 & -- & -0.0178 & 0.0074 & 0.1486 & -- & -0.0028 & 0.0010 & 0.0518 & -- \\
20 & 0.000 & 0.0000 & 0.0000 & 0.0000 & -- & 0.0000 & 0.0000 & 0.0000 & -- & 0.0000 & 0.0000 & 0.0000 & -- \\
21 & 0.000 & 0.0000 & 0.0000 & 0.0000 & -- & 0.0000 & 0.0000 & 0.0000 & -- & 0.0000 & 0.0000 & 0.0000 & -- \\
22 & 0.000 & 0.0000 & 0.0000 & 0.0000 & -- & 0.0000 & 0.0000 & 0.0000 & -- & 0.0000 & 0.0000 & 0.0000 & -- \\
23 & 2.000 & 2.0068 & 0.1585 & 0.1655 & 0.941 & 1.9892 & 0.1583 & 0.1845 & 0.935 & 2.0032 & 0.1121 & 0.1086 & 0.953 \\
24 & 0.000 & 0.0000 & 0.0000 & 0.0000 & -- & 0.0000 & 0.0000 & 0.0000 & -- & 0.0000 & 0.0000 & 0.0000 & -- \\
25 & 0.000 & 0.0000 & 0.0000 & 0.0000 & -- & 0.0000 & 0.0000 & 0.0000 & -- & 0.0000 & 0.0000 & 0.0000 & -- \\
26 & 0.000 & 0.0009 & 0.0003 & 0.0296 & -- & 0.0007 & 0.0003 & 0.0228 & -- & 0.0000 & 0.0000 & 0.0000 & -- \\
27 & 0.000 & 0.0000 & 0.0000 & 0.0000 & -- & 0.0000 & 0.0000 & 0.0000 & -- & 0.0000 & 0.0000 & 0.0000 & -- \\
28 & 0.000 & 0.0000 & 0.0000 & 0.0000 & -- & 0.0000 & 0.0000 & 0.0000 & -- & 0.0000 & 0.0000 & 0.0000 & -- \\
29 & 0.000 & 0.0000 & 0.0000 & 0.0000 & -- & 0.0000 & 0.0000 & 0.0000 & -- & 0.0000 & 0.0000 & 0.0000 & -- \\
30 & 1.667 & 1.6610 & 0.1558 & 0.2121 & 0.956 & 1.6203 & 0.1504 & 0.3571 & 0.927 & 1.6691 & 0.1105 & 0.1102 & 0.951 \\
31 & 0.000 & 0.0000 & 0.0000 & 0.0000 & -- & 0.0000 & 0.0000 & 0.0000 & -- & 0.0000 & 0.0000 & 0.0000 & -- \\
32 & 0.000 & 0.0000 & 0.0000 & 0.0000 & -- & 0.0000 & 0.0000 & 0.0000 & -- & 0.0000 & 0.0000 & 0.0000 & -- \\
33 & 0.000 & 0.0000 & 0.0000 & 0.0000 & -- & 0.0016 & 0.0005 & 0.0366 & -- & 0.0000 & 0.0000 & 0.0000 & -- \\
34 & 0.000 & 0.0000 & 0.0000 & 0.0000 & -- & 0.0000 & 0.0000 & 0.0000 & -- & 0.0000 & 0.0000 & 0.0000 & -- \\
35 & 0.000 & 0.0000 & 0.0000 & 0.0000 & -- & 0.0000 & 0.0000 & 0.0000 & -- & 0.0000 & 0.0000 & 0.0000 & -- \\
36 & 0.000 & 0.0000 & 0.0000 & 0.0000 & -- & 0.0000 & 0.0000 & 0.0000 & -- & 0.0000 & 0.0000 & 0.0000 & -- \\
37 & 2.500 & 2.5155 & 0.1638 & 0.1614 & 0.948 & 2.4969 & 0.1635 & 0.1810 & 0.931 & 2.5051 & 0.1155 & 0.1150 & 0.959 \\
38 & 0.000 & 0.0000 & 0.0000 & 0.0000 & -- & 0.0000 & 0.0000 & 0.0000 & -- & 0.0000 & 0.0000 & 0.0000 & -- \\
39 & 0.000 & 0.0000 & 0.0000 & 0.0000 & -- & 0.0000 & 0.0000 & 0.0000 & -- & 0.0000 & 0.0000 & 0.0000 & -- \\
40 & 0.000 & 0.0011 & 0.0008 & 0.0579 & -- & 0.0044 & 0.0016 & 0.0573 & -- & 0.0000 & 0.0000 & 0.0000 & -- \\
41 & 0.000 & 0.0000 & 0.0000 & 0.0000 & -- & 0.0000 & 0.0000 & 0.0000 & -- & 0.0000 & 0.0000 & 0.0000 & -- \\
42 & 0.000 & 0.0000 & 0.0000 & 0.0000 & -- & 0.0000 & 0.0000 & 0.0000 & -- & 0.0000 & 0.0000 & 0.0000 & -- \\
43 & -2.000 & -2.0039 & 0.1793 & 0.1959 & 0.942 & -1.9989 & 0.1781 & 0.1851 & 0.945 & -1.9987 & 0.1260 & 0.1335 & 0.944 \\
44 & 0.000 & 0.0000 & 0.0000 & 0.0000 & -- & -0.0043 & 0.0010 & 0.0605 & -- & 0.0000 & 0.0000 & 0.0000 & -- \\
45 & 0.000 & 0.0000 & 0.0000 & 0.0000 & -- & 0.0000 & 0.0000 & 0.0000 & -- & 0.0000 & 0.0000 & 0.0000 & -- \\
46 & 0.000 & 0.0000 & 0.0000 & 0.0000 & -- & 0.0000 & 0.0000 & 0.0000 & -- & 0.0000 & 0.0000 & 0.0000 & -- \\
47 & 0.000 & -0.0213 & 0.0096 & 0.1554 & -- & -0.0135 & 0.0060 & 0.1243 & -- & -0.0104 & 0.0046 & 0.0918 & -- \\
48 & 0.000 & 0.0000 & 0.0000 & 0.0000 & -- & 0.0000 & 0.0000 & 0.0000 & -- & 0.0000 & 0.0000 & 0.0000 & -- \\
49 & 0.000 & 0.0000 & 0.0000 & 0.0000 & -- & 0.0000 & 0.0000 & 0.0000 & -- & 0.0000 & 0.0000 & 0.0000 & -- \\
50 & 0.000 & 0.0000 & 0.0000 & 0.0000 & -- & 0.0000 & 0.0000 & 0.0000 & -- & 0.0000 & 0.0000 & 0.0000 & -- \\
51 & 2.400 & 2.4109 & 0.1626 & 0.1698 & 0.940 & 2.3992 & 0.1625 & 0.1818 & 0.921 & 2.4041 & 0.1146 & 0.1115 & 0.951 \\
52 & 0.000 & 0.0000 & 0.0000 & 0.0000 & -- & 0.0000 & 0.0000 & 0.0000 & -- & 0.0000 & 0.0000 & 0.0000 & -- \\
53 & 0.000 & 0.0000 & 0.0000 & 0.0000 & -- & 0.0000 & 0.0000 & 0.0000 & -- & 0.0000 & 0.0000 & 0.0000 & -- \\
54 & 0.000 & 0.0000 & 0.0000 & 0.0000 & -- & 0.0007 & 0.0003 & 0.0235 & -- & 0.0000 & 0.0000 & 0.0000 & -- \\
55 & 0.000 & 0.0000 & 0.0000 & 0.0000 & -- & 0.0000 & 0.0000 & 0.0000 & -- & 0.0000 & 0.0000 & 0.0000 & -- \\
56 & 0.000 & 0.0000 & 0.0000 & 0.0000 & -- & 0.0000 & 0.0000 & 0.0000 & -- & 0.0000 & 0.0000 & 0.0000 & -- \\
57 & 0.000 & 0.0000 & 0.0000 & 0.0000 & -- & 0.0000 & 0.0000 & 0.0000 & -- & 0.0000 & 0.0000 & 0.0000 & -- \\
58 & 0.000 & 0.0000 & 0.0000 & 0.0000 & -- & -0.0045 & 0.0009 & 0.0638 & -- & 0.0000 & 0.0000 & 0.0000 & -- \\
59 & 0.000 & 0.0000 & 0.0000 & 0.0000 & -- & 0.0000 & 0.0000 & 0.0000 & -- & 0.0000 & 0.0000 & 0.0000 & -- \\
60 & 0.000 & 0.0000 & 0.0000 & 0.0000 & -- & 0.0000 & 0.0000 & 0.0000 & -- & 0.0000 & 0.0000 & 0.0000 & -- \\
61 & 0.000 & 0.0000 & 0.0000 & 0.0000 & -- & 0.0000 & 0.0000 & 0.0000 & -- & -0.0007 & 0.0002 & 0.0219 & -- \\
62 & 1.800 & 1.8127 & 0.1551 & 0.1580 & 0.950 & 1.8029 & 0.1547 & 0.1526 & 0.950 & 1.8077 & 0.1092 & 0.1124 & 0.942 \\
63 & 0.000 & 0.0000 & 0.0000 & 0.0000 & -- & 0.0000 & 0.0000 & 0.0000 & -- & 0.0000 & 0.0000 & 0.0000 & -- \\
64 & 0.000 & 0.0000 & 0.0000 & 0.0000 & -- & 0.0000 & 0.0000 & 0.0000 & -- & 0.0000 & 0.0000 & 0.0000 & -- \\
65 & 0.000 & 0.0000 & 0.0000 & 0.0000 & -- & -0.0016 & 0.0003 & 0.0367 & -- & 0.0000 & 0.0000 & 0.0000 & -- \\
66 & 0.000 & 0.0000 & 0.0000 & 0.0000 & -- & 0.0000 & 0.0000 & 0.0000 & -- & 0.0000 & 0.0000 & 0.0000 & -- \\
67 & 0.000 & 0.0000 & 0.0000 & 0.0000 & -- & 0.0000 & 0.0000 & 0.0000 & -- & 0.0000 & 0.0000 & 0.0000 & -- \\
68 & 0.000 & 0.0000 & 0.0000 & 0.0000 & -- & 0.0008 & 0.0003 & 0.0254 & -- & 0.0000 & 0.0000 & 0.0000 & -- \\
69 & 2.500 & 2.5094 & 0.1621 & 0.1629 & 0.950 & 2.4950 & 0.1615 & 0.1655 & 0.941 & 2.5067 & 0.1145 & 0.1115 & 0.953 \\
70 & 0.000 & 0.0000 & 0.0000 & 0.0000 & -- & 0.0000 & 0.0000 & 0.0000 & -- & 0.0000 & 0.0000 & 0.0000 & -- \\
71 & 0.000 & 0.0000 & 0.0000 & 0.0000 & -- & 0.0000 & 0.0000 & 0.0000 & -- & 0.0000 & 0.0000 & 0.0000 & -- \\
72 & 0.000 & 0.0000 & 0.0000 & 0.0000 & -- & 0.0000 & 0.0000 & 0.0000 & -- & 0.0000 & 0.0000 & 0.0000 & -- \\
73 & 0.000 & 0.0000 & 0.0000 & 0.0000 & -- & 0.0000 & 0.0000 & 0.0000 & -- & 0.0000 & 0.0000 & 0.0000 & -- \\
74 & 0.000 & 0.0000 & 0.0000 & 0.0000 & -- & 0.0000 & 0.0000 & 0.0000 & -- & 0.0000 & 0.0000 & 0.0000 & -- \\
75 & 0.000 & 0.0010 & 0.0004 & 0.0326 & -- & 0.0011 & 0.0004 & 0.0344 & -- & -0.0007 & 0.0003 & 0.0229 & -- \\
76 & 6.667 & 6.7243 & 0.2823 & 0.2890 & 0.943 & 6.6786 & 0.2797 & 0.3044 & 0.927 & 6.6911 & 0.1982 & 0.2025 & 0.937 \\
77 & 0.000 & 0.0000 & 0.0000 & 0.0000 & -- & 0.0000 & 0.0000 & 0.0000 & -- & 0.0000 & 0.0000 & 0.0000 & -- \\
\end{longtable}\endgroup

\clearpage

\begingroup\renewcommand{\baselinestretch}{1}\fontsize{8.5}{10}\selectfont\setlength{\tabcolsep}{3pt}

\begin{longtable}{rr*{12}{r}}
\caption{Complete core coefficient results for $\bm\alpha^{(1)}$}\label{tab:sim-alpha-coordinates}\\
\toprule
Comp. & True & \multicolumn{4}{c}{$(5000,250)$} & \multicolumn{4}{c}{$(5000,500)$} & \multicolumn{4}{c}{$(10000,500)$} \\
\cmidrule(lr){3-6}\cmidrule(lr){7-10}\cmidrule(lr){11-14}
 &  & Mean & SE & SD & Cov. & Mean & SE & SD & Cov. & Mean & SE & SD & Cov. \\
\midrule\endfirsthead
\caption[]{Complete core coefficient results for $\bm\alpha^{(1)}$ (continued)}\\
\toprule
Comp. & True & \multicolumn{4}{c}{$(5000,250)$} & \multicolumn{4}{c}{$(5000,500)$} & \multicolumn{4}{c}{$(10000,500)$} \\
\cmidrule(lr){3-6}\cmidrule(lr){7-10}\cmidrule(lr){11-14}
 &  & Mean & SE & SD & Cov. & Mean & SE & SD & Cov. & Mean & SE & SD & Cov. \\
\midrule\endhead
\midrule\multicolumn{14}{r}{Continued on next page}\\\endfoot
\bottomrule\endlastfoot
1 & 3.938 & 3.9640 & 0.1619 & 0.1636 & 0.948 & 3.9454 & 0.1608 & 0.1768 & 0.927 & 3.9547 & 0.1139 & 0.1146 & 0.947 \\
2 & 0.848 & 0.8545 & 0.0814 & 0.0811 & 0.944 & 0.8468 & 0.0808 & 0.0839 & 0.939 & 0.8478 & 0.0572 & 0.0554 & 0.958 \\
3 & 0.000 & 0.0000 & 0.0000 & 0.0000 & -- & 0.0000 & 0.0000 & 0.0000 & -- & 0.0000 & 0.0000 & 0.0000 & -- \\
4 & 0.000 & 0.0000 & 0.0000 & 0.0000 & -- & 0.0000 & 0.0000 & 0.0000 & -- & 0.0000 & 0.0000 & 0.0000 & -- \\
5 & 4.006 & 4.0364 & 0.1714 & 0.1741 & 0.947 & 4.0070 & 0.1696 & 0.1814 & 0.936 & 4.0211 & 0.1202 & 0.1223 & 0.944 \\
6 & 0.944 & 0.9521 & 0.0869 & 0.0855 & 0.951 & 0.9496 & 0.0863 & 0.0868 & 0.949 & 0.9476 & 0.0610 & 0.0610 & 0.953 \\
7 & 0.800 & 0.8039 & 0.0994 & 0.1172 & 0.945 & 0.7981 & 0.0987 & 0.1008 & 0.946 & 0.8035 & 0.0698 & 0.0712 & 0.946 \\
8 & 0.000 & 0.0000 & 0.0000 & 0.0000 & -- & 0.0000 & 0.0000 & 0.0000 & -- & 0.0000 & 0.0000 & 0.0000 & -- \\
9 & 4.072 & 4.1003 & 0.1720 & 0.1736 & 0.947 & 4.0798 & 0.1706 & 0.1817 & 0.936 & 4.0936 & 0.1210 & 0.1211 & 0.951 \\
10 & 1.043 & 1.0513 & 0.0927 & 0.0942 & 0.950 & 1.0438 & 0.0920 & 0.0964 & 0.940 & 1.0450 & 0.0649 & 0.0653 & 0.947 \\
11 & 1.044 & 1.0502 & 0.1033 & 0.1024 & 0.951 & 1.0483 & 0.1024 & 0.1031 & 0.954 & 1.0492 & 0.0728 & 0.0755 & 0.942 \\
12 & 0.000 & 0.0000 & 0.0000 & 0.0000 & -- & 0.0000 & 0.0000 & 0.0000 & -- & 0.0000 & 0.0000 & 0.0000 & -- \\
13 & 4.145 & 4.1750 & 0.1737 & 0.1811 & 0.946 & 4.1527 & 0.1727 & 0.1842 & 0.932 & 4.1586 & 0.1221 & 0.1230 & 0.949 \\
14 & 1.137 & 1.1423 & 0.0978 & 0.1010 & 0.946 & 1.1389 & 0.0973 & 0.1035 & 0.933 & 1.1406 & 0.0686 & 0.0679 & 0.956 \\
15 & 1.289 & 1.2955 & 0.1060 & 0.1113 & 0.939 & 1.2930 & 0.1049 & 0.1085 & 0.945 & 1.2923 & 0.0744 & 0.0769 & 0.945 \\
16 & 0.000 & 0.0000 & 0.0000 & 0.0000 & -- & 0.0000 & 0.0000 & 0.0000 & -- & 0.0000 & 0.0000 & 0.0000 & -- \\
17 & 4.228 & 4.2590 & 0.1771 & 0.1821 & 0.948 & 4.2314 & 0.1757 & 0.1877 & 0.933 & 4.2449 & 0.1245 & 0.1283 & 0.941 \\
18 & 1.221 & 1.2284 & 0.1023 & 0.1046 & 0.953 & 1.2233 & 0.1015 & 0.1074 & 0.936 & 1.2232 & 0.0716 & 0.0729 & 0.937 \\
19 & 1.533 & 1.5392 & 0.1078 & 0.1127 & 0.934 & 1.5403 & 0.1072 & 0.1125 & 0.940 & 1.5352 & 0.0758 & 0.0781 & 0.943 \\
20 & 0.000 & 0.0000 & 0.0000 & 0.0000 & -- & 0.0000 & 0.0000 & 0.0000 & -- & 0.0000 & 0.0000 & 0.0000 & -- \\
21 & 4.324 & 4.3577 & 0.1820 & 0.1886 & 0.946 & 4.3306 & 0.1809 & 0.1899 & 0.931 & 4.3411 & 0.1280 & 0.1290 & 0.949 \\
22 & 1.291 & 1.3004 & 0.1049 & 0.1047 & 0.950 & 1.2946 & 0.1043 & 0.1113 & 0.930 & 1.2964 & 0.0737 & 0.0734 & 0.957 \\
23 & 1.778 & 1.7882 & 0.1095 & 0.1110 & 0.947 & 1.7861 & 0.1091 & 0.1142 & 0.935 & 1.7772 & 0.0770 & 0.0784 & 0.953 \\
24 & 0.000 & 0.0000 & 0.0000 & 0.0000 & -- & 0.0000 & 0.0000 & 0.0000 & -- & 0.0000 & 0.0000 & 0.0000 & -- \\
25 & 4.434 & 4.4673 & 0.1889 & 0.1916 & 0.953 & 4.4412 & 0.1876 & 0.1955 & 0.933 & 4.4539 & 0.1327 & 0.1348 & 0.950 \\
26 & 1.347 & 1.3581 & 0.1067 & 0.1066 & 0.953 & 1.3506 & 0.1058 & 0.1118 & 0.938 & 1.3533 & 0.0746 & 0.0722 & 0.953 \\
27 & 2.022 & 2.0357 & 0.1116 & 0.1166 & 0.946 & 2.0301 & 0.1106 & 0.1143 & 0.936 & 2.0268 & 0.0782 & 0.0790 & 0.957 \\
28 & 0.000 & 0.0000 & 0.0000 & 0.0000 & -- & 0.0000 & 0.0000 & 0.0000 & -- & 0.0000 & 0.0000 & 0.0000 & -- \\
29 & 4.557 & 4.5883 & 0.1967 & 0.2044 & 0.946 & 4.5649 & 0.1958 & 0.2048 & 0.941 & 4.5779 & 0.1385 & 0.1419 & 0.946 \\
30 & 1.391 & 1.4047 & 0.1067 & 0.1111 & 0.952 & 1.3888 & 0.1058 & 0.1084 & 0.940 & 1.3945 & 0.0747 & 0.0781 & 0.945 \\
31 & 2.267 & 2.2831 & 0.1132 & 0.1170 & 0.942 & 2.2759 & 0.1130 & 0.1188 & 0.942 & 2.2715 & 0.0797 & 0.0784 & 0.949 \\
32 & 0.000 & 0.0000 & 0.0000 & 0.0000 & -- & 0.0000 & 0.0000 & 0.0000 & -- & 0.0000 & 0.0000 & 0.0000 & -- \\
33 & 4.688 & 4.7251 & 0.2057 & 0.2085 & 0.949 & 4.6970 & 0.2045 & 0.2131 & 0.941 & 4.7092 & 0.1448 & 0.1472 & 0.942 \\
34 & 1.424 & 1.4351 & 0.1054 & 0.1043 & 0.951 & 1.4272 & 0.1048 & 0.1107 & 0.921 & 1.4298 & 0.0740 & 0.0738 & 0.958 \\
35 & 2.511 & 2.5305 & 0.1164 & 0.1216 & 0.946 & 2.5191 & 0.1157 & 0.1239 & 0.929 & 2.5153 & 0.0818 & 0.0834 & 0.945 \\
36 & 0.000 & 0.0000 & 0.0000 & 0.0000 & -- & 0.0000 & 0.0000 & 0.0000 & -- & 0.0000 & 0.0000 & 0.0000 & -- \\
37 & 4.827 & 4.8606 & 0.2151 & 0.2211 & 0.947 & 4.8333 & 0.2140 & 0.2225 & 0.939 & 4.8437 & 0.1514 & 0.1550 & 0.952 \\
38 & 1.449 & 1.4619 & 0.1035 & 0.1070 & 0.954 & 1.4463 & 0.1029 & 0.1104 & 0.932 & 1.4537 & 0.0727 & 0.0734 & 0.949 \\
39 & 2.756 & 2.7748 & 0.1200 & 0.1248 & 0.943 & 2.7685 & 0.1191 & 0.1264 & 0.938 & 2.7649 & 0.0841 & 0.0867 & 0.936 \\
40 & 0.000 & 0.0000 & 0.0000 & 0.0000 & -- & 0.0000 & 0.0000 & 0.0000 & -- & 0.0000 & 0.0000 & 0.0000 & -- \\
41 & 4.970 & 5.0142 & 0.2251 & 0.2335 & 0.940 & 4.9797 & 0.2237 & 0.2346 & 0.944 & 4.9929 & 0.1582 & 0.1595 & 0.945 \\
42 & 1.468 & 1.4781 & 0.1010 & 0.1012 & 0.948 & 1.4704 & 0.1005 & 0.1081 & 0.926 & 1.4716 & 0.0709 & 0.0700 & 0.954 \\
43 & 3.000 & 3.0189 & 0.1237 & 0.1297 & 0.940 & 3.0123 & 0.1231 & 0.1266 & 0.940 & 3.0091 & 0.0869 & 0.0866 & 0.951 \\
44 & 0.000 & 0.0000 & 0.0000 & 0.0000 & -- & 0.0000 & 0.0000 & 0.0000 & -- & 0.0000 & 0.0000 & 0.0000 & -- \\
\end{longtable}\endgroup

\counterwithin{equation}{section}

\section{Supporting Materials for Empirical Study}\label{A-empi}

Here we present the clinical variables, action-alignment results for alternative policies, and detailed coefficient tables for the three stages. We used the following 47 clinical variables at each stage, consistent with \citet{Komorowski2018}:\begin{itemize}
	\item \textbf{Demographics and Comorbidities:}
	age, gender, elixhauser, re-admission.
	
	\item \textbf{Vital Signs:}
	HR (heart rate), RR (respiratory rate), Temp-C (temperature), SpO2 (oxygen saturation), SysBP (systolic blood pressure), DiaBP (diastolic blood pressure), MeanBP (mean arterial pressure), Shock-Index.
	
	\item \textbf{Laboratory Results:}
	WBC-count, Hb (hemoglobin), Platelets-count, BUN (blood urea nitrogen), Creatinine, Glucose, INR, PT, PTT, Arterial-BE, Arterial-lactate, Arterial-pH, CO2-mEqL, HCO3, paCO2, paO2, PaO2-FiO2 ratio.
	
	\item \textbf{Electrolytes and Other Blood Chemistry:}
	Sodium, Potassium, Calcium, Magnesium, Chloride, Ionised-Ca, SGOT, SGPT, Total-bili, Albumin.
	
	\item \textbf{Organ Function and Treatment:}
	SOFA (Sequential Organ Failure Assessment), GCS (Glasgow Coma Scale), mechvent (mechanical ventilation), output-4hourly, output-total, SIRS (Systemic Inflammatory Response Syndrome), Weight-kg.
\end{itemize}
Regularization parameters \( \lambda^{(i)} \) (\( i=1,2,3 \)) are chosen via cross-validation.  For \( Q^{(3)} \), AUC is the primary metric, with ACC and death recall as supplements; for \( Q_{\mathrm{dfm}}^{(1)} \) and \( Q_{\mathrm{dfm}}^{(2)} \), BIC is used due to the linear model assumption.

\subsection{Detailed Action-Alignment Results for Alternative and Baseline Policies}\label{app:alignment_alternatives}
In addition to our proposed three-stage model with a 16-dimensional action space, we benchmarked the following alternatives:
\begin{itemize}
	\item Fixed policy selecting from the four most frequent actions: 1, 5, 9, or 13.
	\item Two-stage model with 16 discrete actions, splitting time steps into early-mid (4–9) and mid-late (10–15).
	\item Single-stage model pooling all 12 steps and learning a time-invariant policy.
	\item Three-stage binary-action model: actions classified as conservative (both below 1/3 quantile) or aggressive.
\end{itemize}

Table~\ref{tab:alt_policy_comparison} reports stage-wise action-alignment contrasts for several alternative policy constructions, and Table~\ref{tab:baseline_causal_vertical} summarizes the corresponding alignment results for the baseline policies. These auxiliary analyses describe policy-specific concordance patterns and are not used to rank policy values.

\begingroup\small\renewcommand{\baselinestretch}{1}\selectfont\setlength{\tabcolsep}{6pt}
\begin{longtable}{lcccc}
	\caption{Action-Alignment Results for Alternative Policy Constructions} \label{tab:alt_policy_comparison} \\
	\toprule
	\textbf{Model} & \textbf{ATE} & \textbf{DR-ATE} & \textbf{95\% CI (DR)} & \textbf{Max SMD (weighted)} \\
	\midrule
	\endfirsthead
	
	\toprule
	\textbf{Model} & \textbf{ATE} & \textbf{DR-ATE} & \textbf{95\% CI (DR)} & \textbf{Max SMD (weighted)} \\
	\midrule
	\endhead
	
	1-stage, 16-act               & 0.005 & 0.039  & [0.029,\ 0.048]     & 0.280 \\
	2-stage, 16-act (s2)          & 0.015 & -0.008 & [-0.048,\ 0.031]    & 0.121 \\
	2-stage, 16-act (s1)          & 0.067 & 0.018  & [-0.135,\ 0.278]    & 0.289 \\
	3-stage, 2-act (s3)           & 0.017 & 0.035  & [-0.001,\ 0.066]    & 0.032 \\
	3-stage, 2-act (s2)           & 0.015 & 0.009  & [-0.020,\ 0.040]    & 0.026 \\
	3-stage, 2-act (s1)           & 0.015 & 0.019  & [-0.016,\ 0.049]    & 0.028 \\
	\bottomrule
\end{longtable}
\endgroup

\begin{table}[!htbp]
	\centering
	\small\renewcommand{\baselinestretch}{1}\selectfont
	\setlength{\tabcolsep}{6pt}
	\renewcommand{\arraystretch}{1}
	\caption{Action-Alignment Estimation Results for Baseline Strategies}
	\label{tab:baseline_causal_vertical}
	\begin{tabular*}{\linewidth}{@{\extracolsep{\fill}}lccc}
	\toprule
	\textbf{Stage} & \textbf{Action} & \textbf{Metric} & \textbf{Value} \\
	\midrule
	\multirow{12}{*}{Stage 1}
	& \multirow{3}{*}{Action 1}
	& ATE & 0.036 \\
	&  & DR-ATE & -0.009 \\
	&  & 95\% CI (DR) & [-0.121,\ 0.093] \\
	& \multirow{3}{*}{Action 5}
	& ATE & 0.018 \\
	&  & DR-ATE & -0.007 \\
	&  & 95\% CI (DR) & [-0.058,\ 0.042] \\
	& \multirow{3}{*}{Action 9}
	& ATE & -0.046 \\
	&  & DR-ATE & -0.059 \\
	&  & 95\% CI (DR) & [-0.100,\ -0.019] \\
	& \multirow{3}{*}{Action 13}
	& ATE & 0.029 \\
	&  & DR-ATE & 0.040 \\
	&  & 95\% CI (DR) & [-0.025,\ 0.338] \\
	\midrule
	\multirow{12}{*}{Stage 2}
	& \multirow{3}{*}{Action 1}
	& ATE & 0.048 \\
	&  & DR-ATE & 0.009 \\
	&  & 95\% CI (DR) & [-0.138,\ 0.121] \\
	& \multirow{3}{*}{Action 5}
	& ATE & -0.002 \\
	&  & DR-ATE & -0.037 \\
	&  & 95\% CI (DR) & [-0.090,\ 0.007] \\
	& \multirow{3}{*}{Action 9}
	& ATE & -0.029 \\
	&  & DR-ATE & -0.050 \\
	&  & 95\% CI (DR) & [-0.095,\ -0.010] \\
	& \multirow{3}{*}{Action 13}
	& ATE & -0.008 \\
	&  & DR-ATE & -0.020 \\
	&  & 95\% CI (DR) & [-0.080,\ 0.035] \\
	\midrule
	\multirow{12}{*}{Stage 3}
	& \multirow{3}{*}{Action 1}
	& ATE & 0.071 \\
	&  & DR-ATE & 0.053 \\
	&  & 95\% CI (DR) & [-0.086,\ 0.122] \\
	& \multirow{3}{*}{Action 5}
	& ATE & 0.028 \\
	&  & DR-ATE & 0.022 \\
	&  & 95\% CI (DR) & [-0.008,\ 0.044] \\
	& \multirow{3}{*}{Action 9}
	& ATE & -0.056 \\
	&  & DR-ATE & -0.071 \\
	&  & 95\% CI (DR) & [-0.091,\ -0.034] \\
	& \multirow{3}{*}{Action 13}
	& ATE & -0.021 \\
	&  & DR-ATE & -0.018 \\
	&  & 95\% CI (DR) & [-0.489,\ 0.032] \\
	\bottomrule
	\end{tabular*}
\end{table}

\subsection{Selected Coefficient Estimates}
We next present the estimated coefficients and their corresponding confidence intervals for each stage of the multi-stage decision process. These tables report selected variables that are of primary interest, providing insights into the stage-specific effects.

\begin{table}[!htbp]
	\centering
	\caption{Selected coefficient estimates from Stage 3 }
	\small\renewcommand{\baselinestretch}{1}\selectfont
	\setlength{\tabcolsep}{6pt}
	\renewcommand{\arraystretch}{1}
	\begin{tabular*}{\linewidth}{@{\extracolsep{\fill}}cccc}
		\toprule
		\textbf{Value} & \textbf{Action\_Blocks} & \textbf{SubFeature} & \textbf{CI} \\
		\midrule
		-0.0803 & (0,0) & DiaBP-1 & [-0.1416, -0.0189] \\
		0.1319  & (0,0) & FiO2\_1-1 & [0.0711, 0.1926] \\
		-0.1642 & (0,0) & SIRS-1 & [-0.2311, -0.0973] \\
		-0.2643 & (0,0) & SOFA-1 & [-0.3545, -0.1742] \\
		-0.1492 & (0,0) & Sodium-1 & [-0.2338, -0.0647] \\
		-0.5918 & (0,0) & age-1 & [-0.6636, -0.5201] \\
		0.2176  & (0,0) & vaso\_bin-1 & [0.1411, 0.2941] \\
		0.1166  & (0,0) & Albumin-2 & [0.0575, 0.1757] \\
		0.2761  & (0,0) & Creatinine-2 & [0.0366, 0.5155] \\
		0.1184  & (0,0) & GCS-2 & [0.0533, 0.1835] \\
		-0.0858 & (0,0) & paCO2-2 & [-0.1483, -0.0234] \\
		-0.1184 & (0,0) & SIRS-3 & [-0.1925, -0.0443] \\
		-0.3717 & (0,0) & SOFA-3 & [-0.4605, -0.2830] \\
		0.1291  & (0,0) & SpO2-3 & [0.0624, 0.1958] \\
		-0.0898 & (0,0) & paO2-3 & [-0.1574, -0.0223] \\
		0.1160  & (1,0) & SGPT-1 & [0.0159, 0.2162] \\
		-0.1694 & (1,0) & SpO2-2 & [-0.2792, -0.0595] \\
		0.2682  & (1,0) & mechvent-2 & [0.1535, 0.3829] \\
		0.1828  & (2,0) & paCO2-1 & [0.0536, 0.3119] \\
		-0.1586 & (3,0) & Magnesium-1 & [-0.3051, -0.0122] \\
		0.2349  & (3,0) & SIRS-1 & [0.0585, 0.4113] \\
		0.2581  & (3,0) & SOFA-2 & [0.0788, 0.4374] \\
		\bottomrule
	\end{tabular*}
	\label{table:stage3_selected_cleaned}
\end{table}

\begin{table}[!htbp]
	\centering
	\caption{Selected coefficient estimates from Stage 2}
	\small\renewcommand{\baselinestretch}{1}\selectfont
	\setlength{\tabcolsep}{6pt}
	\renewcommand{\arraystretch}{1}
	\begin{tabular*}{\linewidth}{@{\extracolsep{\fill}}rrrr}
		\toprule
		Value & Action\_Blocks & SubFeature &                    CI \\
		\midrule
		-0.4256 &       (0,0) &          BUN-2 &  [-0.4633, -0.3880] \\
		0.1574 &       (0,0) &           GCS-2 &  [0.0926, 0.2222] \\
		0.1456 &       (1,0) &          HCO3-2 &  [0.0917, 0.1995] \\
		-0.1185 &       (1,0) &           INR-2 &  [-0.1584, -0.0786] \\
		-0.2123 &       (1,0) &          SOFA-2 &  [-0.2594, -0.1653] \\
		-0.1685 &       (1,0) &   Shock\_Index-2 &  [-0.1908, -0.1461] \\
		-0.1680 &       (1,0) &       mechvent-2 &  [-0.2435, -0.0926] \\
		-0.5707 &       (2,0) &          BUN-2 &  [-0.6114, -0.5300] \\
		0.1153 &       (2,0) &          HCO3-2 &  [0.0562, 0.1743] \\
		-0.1109 &       (2,0) &           INR-2 &  [-0.1506, -0.0713] \\
		-0.2081 &       (2,0) &          SOFA-2 &  [-0.2611, -0.1550] \\
		-0.1821 &       (2,0) &   Shock\_Index-2 &  [-0.2061, -0.1580] \\
		-0.2195 &       (2,0) &       mechvent-2 &  [-0.3003, -0.1387] \\
		-0.1518 &      (3,0) &    output\_total-1 &  [-0.2275, -0.0761] \\
		-0.4609 &      (3,0) &          BUN-2 &  [-0.5012, -0.4206] \\
		0.1564 &      (3,0) &          HCO3-2 &  [0.1109, 0.2019] \\
		-0.1149 &      (3,0) &            PT-2 &  [-0.1429, -0.0868] \\
		-0.2046 &      (3,0) &       mechvent-2 &  [-0.3316, -0.0777] \\
		-0.1796 &      (3,1) &    output\_total-1 &  [-0.3265, -0.0328] \\
		-0.4724 &      (3,1) &          BUN-2 &  [-0.6194, -0.3255] \\
		-0.2185 &      (3,2) &    output\_total-1 &  [-0.3707, -0.0663] \\
		-0.4882 &      (3,3) &          BUN-2 &  [-0.5930, -0.3833] \\
		-0.2466 &      (3,3) &   Shock\_Index-2 &  [-0.3210, -0.1722] \\
		\bottomrule
	\end{tabular*}
	\label{table:stage2_selected_cleaned}
\end{table}

\begin{table}[!htbp]
	\centering
	\caption{Selected coefficient estimates from Stage 1}
	\small\renewcommand{\baselinestretch}{1}\selectfont
	\setlength{\tabcolsep}{6pt}
	\renewcommand{\arraystretch}{1}
	\begin{tabular*}{\linewidth}{@{\extracolsep{\fill}}cccc}
		\toprule
		\textbf{Value} & \textbf{Action\_Blocks} & \textbf{SubFeature} & \textbf{CI} \\
		\midrule
		0.1130 & (0,0) & Albumin & [0.0527, 0.1733] \\
		-0.2420 & (0,0) & BUN & [-0.2996, -0.1843] \\
		0.2261 & (0,0) & Creatinine & [0.2015, 0.2507] \\
		0.1899 & (0,0) & GCS & [0.1666, 0.2132] \\
		-0.0729 & (0,0) & RR & [-0.0990, -0.0469] \\
		-0.1370 & (0,0) & elixhauser & [-0.2104, -0.0636] \\
		0.0839 & (1,0) & Albumin & [0.0433, 0.1244] \\
		-0.2365 & (1,0) & BUN & [-0.2770, -0.1960] \\
		-0.0675 & (1,0) & CO2\_mEqL & [-0.0857, -0.0492] \\
		0.2610 & (1,0) & Creatinine & [0.2455, 0.2765] \\
		0.1609 & (1,0) & GCS & [0.1451, 0.1767] \\
		0.0677 & (1,0) & Hb & [0.0503, 0.0852] \\
		-0.0597 & (1,0) & RR & [-0.0754, -0.0439] \\
		-0.1136 & (1,0) & SOFA & [-0.1343, -0.0929] \\
		-0.0918 & (1,0) & Shock\_Index & [-0.1114, -0.0722] \\
		-0.1207 & (1,0) & elixhauser & [-0.1641, -0.0773] \\
		0.0823 & (2,0) & Albumin & [0.0414, 0.1232] \\
		-0.2714 & (2,0) & BUN & [-0.3117, -0.2311] \\
		0.3259 & (2,0) & Creatinine & [0.3102, 0.3417] \\
		0.1616 & (2,0) & GCS & [0.1463, 0.1769] \\
		0.0723 & (2,0) & Hb & [0.0541, 0.0905] \\
		-0.1130 & (2,0) & SOFA & [-0.1338, -0.0923] \\
		-0.0870 & (2,0) & Shock\_Index & [-0.1060, -0.0681] \\
		-0.1196 & (2,0) & elixhauser & [-0.1593, -0.0799] \\
		0.0964 & (3,0) & Albumin & [0.0563, 0.1364] \\
		-0.2399 & (3,0) & BUN & [-0.2792, -0.2006] \\
		0.3027 & (3,0) & Creatinine & [0.2874, 0.3179] \\
		0.1182 & (3,0) & GCS & [0.1029, 0.1335] \\
		-0.1602 & (3,0) & SOFA & [-0.1793, -0.1411] \\
		-0.0983 & (3,0) & Shock\_Index & [-0.1149, -0.0817] \\
		-0.1099 & (3,0) & elixhauser & [-0.1501, -0.0697] \\
		0.1419 & (3,3) & SOFA & [0.0902, 0.1935] \\
		\bottomrule
	\end{tabular*}
\end{table}

\FloatBarrier

\end{document}